\documentclass[useAMS,usenatbib]{mn2e}
\usepackage{graphicx,amssymb, txfonts,color}
\usepackage{pdflscape}
\usepackage{longtable}

\def\kms {$\rm km\,s^{-1}$}

\title[The $\sigma-$discrepancy in spiral galaxies]{Differences between CO- and calcium triplet-derived velocity dispersions in spiral galaxies: evidence for central star formation?}
\author[Riffel et al.]{Rogemar A. Riffel$^{1}$\thanks{E-mail:
rogemar@ufsm.br}, Luis C. Ho$^{2,3}$, Rachel Mason$^{4}$, Alberto Rodr\'iguez-Ardila$^{5}$
\newauthor Lucimara Martins$^{6}$, Rog\'erio Riffel$^{7}$, Ruben Diaz$^8$, Luis Colina$^9$, 
\newauthor Almudena Alonso-Herrero$^{10}$, Helene Flohic$^{11}$, Omaira Gonzalez Martin$^{12,13}$,
\newauthor Paulina Lira$^{14}$, Richard McDermid$^{4,15}$, Cristina Ramos Almeida$^{12,13}$,
\newauthor Ricardo Schiavon$^{2,16}$, Karun Thanjavur$^{17}$,  Daniel Ruschel Dutra$^{7,12}$,  
\newauthor Claudia Winge$^{8}$, Eric Perlman$^{18}$ 
\\
$^{1} $ Universidade Federal de Santa Maria, Departamento de F\'\i sica/CCNE, 97105-900, Santa Maria, RS, Brazil.\\
$^{2} $ Kavli Institute for Astronomy and Astrophysics, Peking University, Beijing, China. \\
$^{3} $ Department of Astronomy, School of Physics, Peking University, Beijing, China. \\
$^{4} $  Gemini Observatory, Northern Operations Center, 670 N. A’ohoku Place, Hilo, HI 96720, USA.\\
$^{5} $ Laborat\'orio Nacional de Astrof\'isica/MCT, Rua dos Estados Unidos 154, Itajub\'a, MG, Brazil.\\
$^{6} $ NAT -- Universidade Cruzeiro do Sul, Rua Galv\~ao Bueno, 868, S\~ao Paulo, SP, Brazil.\\
$^{7} $ Universidade Federal do Rio Grande do Sul, Instituto de F\'\i sica, CP 15051, Porto Alegre 91501-970, RS, Brazil.\\
$^{8} $ Gemini Observatory, Southern Operations Center, Casilla 603, La Serena, Chile; \\
$^{9} $ Astrophysics Department, Center for Astrobiology (CSIC-INTA), Torrejon de Ardoz, 28850 Madrid, Spain.  \\
$^{10} $ Instituto de Fisica de Cantabria, CSIC-UC, 39005 Santander, Spain. \\
$^{11} $ University of the Pacific, Department of Physics, 3601 Pacific Avenue, Stockton, CA 95211, USA \\
$^{12} $ Instituto de Astrof\'isica de Canarias, Calle V\'ia L\'actea, s/n, E-38205, La Laguna, Tenerife, Spain. \\
$^{13} $ Departamento de Astrof\'isica, Universidad de La Laguna, E-38205, La Laguna, Tenerife, Spain. \\
$^{14} $ Departamento de Astronom\'ia, Universidad de Chile, Casilla 36-D, Santiago, Chile. \\
$^{15} $ Department of Physics and Astronomy, Macquarie University, NSW 2109, Australia. \\
$^{16} $  Astrophysics Research Institute, Liverpool John Moores University, IC2, Liverpool Science Park 146 Brownlow Hill, Liverpool
L3 5RF, United Kingdom.\\
$^{17} $ Departament of Physics \& Astronomy, University of Victoria, PO Box 1700, STN CSC Victoria, BC, V8W 2Y2, CANADA   \\
$^{18} $ Department of Physics and Space Sciences, Florida Insitute of Technology, 150 West University Boulevard, Melbourne, FL 32901, USA
}

\begin{document}

%\date{Accepted 1988 December 15. Received 1988 December 14; in original form 1988 October 11}

\pagerange{\pageref{firstpage}--\pageref{lastpage}} \pubyear{2014}

\maketitle

\label{firstpage}

\begin{abstract}

We examine the stellar velocity dispersions ($\sigma$) of a sample of 48 galaxies, 35 of which are spirals, from the Palomar nearby galaxy survey. It is known that for ultra-luminous infrared galaxies (ULIRGs) and merger remnants the $\sigma$ derived from the near-infrared CO band-heads is smaller than that measured from optical lines, while no discrepancy between these measurements is found for early-type galaxies. No such studies are available for spiral galaxies -- the subject of this paper. We used cross-dispersed spectroscopic data obtained with the Gemini Near-Infrared Spectrograph (GNIRS), with spectral coverage from 0.85 to 2.5~$\mu$m, to obtain $\sigma$ measurements from the 2.29 $\mu$m CO band-heads ($\sigma_{CO}$), and the 0.85 $\mu$m calcium triplet ($\sigma_{CaT}$). For the spiral galaxies in the sample, we found that $\sigma_{\rm CO}$ is smaller than $\sigma_{\rm CaT}$,  with a mean fractional difference of 14.3\,\%.
The best fit to the data is given by  $\sigma_{\rm opt} = (46.0\pm18.1) + (0.85\pm0.12)\sigma_{\rm CO}$. This  ``$\sigma$ discrepancy'' may be related to the presence of warm dust, as suggested by a slight correlation  between the discrepancy and the infrared luminosity. This is consistent with studies that have found no $\sigma-$discrepancy in dust-poor early-type galaxies, and a much larger discrepancy in dusty merger remnants and ULIRGs.  That $\sigma_{\rm CO}$ is lower than $\sigma_{\rm opt}$ may also indicate the presence of a dynamically cold stellar population component. This would agree with the spatial correspondence between low $\sigma_{\rm CO}$ and young/intermediate-age stellar populations that has been observed in spatially-resolved spectroscopy of a handful of galaxies.
%It may in turn indicate the presence of a dynamically cold young stellar population component, in agreement with the spatial correspondence between low $\sigma_{\rm CO}$ and young/intermediate-age stellar populations observed in integral field unit spectroscopy of a handful of galaxies. }
%Finally, we find that the $\sigma$-discrepancy results in an additional logarithmic uncertainty of $-0.29\pm0.12$ (in units of M$_\odot$) that should be taken into account when estimating black hole masses for late-type galaxies based on the black hole mass-$\sigma$ relationship, {\bf and the uncertainties in mass by using .$\sigma_{\rm CO}$ might be higher than those from $\sigma_{\rm opt}$.}

\end{abstract}
\begin{keywords}
galaxies: active -- galaxies: stellar populations -- infrared: galaxies --  galaxies: kinematics
\end{keywords}

\section{Introduction} \label{intro}

The empirical relationship between the stellar velocity dispersion ($\sigma$) of the spheroidal component of galaxies and the mass of the super-massive black hole ($M_{\bullet}$) at their center  \citep[e.g.][]{ferrarese00,gebhardt00} has been extensively used to estimate  $M_{\bullet}$ in active and inactive galaxies. More direct determinations of $M_{\bullet}$, through stellar kinematics within the black hole's sphere of influence, or broad emission line measurements, are only feasible for a limited number of objects, making the $M_{\bullet}-\sigma$ relation a very useful alternative. Cosmological simulations suggest that the central super-massive black hole (SMBH) evolves together with the host galaxy and plays a fundamental role in its evolution \citep{dimateo05,springel05,bower06,nemmen07}, and this co-evolution may be the mechanism that leads to the $M_{\bullet}-\sigma$ relation. The intrinsic scatter in the relation (i.e., the range of $M_{\bullet}$ found for a given $\sigma$) therefore contains information about the processes of galaxy and black hole evolution \citep[e.g.][]{gultekin09}.  Besides the $M_{\bullet}-\sigma$ relation, stellar velocity dispersion measurements are also relevant for many other astrophysical applications, including the galaxy fundamental plane \citep[e.g.][]{djorgovski87,dressler87,falcon-barroso02,gebhardt03,bernardi03,valluri04}, the metallicity-$\sigma$ relation \citep[e.g.][]{terlevich81,dressler84a,bender93}, the $V/\sigma$ ratio that is an important criterium to determine the dynamical state of early type galaxies \citep[e.g.][]{cox06,cappellari07,naab13}, etc.   For all of these reasons, understanding the factors affecting $\sigma$ measurements is an important issue.

Measurements of $\sigma$ in galaxies have traditionally been done at wavelengths $<$1 $\mu$m, often using the  ``Mg b'' line at 0.52 $\mu$m, or the 0.85 $\mu$m calcium triplet absorption  \citep[e.g.][]{ho09}. 
%I added the Mg b line because it should be sensitive to extinction even if the CaT isn't.
Measurements based on stellar absorption lines in the infrared (IR), on the other hand, 
can also probe regions that are obscured by dust at optical wavelengths, or different stellar populations from those revealed in the optical. For these reasons, several recent studies have compared $\sigma$ values obtained from the fitting of stellar absorptions in the optical and in the near-IR spectral regions, generally using the CO absorption band heads in the H and K bands \citep{silge03,rothberg10,vanderbeke11,rothberg13,kang13}.

\citet{silge03} presented stellar velocity dispersions measured from the 2.29$\mu$m CO band head in a sample composed of 18 lenticular (S0) and 7 elliptical galaxies. The resulting sigma values were compared with literature values derived primarily from fitting the Ca II lines near 4000 A, the Mgb lines near 5175 A, and the 8500\AA\ Ca II triplet (hereafter CaT). The $\sigma_{\rm CO}$ values were found to be up to 30\% smaller than $\sigma_{\rm opt}$, with a median difference of 11\%. However, this difference was observed only in the lenticular galaxies; in the elliptical galaxies, the optical and IR measurements were consistent. The authors speculate that the difference may be related to the presence and distribution of dust in the S0 objects: if the dust is mainly located in the stellar disk, optical $\sigma$ measurements will be biased towards the less-obscured, dynamically hotter bulge component. 
\footnotetext{We use  $\sigma_{\rm CO}$ to denote velocity dispersions derived from the 2.29 $\mu$m CO bands; $\sigma_{\rm CaT}$ for those based on the CaT lines; and $\sigma_{\rm opt}$ as a general term to include all measurements based on lines at $\lambda$} $<$1 $\mu$m.

\citet{vanderbeke11} measured $\sigma_{\rm CO}$ in a sample of 19 Fornax cluster members, composed of roughly equal numbers of elliptical and lenticular galaxies. These measurements were compared with $\sigma_{\rm opt}$ from \citet{kuntschner00} and found to be consistent, with  $\sigma_{\rm frac}=\frac{\sigma_{\rm CO} - \sigma_{\rm opt}}{\sigma_{\rm opt}} = 6.4\%$. The lack of a discernible difference between $\sigma_{\rm CO}$ and $\sigma_{\rm opt}$ in the lenticular galaxies does not agree with the findings of \citet{silge03}. It is, however, consistent with a previous study of velocity dispersions in the Fornax cluster by \citet{silva08}.

Optical and IR velocity dispersions were also compared by \citet{kang13}. This work used the CO absorption bands around 1.6~$\mu$m, rather than those near 2.3~$\mu$m, to derive $\sigma_{\rm CO}$ for 31 galaxies: 19 elliptical, 9 lenticular, and 3 spiral. No significant difference was found between $\sigma_{\rm CO}$ and $\sigma_{\rm opt}$ (based mainly on the CaT). This suggests that both sets of lines probe roughly the same stars, and provides no evidence for a dynamically cold, obscured population.

\citet{rothberg10} also compared $\sigma_{\rm CO}$ and $\sigma_{\rm CaT}$ for a set of galaxies. This time, however, they studied 14 galaxy mergers, accompanied by a control sample of 23 elliptical galaxies. The measurements for the control sample were taken mostly from the literature, and $\sigma_{\rm CO}$ and $\sigma_{\rm CaT}$ were found to be similar, in agreement with the studies above. On the other hand, large differences are found for the mergers. In particular, for luminous infrared galaxies (LIRGS, 6 objects in their sample) the $\sigma$ derived in the optical is up to twice the value obtained for the near-IR. For the remaining 8 non-LIRG mergers, $\sigma_{\rm CaT}$ is slightly larger than  $\sigma_{\rm CO}$. Even larger discrepancies have been found for single nucleus Ultra-Luminous Infrared Galaxies (ULIRGs), for which $\sigma_{\rm CaT}$  can be three times larger  than $\sigma_{\rm CO}$ \citep{rothberg13}. The discrepancies arise because $\sigma_{\rm CO}$, although conveniently insensitive to dust absorption, probes a luminous, young stellar disk, whereas $\sigma_{\rm CaT}$ gives information about an older stellar population that is more representative of the galaxy's overall dynamical mass. CO-based $\sigma$ measurements imply that ULIRGs cannot be the progenitors of giant elliptical galaxies, whereas CaT-based values are consistent with a range of final galaxy masses. 

The above studies have compared optical and  CO-based $\sigma$ measurements for early-type (E and S0) galaxies, and for galaxy mergers 
%Changed near-IR to CO-based because the difference seems to be related to the spectral line rather than the wavelength range
and ULIRGs. Little information is available, though, for late-type (spiral) objects,  so the range of morphological types in which the large differences between $\sigma_{\rm CO}$ and $\sigma_{\rm CaT}$ occur is not yet known. We aim to rectify this situation by measuring  $\sigma_{\rm CO}$ and $\sigma_{\rm CaT}$ in a sample of 48 nearby galaxies: 35 spirals, 7 lenticulars, and 6 ellipticals. We do this using  slit spectroscopy covering  simultaneously the  CaT and 2.29 $\mu$m CO spectral regions. 
%Changed long-slit to "slit" throughout, as the XD mode uses a very short slit and shouldn't really be confused with classical long-slit spectroscopy. Also, calcium triplet is now defined earlier as CaT, so we can use this notation throughout. Also explicitly stated 2.29 um CO bands; may need to change this if we say anything about the H-band CO lines.
In Section~2, we describe the sample and the observational data, and in Section~3  we discuss the methods used to measure the stellar dispersion in the optical and in the near-IR. The results are presented in Section~4, while their implications are discussed in Section~5. The conclusions of the present paper are given in Section~6.

\section{Observations and data reduction}
\label{obsdr}

The sample of galaxies used in this work comprises 48 objects selected from  the 
 Palomar spectroscopic survey of nearby galaxies \citep{ho95,ho97}, covering a wide
 range of luminosity and nuclear activity type. Some properties of the sample are shown in Table~\ref{table}, while full details of the overall program, sample, observations and data reduction are given in Mason et al. (submitted).
 
Briefly, the spectroscopic data were obtained in queue mode\footnotemark\ with the Gemini Near-Infrared Spectrograph (GNIRS) on the Gemini North telescope. The cross-dispersed mode was used with the 32 l/mm grating, providing simultaneous spectral coverage from approximately 0.85 to 2.5~$\mu$m. We used the  $0\farcs3 \times 7\arcsec\ $ slit, generally oriented along the mean parallactic angle  at the time of the observations to minimize differential refraction effects. The data were obtained between  October 2011 and May 2013. Due to work done to address an issue with the 32 l/mm grating mount in 2012, the spectral resolution of the spectra obtained with the 0\farcs3 slit after November 2012 differs from that achieved previously. The spectral resolution before November 2012 is 11.6~\AA\ and 4.4~\AA\  for the K-band CO band heads and the CaT regions, respectively, obtained from the full width at half maximum (FWHM) of the arc lamp calibration spectra. These values correspond to a resolution in velocity dispersion of  $\sim$65\,\kms\ for both regions. After November 2012, the resulting spectral resolutions are 18.2~\AA\ and 7.4~\AA\, corresponding to  $\sim$100\,\kms\ and $\sim$110\,\kms.
\footnotetext{Programs GN-2011B-Q-111, GN-2012A-Q-23, GN-2012B-Q-80, GN-2012B-Q-112, GN-2013A-Q-16.}

The data were reduced using standard procedures\footnotemark. To summarize, the raw frames are first cleaned of any electronic striping and cosmic ray-like features arising from $\alpha$ particles in the camera lens coatings. The files are divided by a flat field and sky-subtracted using
blank sky exposures made between the on-source observations, then rectified using pinhole images. Wavelength calibration is achieved using argon arc spectra, then a spectrum of each order is extracted, divided by a standard star to cancel telluric absorption lines, and roughly flux-calibrated using the telluric standard star spectrum. The individual orders are then combined to produce the final spectrum.  The extraction aperture used for this work was 1\farcs8 along the 0\farcs3 slit, corresponding to a few tens to a few hundreds of parsecs at the distances of these galaxies.

\footnotetext{ Described at www.gemini.edu/sciops/instruments/gnirs/data-format-and-reduction/reducing-xd-spectra.}

\section{Methods}
\label{met}

%\begin{figure*}
%\centering
%  \begin{tabular}{ccc}
%    \includegraphics[scale=0.28]{sample-co/NGC205.ps}&
%    \includegraphics[scale=0.28]{sample-co/NGC2273.ps}&
%    \includegraphics[scale=0.28]{sample-co/NGC3147.ps}\\    
%    \includegraphics[scale=0.28]{sample-co/NGC4235.ps}&
%    \includegraphics[scale=0.28]{sample-co/NGC5005.ps}&
%    \includegraphics[scale=0.28]{sample-co/NGC7217.ps}\\
%  \end{tabular}
%  \caption{Examples of fits of the CO absorption band-heads in the K band for NGC\,205, NGC\,2273, NGC\,3147, NGC\,4235, NGC\,5005 and NGC\,7217. The observed spectra are shown as dotted lines and the best-fit model as continuous lines.}
%  \label{fits-co}
%\end{figure*}

%\begin{figure*}
%\centering
%  \begin{tabular}{ccc}
%    \includegraphics[scale=0.28]{sample-cat/NGC205.ps}&
%    \includegraphics[scale=0.28]{sample-cat/NGC2273.ps}&
%    \includegraphics[scale=0.28]{sample-cat/NGC3147.ps}\\    
%    \includegraphics[scale=0.28]{sample-cat/NGC4235.ps}&
%    \includegraphics[scale=0.28]{sample-cat/NGC5005.ps}&
%    \includegraphics[scale=0.28]{sample-cat/NGC7217.ps}\\
%  \end{tabular}
%  \caption{Same as Fig.~\ref{fits-co} for the CaT spectral region.}
%  \label{fits-cat}
%\end{figure*}

In order to obtain the line-of-sight velocity distribution (LOSVD) we have
used the penalized Pixel-Fitting (pPXF) method of
\citet{ppxf} to fit the CO absorption band heads around 2.3~$\mu$m in the K-band and the Ca\,{\sc ii}$\lambda\lambda$8500,8544,8665 (the CaT) stellar absorptions in the Z band. The best fit of the galaxy spectrum is obtained by convolving template stellar spectra with the corresponding LOSVD, 
assumed to be well represented by Gauss-Hermite series.  The pPXF method outputs the radial velocity, velocity dispersion ($\sigma$) and higher order 
Gauss-Hermite moments ($h_3$ and $h_4$), as well as the uncertainties for each parameter.  The $h_3$ and $h_4$ moments measure deviations of the line of sight velocity distribution from  a  Gaussian curve: the parameter $h_3$ measures asymmetric deviations (e.g. wings) and the $h_4$ quantifies the peakiness of the profile, with $h_4 > 0$ and $h_4 < 0$ indicating narrower and broader profiles than Gaussian, respectively \citep{marel93,profit}. 

The dominant source of error in the velocity dispersion is usually related to the  choice of stellar template used to fit the galaxy spectrum \citep[e.g.][]{barth02,n4051,winge09}. This can be minimized by using a large stellar template library, instead of the spectrum of a single star. The pPXF allows the use of several stellar template spectra, and varies the weighting of each one to obtain the best match to the observed spectrum, minimizing issues arising from template mismatches. The set of templates must include stars that closely match the fitted galaxy spectrum \citep[e.g.][]{silge03,emsellem04,n4051}. 
For the fitting of the CO absorptions, we used as template spectra those of the Gemini Near-IR Late-type stellar library \citep{winge09}, which contains spectra of 60 stars with spectral types ranging from F7 III to M5 III, observed  in the K band at a spectral resolution of $\sim$3.2~\AA\ (FWHM). 

As template spectra for the CaT region, we used selected spectra of stars from the CaT stellar library of \citet{cenarro01} with a spectral resolution of 1.5~\AA\ (FWHM). This library contains spectra of 706 stars with all spectral types and is part of the Medium-resolution Isaac Newton Telescope library \citep[MILES; ][]{miles}. In this work, we used only the spectra of stars with S/N ratio larger than 50 in the CaT region in order to avoid the selection of noisy spectra by the pPXF code. The final template library contains 190 spectra, including several spectral types.  We also tested the NASA InfraRed Telescope Facility (IRTF) stellar spectral library \citep{irtf05,irtf09} that presents spectra of late-type stars ranging from 0.8 to 5.5~$\mu$m at a spectral resolving power of $R = \lambda/\Delta\lambda\sim2000$, similar to the spectral resolution used in this work. The comparison of the kinematic parameters obtained with the IRTF spectral library are similar to the ones obtained with the two libraries mentioned above.   However, the standard deviation of the residuals (defined as the difference between the galaxy spectrum and the best fit model) and the uncertainties are much larger using the IRTF library. This may be related to the S/N ratio of the IRTF spectra, and/or their lower spectral resolution compared to the MILES library. We therefore decided to use the Gemini and MILES libraries for this work.

Since the spectral resolution of the template stellar spectra for both spectral regions is better than the resolution of our observations, we degraded the stellar templates to the same resolution as the spectra of the galaxies before running the pPXF to measure the LOSVD.  In order to properly fit the continuum emission we allowed pPXF to fit multiplicative Legendre polynomials of order 1  to account for any slope introduced by dust/AGN emission. Since the order of the polynomials included is small, they do not introduce any bias in the derived stellar kinematics.

  Although pPXF outputs the errors of the measurements, we also performed 100 iteration of Monte Carlo simulations where random noise was introduced to the galaxy spectrum, keeping constant the signal-to-noise ratio and the standard deviation of the 100 measurements for each galaxy. The errors obtained from the simulations are similar to the uncertainties that pPXF outputs. 

Figure~\ref{fits-co} shows the fits of the galaxy spectra at the region of the CO band heads (2.25--2.41~$\mu$m) for all the galaxies of our sample. The fits reproduce the observed spectra very well for all objects, with the residuals being smaller than 3 times the standard deviation of the continuum emission next to the CO absorptions. Two objects (NGC~2273 and NGC~4388) present strong [Ca{\sc viii}]  2.322 $\mu$m line emission superimposed on the second $^{12}$CO absorption band head, so this band was excluded from the fitted region. 

The fits for the CaT spectral region (8480--8730~\AA) are shown in Figure~\ref{fits-ca}. They also reproduce the observed spectra well, with the residuals again generally being smaller than 3 times the standard deviation of the continuum emission next to the CaT. However, for some objects the residuals are somewhat larger (e.g. NGC\,5194, NGC\,5371, NGC\,7743, and NGC\,7332 -- see bottom panels of Fig.~\ref{fits-ca}) due to the lower S/N ratio of those spectra in this region. %Although the fits of the CaT are satisfactory, they are in general worser than those for the CO region, probably due to a lower S/N ratio at this spectral region. 
This results in uncertainties of up to 25\,\kms\ in the resulting velocity dispersions.

For the CO spectral region, we  find that $\sigma_{CO}$ is smaller than the spectral resolution of the spectra for 3 objects (NGC\,205, NGC\,404 and NGC\,5194), while for the CaT region only the dwarf elliptical NGC\,205 has $\sigma$ smaller than the resolution.  These values are marked by $^*$ in Table~\ref{table} 
and should be considered highly uncertain. In particular, for the galaxies NGC\,205 and NGC\,404, \citet{ho09} measured values lower than our resolution, using higher-resolution data.

\section{Results}

The major difference and main advantage of the present work, compared to previous studies, is that we measure both $\sigma_{CO}$ and $\sigma_{CaT}$ from the same
 spectra, observed in the same way, with the same aperture and using the same method, while previous studies used their own $\sigma_{CO}$ measurements and compared them 
with optical $\sigma$ values from the literature.  Resulting measurements of the stellar velocity dispersion and Gauss-Hermite moments for our sample are shown in Table~\ref{table}.

\begin{table*}
\caption{Properties of the galaxies from our sample. Col. 1: Name of the object. Col. 2: Morphological type from \citet{ho97}. Col. 3: Nuclear Activity from \citet{ho97} -  S: Seyfert nucleus, L: LINER, T: Transition object and H: H\,{\sc ii} nucleus. ``:" means that the classification is uncertain.  Cols 4 -- 9: Stellar kinematic parameters. Col 10: The instrumental configuration used for the observations. Configuration ``a" corresponds to observations done before November 2012 and has an instrumental $\sigma$ of  65 \kms\ for both spectral regions.  Configuration ``b" corresponds to observations done after November 2012 and has an instrumental $\sigma$ of  100 and 110 \kms\ for the CO and CaT regions. The uncertainties included for the kinematic parameters are those that pPXF outputs. The dashes in the table are due to the fact that for a few objects we were not able to get a good fit of the spectrum in one of the spectral regions due to a low signal-to-noise ratio or non-detection of the absorption lines.}
%Col 10: Distance to the object and Col 11 logarithm of the infrared flux in W m$^{-2}$ from \citet{ho97}. }
\label{table}
\begin{tabular}{l l l c c c c c c c }
\hline
Object     &Hubble Type & N. A.  & $\sigma_{CO}$(km/s) &$\sigma_{CaT}$(km/s)&  $h_{3CO}$	   &  $h_{3CaT}$       &  $h_{4CO}$	   &  $h_{4CaT}$    & Conf. \\ % $ \sigma_{\rm inst}^{\rm CO}$  & $\sigma_{\rm inst}^{\rm CaT}$  $D$  & log $F_{\rm IR}$   \\
\hline
    NGC\,205 & dE5 pec     & --        &   40.7$\pm$ 35.3$^*$&  98.3$\pm$5.3$^*$  &    0.00$\pm$ 0.15 &   0.00$\pm$ 0.05   &   0.01$\pm$ 0.12   &  -0.13$\pm$ 0.03    & b \\ %100   & 109 \\	   &  0.70  &  -13.17	
    NGC\,266 & SB(rs)ab    & L1.9      &   204.4$\pm$  8.2   &  248.5$\pm$ 26.3   &    0.00$\pm$ 0.02 &   0.04$\pm$ 0.07   &   0.07$\pm$ 0.03   &   0.01$\pm$ 0.08    & a  \\ %64   &  65 \\	   & 62.4   & -13.17	
    NGC\,315 & E+:	   & L1.9      &   322.9$\pm$  5.5   &  362.0$\pm$ 15.0   &    0.02$\pm$ 0.02 &   0.03$\pm$ 0.04   &  -0.08$\pm$ 0.02   &  -0.20$\pm$ 0.03    & a \\ %64   &  65 \\	   & 65.8   & -13.83	
    NGC\,404 & SA(s)0-:    & L2        &   55.0$\pm$ 18.4$^*$&   74.5$\pm$ 22.5   &   -0.02$\pm$ 0.13 &  -0.02$\pm$ 0.19   &   0.03$\pm$ 0.20   &  -0.03$\pm$ 0.16    & a\\ %64   &  65 \\	   & 2.40   & -12.90	
    NGC\,410 & E+:	   & T2        &   276.4$\pm$ 20.0   &       --           &   -0.05$\pm$ 0.05 &   --	           &  -0.03$\pm$ 0.06   &    --               & a \\  %64   &  65 \\	& 70.6   & -14.23$^u$
    NGC\,474 &(R')SA(s)0   & L2:       &   164.8$\pm$  5.4   &  178.7$\pm$  7.5   &    0.03$\pm$ 0.02 &   0.06$\pm$ 0.04   &   0.06$\pm$ 0.02   &  -0.05$\pm$ 0.04    & a \\%64 &  65 \\	      & 70.6   & -14.23$^u$
    NGC\,660 & SB(s)a pec  & T2/H      &   164.7$\pm$17.3    &       --           &   -0.05$\pm$0.06  &     --   	   &   0.12$\pm$ 0.07   &   --		      & b \\ % 100 & 109 \\   & 11.8	& -11.45    
   NGC\,1052 & E4	   & L1.9      &   220.3$\pm$  4.1   &  250.6$\pm$ 22.3   &   -0.01$\pm$ 0.01 &  -0.07$\pm$ 0.05   &   0.03$\pm$ 0.01   &   0.04$\pm$ 0.06    & a\\ %64  &  65 \\	      & 17.8   & -13.33    
   NGC\,1167 & SA0-	   & S2        &   179.0$\pm$  6.2   &  160.6$\pm$ 41.7   &    0.01$\pm$ 0.02 &   0.03$\pm$ 0.06   &   0.00$\pm$ 0.03   &   0.00$\pm$ 0.21    & a\\  %64 &  65 \\	      & 65.3   & -13.81    
   NGC\,1358 & SAB(r)0/a   & S2        &   182.3$\pm$  6.8   &  174.3$\pm$ 21.2   &   -0.03$\pm$ 0.03 &   0.08$\pm$ 0.10   &  -0.03$\pm$ 0.03   &  -0.09$\pm$ 0.09    & a \\ %64   &  65 \\	   & 53.6   & -13.62	
   NGC\,1961 & SAB(rs)c    & L2        &   189.9$\pm$  9.2   &  249.7$\pm$ 13.9   &    0.01$\pm$ 0.02 &   0.03$\pm$ 0.05   &   0.11$\pm$ 0.03   &  -0.07$\pm$ 0.04    & b \\ %100  & 109 \\	   & 53.1   & -12.32	
NGC\,2273$^{**}$ & SB(r)a: &S2         &   105.7$\pm$ 14.4   &  142.4$\pm$  8.3   &   -0.11$\pm$ 0.06 &   0.05$\pm$ 0.04   &   0.03$\pm$ 0.09   &   0.02$\pm$ 0.05    & b \\  %100  & 109 \\	   & 28.4   & -12.49	
   NGC\,2639 &(R)SA(r)a?   & S1.9      &   160.4$\pm$  5.2   &         --	  &    0.02$\pm$ 0.02 &        --	   &  -0.04$\pm$ 0.03   &	 --	      & a\\  %64 &  65 \\	      & 42.6   & -12.81    
   NGC\,2655 & SAB(s)0/a   & S2        &   145.8$\pm$  6.5   &  181.1$\pm$  7.1   &    0.02$\pm$ 0.02 &   0.03$\pm$ 0.03   &   0.08$\pm$ 0.03   &   0.01$\pm$ 0.03    & b\\ %100  & 109 \\	  & 24.4   & -12.92    
   NGC\,2768 & E6:         & L2        &   172.8$\pm$  3.8   &  177.3$\pm$  6.7   &    0.01$\pm$ 0.01 &  -0.05$\pm$ 0.03   &   0.02$\pm$ 0.02   &   0.04$\pm$ 0.03    & a\\ %64  &  65 \\	      & 23.7   & -13.55    
   NGC\,2832 & E+2:	   & L2:       &   328.7$\pm$  7.1   &  254.2$\pm$ 16.9   &    0.06$\pm$ 0.02 &   0.02$\pm$ 0.05   &  -0.01$\pm$ 0.02   &  -0.03$\pm$ 0.05    & b\\ %100  & 109 \\	  & 91.6   & -13.51    
   NGC\,3031 & SA(s)ab     & S1.5      &   182.5$\pm$  3.8   &  149.8$\pm$  7.1   &    0.00$\pm$ 0.01 &   0.04$\pm$ 0.03   &   0.08$\pm$ 0.01   &   0.03$\pm$ 0.04    & a\\ %64   &  65 \\	  & 1.4    & -11.44    
   NGC\,3079 &SB(s)c spin  & S2        &   143.6$\pm$  4.7   &  --		  &   -0.01$\pm$0.02  &   --		   & 	0.05$\pm$0.02   &   --  	      & a\\  %64  --   \\	      & 20.4   & -11.53    
   NGC\,3147 & SA(rs)bc    & S2        &   229.1$\pm$  4.2   &  250.2$\pm$ 11.7   &    0.00$\pm$ 0.01 &   0.02$\pm$ 0.03   &   0.06$\pm$ 0.01   &   0.03$\pm$ 0.03    & b\\  %100  & 109 \\	  & 40.9   & -12.19    
   NGC\,3169 &SA(s)a pec   & L2        &   169.4$\pm$  4.0   &  191.6$\pm$ 17.5   &    0.01$\pm$ 0.02 &   0.13$\pm$ 0.04   &   0.01$\pm$ 0.02   &   0.04$\pm$ 0.06    & a\\  %64   &  65 \\	  & 19.7   & -12.27    
   NGC\,3190 &SA(s)a pec spin& L2      &   189.2$\pm$  3.9   &  202.9$\pm$ 13.6   &    0.02$\pm$ 0.01 &   0.00$\pm$ 0.06   &   0.06$\pm$ 0.01   &  -0.07$\pm$ 0.05    & a\\  %64   &  65 \\	  & 22.4   & -12.64    
   NGC\,3607 &SA(s)0:	   & L2        &   213.5$\pm$  3.6   &  201.0$\pm$ 13.9   &    0.01$\pm$ 0.01 &   0.00$\pm$ 0.07   &   0.06$\pm$ 0.01   &  -0.09$\pm$ 0.06    & a\\  %64   &  65 \\	  & 19.9   &  --	   
   NGC\,3718 &SB(s)a pec   & L1.9      &   192.7$\pm$  5.1   &  224.0$\pm$  8.4   &   -0.03$\pm$ 0.01 &  -0.04$\pm$ 0.03   &   0.08$\pm$ 0.02   &  -0.01$\pm$ 0.03    & b\\  %100  & 109 \\	  & 17.0   & -13.25    
   NGC\,3998 &SA(s)ab	   & L1.9      &   346.9$\pm$  5.9   &  331.5$\pm$ 17.7   &    0.02$\pm$ 0.01 &  -0.03$\pm$ 0.04   &  -0.02$\pm$ 0.02   &   0.03$\pm$ 0.04    & b\\  %100  & 109 \\	  & 21.6   & -13.50    
   NGC\,4203 & SAB0-:      & L1.9      &   176.4$\pm$  5.9   &  182.2$\pm$  8.4   &    0.00$\pm$ 0.02 &  -0.02$\pm$ 0.04   &   0.03$\pm$ 0.02   &  -0.05$\pm$ 0.04    & b\\ %100   & 109	      & 9.7    & -13.36    
   NGC\,4235 & SA(s)a spin & S1.2      &   209.6$\pm$ 13.5   &  156.4$\pm$ 12.3   &   -0.10$\pm$ 0.03 &   0.01$\pm$ 0.05   &   0.12$\pm$0.03    &   0.06$\pm$ 0.05    & b\\  %100  & 109 \\	  & 35.1   & -13.73   
   NGC\,4258 & SAB(s)b     & S1.9      &   129.6$\pm$  3.2   &  132.4$\pm$  6.4   &   -0.02$\pm$ 0.01 &  -0.02$\pm$ 0.04   &   0.02$\pm$ 0.02   &  -0.05$\pm$ 0.04    & a\\  %64   &  65 \\	  &  6.8   & -11.77    
   NGC\,4346 & SA0 spin    & L2:       &   124.1$\pm$  7.4   &  154.4$\pm$  6.5   &    0.00$\pm$ 0.02 &   0.06$\pm$ 0.02   &   0.07$\pm$ 0.04   &   0.00$\pm$ 0.03    & b\\  %100  & 109 \\	  & 17.0   &  --	   
NGC\,4388$^{**}$ & SA(s)b: spin& S1.9  &   103.3$\pm$ 12.4   &  165.3$\pm$ 20.8  &    -0.01$\pm$ 0.05 &   0.01$\pm$ 0.09   &   0.01$\pm$ 0.08   &   0.01$\pm$ 0.09    & b\\  %100  & 109 \\	  & 16.8   & -12.22    
   NGC\,4450 & SA(s)ab     & L1.9      &   118.8$\pm$  4.6   &  136.4$\pm$ 10.9   &    0.00$\pm$ 0.02 &  -0.07$\pm$ 0.05   &   0.08$\pm$ 0.03   &  -0.05$\pm$ 0.07    & a\\  %64   &  65 \\	  & 16.8   & -12.80    
   NGC\,4548 & SB(rs)b     & L2        &   104.0$\pm$  8.8   &  131.6$\pm$  6.8   &    0.00$\pm$ 0.03 &   0.00$\pm$ 0.04   &   0.04$\pm$ 0.06   &   0.02$\pm$ 0.04    & b\\  %100  & 109 \\	  & 16.8   & -12.64    
   NGC\,4565 & SA(s)b? spin& S1.9      &   151.6$\pm$  4.0   &  180.0$\pm$  5.0   &    0.03$\pm$ 0.01 &  -0.01$\pm$ 0.02   &   0.02$\pm$ 0.02   &  -0.04$\pm$ 0.02    & b\\  %100  & 109 \\	  &  9.7   & -12.04    
   NGC\,4569 & SAB(rs)ab   & T2        &   106.4$\pm$  7.6   &   178.2$\pm$7.8    &   -0.05$\pm$ 0.03 &   0.08$\pm$0.02    &   0.03$\pm$ 0.05   &   0.05$\pm$0.03     & b\\  %100  & 109 \\	  & 16.8   & -12.19    
   NGC\,4579 & SAB(rs)b &S1.9/L1.9     &   177.5$\pm$  5.3   &  174.9$\pm$ 10.8   &    0.05$\pm$ 0.02 &   0.03$\pm$ 0.04   &   0.06$\pm$ 0.02   &   0.01$\pm$ 0.05    & a\\  %64   &  65 \\	  & 16.8   & -12.34    
   NGC\,4594 & SA(s)a spin & L2        &   253.9$\pm$  3.6   &  271.4$\pm$  6.3   &    0.00$\pm$ 0.01 &   0.03$\pm$ 0.02   &   0.06$\pm$ 0.01   &  -0.01$\pm$ 0.02    & b\\  %100  & 109 \\	  & 20.0   & -12.37    
   NGC\,4725 & SAB(r)ab pec& S2:       &   133.5$\pm$  3.3   &  162.4$\pm$ 11.3   &   -0.05$\pm$ 0.02 &   0.27$\pm$ 0.08   &  -0.02$\pm$ 0.02   &  -0.27$\pm$ 0.07    & a\\  %64   &  65 \\	  & 12.4   & -12.40    
   NGC\,4736 & (R)SA(r)ab  & L2        &   120.2$\pm$  6.9   &  135.4$\pm$ 10.0   &    0.13$\pm$ 0.04 &   0.01$\pm$ 0.05   &  -0.05$\pm$ 0.04   &   0.01$\pm$ 0.06    & b\\  % 100 & 109 \\	  & 4.3    & -11.43    
   NGC\,4750 & (R)SA(rs)ab & L1.9      &   105.9$\pm$ 11.5   &  155.9$\pm$  7.3   &   -0.01$\pm$ 0.03 &  -0.05$\pm$ 0.04   &   0.08$\pm$ 0.07   &  -0.08$\pm$ 0.04    & b\\  %100  & 109 \\	  & 26.1   & -12.49    
   NGC\,5005 & SAB(rs)bc   & L1.9      &   153.3$\pm$  6.4   &  183.1$\pm$  7.5   &   -0.01$\pm$ 0.02 &  -0.01$\pm$ 0.02   &   0.08$\pm$ 0.03   &   0.06$\pm$ 0.03    & b\\  %100  & 109 \\	  & 21.3   & -11.81    
   NGC\,5033 & SA(s)c	   & S1.5      &   151.0$\pm$ 10.7   &  147.0$\pm$ 32.7   &   -0.12$\pm$ 0.04 &   0.04$\pm$ 0.16   &   0.15$\pm$ 0.05   &  -0.01$\pm$ 0.22    & a\\   %64	&  65 \\      &  18.7  &  -11.92   
   NGC\,5194 & SA(s)bc pec & S2        &  56.3$\pm$  8.0$^*$ &   91.8$\pm$  7.0   &    -0.02$\pm$ 0.0 &  -0.04$\pm$ 0.05   & 	0.03$\pm$ 0.09  &  -0.01$\pm$ 0.05    & a\\  %64   &  65 \\	  &  7.7   & -11.14    
   NGC\,5371 & SAB(rs)bc   & L2        &   142.8$\pm$  6.6   &  159.3$\pm$  6.8   &   -0.01$\pm$ 0.02 &   0.01$\pm$ 0.03   &   0.04$\pm$ 0.03   &  -0.03$\pm$ 0.04    & b\\ %100  & 109 \\	  & 37.8   & -12.39    
   NGC\,5850 & SB(r)b	   & L2        &   118.0$\pm$  8.5   &  179.3$\pm$  7.2   &    0.05$\pm$ 0.03 &   0.02$\pm$ 0.03   &   0.07$\pm$ 0.05   &  -0.05$\pm$ 0.03    & b\\ %100  & 109 \\	  & 28.5   & -13.10    
   NGC\,6500 & SAab:	   & L2        &   177.5$\pm$  5.6   &         --	  &    0.00$\pm$ 0.02 &        --	   &   0.04$\pm$ 0.02   &	 --	      & a\\  %64   &  65 \\	  & 39.7   & -13.28    
   NGC\,7217 & (R)SA(r)ab  & L2        &   125.8$\pm$  5.9   &  157.6$\pm$  5.6   &   -0.01$\pm$ 0.02 &   0.04$\pm$ 0.02   &   0.03$\pm$ 0.03   &   0.00$\pm$ 0.03    & b\\  %100  & 109 \\	  & 16.0   & -12.40    
   NGC\,7331 & SA(s)b	   & T2        &   137.3$\pm$  3.2   &  141.3$\pm$  6.6   &    0.03$\pm$ 0.01 &  -0.03$\pm$ 0.04   &   0.01$\pm$ 0.02   &  -0.05$\pm$ 0.04    & a\\  %64   &  65 \\	  & 14.3   & -11.59    
%   NGC\,7469 & (R')SAB(rs)a& S1        &   -10.0$\pm$ 10.0   &         --	  &  -10.00$\pm$10.00&        --	   & -10.00$\pm$10.00   &	 --	      & b\\  %100  & 109 \\	    & 56.7$^\dagger$&-- 
   NGC\,7743 & (R)SB(s)0+  & S2        &    90.2$\pm$  4.8   &  133.4$\pm$  4.8   &    0.04$\pm$ 0.02 &  -0.13$\pm$ 0.04   &   0.05$\pm$ 0.04   &  -0.20$\pm$ 0.03    & a\\   %64	&  65 \\      & 24.4   & -13.16  
\hline
\multicolumn{9}{|l|}{$^*$ The measured $\sigma$ is smaller than the instrumental $\sigma$.} \\
\multicolumn{9}{|l|}{$^{**}$ The second CO band was excluded from the fitting due to contamination by the [Ca{\sc viii}]\,2.322$\mu$m emission.} \\
%\multicolumn{}{|9|}{$^u$ Upper limit.} \\
%\multicolumn{11}{9}{$^\dagger$ Distance from NED.} \\

\end{tabular}
\end{table*}

\begin{figure}
\centering
    \includegraphics[scale=0.62]{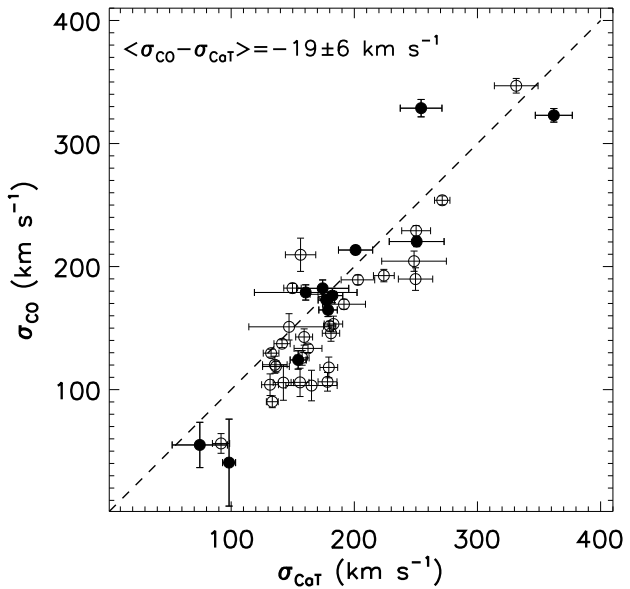}
     \caption{Comparison between the stellar velocity dispersion obtained from the fitting of the CO band heads ($y$-axis) and from the fitting of the CaT ($x$-axis). The dashed line shows a one-to-one relation. Filled circles are for the elliptical and lenticular galaxies of our sample and open circles represent the spiral galaxies.
% {\bf Why have you indicated N205 and N2832? Should this figure indicate the 3 objects with sigma smaller than the resolution?}
%\textcolor{red}{I excluded this now. The referee asked to identify the outliers, but there isn't any clear outlier. I think it's better do not identify any object here.}
}
  \label{sigma}
\end{figure}

In Figure \ref{sigma} we show $\sigma_{\rm CO}$ vs $\sigma_{\rm CaT}$, excluding objects  for which good fits could not be obtained for one of the spectral regions (marked by dashes in the Table~\ref{table}). We find that $\sigma_{\rm CaT}$ tends to be higher than $\sigma_{\rm CO}$, with an average difference of $\sigma_{\rm CO} - \sigma_{\rm CaT} = -19.2\pm5.6$~\kms\ (for all morphological types). The error in this value was obtained using Monte Carlo simulations and the bootstrap technique \citep[e.g.][]{bootstrap} as follows. First, 10000 Monte Carlo iterations were run to determine the effect to the uncertainties in $\sigma$ to the mean sigma difference ($<\sigma_{\rm CO}-\sigma_{\rm CaT}>$). At each run, ramdom values for $\sigma_{\rm CO}$ and $\sigma_{\rm CaT}$ constrained to be within their measured uncertainties were generated and then the $<\sigma_{\rm CO}-\sigma_{\rm CaT}>$ was calculated. The standard deviation of the 10000 simulations of $<\sigma_{\rm CO}-\sigma_{\rm CaT}>$ ($\epsilon_{\sigma_u}$) represents the effects of the uncertainties in $\sigma$ to the $<\sigma_{\rm CO}-\sigma_{\rm CaT}>$. Then, to evaluate the completeness of the sample and its effect to the mean $\sigma$ difference, we run a bootstrap with 10000 realizations in which for each iteration the $<\sigma_{\rm CO}-\sigma_{\rm CaT}>$ is calculated for a sample selected randomly amoung the galaxies of our sample. The standard deviation  of the simulated $<\sigma_{\rm CO}-\sigma_{\rm CaT}>$ ($\epsilon_{\sigma_s}$) represents the intrinsic scatter of our sample. Finally, the uncertainty $\epsilon_{<\sigma_{\rm CO}-\sigma_{\rm CaT}>}$ is obtained by the sum of $\epsilon_{\sigma_u}$ and $\epsilon_{\sigma_s}$ in quadrature, as 
$\epsilon_{<\sigma_{\rm CO}-\sigma_{\rm CaT}>}=\sqrt{(\epsilon_{\sigma_u}^2 + \epsilon_{\sigma_s}^2)}=\sqrt{(2.6\,{\rm km\,s^{-1}})^2 + (4.6\,{\rm km\,s^{-1}})^2}=5.6\,{\rm km\,s^{-1}}$.  This value is similar to the standard error. If we exclude also the 3 objects with $\sigma$ values smaller than the instrumental broadening, we find $\sigma_{\rm CO} - \sigma_{\rm CaT} = -17.7\pm5.7$~\kms. 

We also performed a  Kolmogorov-Smirnov  test to determine if $\sigma_{\sc CaT}$ and $\sigma_{\rm CO}$ differ significantly. We found a statistic significance $P=0.2$ for our sample, where $P$ ranges from 0 to 1 and small values mean that the two data sets are significantly different. The value of $P$ obtained for our sample indicates that there is a reasonable probability (80\,\%) that $\sigma_{\rm CaT}$ and $\sigma_{\rm CO}$ present discrepant values.

%The comparisons of higher order Gauss-Hermite moments $h_3$ and $h_4$ obtained from the fitting of the CaT with those obtained from the fitting of the CO band heads are presented in Figures~\ref{h3} and \ref{h4}. Fig.~\ref{h3} shows that $h_3$ presents values ranging from  $\sim -0.2$ to $\sim0.2$ derived from the CaT, with similar values for  the CO bands. The average difference between the $h_3$ values measured for the CaT and CO region is $h_{3CO} - h_{3CaT} = -0.01\pm0.08$. The comparison between $h_{\rm 4CO}$ and $h_{\rm 4CaT}$ is shown in Fig.~\ref{h4} and shows a similar scatter as the one observed for $h_3$, but now with the values derived from the CO region being on average larger than the ones obtained from the CaT. The average difference for $h_4$ is $h_{\rm 4CO} - h_{\rm 4CaT} = 0.07\pm0.07$. For both the $h_3$ and $h_4$ parameters there is no correlation between the values found from the fitting of the CaT and the ones from the CO band heads.
%The values found for $h_3$ and $h_4$ are small for most of the objects in our sample, indicating that the LOSVD of the stars for the nucleus of these galaxies is reasonably well reproduced by a Gaussian velocity distribution.   

 The comparisons of higher order Gauss-Hermite moments $h_3$ and $h_4$ obtained from the fitting of the CaT with those of the CO band heads show an average differences of  $h_{3CO} - h_{3CaT} = -0.01\pm0.03$ and $h_4$ is $h_{\rm 4CO} - h_{\rm 4CaT} = 0.07\pm0.03$.  For both the $h_3$ and $h_4$ parameters there is no correlation between the values found from the fitting of the CaT and the ones from the CO band heads. The values found for $h_3$ and $h_4$ are small for most of the objects in our sample, indicating that the LOSVD of the stars for the nucleus of these galaxies is reasonably well reproduced by a Gaussian velocity distribution.

%Finally, we can also compare the differences in the resulting radial velocities. The velocities -- after correction of the spectra by Doppler shift -- derived from the CaT region are on average 12 \kms\ larger than the values obtained from the CO region, but a scatter of 36 \kms\ is observed between both measurements. We do not show a plot for the radial velocity and their values in Table~\ref{table} because the correction for Doppler shift was done using the redshift values from NASA/IPAC Extragalactic Database (NED\footnote{http://ned.ipac.caltech.edu/}), which may be uncertain and the main goal of this paper is the comparison of the $\sigma$ values obtained in the optical and near-IR. 

%\begin{figure}
%\centering
%    \includegraphics[scale=0.62]{figs/fig-co-ca-h3.ps}
%     \caption{Same as Fig.~\ref{sigma} for the $h_3$ Gauss-Hermite moment.}
%  \label{h3}
%\end{figure}

%\begin{figure}
%\centering
%    \includegraphics[scale=0.62]{figs/fig-co-ca-h4.ps}
%     \caption{Same as Fig.~\ref{sigma} for the $h_4$ Gauss-Hermite moment.}
%  \label{h4}
%\end{figure}

\section{Analysis of the results}

 To understand the observed difference between $\sigma_{\rm CO}$ and $\sigma_{\rm CaT}$ in the galaxies, we performed several tests to  identify possible systematic effects in the data or introduced by the fitting procedure. 

 As discussed in Sec.~\ref{met}, we were very careful in the estimation of the instrumental broadening, taking into account the facts that (i) our observations present distinct spectral resolutions depending on the date of the observations, and (ii) the spectral resolution for observations performed after November 2012 differs between the CaT and CO regions  (\S\ref{obsdr}). The uncertainty (rms) in the FWHM of the arc lamp lines is smaller than 10~\kms\ for both spectral regions and in order to ``correct" the offset from the one-to-one relation observed in $\sigma$ in Fig.~\ref{sigma} the FWHM in the CaT region would have to be underestimated by more than 30~\kms.
% ({\bf Why $>$40 km/s, if sigma\_CO - sigma\_CaT is only 20 km/s?}) \textcolor{red}{The instrumental sigma must be subtracted in quadrature. This value is $\sigma=\sqrt{\sigma_{obs}^2-\sigma_{inst}^2}$. We quote an resolution of 100\,\kms\ for one instrumental configuration and 100\,\kms\ for the other.  If $\sigma_{obs}\sim200$\kms\ (typical value for our sample), then $\sigma=\sqrt{200^2-65^2}=189$\kms\. If the $\sigma_inst$ is undestimated by 10 km/s, then $\sigma=185$\,\kms (difference of only 4 km/s, much smaller than the discrepancy found), for an undesprediction of sigma of 40\kms, the resulting sigma would be 170\kms (similar to the discrepancy). For the second set of observations, an uncertainty of 30 km/s could reproduce the discrepancy. Thus, I modified the "40" above by "30" \kms}.
 Thus, uncertainties in the instrumental broadening cannot account for the observed differences in CaT- and CO-based $\sigma$ values in our sample.

%The template mismatch could also introduce systematic uncertainties in the $\sigma$ measurements \citep[e.g.][]{silge03,n4051}. Usually fewer than 10 template spectra are used by pPXF to reproduce the CO absorption band heads and the CaT. Although the libraries used here are large and contain spectra of stars with several spectral types, most of the contribution to the fit of the CaT region is from F/G dwarf stars, while for the CO band heads the pPXF choice is dominated by K giants/super-giants stars, although some contribution of dwarf stars is also present.  We measured $\sigma_{CaT}$ for some stars from the library employed and found that the intrinsic difference between dwarf and giant stars is about 10~\kms, with the K giants/super-giants having smaller values of $\sigma$ than the F/G dwarf stars. 
 As noted in \S\ref{met}, template mismatch could also introduce systematic uncertainties in the $\sigma$ measurements \citep[e.g.][]{silge03,n4051}. Usually fewer than 10 template spectra are needed by pPXF to successfully reproduce the CO absorption band heads and the CaT. Although the libraries used here are large and contain spectra of stars with several spectral types, the fits to both spectral regions are dominated by giant and super-giant stars, with a slightly larger contribution from super-giants in the K band. These are the spectral types that are expected to dominate the emission in the near-IR, suggesting that pPXF is selecting the appropriate stars with which to fit the spectra. For most galaxies M-stars are the dominant contributor to the fits in the K-band, while K-stars dominate in the region of the CaT. The difference between the intrinsic widths of the absorption lines in these stars is $\sim10$\,\kms, which is negligible compared to the overall difference between $\sigma_{\rm CaT}$ and $\sigma_{\rm CO}$ in the sample. In Appendix \ref{appendix} we show the stellar templates used to fit the spectrum of each galaxy. 
%{\bf I don't understand why the relative contributions of those types of star are evidence that template mismatch is not important. Can you state the reason for that?} 
%\textcolor{red}{Giants and supergiants are expected to dominate the emission in the near-IR (including the CaT region) and thus, if these types of stars are selected by pPXF means that the choice is ok. Differences in the choice of an M or and K stars (or even a giant or supergiant) are small due to intrinsic differences in the width of the absorption lines in the star's spectrum.}
Additionally, there is no difference in the $\chi^2$ values of the fits between early and late-type galaxies, suggesting that there is no bias in the choice of template for each type of galaxy. 
%{\bf Are the chi-sq values stated somewhere? If not, should we state  the mean value for early- and late-type objects, or something like that?} \textcolor{red}{I'm avoiding to quote $\chi^2$ values in the paper, since pPXF does not compute a reduced $\chi^2$ value if the noise of the spectra is not well determined and this is our case. The noise is determined as the standard deviation of each spectrum. I think that this is too simplistic. Vaules of $\chi^2$ are $10^{-2}$ for the CO region and $10^{-1}$ for the CaT region. These values can be used only to compare the fits but the absolute values make no sense.}

As several galaxies of our sample have a Seyfert nucleus, hot dust emission may play an important role in the K-band continuum. The CO band-heads can be ``diluted" by this emission, which might introduce an uncertainty in the measurements of $\sigma_{\rm CO}$ \citep[e.g.][]{ivanov00,rogerio06,kotilainen12}. To test whether the pPXF accounts correctly for variations of the continuum shape, we have simulated contributions of Planck functions with temperatures ranging from 700 to 2000~K  to the continuum emission at the K-band. No significant difference in $\sigma_{\rm CO}$ was found  when including dust emission ranging from 1\% to 70\% of the total K-band emission. Furthermore, for the objects in which we could measure the $\sigma$ from the H-band CO lines \citep[where the contribution from AGN-heated dust is smaller;][]{origlia93}, it agrees to within 10\% with that measured from the K-band and no systematic difference is found between H and K band measurements. Thus, we conclude that the hot dust emission plays a negligible role in the $\sigma$ measurements. 
%{\bf Actually, isn't the sigma-discrepancy roughly a 10\% effect? So shouldn't we say that there was no systematic offset in any direction, if that's true?}
%\textcolor{red}{Yes, that's true. I modified the sentence above to make sure that the 10\% are not systematic.}

\begin{figure}
\centering
    \includegraphics[scale=0.62]{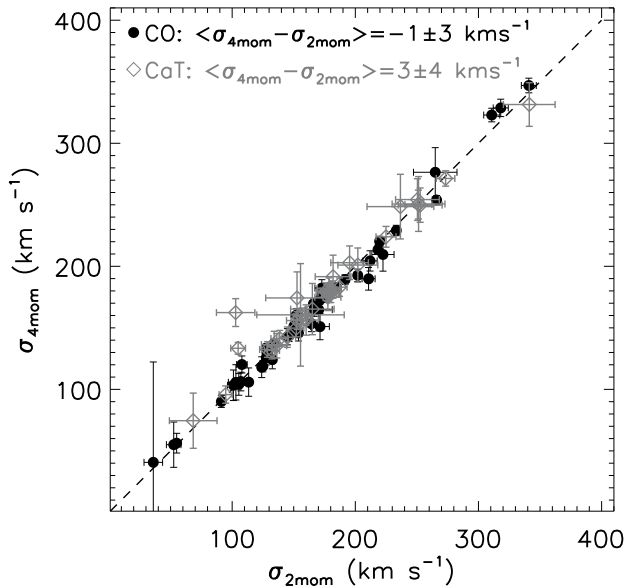}
     \caption{Comparison of $\sigma$ values obtained from the fit of 2 and 4 moments in the LOSVD.}
  \label{moments}
\end{figure}

 In galaxies with low values of $\sigma$, the fitting of higher order Gauss-Hermite moments can introduce uncertainties in the $\sigma$ measurements \citep[see the {\it bias} parameter of the pPXF program; ][]{ppxf}. To test this, we fitted the spectra of the galaxies assuming that the LOSVD is well described by a Gaussian, by fitting only the first 2 moments.  In Fig.~\ref{moments} we present the comparison of $\sigma$ values obtained from the fitting of 4 moments with those obtained from the fitting of 2 moments. This figure shows that the resulting $\sigma$ for both the CO and CaT spectral regions are very similar to those obtained when allowing the pPXF to include the $h_3$ and $h_4$ parameters, suggesting that the inclusion of these parameters does not affect significantly the $\sigma$ measurements for our sample.

\begin{figure}
\centering
    \includegraphics[scale=0.62]{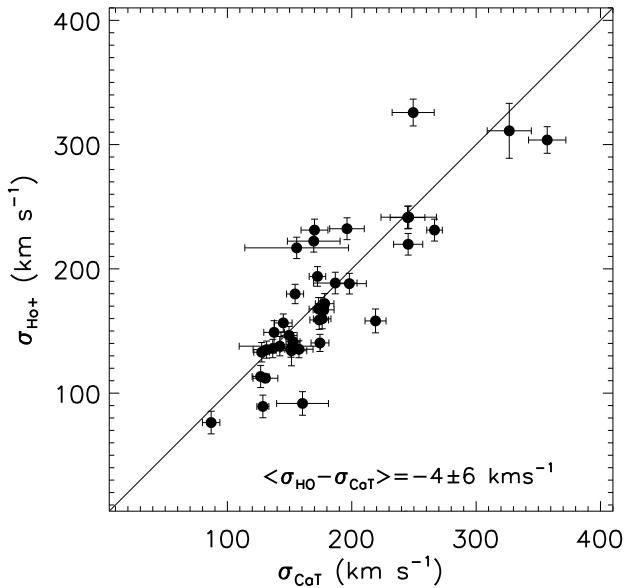}
     \caption{Comparison of $\sigma_{\rm CaT}$ with those found by \citet{ho09} using various optical lines.}
  \label{ourho}
\end{figure}

Finally, we compare our $\sigma_{\rm CaT}$ measurements with the optical $\sigma$ values of  \citet{ho09} for the same galaxies. %although some differences may be expected as they used a much larger aperture (2$^{\prime\prime}\times$4$^{\prime\prime}$) than ours. 
\citet{ho09} used two spectral ranges to measure $\sigma$: a blue region, from 4200 to 5000\,\AA\ that includes several Fe lines, and a red region, covering the range 6450--6550\,\AA\, where Ca+Fe lines are present. They found that both values are in good agreement.
% {\em Rachel: Ho et al. 2009 didn't measure $sigma$ from the CaT. Something needs clarifying here. Rogemar: Done in the sentence above.}.
 In Fig.~\ref{ourho} we show our $\sigma_{\rm CaT}$ vs. the $\sigma$ values presented in \citet{ho09}.  This comparison shows that 50\% of the objects have $\sigma$ differences smaller than 10\% and for about 90\% of the objects the differences are smaller than 25\%, indicating that our measurements are in  agreement with those from \citet{ho09}. Differences between the measurements may be due to the larger aperture (2\arcsec $\times$ 4\arcsec) used by \citet[][]{ho09}, as well as differences in the S/N ratio of the spectra and the exact measurement procedures used.

We therefore conclude that our $\sigma$ measurements are robust, and that the observed difference between $\sigma_{\rm CaT}$ and $\sigma_{\rm CO}$ is not due to measurement error.

\section{Discussion}

%\citet{barth02a} presented measurements of the central stellar velocity dispersions of 33 nearby spiral and elliptical galaxies based on spectra from an aperture of 2$^{\prime \prime}\times$3\farcs7. They derived $\sigma$ from the CaT lines and from the Mg{\sc i} b lines at 5200~\AA, and concluded that the $\sigma$ values  obtained from the CaT lines are not sensitive to the choice of the template star.  The Mg~{\sc i} b region is subject to larger uncertainties. However, both $\sigma$ values are consistent to each other within the uncertainties for 65\,\% of the sample. 
%This was already found 30 years ago by \citet{dressler84b} -- see also \citet{barth02}. Thus, since the CaT region is not very sensitive to stellar population variations, we consider the $\sigma_{\rm CaT} $ as the ``reference" value for the velocity dispersions for our sample. {\bf On the other hand, the CO based $\sigma$ values can be strongly affected by the choice of the template stellar spectra \citep[e.g.][]{winge09,silge03}. }

%{\bf Haven't we just shown that template mismatch isn't noticeably affecting our results? And don't we conclude that we don't yet understand the effect of the stellar population? I think we should just remove the previous paragraph as it's confusing and doesn't add anything to our understanding of the problem, if I'm following correctly. I'm adding a sentence to the start of the next paragraph that can open this section if we delete the current first paragraph.}

We have found a systematic offset 
%-- albeit with large scatter -- 
between $\sigma_{\rm CO}$ and $\sigma_{\rm CaT}$ in the galaxies in our sample, which are primarily spirals. 
In order to further investigate the $\sigma$-discrepancy in late-type galaxies and compare the results with the available studies of other galaxy types, we compiled values for the $\sigma_{\rm CO}$ and for optical measurements (most of them obtained from the CaT region) from the literature for distinct classes of objects. Table~\ref{literature} presents the resulting $\sigma_{\rm opt}$ and $\sigma_{\rm CO}$ values for elliptical, lenticular and spiral galaxies and merger remnants. Since there is good agreement between the $\sigma$ values obtained from CaT and those from the fitting of other optical lines \citep[e.g.][]{barth02a,barth02} and as not all the optical values from the literature were obtained from the CaT region, we will use the nomenclature ``optical" velocity dispersion ($\sigma_{\rm opt}$) to refer to measurements of $\sigma$ obtained from the CaT or other optical lines.   

%{\em Rachel: Some of the galaxies in table 2 are also in our sample. They haven't been included twice in the figures, have they? Rogemar: Yes, they were. I couldn't decide what measurements should I take as the correct and thus I included all - this is certainly wrong, but what is better?}

\begin{table*}
\caption{Velocity dispersions compiled from the literature.}
\label{literature}
\centering
\begin{tabular}{l l l l  l l l l }
\hline
Object    & $\sigma_{\rm CO}$(\kms) &$\sigma_{\rm opt}$(\kms) & Ref.  & 	      Object	& $\sigma_{\rm CO}$(\kms) &$\sigma_{\rm opt}$(\kms) & Ref.  \\
\hline										          		 
\multicolumn{4}{|c |}{\bf ELLIPTICAL GALAXIES} 	              &	                    	 \multicolumn{4}{|c|}{\bf LENTICULAR GALAXIES} \\
NGC\,221      & 71$\pm$8        & 75$\pm$4      & [1]         & 		         NGC\,1023    & 152$\pm$11    & 205$\pm$10  & [1] \\
              & 70$\pm$2        & 75$\pm$3      & [2]         & 		         	      & 217$\pm$5     & 205$\pm$10  & [2] \\
              & 60$\pm$8        & 69$\pm$2      & [4]         & 		         NGC\,1161    & 274$\pm$19    & 297$\pm$17  & [1] \\
NGC\,315      & 321$\pm$59      & 310$\pm$16    & [1]         & 		         NGC\,1375    & 64$\pm$4      & 56$\pm$10   & [3] \\ 
              & 324$\pm$59      & 351$\pm$16    & [4]         & 		         NGC\,1380    & 190$\pm$17    & 219$\pm$11  & [3] \\ 
NGC\,821      & 195$\pm$17      & 209$\pm$10    & [1]         & 		         NGC\,1380A   & 60$\pm$9      & 55$\pm$9    & [3] \\ 
              & 208$\pm$5       & 209$\pm$10    & [2]         & 		         NGC\,1381    & 155$\pm$6     & 153$\pm$8   & [3] \\
              & 188$\pm$17      & 197$\pm$20    & [4]         & 		         NGC\,1400    & 212$\pm$12    & 264$\pm$26  & [1] \\
NGC\,1052     & 211$\pm$20      & 196$\pm$4     & [4]         & 		         NGC\,2110    & 224$\pm$49    & 220$\pm$25  & [1] \\
NGC\,1316     & 212$\pm$20      & 243$\pm$9     & [4]         & 		         NGC\,2293    & 255$\pm$44    & 254$\pm$13  & [1] \\
NGC\,1336     & 119$\pm$8       & 96$\pm$5      & [3]         & 		         NGC\,2380    & 164$\pm$31    & 189$\pm$9   & [1] \\
NGC\,1339     & 182$\pm$9       & 158$\pm$8     & [3]         & 		         NGC\,2681    & 82$\pm$9      & 111$\pm$22  & [1] \\
NGC\,1344     & 158$\pm$20      & 166$\pm$7     & [4]         & 		         NGC\,2768    & 235$\pm$51    & 198$\pm$28  & [1] \\
NGC\,1351     & 153$\pm$7       & 157$\pm$8     & [3]         & 		         NGC\,2787    & 153$\pm$8     & 210$\pm$12  & [1] \\
NGC\,1373     & 80$\pm$5        & 75$\pm$4      & [3]         & 		         	      & 186$\pm$3     & 189$\pm$9   & [2] \\
NGC\,1374     & 207$\pm$10      & 185$\pm$9     & [3]         & 		         NGC\,3115    & 272$\pm$12    & 230$\pm$11  & [2] \\
              & 181$\pm$20      & 180$\pm$8     & [4]         & 		         NGC\,3245    & 206$\pm$7     & 205$\pm$10  & [2] \\
NGC\,1379     & 130$\pm$7       & 130$\pm$7     & [3]         & 		         NGC\,3384    & 151$\pm$3     & 143$\pm$7   & [2] \\
              & 126$\pm$20      & 127$\pm$5     & [4]         & 		         NGC\,3998    & 205$\pm$16    & 297$\pm$15  & [1] \\
NGC\,1399     & 406$\pm$33      & 375$\pm$19    & [3]         & 		         NGC\,4150    & 113$\pm$18    & 132$\pm$10  & [1] \\
              & 336$\pm$20      & 325$\pm$15    & [4]         & 		         NGC\,4342    & 224$\pm$5     & 225$\pm$11  & [2] \\
NGC\,1404     & 247$\pm$22      & 260$\pm$13    & [3]         & 		         NGC\,4564    & 175$\pm$7     & 162$\pm$8   & [2] \\
              & 204$\pm$20      & 230$\pm$10    & [4]         & 		         NGC\,4596    & 139$\pm$3     & 136$\pm$6   & [2] \\
NGC\,1407     & 297$\pm$40      & 283$\pm$13    & [4]         & 		         NGC\,5195    & 95$\pm$6      & 175$\pm$30  & [1] \\
              & 306$\pm$40      & 285$\pm$40    & [1]         & 		         NGC\,5866    & 186$\pm$14    &  139$\pm$7  & [1] \\
NGC\,1419     & 125$\pm$5       & 117$\pm$6     & [3]         & 		         NGC\,6548    & 225$\pm$47    &  307$\pm$23 & [1] \\
              & 116$\pm$20      & 110$\pm$6     & [4]         & 		         NGC\,6703    & 146$\pm$42    &  186$\pm$9  & [1] \\
NGC\,1427     & 155$\pm$18      & 175$\pm$9     & [3]         & 		         NGC\,7332    & 148$\pm$13    &  130$\pm$10 & [1] \\
              & 174$\pm$20      & 172$\pm$8     & [4]         & 		         NGC\,7457    & 63$\pm$2      & 67$\pm$3    & [2] \\
NGC\,2778     & 161$\pm$4       & 175$\pm$8     & [2]         & 		         NGC\,7743    & 66$\pm$12     & 83$\pm$20   & [1] \\ 
NGC\,2974     & 272$\pm$19      & 262$\pm$13    & [1]         & 		         IC\,1963     & 49$\pm$6      & 58$\pm$10   & [3] \\
              & 262$\pm$19      & 255$\pm$12    & [4]         & 		         ESO\,358-G06 & 55$\pm$25     & 58$\pm$11   & [3] \\ 
NGC\,3377     & 144$\pm$20      & 145$\pm$7     & [1]         & 		         ESO\,358-G59 & 70$\pm$20     & 54$\pm$9    & [3] \\
              & 147$\pm$4       & 145$\pm$7     & [2]         & 		         \multicolumn{4}{|c|}{\bf SPIRAL GALAXIES}		\\ 
              & 134$\pm$20      & 135$\pm$4     & [4]         & 		          NGC\,1068    & 129$\pm$3     &151$\pm$7	& [2]	 \\
NGC\,3379     & 235$\pm$20      & 185$\pm$2     & [4]         & 		          NGC\,3031    & 157$\pm$3     &143$\pm$7	& [2]	 \\
NGC\,3607     & 210$\pm$8       & 229$\pm$11    & [2]         & 		          NGC\,4258    & 111$\pm$2     &115$\pm$10	& [2]	 \\
NGC\,3608     & 187$\pm$4       & 182$\pm$9     & [2]         & 		          \multicolumn{4}{|c|}{ \bf MERGER REMNANTS/ULIRGS} \\
NGC\,4261     & 286$\pm$6       & 315$\pm$15    & [2]         & 		          NGC\,1614	  & 133$\pm$3	 & 219$\pm$3	  & [4] \\
NGC\,4291     & 248$\pm$7       & 242$\pm$12    & [2]         & 		          NGC\,2418	  & 245$\pm$7	 & 282$\pm$3	  & [4] \\
NGC\,4365     & 262$\pm$20      & 240$\pm$3     & [4]         & 		          NGC\,2623	  & 152$\pm$4	 & 174$\pm$3	  & [4] \\
NGC\,4374     & 290$\pm$8       & 296$\pm$14    & [2]         & 		          NGC\,2914	  & 179$\pm$6	 & 178$\pm$2	  & [4] \\
NGC\,4459     & 164$\pm$6       & 167$\pm$8     & [2]         & 		          NGC\,3256	  & 111$\pm$20   & 239$\pm$4	  & [4] \\
NGC\,4472     & 291$\pm$20      & 269$\pm$3     & [4]         & 		          NGC\,4194	  & 98$\pm$25	 & 103$\pm$2	  & [4] \\
NGC\,4473     & 186$\pm$3       & 190$\pm$9     & [2]         & 		          NGC\,5018	  & 243$\pm$7	 & 222$\pm$2	  & [4] \\
NGC\,4486     & 310$\pm$20      & 361$\pm$37    & [4]         & 		          NGC\,7252	  & 119$\pm$19   & 160$\pm$3	  & [4] \\
              & 331$\pm$11      & 375$\pm$18    & [2]         & 		          Arp\,193	& 143$\pm$5    & 229$\pm$4	& [4] \\
NGC\,4649     & 327$\pm$11      & 385$\pm$19    & [2]         & 		          IC\,5298	& 150$\pm$28   & 187$\pm$4	& [4] \\
NGC\,4697     & 172$\pm$4       & 177$\pm$8     & [2]         &         		  AM\,0612-373  & 240$\pm$9    & 286$\pm$9	& [4] \\
NGC\,4742     & 104$\pm$3       & 90$\pm$5      & [2]         &         		  AM\,1419-263  & 262$\pm$6    & 258$\pm$3	& [4] \\
NGC\,5128     & 190$\pm$13      & 145$\pm$6     & [4]         &         		  AM\,2038-382  & 207$\pm$4    & 256$\pm$5	& [4] \\
NGC\,5812     & 230$\pm$6       & 248$\pm$2     & [4]         &         		  AM\,2055-425  & 137$\pm$15   & 207$\pm$7	& [4] \\
NGC\,5845     & 237$\pm$4       & 234$\pm$11    & [2]         &         		  IRAS\,02021-2103& 143$\pm$21 & 209$\pm$8	& [5] \\
NGC\,6251     & 290$\pm$8       & 290$\pm$14    & [2]         &         		  IRAS\,05189-2524& 131$\pm$16 & 265$\pm$7	& [5] \\
NGC\,7052     & 327$\pm$13      & 266$\pm$13    & [2]         &         		  IRAS\,12540-5708& 117$\pm$10 & 346$\pm$9	& [5] \\
NGC\,7619     & 246$\pm$47      & 296$\pm$15    & [1]         &         		  IRAS\,17208-0014& 223$\pm$15 & 261$\pm$5	 & [5] \\
              & 246$\pm$47      & 296$\pm$11    & [4]         &         		  IRAS\,23365-3604& 143$\pm$15 & 221$\pm$6	 & [5] \\
NGC\,7743     & 66$\pm$12       & 83$\pm$20     & [1]         &         		      & 	      & 	    &	   \\
NGC\,7626     & 313$\pm$20      & 265$\pm$10    & [4]         &         		      & 	      & 	    &	   \\
IC\,2006      & 125$\pm$10      & 136$\pm$7     & [3]         &         		      & 	      & 	    &	   \\	
\hline	      
\multicolumn{8}{|l|}{[1] - \citet{silge03}; [2] - \citet{kang13}; [3] - \citet{vanderbeke11}} \\
\multicolumn{8}{|l|}{ [4] - \citet{rothberg10}, [5] - \citet{rothberg13} } \\

\end{tabular} 
\end{table*}

\begin{figure*}
\centering
    \includegraphics[scale=0.585]{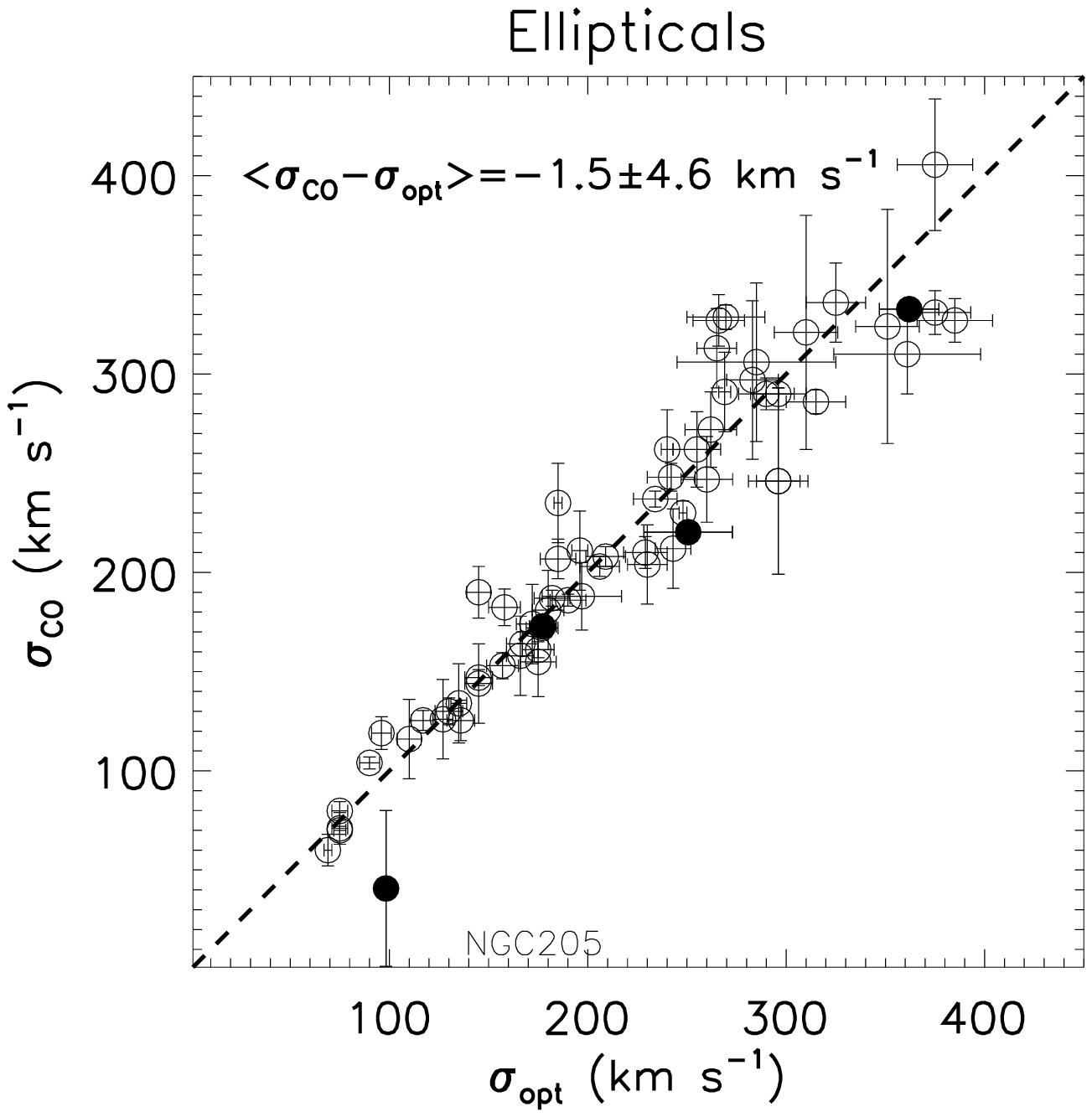}
    \includegraphics[scale=0.585]{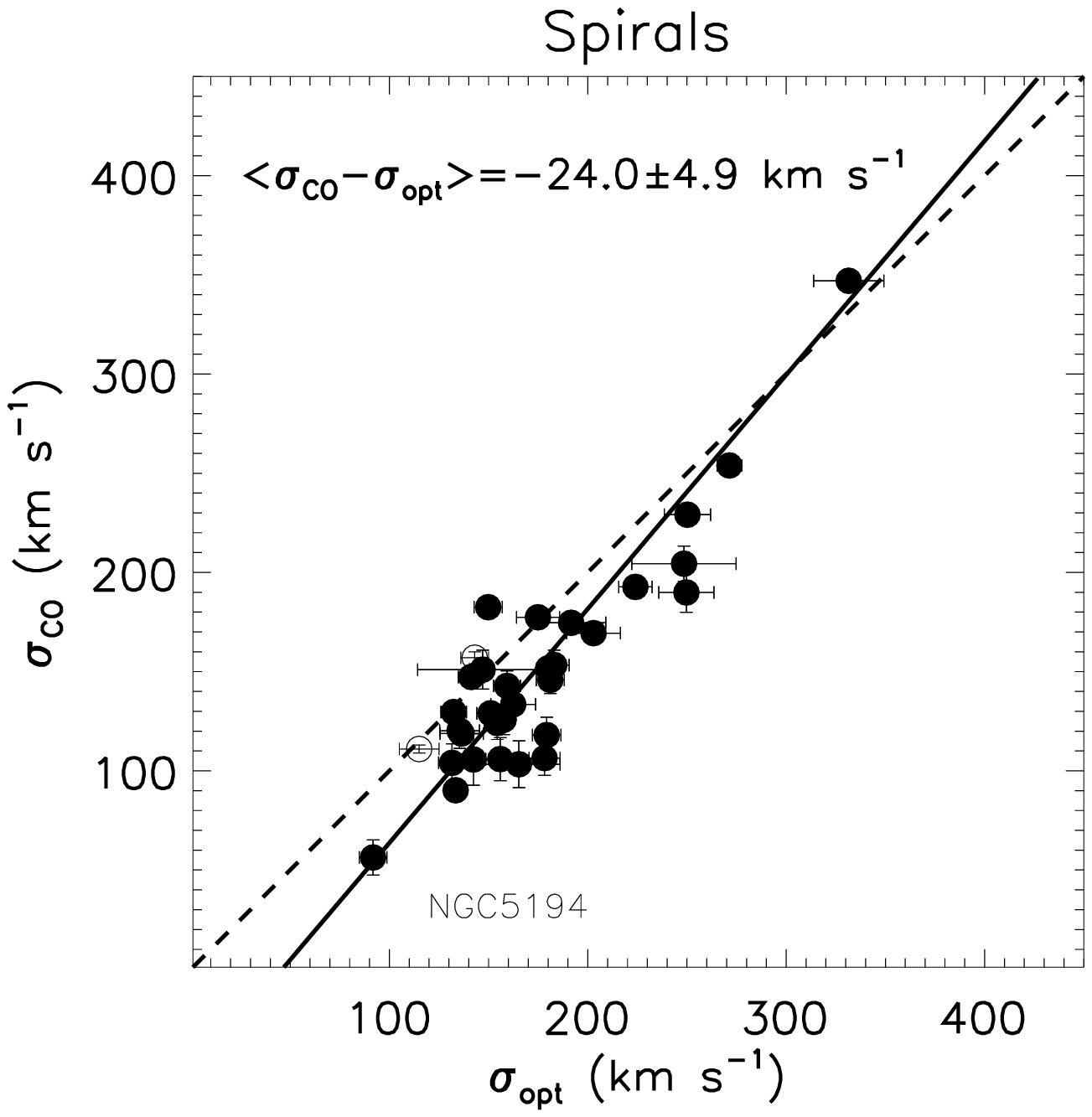}
    \includegraphics[scale=0.585]{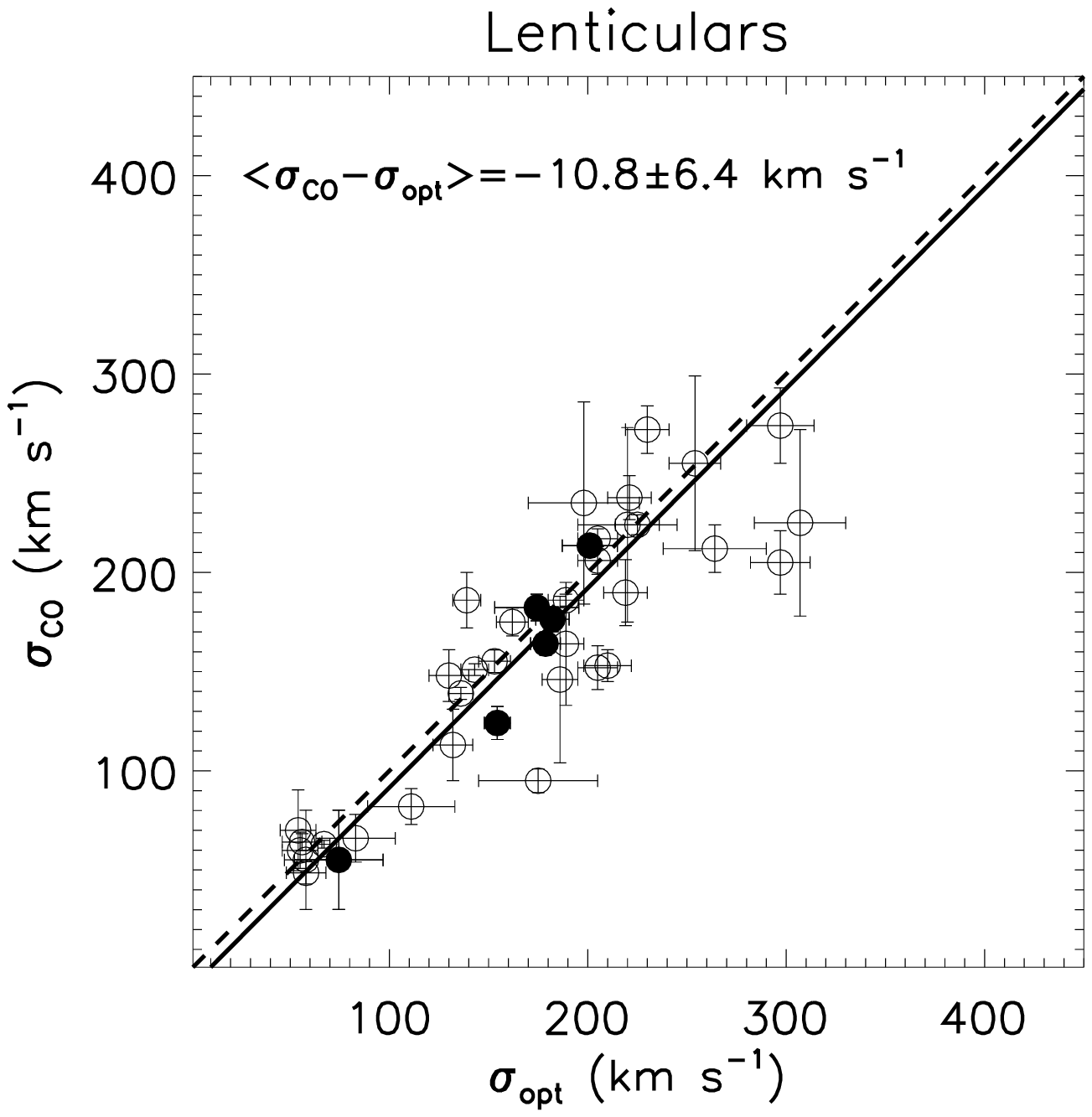}
    \includegraphics[scale=0.585]{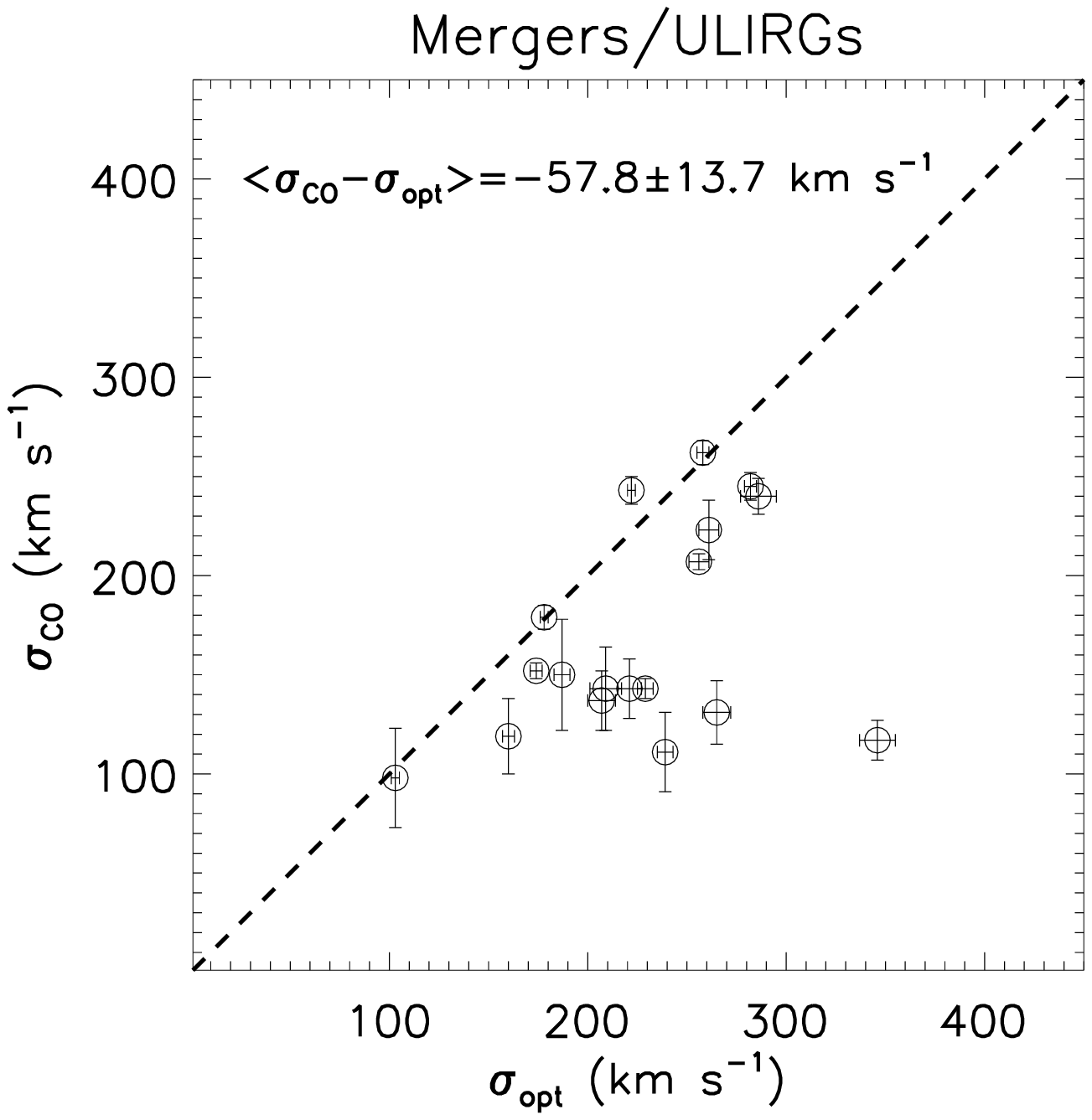}
     \caption{Comparison of the $\sigma$ values in the near-IR  with the optical values  for distinct morphological types. 
%Top-left panel: Elliptical galaxies; Bottom-left panel: Lenticular galaxies; Top-right panel: Spiral galaxies and bottom-right panel: Merger remnants and ULIRGs. 
The objects from our sample are shown as filled circles and open circles are for measurements from the literature. The dashed line shows a one-to-one relation and the continuous line is the best linear fit of the data.  NGC~205 and NGC~5194, which have $\sigma$ smaller than the instrumental resolution, are identified in the plots}.
  \label{types}
\end{figure*}

%The differences in sigma seen in Fig.~\ref{histogram} indicate a dependency on the morphological type of the galaxies. 
In order to investigate the relation between $\sigma$ discrepancy and morphological classification, we plotted in Figure~\ref{types} $\sigma_{\rm opt}$ vs. $\sigma_{\rm CO}$ for distinct classes of objects. %The corresponding plot for elliptical galaxies is shown in the top-left panel, for lenticulars in the bottom-left panel, spirals are shown in the top-right panel and merger remnants and ULIRGs are presented in the bottom-right panel. 
The largest $\sigma$ difference is observed for the merger remnants and ULIRGs, for which no correlation is found between $\sigma_{\rm opt}$ and $\sigma_{\rm CO}$ and the mean difference is $\sigma_{\rm CO} - \sigma_{\rm opt} = -57.8\pm13.7$~\kms. 
%{\bf The standard deviation of the mean ($SD$) is 59.8\,\kms.} 
 Elliptical galaxies follow the one-to-one relation and no discrepancy between optical and near-IR measurements is found. The mean difference is $\sigma_{\rm CO} - \sigma_{\rm opt} = -1.5\pm4.6$~\kms.
%, {\bf with $SD=25.9\,$\kms}. 
For lenticulars, the mean difference is $\sigma_{\rm CO} - \sigma_{\rm opt} = -10.8\pm6.4$~\kms\ and it can be seen from the figure that most points are distributed around the one-to-one relation.
%, with a scatter of $SD=31.8$\,\kms. 
The best linear fit for lenticular galaxies is given by 
\begin{equation}
\sigma_{\rm opt} = (9.1\pm13.2) + (0.99\pm0.08)\sigma_{\rm CO},
\end{equation}
shown as a dashed line in the bottom-left panel of Fig.~\ref{types}.

For spiral galaxies we found a mean difference of $\sigma_{\rm CO} - \sigma_{\rm opt} =  -24.0\pm4.9$~\kms.
%,{\bf with $SD=23.9$\,\kms. 
Excluding NGC\,5194, which presents a sigma value smaller than the spectral resolution of the data, we find the same relation, with $\epsilon_{<\sigma_{\rm CO}-\sigma_{\rm CaT}>}=5.0$\,\kms.
%$SD=24.2\,$\kms.} 
Most objects have $\sigma$ smaller than 200~\kms\ and for these objects   $\sigma_{\rm CO}$ is clearly smaller than $\sigma_{\rm opt}$.

%We computed the Pearson Correlation coefficient ($r$) between $\sigma_{\rm CO}$ and $\sigma_{\rm opt}$ using the $\sc idl$ routine $\sc correlate.pro$ and found $r\approx0.9$, meaning that a good correlation is present ($r$ ranges from $-1$ to $1$: anti-correlation to perfect positive correlation). 
The best linear equation for spiral galaxies is 
\begin{equation}\label{eqs}
\sigma_{\rm opt} = (46.0\pm18.1) + (0.85\pm0.12)\sigma_{\rm CO},
\end{equation}
where we excluded from the fit the galaxy NGC\,5194 (identified in Fig.~\ref{types}). The main cause of uncertainty in Eq.~\ref{eqs} is the small range of $\sigma$ probed by the observations. Further observations are needed to cover the high $\sigma$ region ($\sigma \gtrsim$220~\kms), and higher spectral resolution observations of objects with $\sigma \lesssim$100~\kms, in order to improve the calibration of the equation above. 

 We therefore observe that $\sigma_{\rm CO}$ and $\sigma_{\rm opt}$ become both more similar and more correlated in early-type galaxies compared with spirals and ULIRGs/merger remnants. 
%In Table~\ref{sigfrac} we present the $\sigma_{\rm CO} - \sigma_{\rm opt}$ observed for distinct classes of objetcs. 

%\begin{table}
%\caption{Sigma difference for distinct classes of objects.}
%\label{sigfrac}
%\centering
%\begin{tabular}{l c}
%\hline
%Object Type                 &  $\sigma_{\rm CO} - \sigma_{\rm opt}$ \\
%\hline
%Ellipticals                 & $-1.5\pm25.9$~\kms                    \\
%Lenticulars                 & $ -10.8\pm31.8$~\kms                  \\
%Spirals                     & $ -24.3\pm23.9$~\kms                   \\
%Merger Remnants             & $-57.8\pm59.8$~\kms                   \\
%\hline
%\end{tabular}
%\end{table}

\subsection{What is the origin of the sigma discrepancy for late-type galaxies?}

As discussed above and in \S\ref{intro}, the discrepancy between the stellar velocity dispersion obtained from optical bands and  that obtained from the  near-IR CO absorption band heads is  larger  for mergers of galaxies and ULIRGs than for early-type galaxies. \citet{rothberg10} found a correlation between the infrared luminosity ($L_{\rm IR}$) and $\sigma_{\rm frac}$ for merger remnants, while no correlation is found for elliptical galaxies. \citet{rothberg13} showed that the correlation found for merger remnants extends to ULIRGs, suggesting that dust might  play an important role in the $\sigma_{\rm CO}$ values for this kind of object. 

\begin{figure}
\centering
    \includegraphics[scale=0.62]{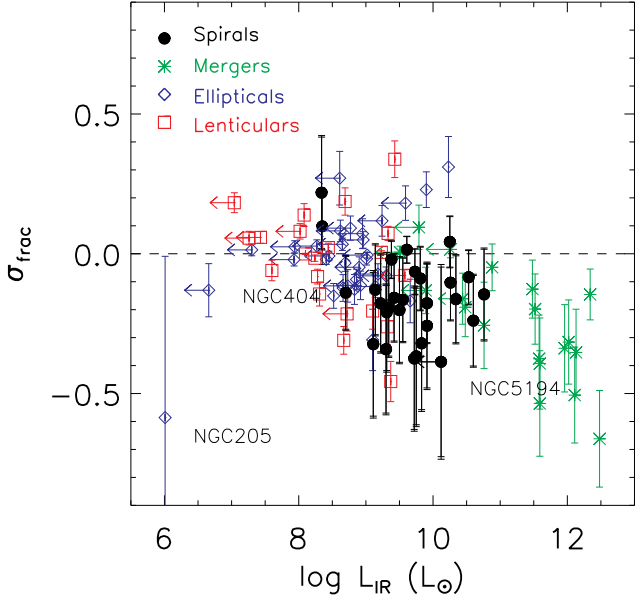}
     \caption{Fractional difference between $\sigma_{\rm CO}$ and $\sigma_{\rm CaT}$, $\sigma_{\rm frac}=\frac{\sigma_{\rm CO} - \sigma_{\rm CaT}}{\sigma_{\rm CaT}}$) vs. $L_{\rm IR}$. The three objects with $\sigma$ values smaller then the spectral resolution of our data are identified in the plot. }
  \label{lir}
\end{figure}

Figure~\ref{lir} shows a plot of $\sigma_{\rm frac}=\frac{\sigma_{\rm CO} - \sigma_{\rm opt}}{\sigma_{\rm opt}}$ vs. log $L_{\rm IR}$ for the galaxies with available infrared luminosities (from \citet{ho97}, \citet{rothberg13} and \citet{rothberg10}).
%, in which the filled circles represent the spiral galaxies, diamonds are for ellipticals, squares are for lenticular galaxies and the asterisks are for the merger remnants and ULIRGs. 
  The values of  $L_{\rm IR}$ were estimated using their infrared fluxes $F_{\rm IR}$ from \citet{ho97}, who defined it as  $F_{\rm IR}=1.26\times10^{-14}(2.58 S_{60}+S_{100})$ W m$^2$, $S_{60}$ and $S_{100}$ being the flux densities at 60 and 100 $\mu$m, respectively. Although data with higher angular resolution and wider wavelength coverage are now available from Herschel and Spitzer Telescopes, we use the IRAS fluxes since they are available for most of our objects, while Herschel and Spitzer data are still not available for most of them of them \citep[e.g.][]{marleau06,sauvage10,looze12, ciesla12, auld13}.
%The dashed line marks  $\sigma_{\rm CO}=\sigma_{\rm CaT}$.
Figure \ref{lir} shows that galaxies with higher $L_{\rm IR}$ also have higher negative values of $\sigma_{\rm frac}$, and that the spiral galaxies fill the gap between early-type objects and merger remnants. 
%For ULIRGs and merger remnants \citet{rothberg13} and \citet{rothberg10} found a good correlation between $\sigma_{\rm frac}$ and log $L_{\rm IR}$ with a Pearson Correlation coefficient ($r=0.75$), while no trend is found for elliptical and lenticular galaxies {\bf you already said this in the previous paragraph, just remove this sentence}. 
This smooth trend suggests that dust might play some role in the observed $\sigma-$discrepancy, as more warm dust is expected in spiral galaxies than in elliptical galaxies, and less than in merger remnants and ULIRGs.

\begin{figure}
\centering
    \includegraphics[scale=0.62]{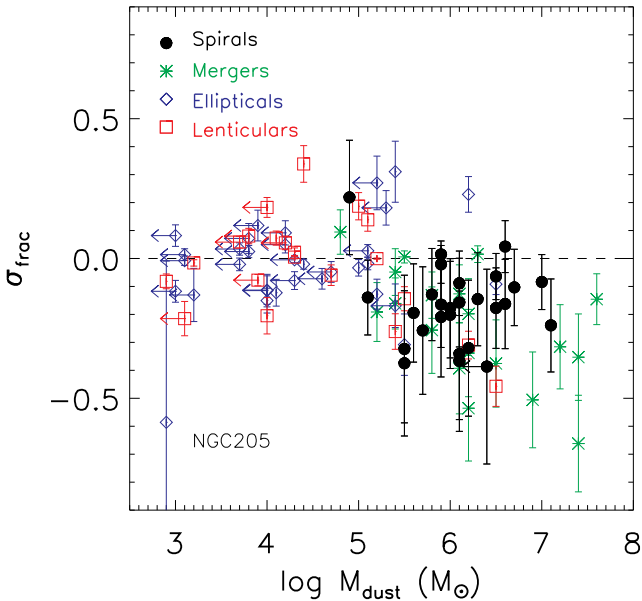}
     \caption{$\sigma_{\rm frac}$ vs. log$M_{\rm dust}$ for distinct types of objects.}
  \label{mdust}
\end{figure}

 Following \citet{rothberg10} we estimate the mass of dust by

\begin{equation} 
\frac{M_{\rm dust}}{\rm M_\odot} =0.959S_{100}D^2\left[\left(9.96\frac{S_{100}}{S_{60}}\right)^{1.5}-1 \right], 
\end{equation}
where $S_{60}$ and $S_{100}$ are the IRAS flux densities at 60 and 100$\mu$m
in Jy, respectively, and D is the distance to the galaxy in Mpc \citep[see also][]{hildebrand83,thuan92}. \citet{rothberg10} found that $\sigma_{\rm frac}$ correlates with M$_ {dust}$ for merger remnants, while no correlation is found for elliptical galaxies. Figure~\ref{mdust} shows the plot of $\sigma_{\rm frac}$ vs. M$_ {dust}$ for the galaxies studied here. A similar trend to that seen in Fig.~\ref{lir} is observed in this plot, suggesting that dust plays a role in the observed $\sigma$-discrepancy for spirals and merger remnants.

%{\bf I know the referee asked for this, but the dust mass calculation is based on exactly the same data as L\_FIR. I don't see that the dust mass gives any new information, so I don't think it's interesting to show it here/}
%\textcolor{red}{I agree with you. But, I think that we could include this figure just to please the referee. Although the M is estimated from the flux densities, the relation is not just linear.}

 Dust may be relevant to the $\sigma$-discrepancy in two ways. First, as extinction is lower in the K band, the $\sigma_{\rm CO}$ measurements could probe a dynamically cold, disk-like component that is more obscured than the dynamically hot bulge stars. Indeed, near-IR studies of nearby galaxies show that the reddening obtained from near-IR lines is larger than that obtained from optical lines, indicating that the near-IR samples an obscuring column larger than
the optical spectral region \citep[e.g.][]{morwood88,heisler99,martins13b}.   Assuming a standard Galactic extinction curve \citep{weingartner01,draine03}, the extinction at 0.85 $\mu$m (CaT) is about a factor of 5 larger than that at 2.3 $\mu$m (CO), and the extinction at 0.52 $\mu$m (Mgb) is about a factor of 2 greater again.  If the difference between $\sigma_{CaT}$ and $\sigma_{CO}$ is due to extinction, we may also expect a difference between $\sigma_{CaT}$ and $\sigma_{Mgb}$. However, \citet{barth02a} compared $\sigma_{CaT}$ and $\sigma_{Mgb}$ and did not find any systematic difference between them.

% Rogemar: would such a difference be visible in the Ho (2009) data? Or is it within the uncertainties? I think we should at least comment on this, as extinction is an obvious possible explanation.

%However, in this case some differences should aslo be found in the $\sigma$ values derived from the CaT and Mgb  lines due to differences in the extinction at  8500\,\AA\ and  5200\,\AA.
%In this case, as the difference in extinction between 2.3 $\mu$m and 0.85 $\mu$m is smaller than that between 0.85 $\mu$m and 0.52 $\mu$m (where velocity dispersions are commonly measured from the Mgb band), larger discrepancies should be found between $\sigma_{CaT}$ and $\sigma_{Mgb}$ than between $\sigma_{CO}$ and $\sigma_{CaT}$. 
%In fact, we find good agreement between our CaT values and the $\sigma_{\rm opt}$ measured by \citet{ho09} -- for about 90\% of the objects the differences in $\sigma_{\rm Ca}$ and $\sigma_{\rm opt}$ are smaller than 25\%, implying that extinction might not be the principal cause of this effect.

%\textcolor{green}{Luis Ho, can you answer the Rachel's question? I think you didn't find any sistematic difference between  $\sigma_{CaT}$ and $\sigma_{Mgb}$ (even not a small difference). Am I right?}

Secondly, warm ($T\sim50$~K), FIR-emitting dust may be associated with star formation in these galaxies. Indeed, the emission of the warm dust is directly correlated with the star formation rate, as it is heated by young stars \citep{kennicutt98,kennicutt12}. 
%Regarding the amount of dust present in late-type galaxies, \citet{bourne13} used a sample of local galaxies $(z< 0.05)$ selected from the 500 $\mu$m-Herschel Astrophysical Terahertz Large Area Survey (H-ATLAS), including dust-rich late-type spirals, starbursts, merging/interacting galaxies, low-surface-brightness disks and dusty early-type galaxies,  and found masses of dust of the order of $10^7$\, M$_\odot$.

%I'm not sure this is useful, as it covers so many galaxy types. Also, it sort of contradicts our dust mass figure. And anyway, what does a dust mass of 10$^7$ Mo imply for star formation?

%\textcolor{red}{I agreee. I was trying to answer this question:   Referee: "If dust really plays an important role in the sigma-discrepancy, please give a short introduction to recent progress in that field (e.g. regarding the dust mass determination)".  Corrently our answer is: "In the introduction, we concentrate on reviewing the previous literature on this subject. Rather than add this material to the introduction, we have added more information and discussion about extinction, dust masses, and star formation to section 6." Maybe I should exclude the sentence "Regarding the amount...." and the "dust masses" from the response to the referee. What do you think? }

 Young stars form in disks, and in the absence of major perturbations generally remain dynamically cold. 
Indications of young stars in disks have been found in recent spatially resolved spectroscopy of galaxy nuclei with the Near-IR IFU Spectrograph (NIFS) on Gemini North. \citet{mrk1066-pop,mrk1157-pop} and \citet{sb12} carried out stellar population synthesis and found a spatial relation between low $\sigma_{\rm CO}$ and the young/intermediate age stellar population, confirming that $\sigma_{\rm CO}$ is affected by the presence of young/intermediate age stars. The presence of a young stellar population has also been proposed as an explanation of the $\sigma$-drop observed in some galaxies \citep[e.g.][]{emsellem01,marquez03}. However, the comparison of our results with spatially resolved measurements should be taken with caution. While the NIFS data resolve a dynamically cold structure, our single-aperture measurements probe the second moment of the LOSVD, which does not necessarily imply a direct link between the two sets of results. New spatially resolved measurements of both the CaT and CO lines would show whether the effect we are observing here is related to the low-$\sigma$ regions observed in the works cited above.

If the difference between $\sigma_{CO}$ and $\sigma_{CaT}$ is due to the presence of a young stellar population in the disk of the galaxies, young stars must contribute more to the CO absorption features than do older stars, and the effect of this population on the $\sigma$ measured from the CaT and optical lines must be negligible. The connection between the CO and CaT bands and the age of the stellar population is not straightforward, though. For example, the CO bands are relatively strong in both young/intermediate-age stars (AGB/TP-AGB stars) and old ones (M stars), while hotter, younger stars produce strong CaT bands along with weak CO features \citep[e.g.][]{maraston05,rogerio07}. On the other hand, previous work has shown that $\sigma_{CaT}$ is not very sensitive to the stellar population \citep{barth02,barth02a}. Full stellar population synthesis would help to resolve these issues, although different models currently make very different predictions for the NIR spectral region \citep[e.g.][]{bruzual03,maraston05}.

%In order to evaluate if the rotating disk component, which may be present in most galaxies of our sample, has some effect on the observed $\sigma$ discrepancy we constructed a plot of $\sigma_{\rm frac}$ vs. $i$ (the inclination of the disk). The rotational component is expected to be more important in highly inclined galaxies \citep[e.g.][]{bellovary14} and thus, if rotation affects the $\sigma$ in a different way in the CO and CaT region a larger $\sigma$ discrepancy would be expected for highly-inclined galaxies compared to face-on disks.  However, we found no correlation between $\sigma_{\rm frac}$ and $i$ suggesting that the rotational component is not playing any role in the $\sigma$ discrepancy  for spiral galaxies.

\subsection{Implications for the $M_{\bullet}-\sigma$ relationship}

As discussed above, the $\sigma $ discrepancy observed in our sample of
mostly late-type galaxies   does not appear in previous studies of early-type galaxies. We can use the measured $\sigma$ values to evaluate the impact of the $\sigma$ discrepancy on determinations of the mass of the central SMBH using the $M_{\bullet}-\sigma$ relationship. Several studies have aimed at properly calibrating the  $M_{\bullet}-\sigma$ relation for distinct classes of objects. \citet{xiao11} investigated the $M_{\bullet}-\sigma$ relation using a sample of 93 late-type galaxies with a Seyfert 1 nucleus. They found no difference in the slope for subsamples of barred and unbarred galaxies, but they found a small offset in the relation between low- and high-inclination disk galaxies, with the latter having a larger $\sigma$ value for a given  black hole mass. % \citet{graham11} used a sample of 64 galaxies  with distinct Hubble types to investigate the $M_{\bullet}-\sigma_\star$ relationship and found that the  $M_{\bullet}-\sigma$ relation obtained when the sample is restricted to barred galaxies only lies $\approx$\,0.45 dex bellow the relation obtained for elliptical and non-barred galaxies. 
 For a review of calibrations of the $M_{\bullet}-\sigma$ relation, see \citet{msigma}. 

Actually, \citet{kormendy11} show that the physically relevant  parameter in black hole correlations
with host galaxy type is not early-type vs. late-type  objects, but rather classical bulges versus pseudo bulges. The latter is
defined as the buildup of dense central
components that look like classical merger-built bulges but that were in fact formed slowly by disks out of disk material \citep{kormendy04}.

Since the aim of the present paper is to evaluate the
impact of the use of  CO-based measurements of the stellar velocity
dispersion on the derived mass of the super-massive black hole, and not to
calibrate the $M_{\bullet}-\sigma$ relationship, we use the same calibration
 for all objects of our sample, given by \citep{msigma} as:

\begin{equation}
{\rm log}\left(\frac{M_\bullet}{10^9\,M_\odot}\right) = -(0.500\pm0.049)+(4.420\pm0.295){\rm log}\left(\frac{\sigma}{200\,\rm km s^{-1}}\right). 
\end{equation}

We estimated the mass of the SMBH for all spiral galaxies of our sample using $\sigma_{\rm CaT}$ and $\sigma_{\rm CO}$ in the equation above. 
%Figure~\ref{mass} shows a plot of the resulting mass obtained using $\sigma_{\rm CO}$  ($y$-axis) vs. the one from $\sigma_{\rm CaT}$ ($x$-axis). The solid line represents a one-to-one relation and uncertainty in $\sigma$. The masses derived using $\sigma_{\rm CO}$ are smaller, as expected (Figure~\ref{sigma}) the masses derived using $\sigma_{\rm CO}$ are smaller than the values obtained from $\sigma_{\rm CaT}$, as expected. 
The mean logarithmic difference is $-0.29\pm0.12$, which may be taken as a systematic error in the $M_{\bullet}-\sigma$ relation when using  CO-based estimates  of the stellar velocity dispersion. %{\bf However, it should be noticed that the scatter of this difference is high, compared to the difference itself.}

%\begin{figure}
%\centering
%    \includegraphics[scale=0.62]{figs/fig-co-ca-mass.ps}
%     \caption{Comparison between the mass of the super-massive black hole obtained with the $M_*-\sigma$ relationship from \citet{msigma} using the  velocity dispersion obtained from the fitting of the CO band heads ($y$-axis) and from the fitting of the CaT ($x$-axis).}
%  \label{mass}
%\end{figure}

\section{Conclusions}

We have used 0.85 -- 2.5~$\mu$m spectroscopy of a sample of  48 galaxies  (35 spirals, 7 lenticulars and 6 ellipticals) obtained with the Gemini Near-Infrared Spectrograph (GNIRS) on Gemini North telescope to measure the stellar kinematics by fitting the K-band CO absorption band heads and the CaT at 8550~\AA. This work is aimed at determining whether the difference in $\sigma_{CO}$ and $\sigma_{CaT}$ (the ``$\sigma-$ discrepancy'') reported for ULIRGs and merger remnants persists in the hitherto unexplored regime of late-type galaxies. Our main conclusions are:

\begin{itemize}

\item The velocity dispersion obtained from the 2.29 $\mu$m CO band-heads is slightly smaller than the one from fitting the CaT, with an average difference of $\sigma_{\rm CO} - \sigma_{\rm CaT} = -19\pm6$~\kms\  for the complete sample (all morphological types). 
%The higher order Gauss-Hermite moments are small for most objects of our sample, but the scatter between optical and near-IR estimates is very high, with an average difference of  $h_{3CO} - h_{3CaT} = -0.01\pm0.08$ and $h_{\rm 4CO} - h_{\rm 4CaT} = 0.07\pm0.07$ for $h_3$ and $h_4$.

\item We compiled the available $\sigma$ values from literature and found an almost one-to-one relation between optical (CaT, Mgb, etc.) and CO-based estimates for early-type galaxies. For spiral galaxies the discrepancy is higher, but still much lower than for merger remnants. The best fit  for spiral galaxies is  $\sigma_{\rm opt} = (46.0\pm18.1) + (0.85\pm0.12)\sigma_{\rm CO}$, but more observations covering the $\sigma$ ranges $\sigma < 100$~\kms and  $\sigma > 200$~\kms  are needed to properly calibrate this relation. 

\item The fractional $\sigma$ difference correlates with  the infrared luminosity, which may suggest that the $\sigma$-discrepancy is related to the presence of warm dust. In this scenario, the CO absorption band heads would be dominated by young stars located in the disc of the galaxy and thus result in smaller $\sigma$ values, while the optical estimates are less sensitive to variations in the stellar population.   %{\bf Less sensitive to this specific stellar population, no? 
However, the detailed spectral synthesis that would be needed to test this interpretation requires high spectral resolution SSP models, which are not yet available.

\item We investigated the impact of the $\sigma-$discrepancy on the mass of the SMBH obtained via the $M_\bullet-\sigma$ relation and found a mean logarithmic difference of $-0.29\pm0.12$, that must be considered as a systematic error in the SMBH mass when using  $\sigma_{\rm CO}$ for spiral galaxies. However, this uncertainty is dominated by scatter of the relation and the conversion from $\sigma_{\rm CO}$ to $\sigma_{\rm opt}$ may introduce a even larger uncertainty in the derived $M_\bullet$. 

 Although the ``$\sigma-$ discrepancy'' has already been discussed for ULIRGs and merger remnants \citep[e.g.][]{rothberg10,rothberg13} and no discrepancy was found for early-type galaxies \citep[e.g.][]{silge03,vanderbeke11,kang13}, this is the first time that this comparison is done for a sample of mostly late-type galaxies.

\end{itemize}

\section*{Acknowledgements}
We thank an anonymous referee for useful suggestions which helped to improve the paper and S. Rembold for help with the bootstrap technique.  
This work is  based on observations obtained at the Gemini Observatory, which is operated by the Association of Universities for Research in Astronomy, Inc., under a cooperative agreement
with the NSF on behalf of the Gemini partnership: the National Science Foundation (United
States), the Science and Technology Facilities Council (United Kingdom), the
National Research Council (Canada), CONICYT (Chile), the Australian Research Council
(Australia), Minist\'erio da Ci\^encia, Tecnologia e Inova\c c\~ao (Brazil) 
and Ministerio de Ciencia, Tecnologia e Innovaci\'on Productiva  (Argentina). 
{\it R.A.R.} acknowledges support from FAPERGS (project N0. 12/1209-6) and CNPq (project N0. 470090/2013-8).
{\it L.C.H.} acknowledges support from the Kavli Foundation, Peking University, and grant N0. XDB09030102 (Emergence of Cosmological Structures) from the Strategic Priority Research Program of the Chinese Academy of Sciences.
 {\it A.R.A.} acknowledges CNPq for partial support to this work through grant 307403/2012-2.
 {\it L.M.} thanks CNPq through grant 305291/2012-2. 
{\it L.C.} acknowledges support from the Special Visiting Researcher Fellowship (PVE 313945/2013-6) under the Brazilian Scientific Mobility Program ``Ci\^encias sem Fronteiras".
{\it R.R.} acknowledges funding from FAPERGs (ARD 11/1758-5) and CNPq (PeP 304796/2011-5). 
{\it C.R.A.} is supported by a Marie Curie Intra European Fellowship within the 7th European Community Framework Programme (PIEF-GA-2012-327934) and by the Spanish Ministry of Science and Innovation (MICINN) through project PN AYA2010-21887-C04.04.

\appendix
\section{Fits of the spectra}

Figures \ref{fits-co} and \ref{fits-ca} show the resulting fit of the galaxy spectra for the CO and Ca triplet spectral regions, respectively.

\begin{figure*}
\centering
  \begin{tabular}{cccc}
    \includegraphics[scale=0.2]{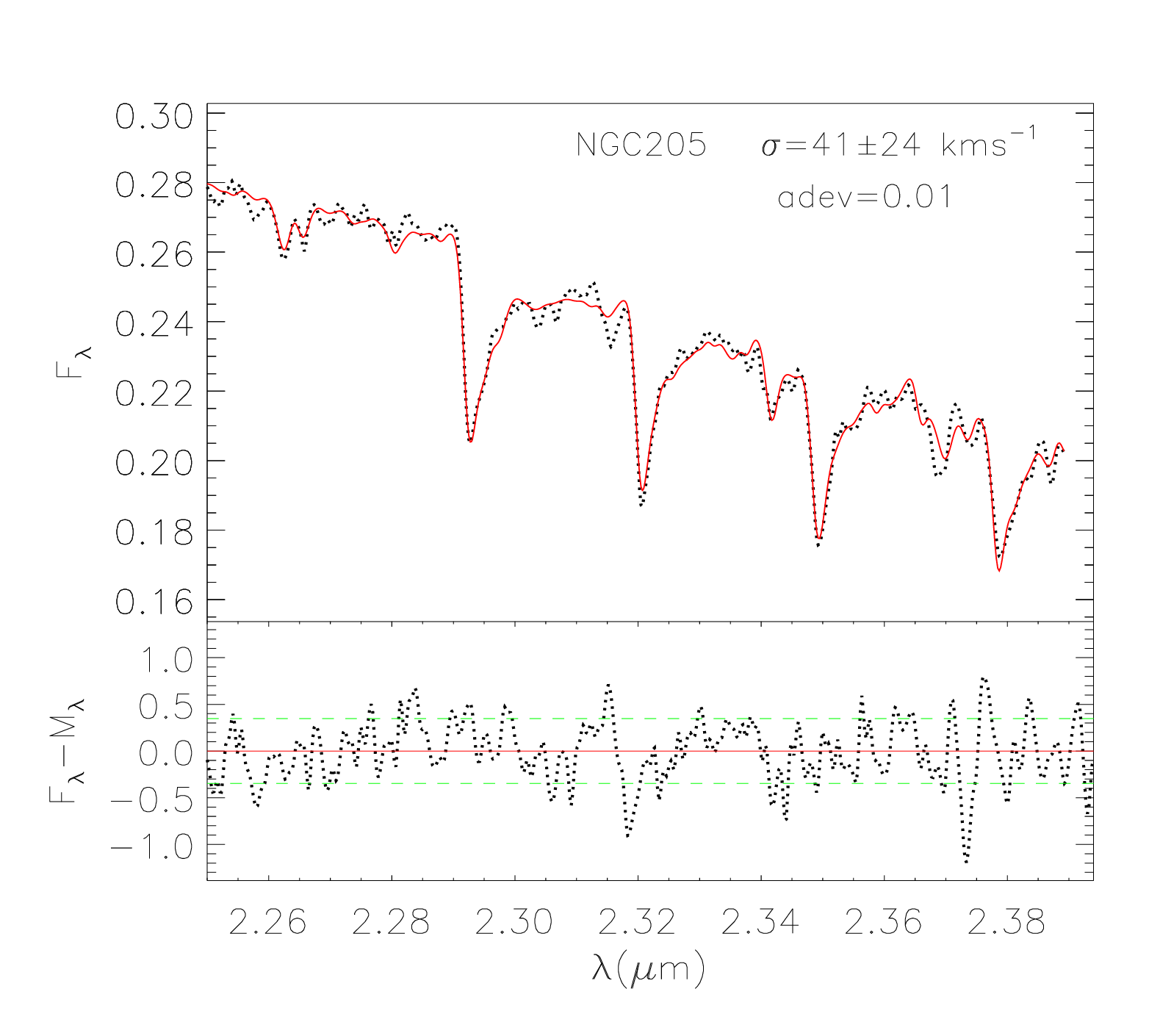}&
    \includegraphics[scale=0.2]{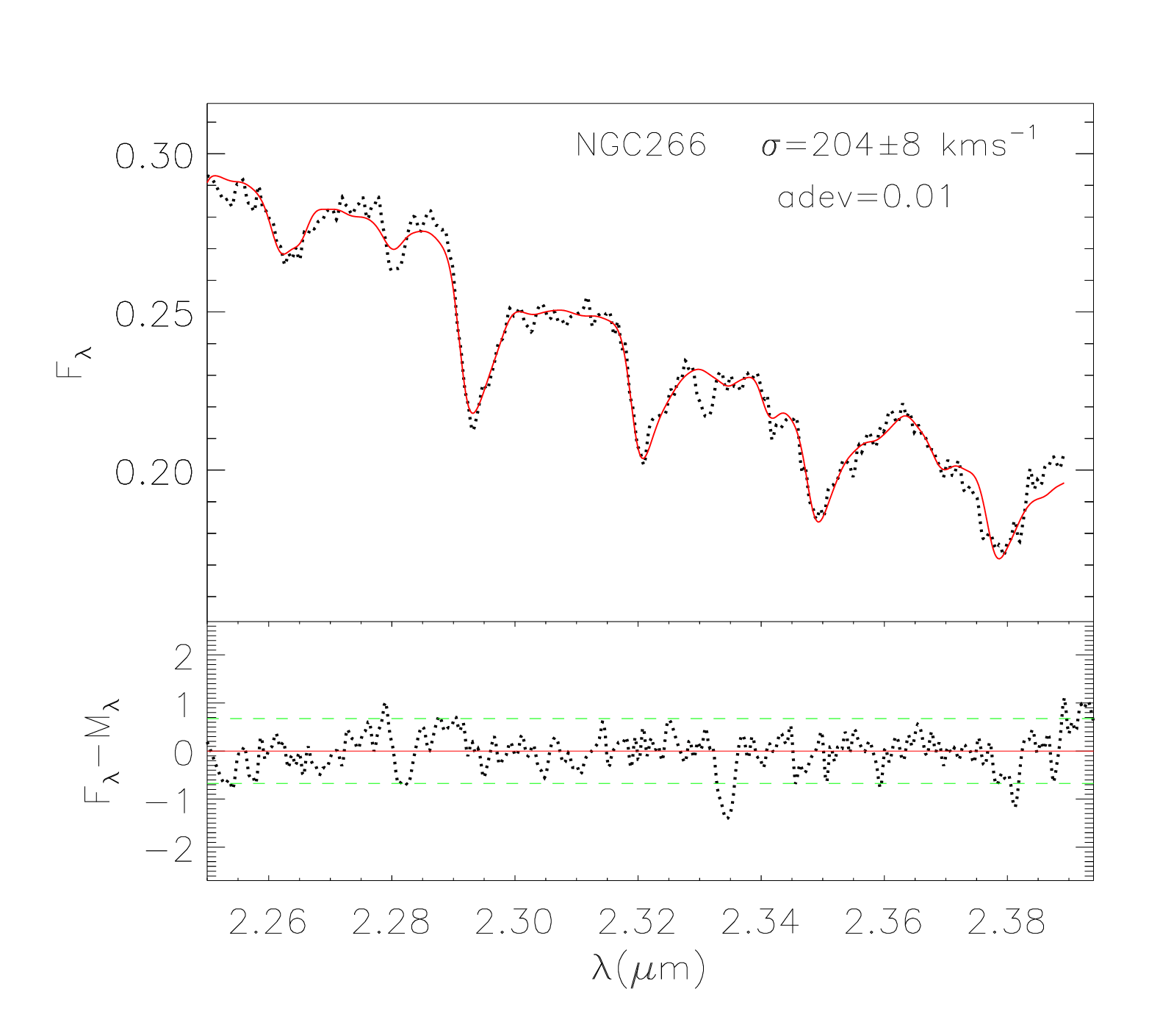}&
    \includegraphics[scale=0.2]{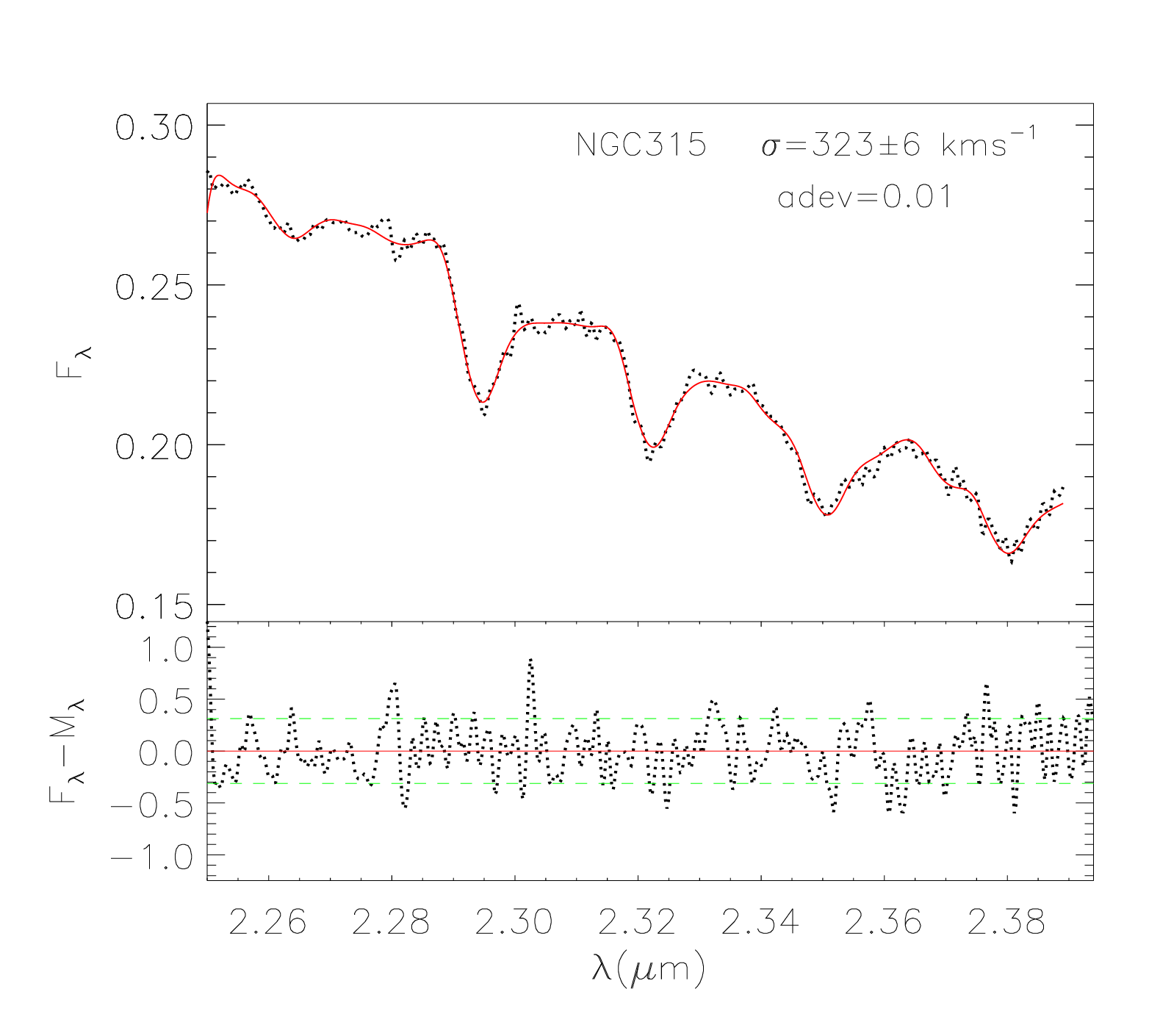}&    
    \includegraphics[scale=0.2]{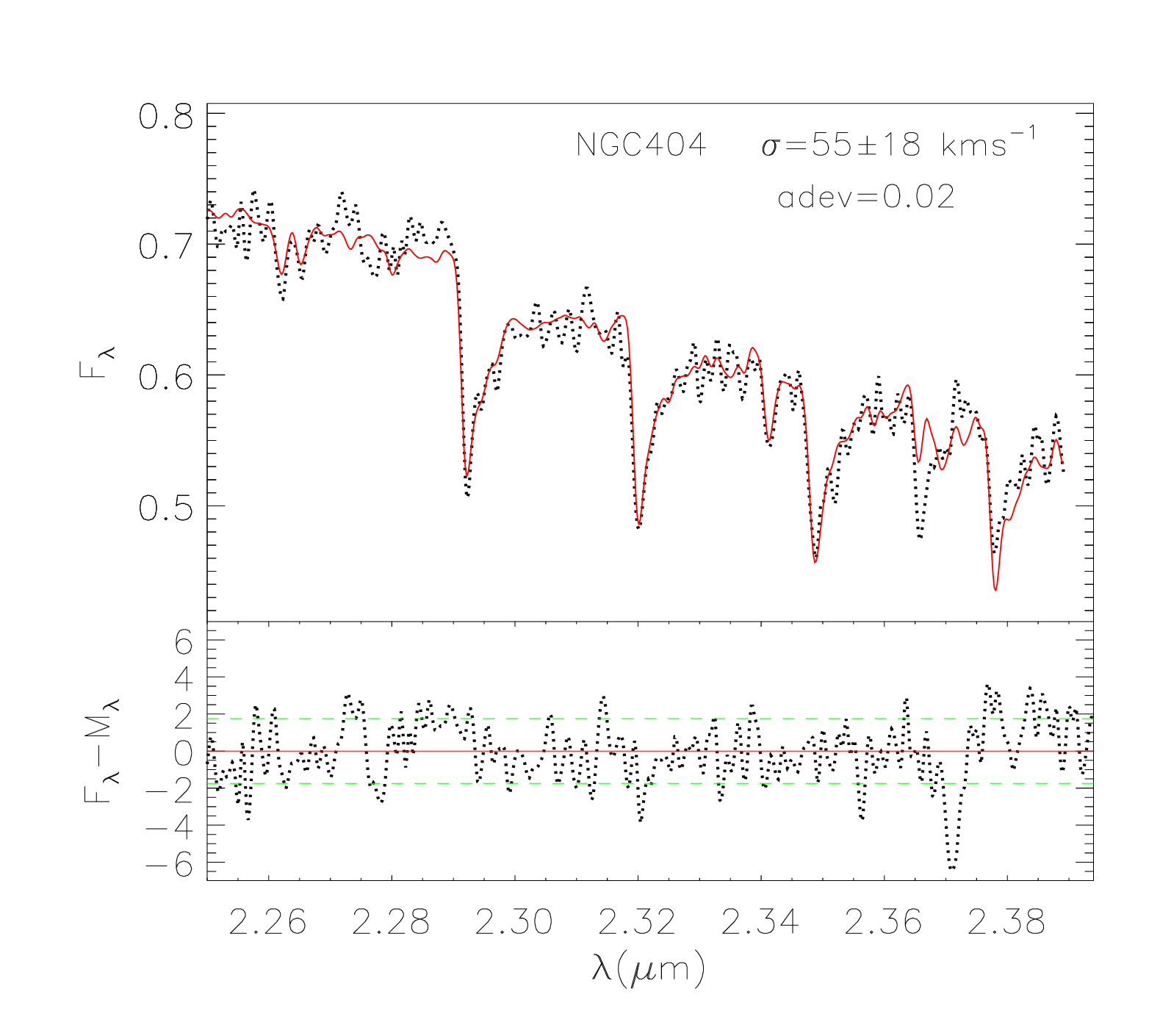} \\

   \includegraphics[scale=0.2]{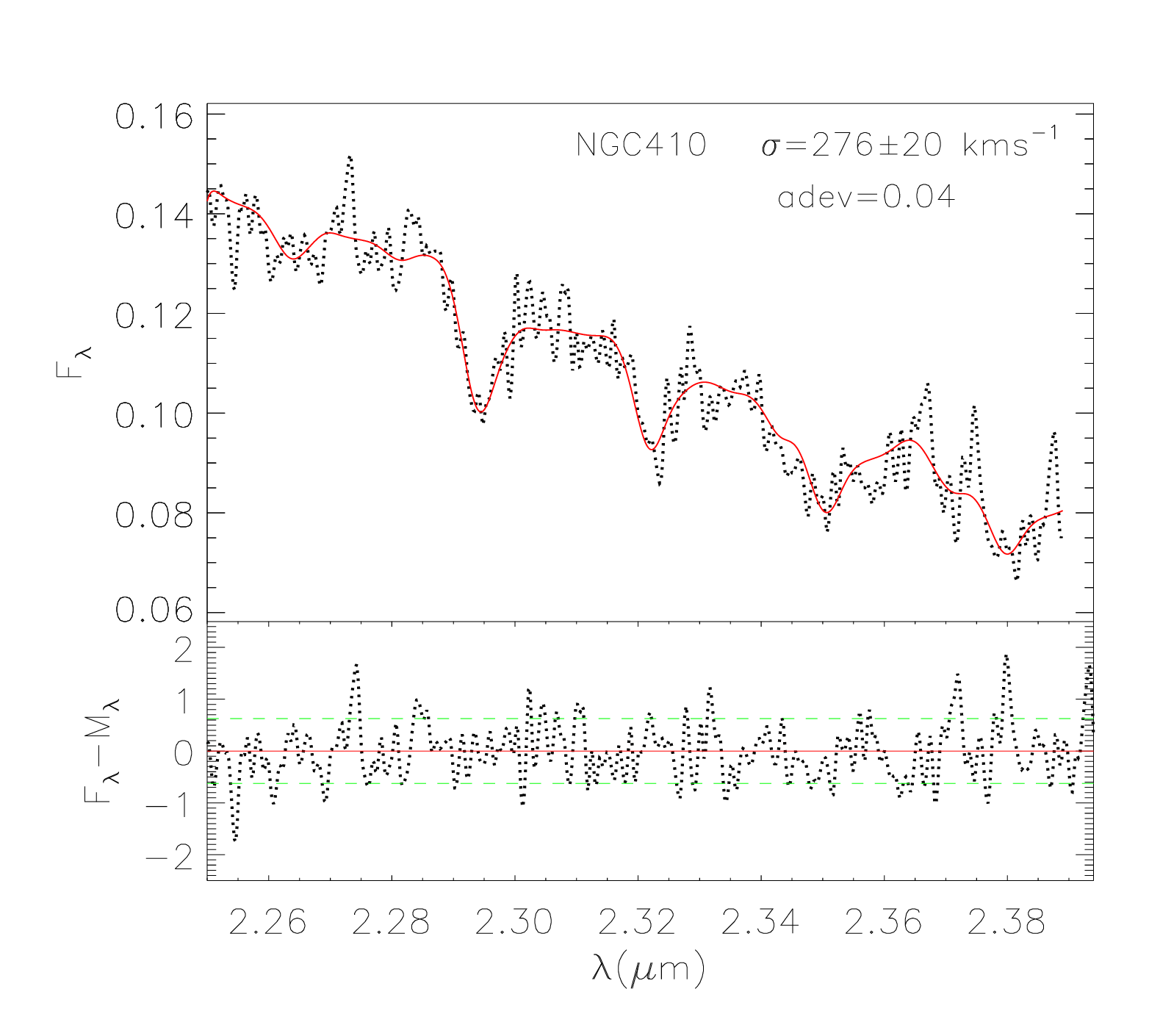}&
    \includegraphics[scale=0.2]{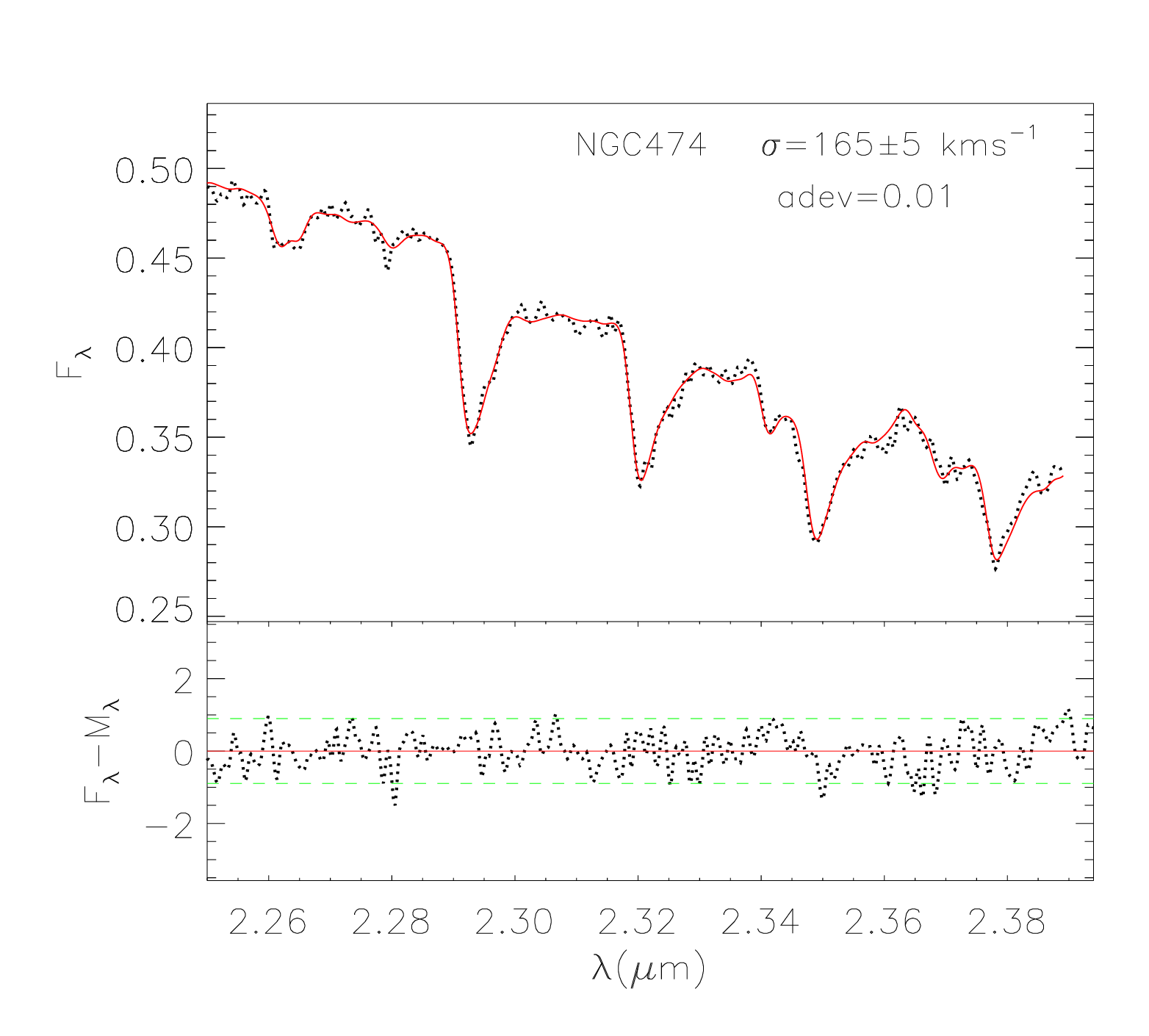}&
    \includegraphics[scale=0.2]{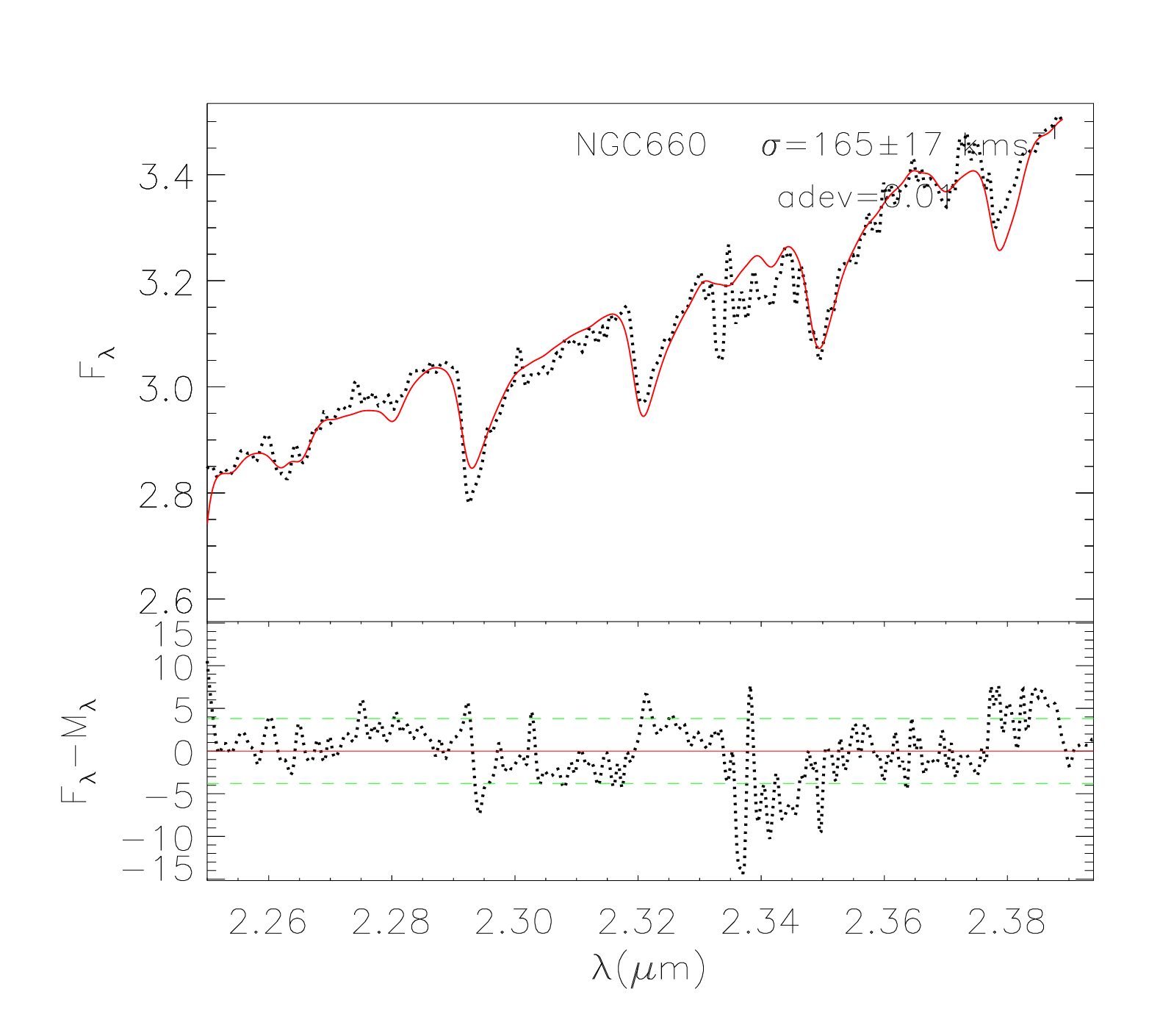}&    
    \includegraphics[scale=0.2]{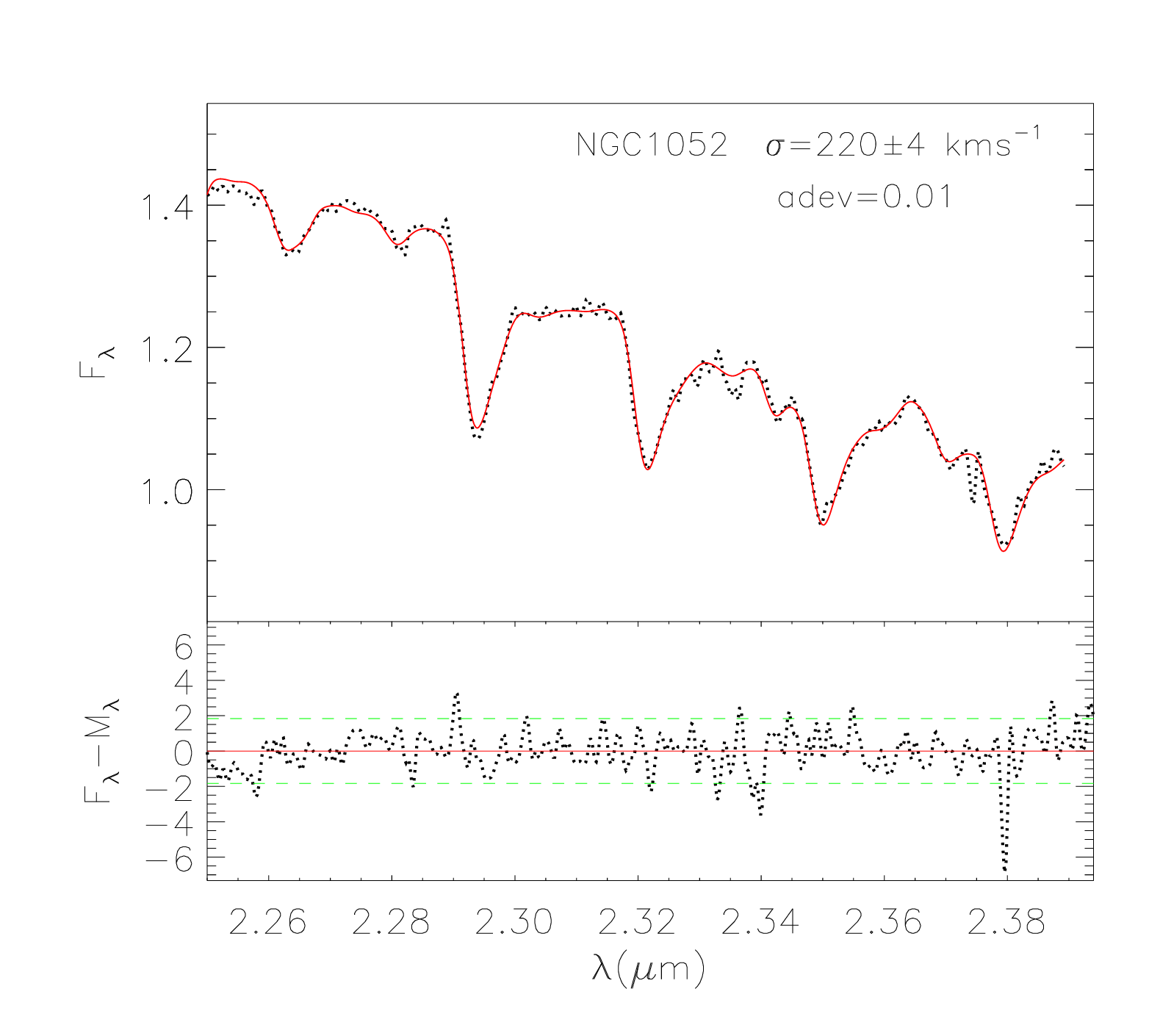} \\

   \includegraphics[scale=0.2]{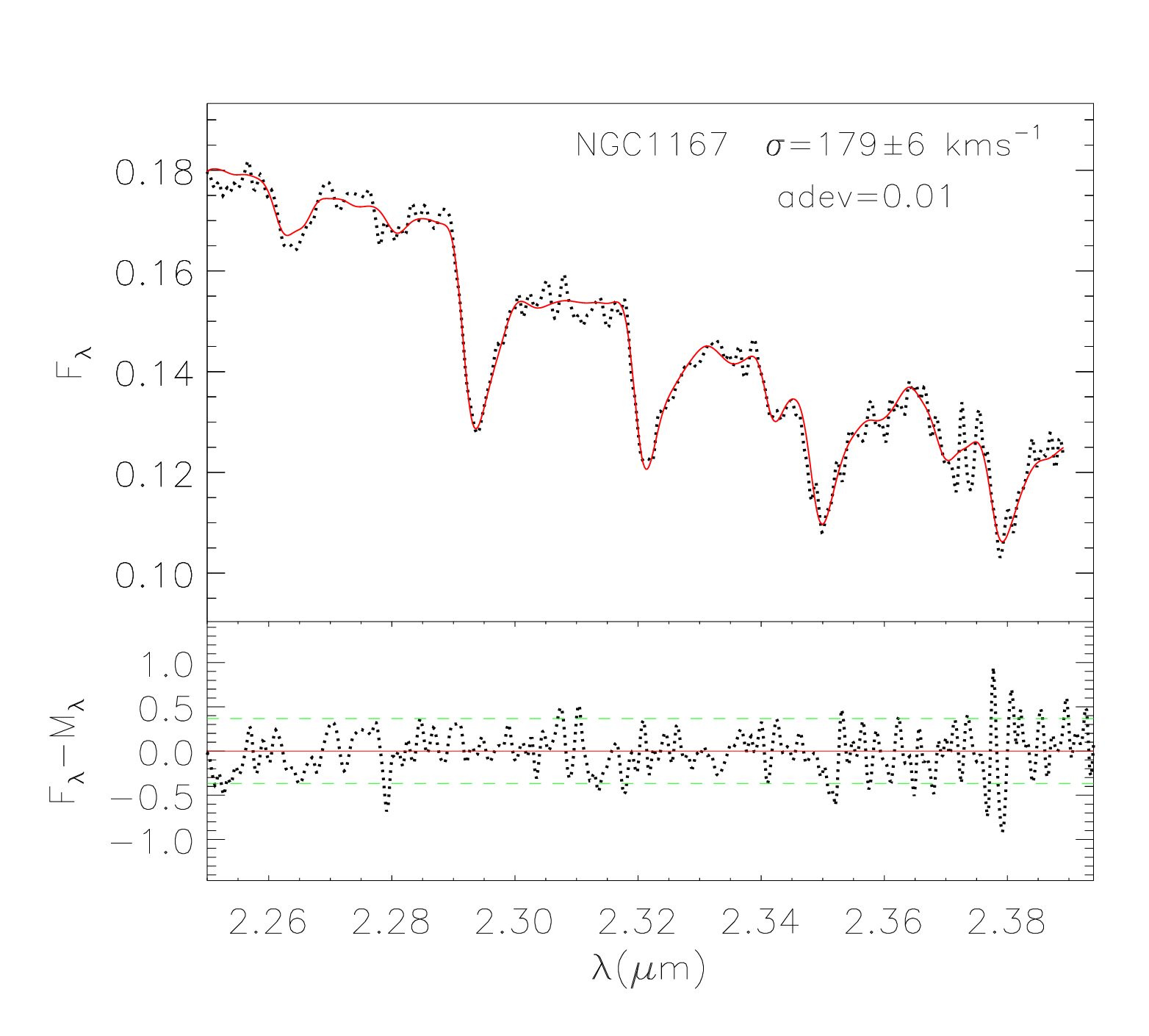}&
    \includegraphics[scale=0.2]{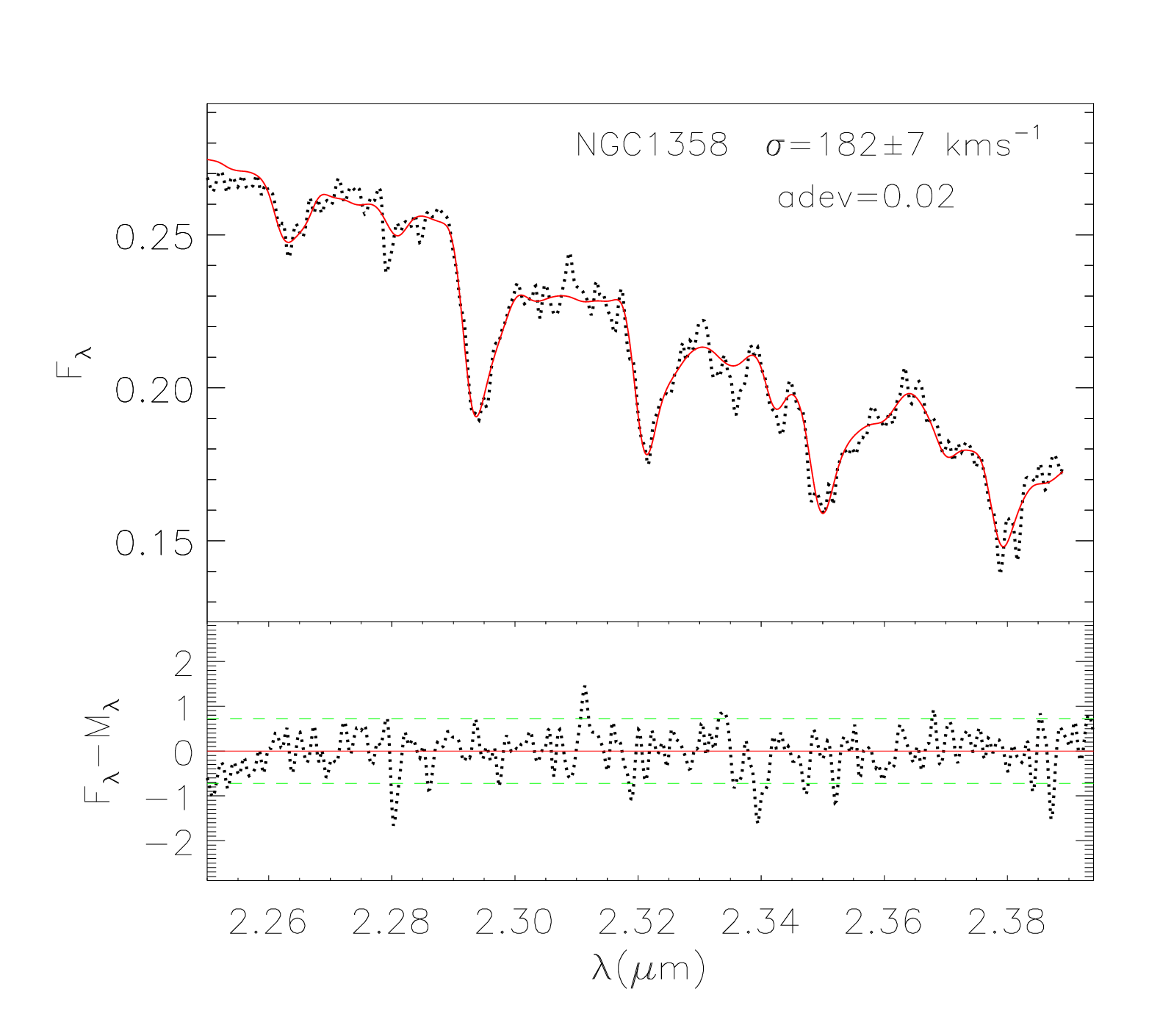}&
    \includegraphics[scale=0.2]{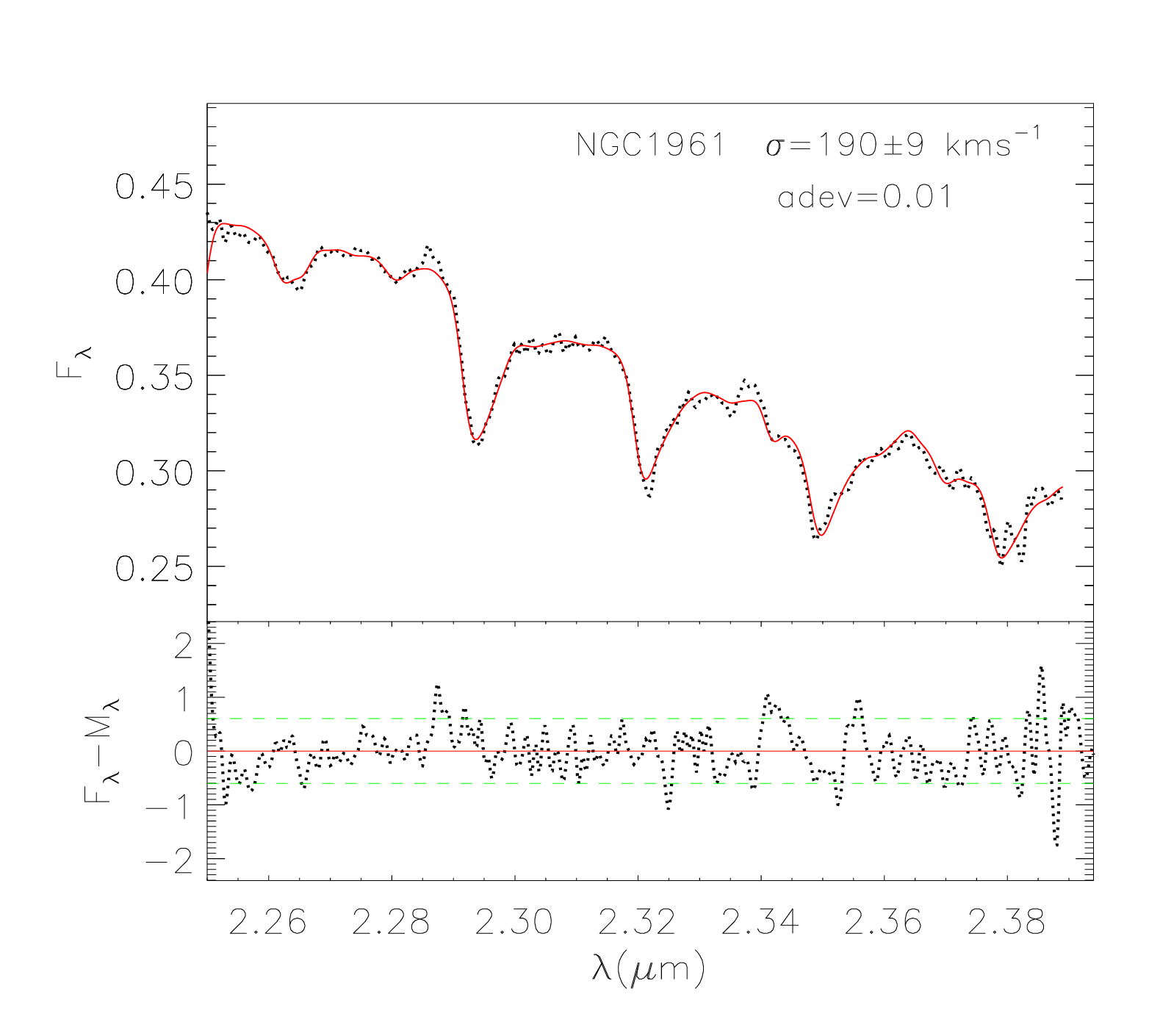}&    
    \includegraphics[scale=0.2]{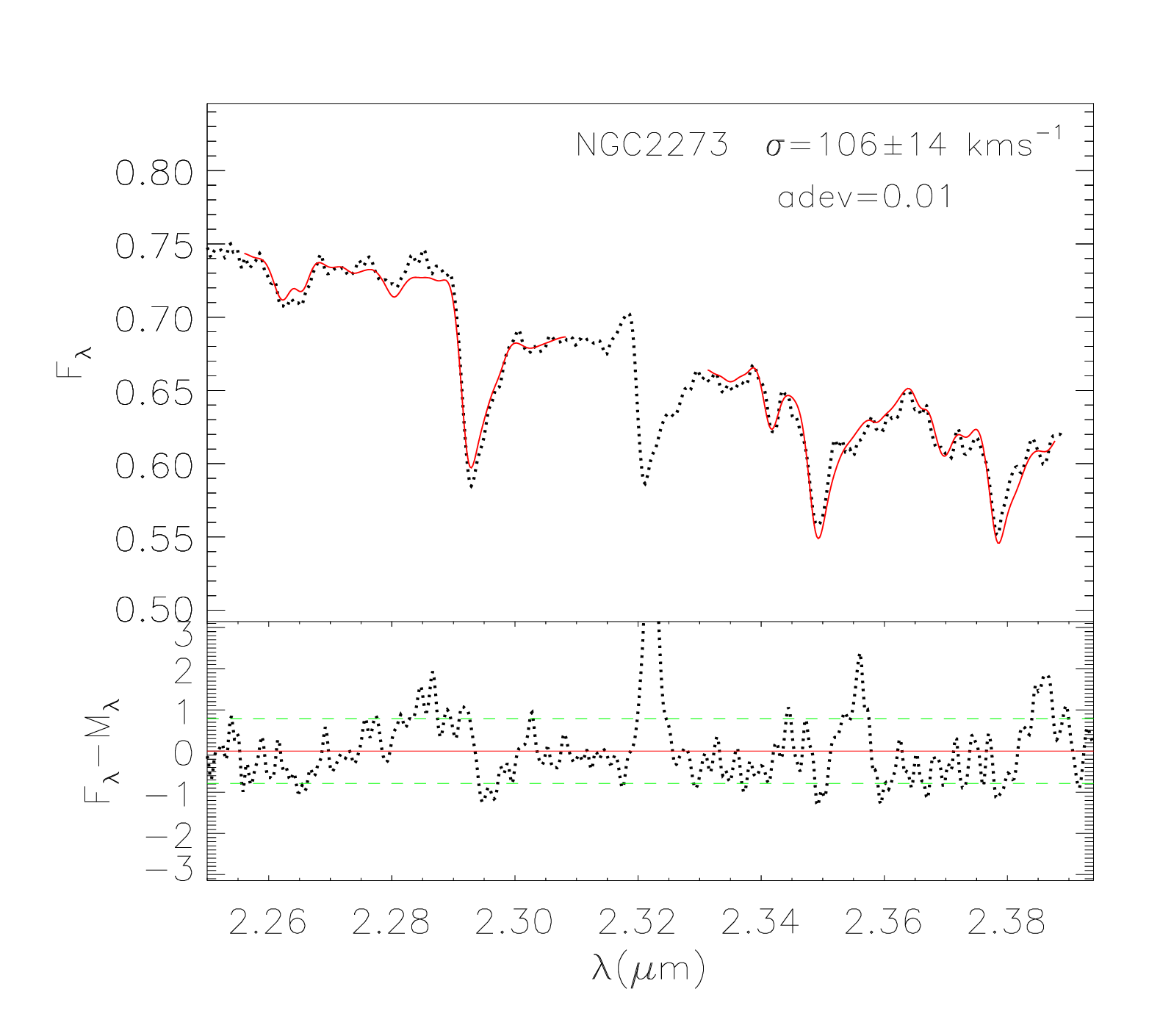} \\

   \includegraphics[scale=0.2]{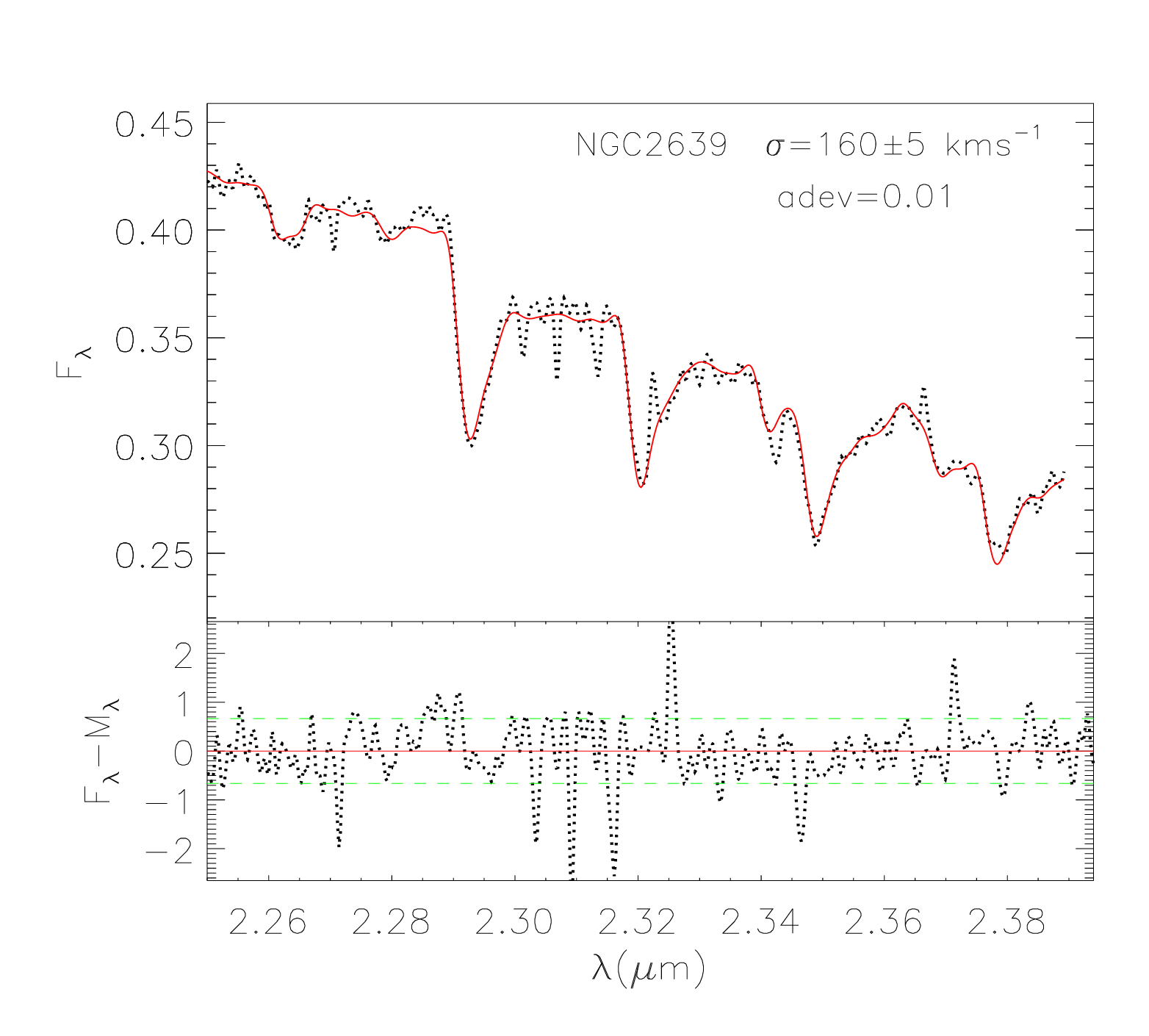}&
    \includegraphics[scale=0.2]{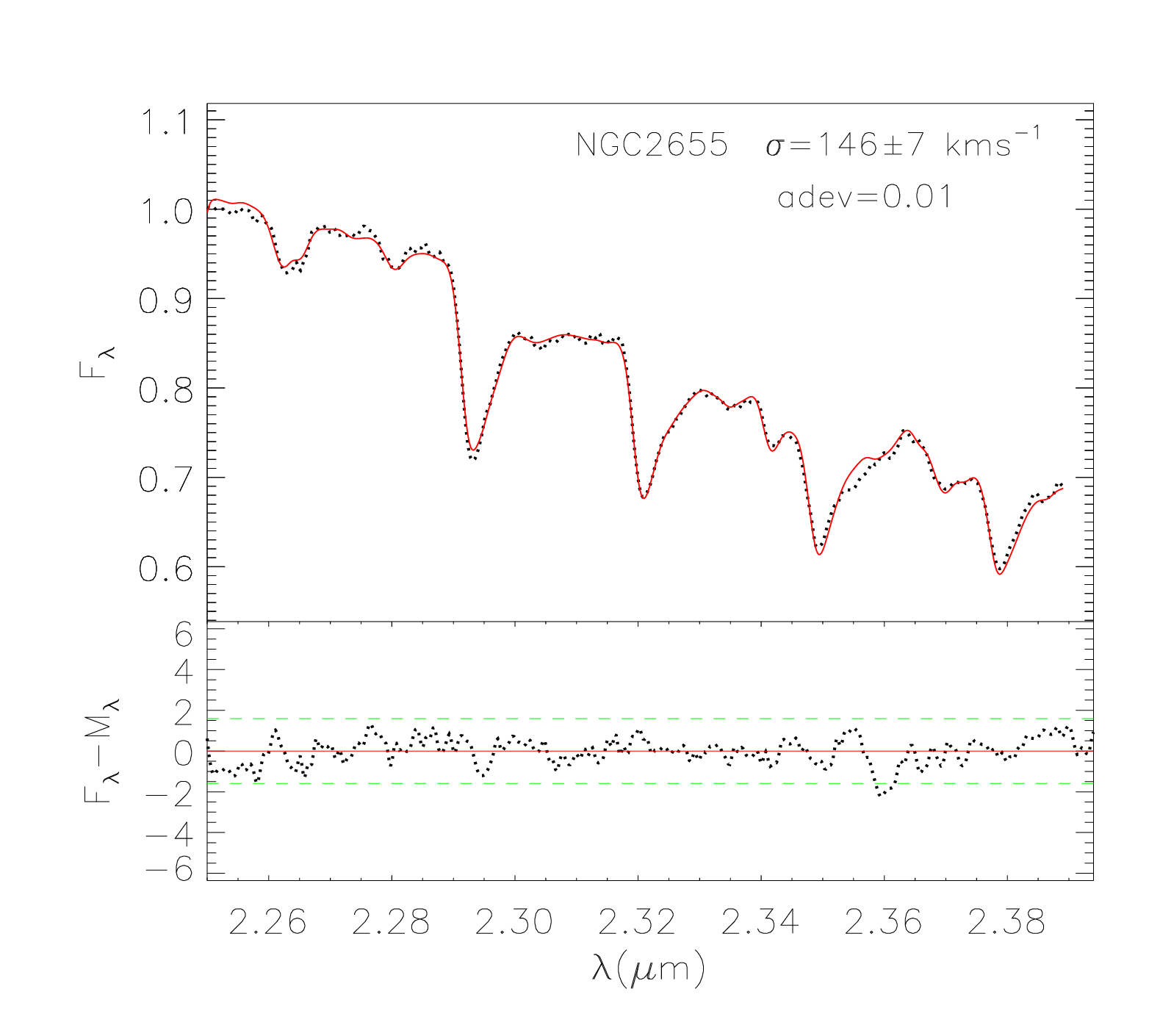}&
    \includegraphics[scale=0.2]{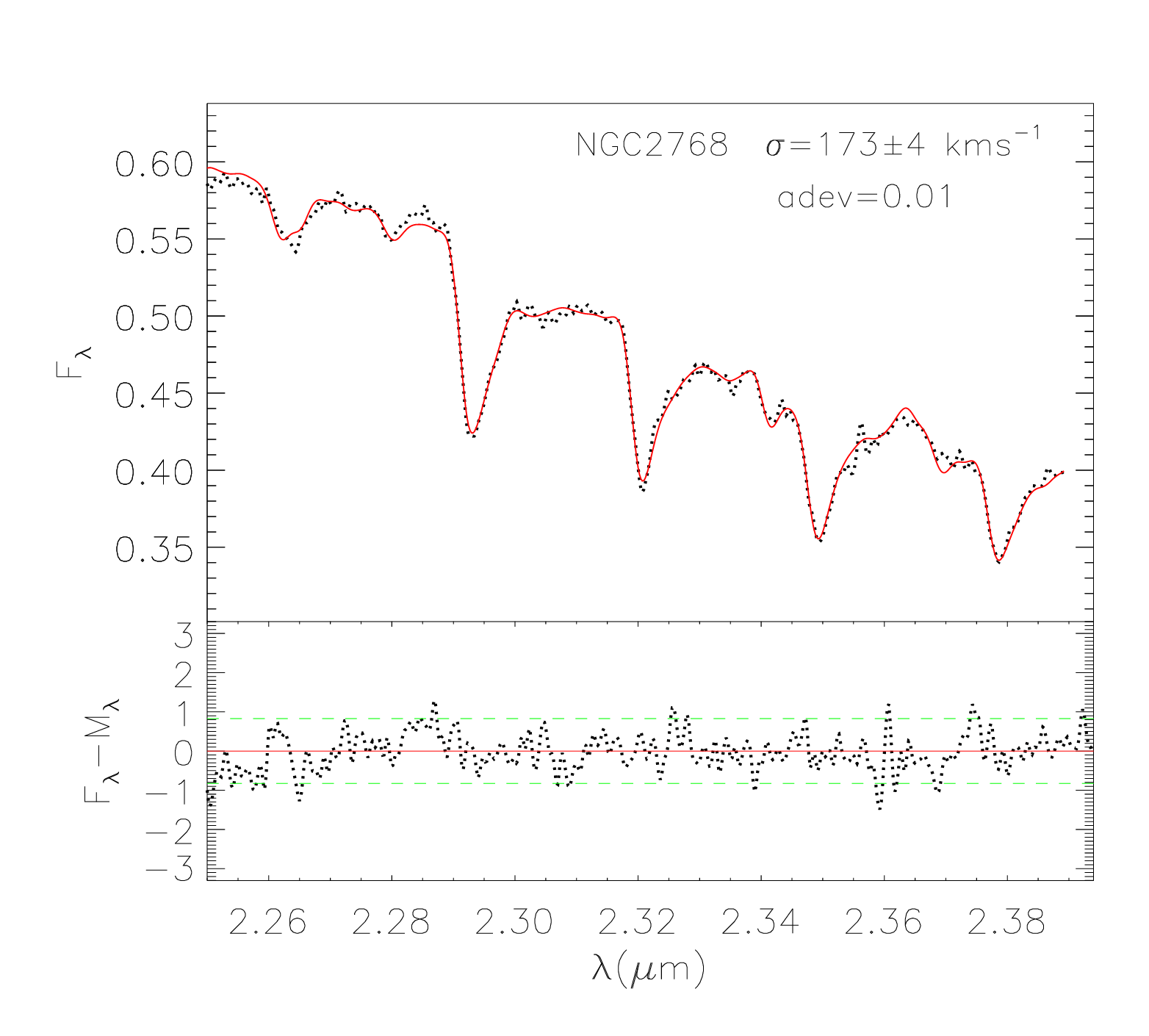}&    
    \includegraphics[scale=0.2]{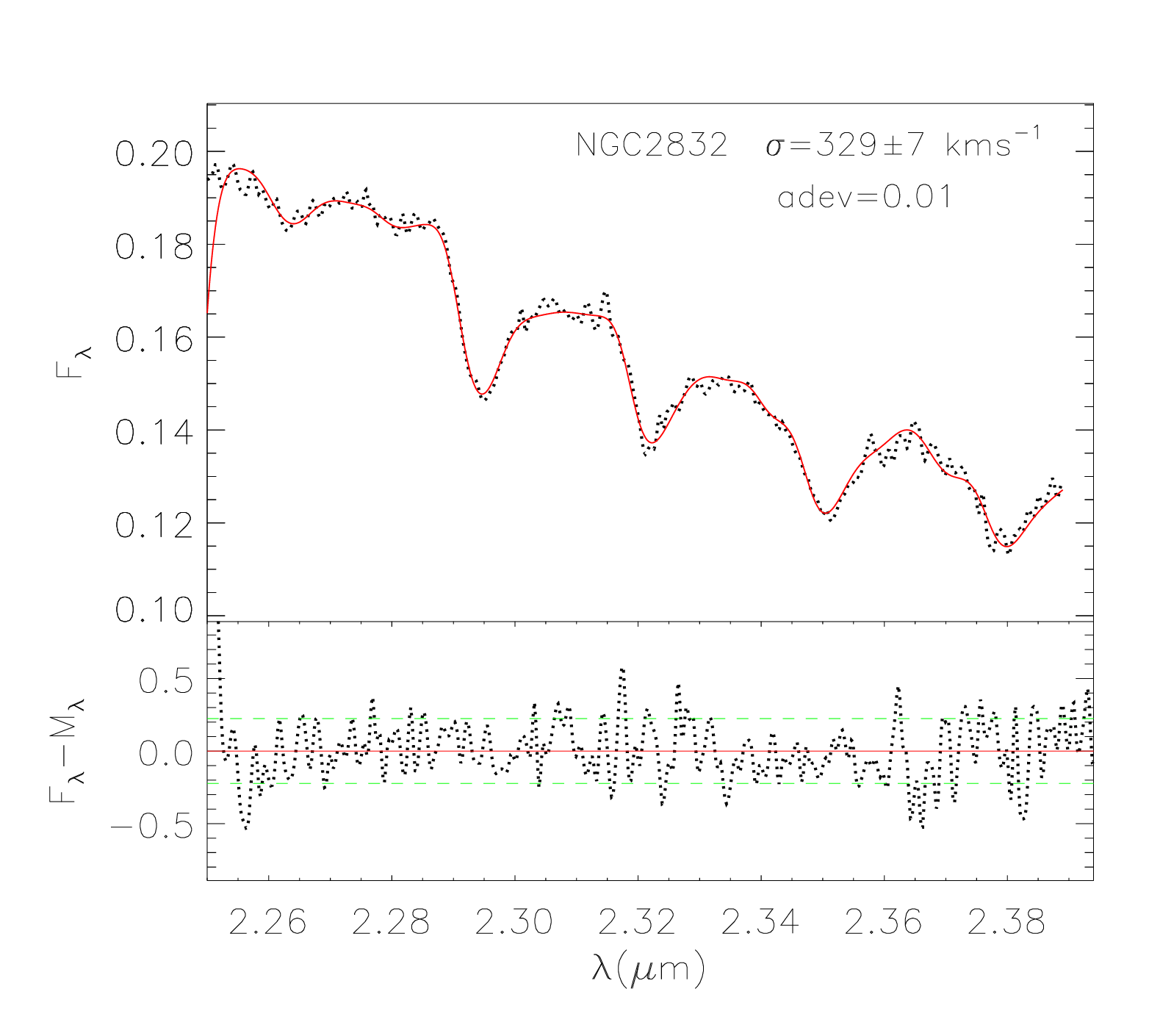} \\

    \includegraphics[scale=0.2]{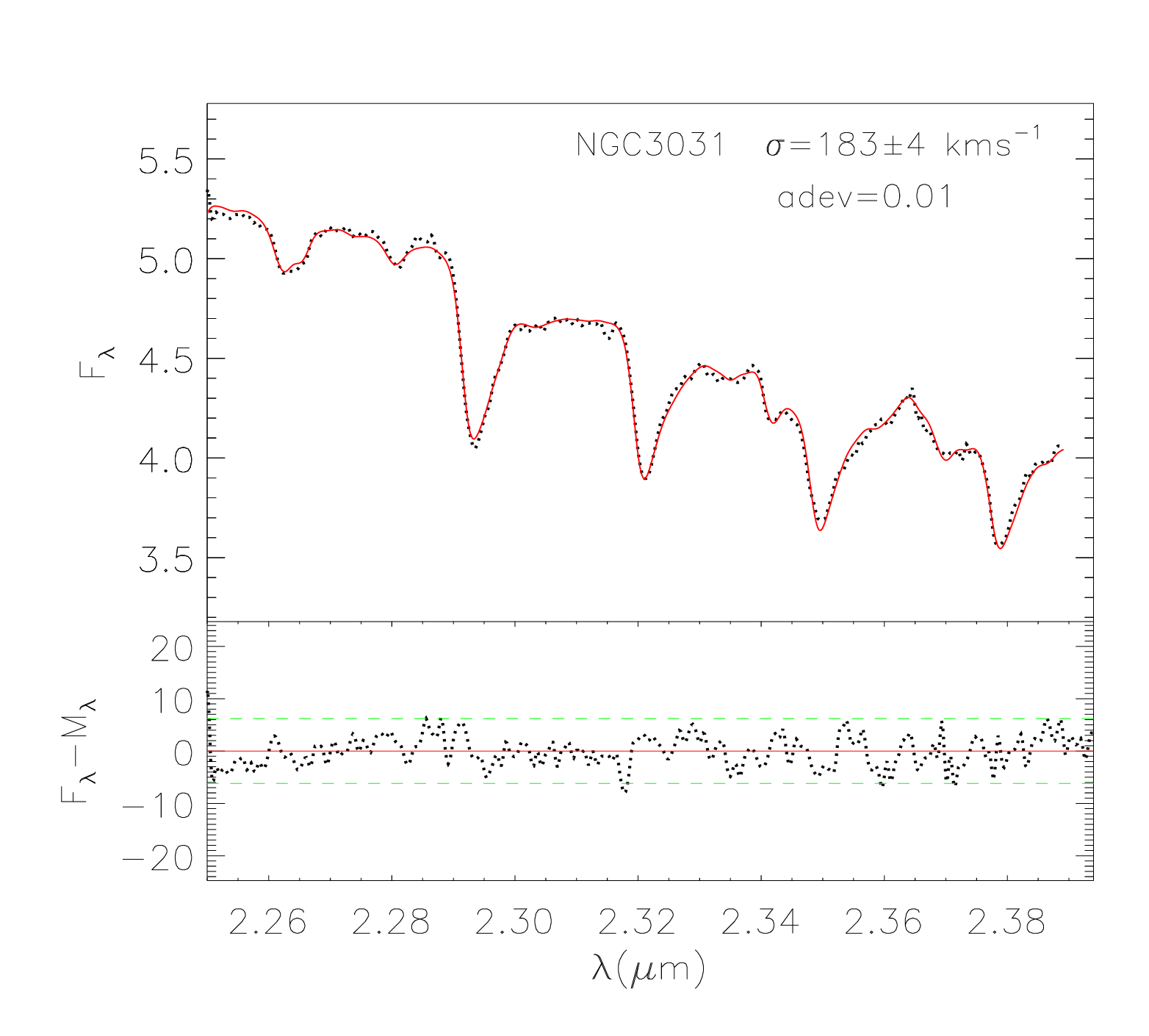}&
    \includegraphics[scale=0.2]{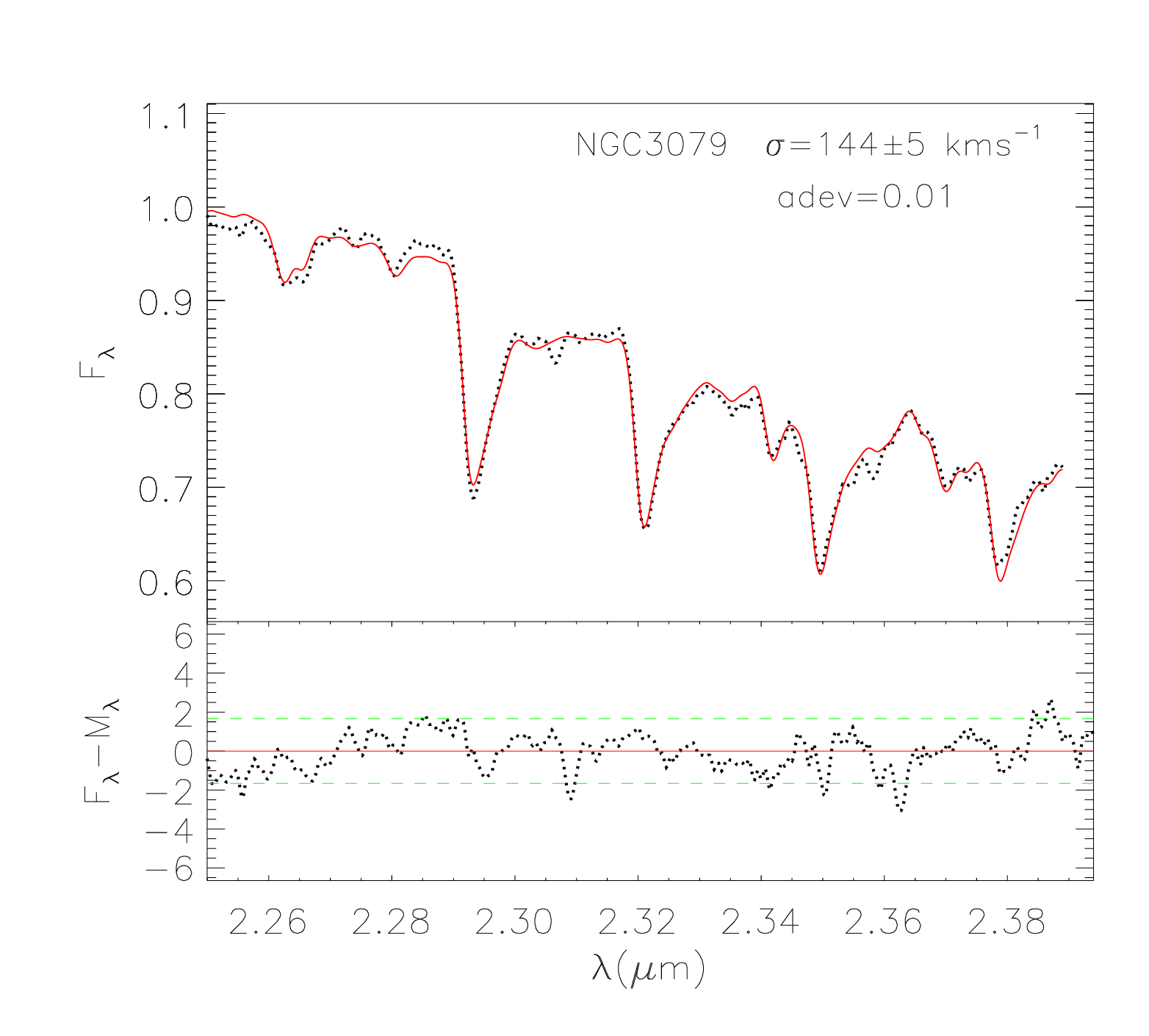}&
    \includegraphics[scale=0.2]{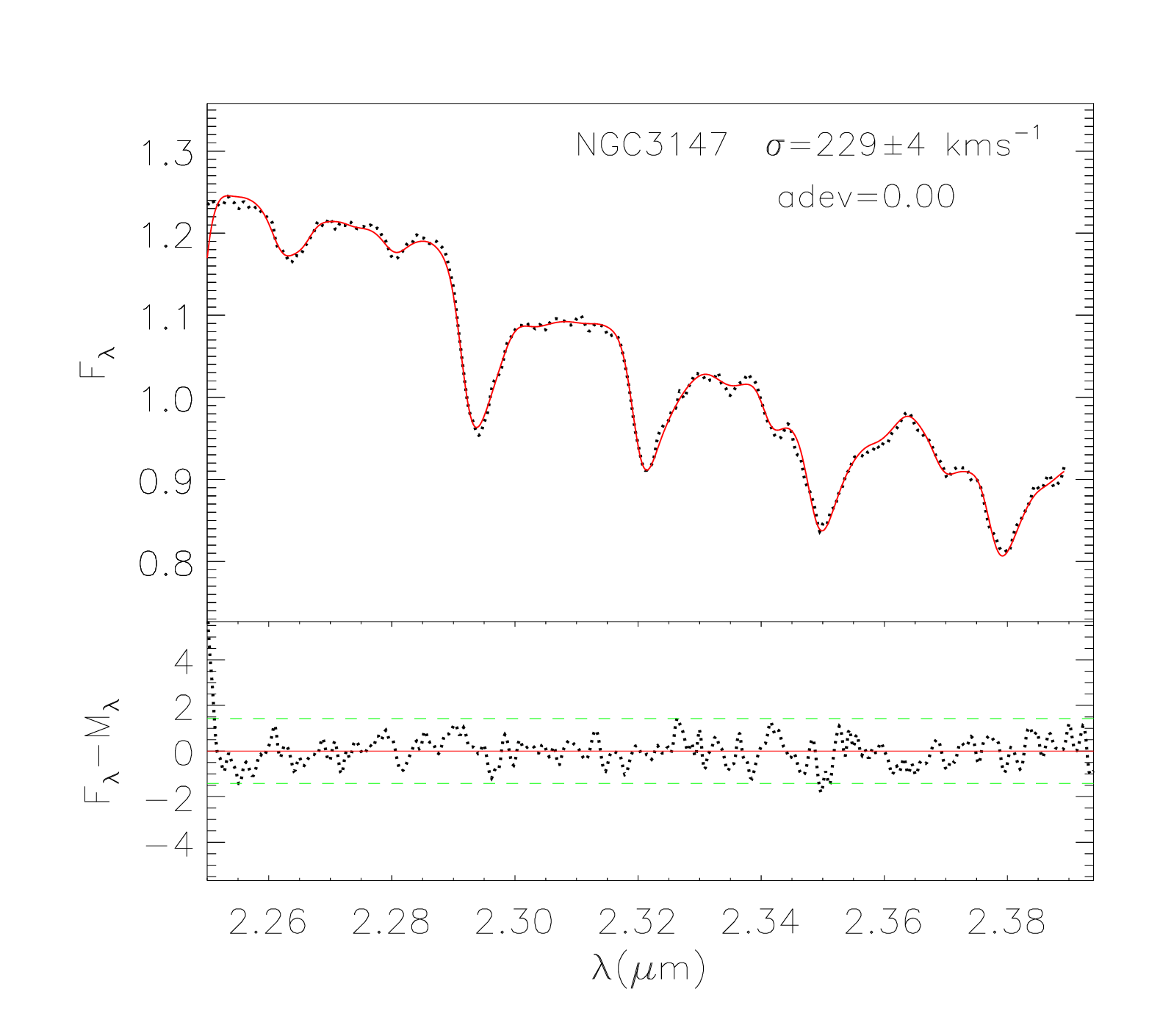}&    
    \includegraphics[scale=0.2]{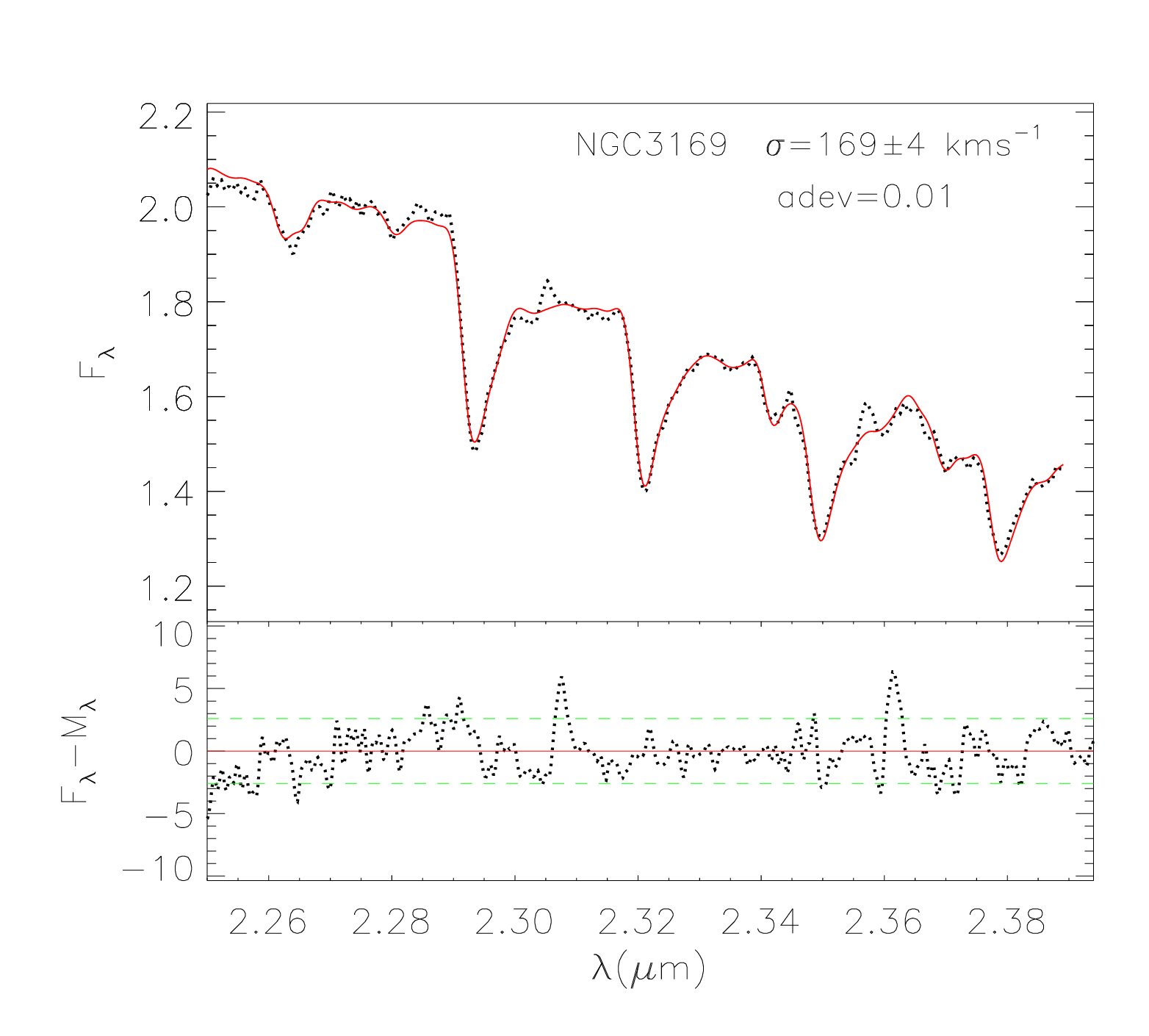} \\

    \includegraphics[scale=0.2]{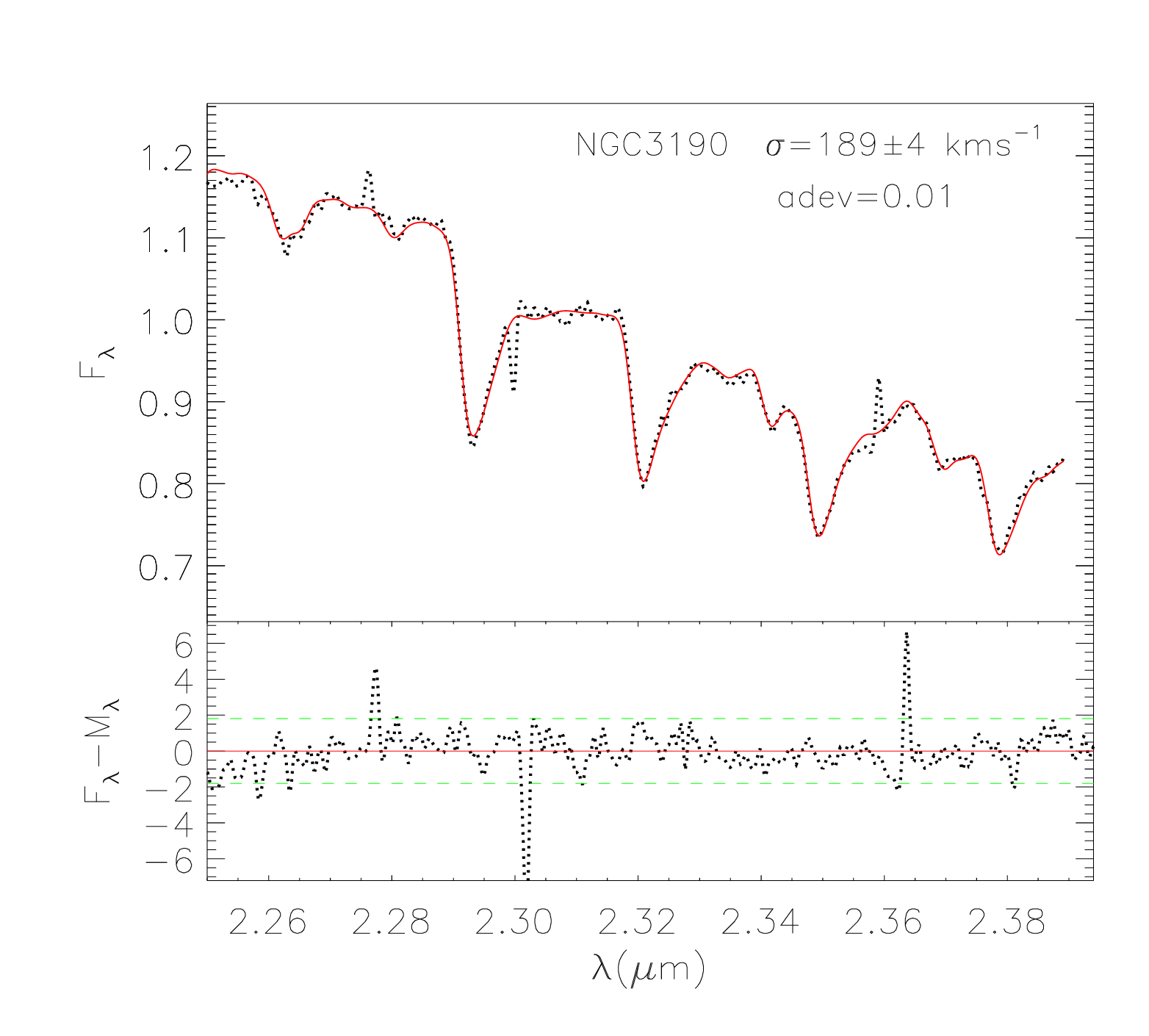}&
    \includegraphics[scale=0.2]{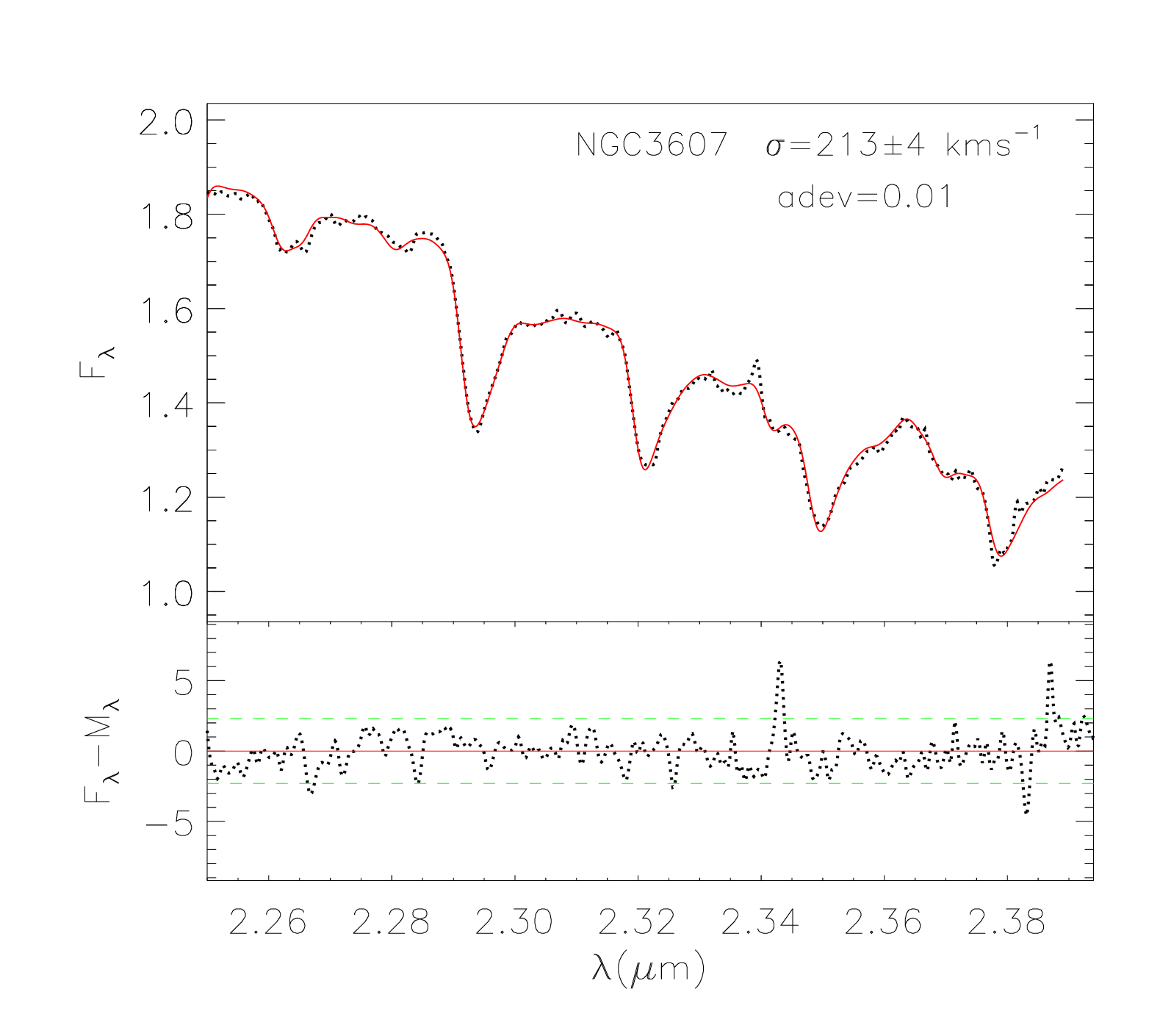}&
    \includegraphics[scale=0.2]{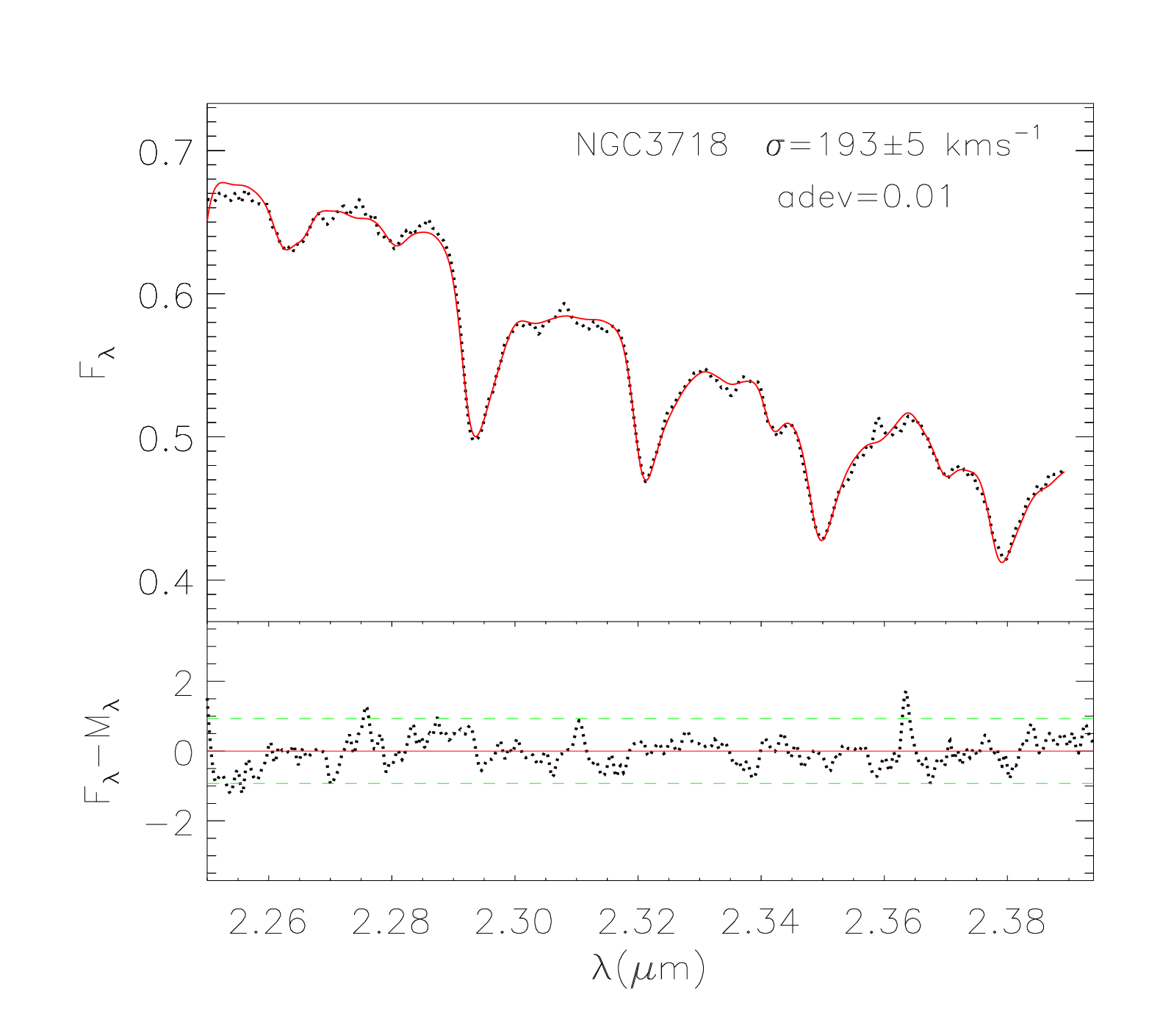}&    
    \includegraphics[scale=0.2]{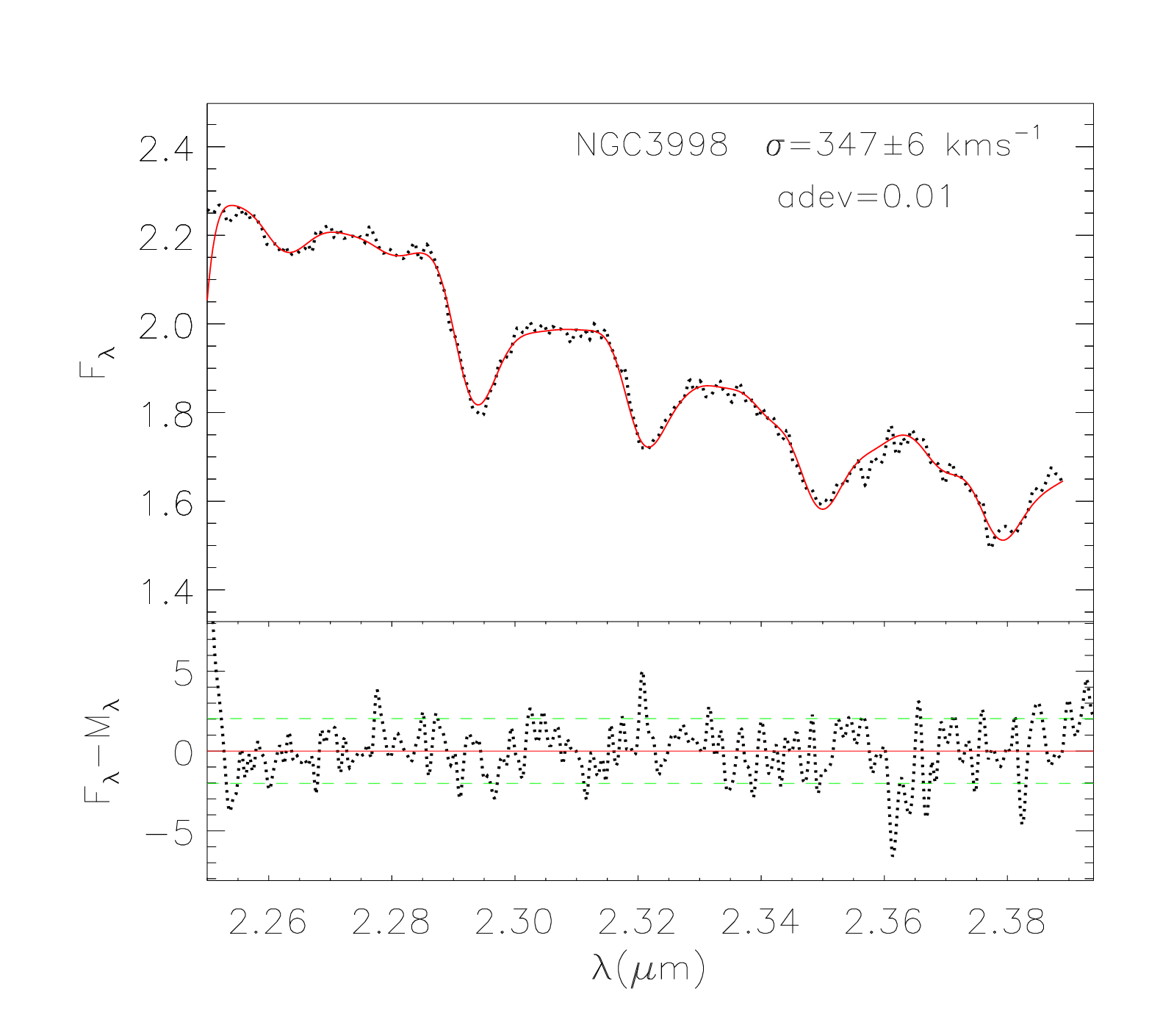} \\

  \end{tabular}
  \caption{Fits of the CO absorption band-heads in the K band. The observed spectra are shown as dotted lines and the best-fit model as continuous lines. In the bottom panels the residuals of the fits (observed $-$ model) are shown as dotted lines, where the dashed lines show the $1-\sigma$ level of the continuum. Fluxes are shown in units of ${ \rm 10^{-15}\,erg\,s^{-1}\,cm^{-2}\,\AA^{-1}}$ and the residuals in units of ${ \rm 10^{-17}\,erg\,s^{-1}\,cm^{-2}\,\AA^{-1}}$. The $adev$ parameter  shown at each panel gives the
percentage mean $|O−M|/O$ deviation over all fitted pixels, where $O$ is the observed spectrum and $M$ is the model.}
  \label{fits-co}
\end{figure*}

\setcounter{figure}{0}
\begin{figure*}
\centering
  \begin{tabular}{cccc}

    \includegraphics[scale=0.2]{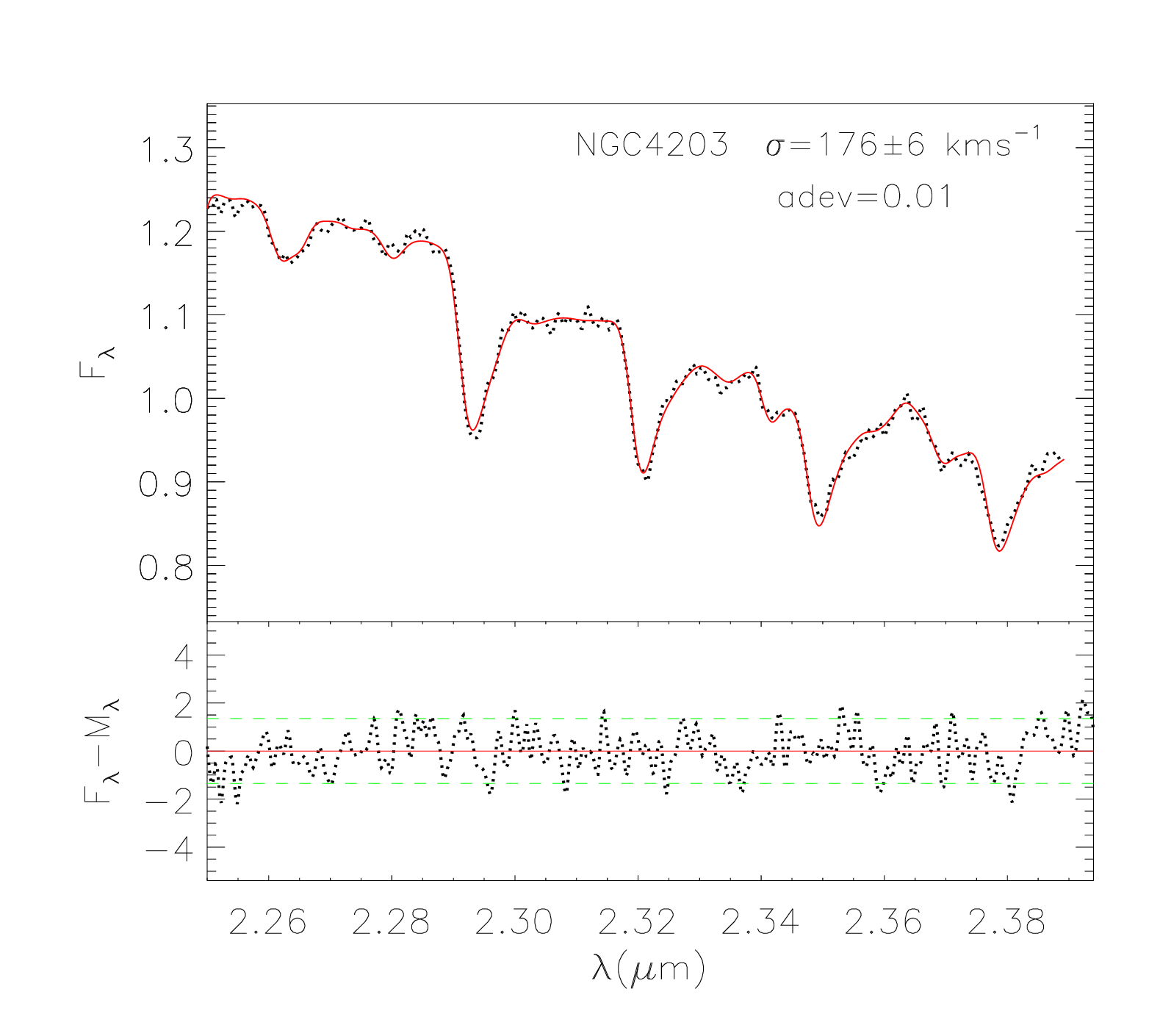}&
    \includegraphics[scale=0.2]{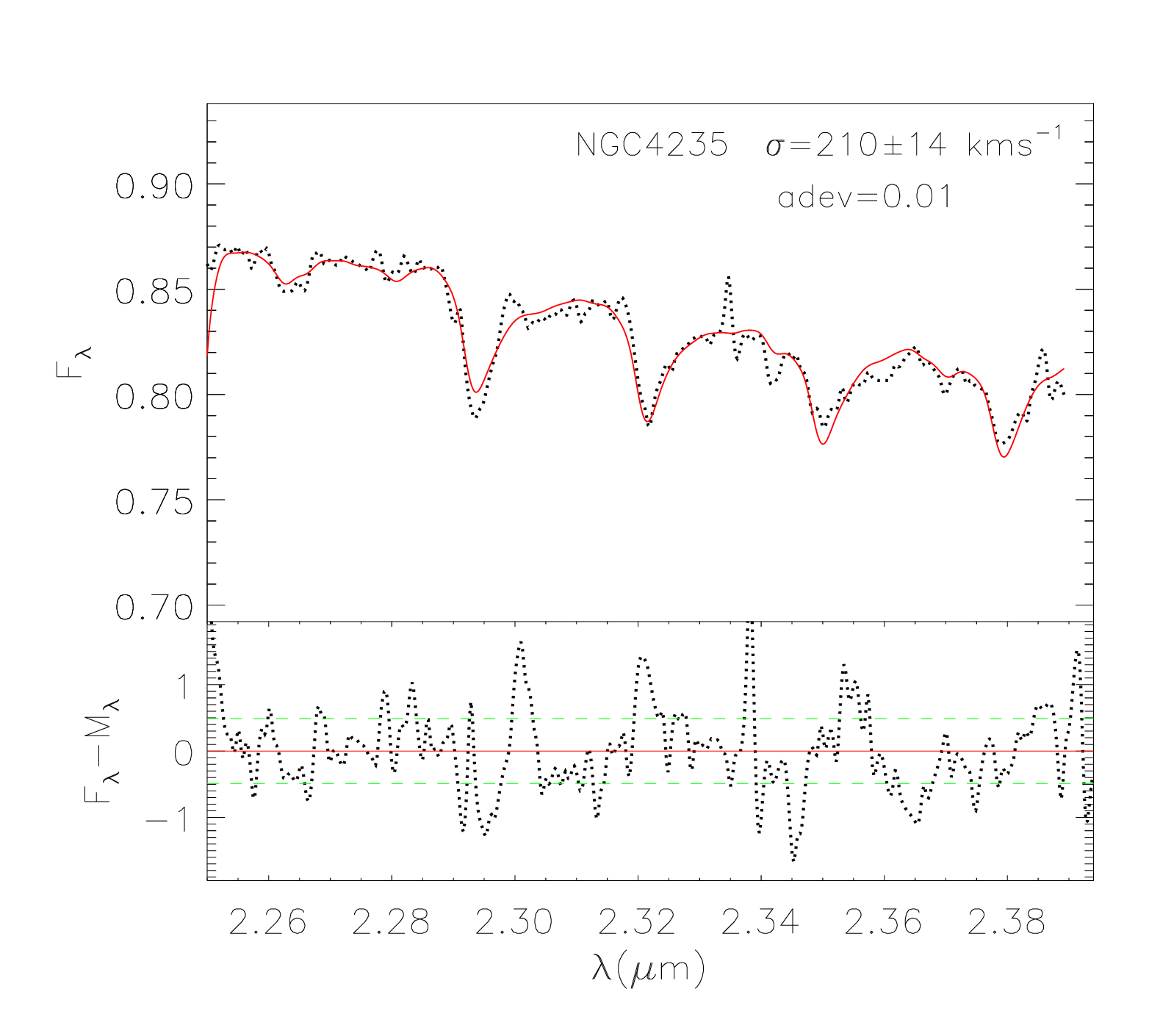}&
    \includegraphics[scale=0.2]{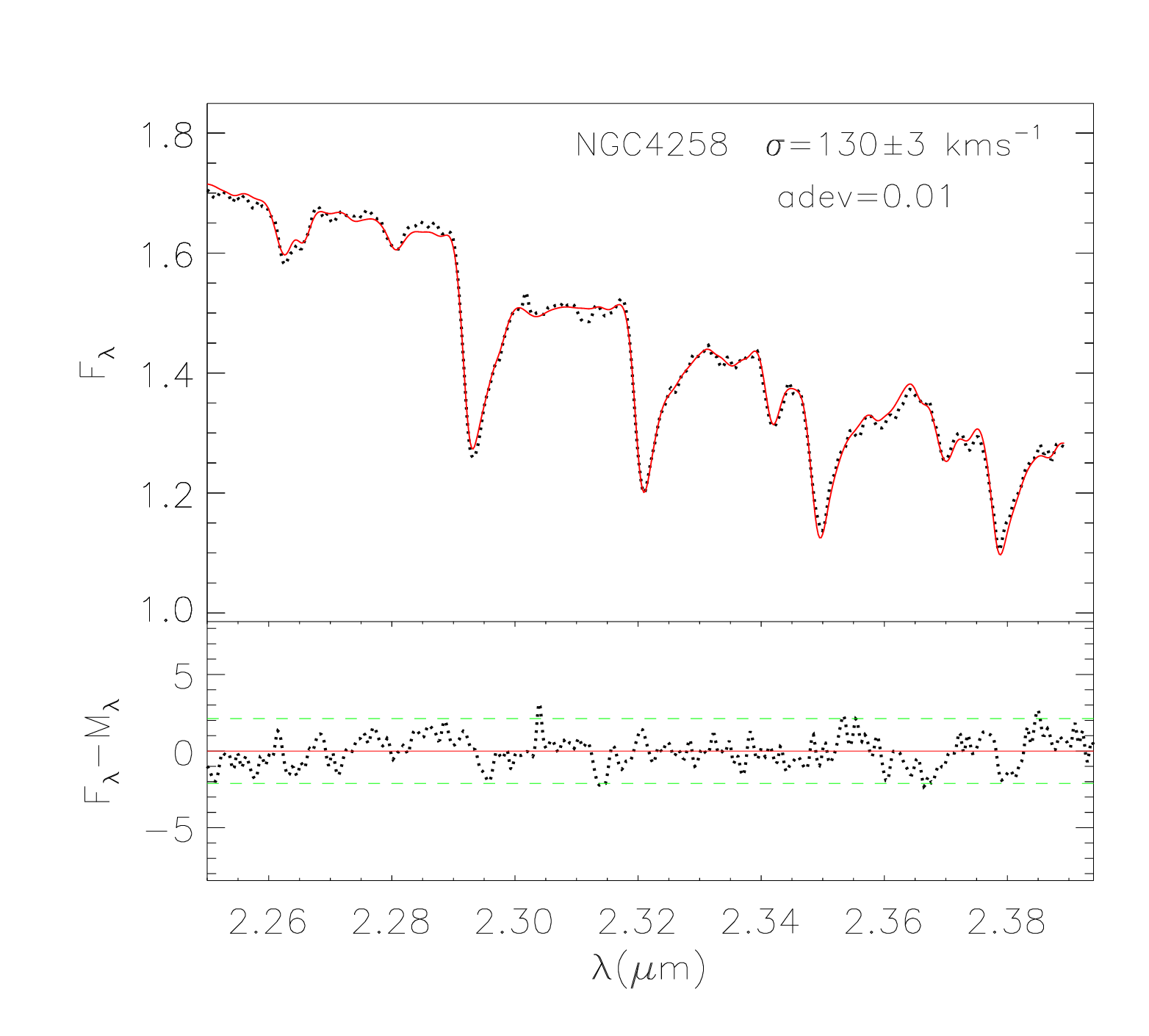}&    
    \includegraphics[scale=0.2]{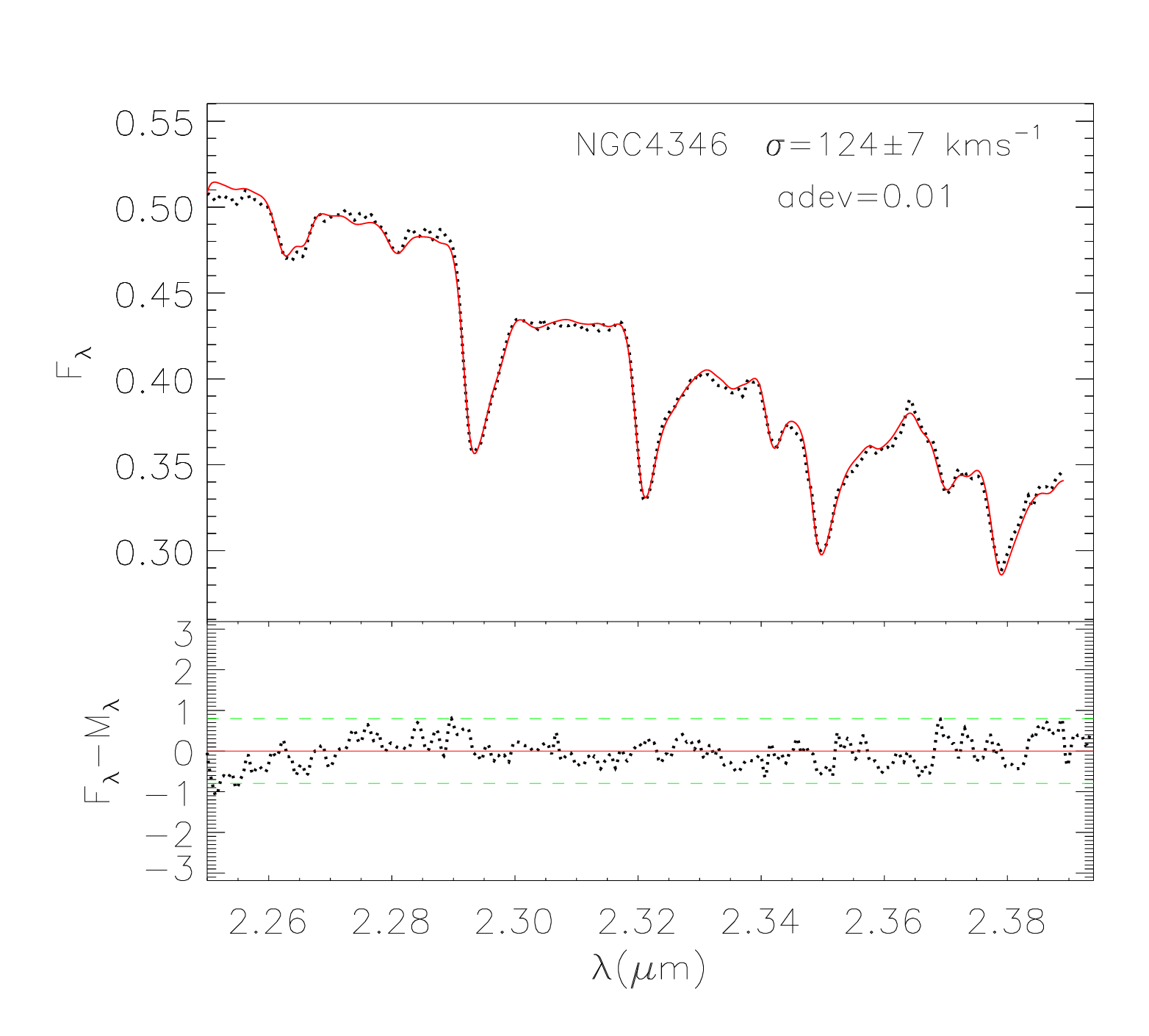} \\

    \includegraphics[scale=0.2]{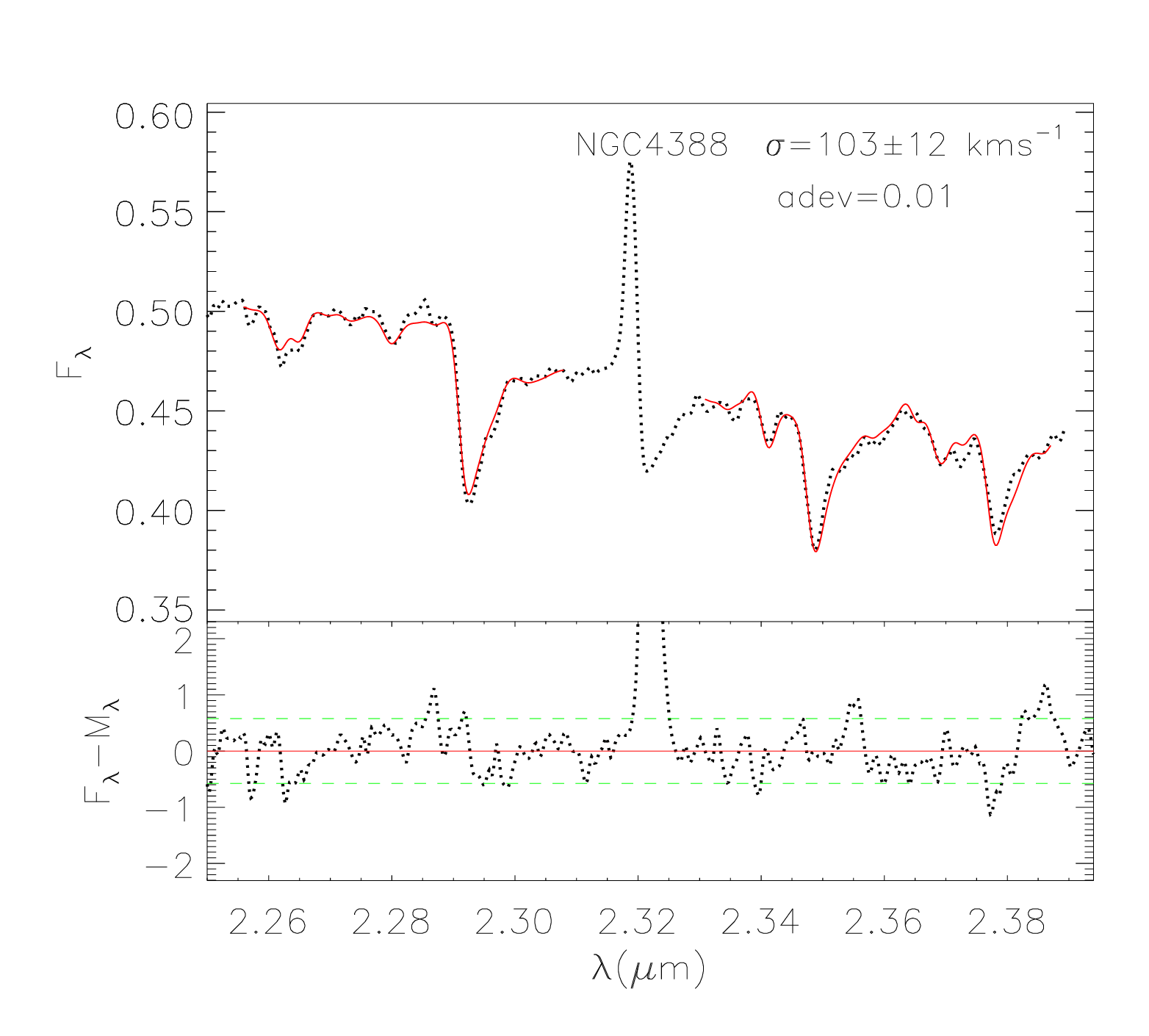}&
    \includegraphics[scale=0.2]{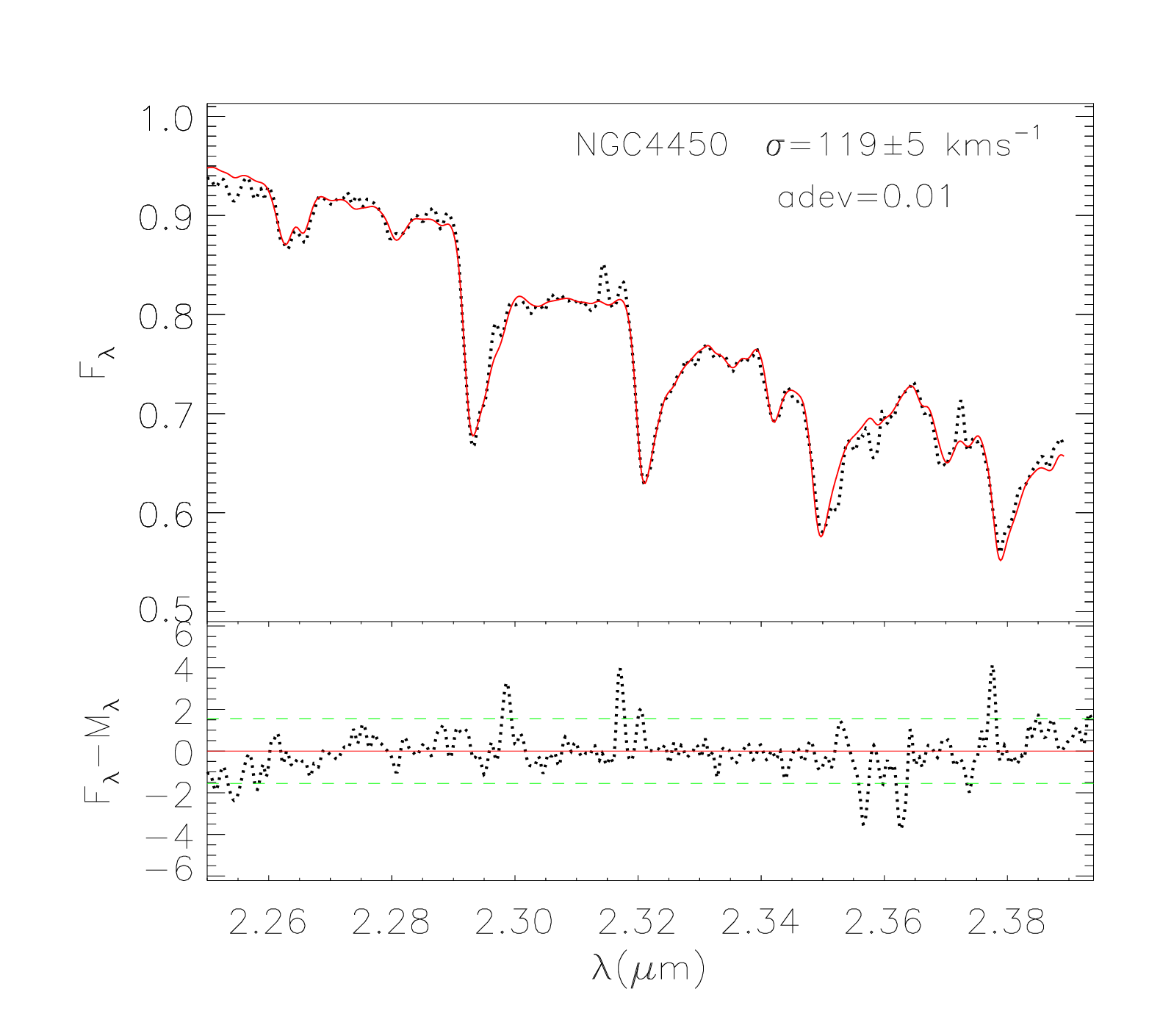}&
    \includegraphics[scale=0.2]{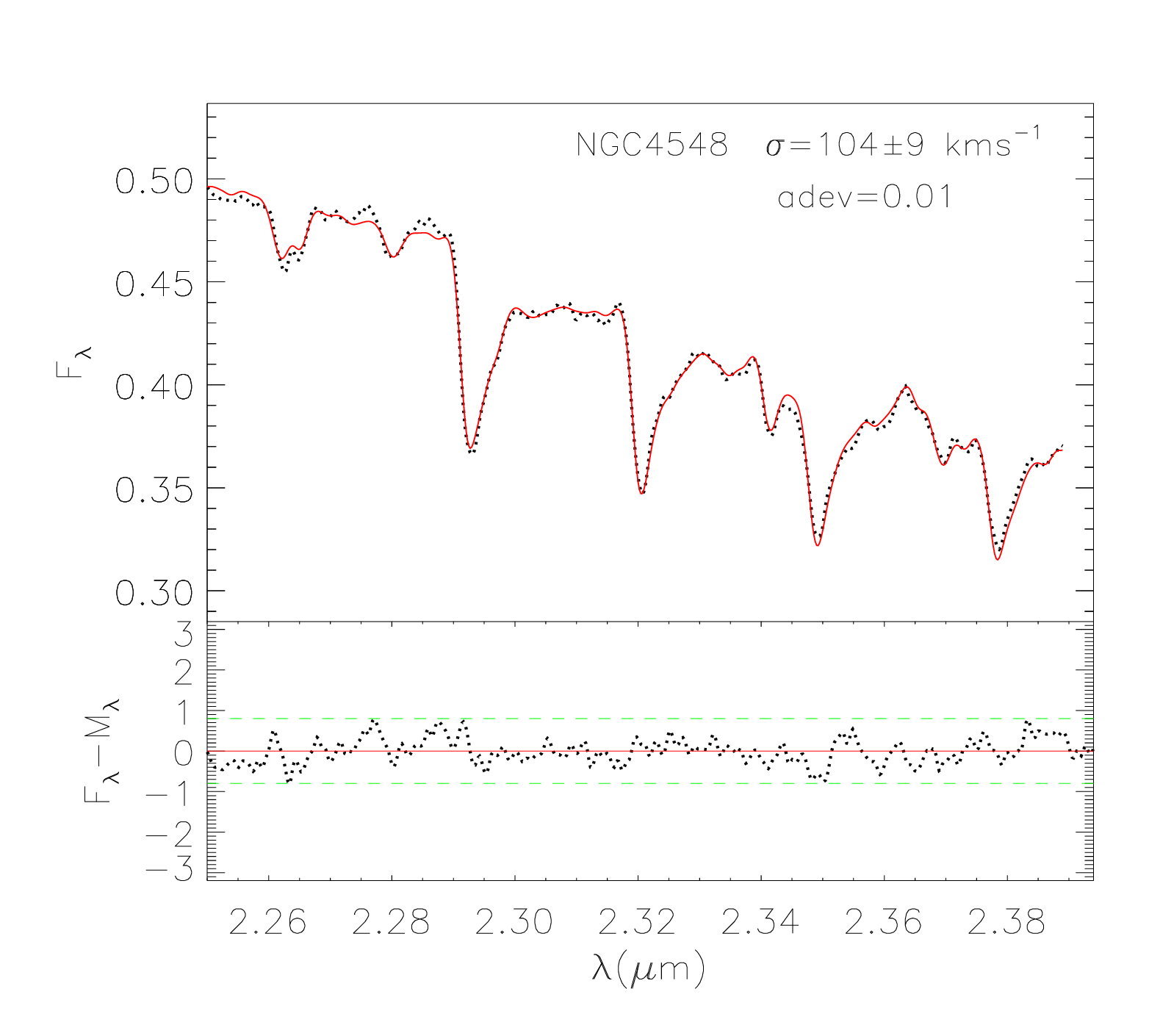}&    
    \includegraphics[scale=0.2]{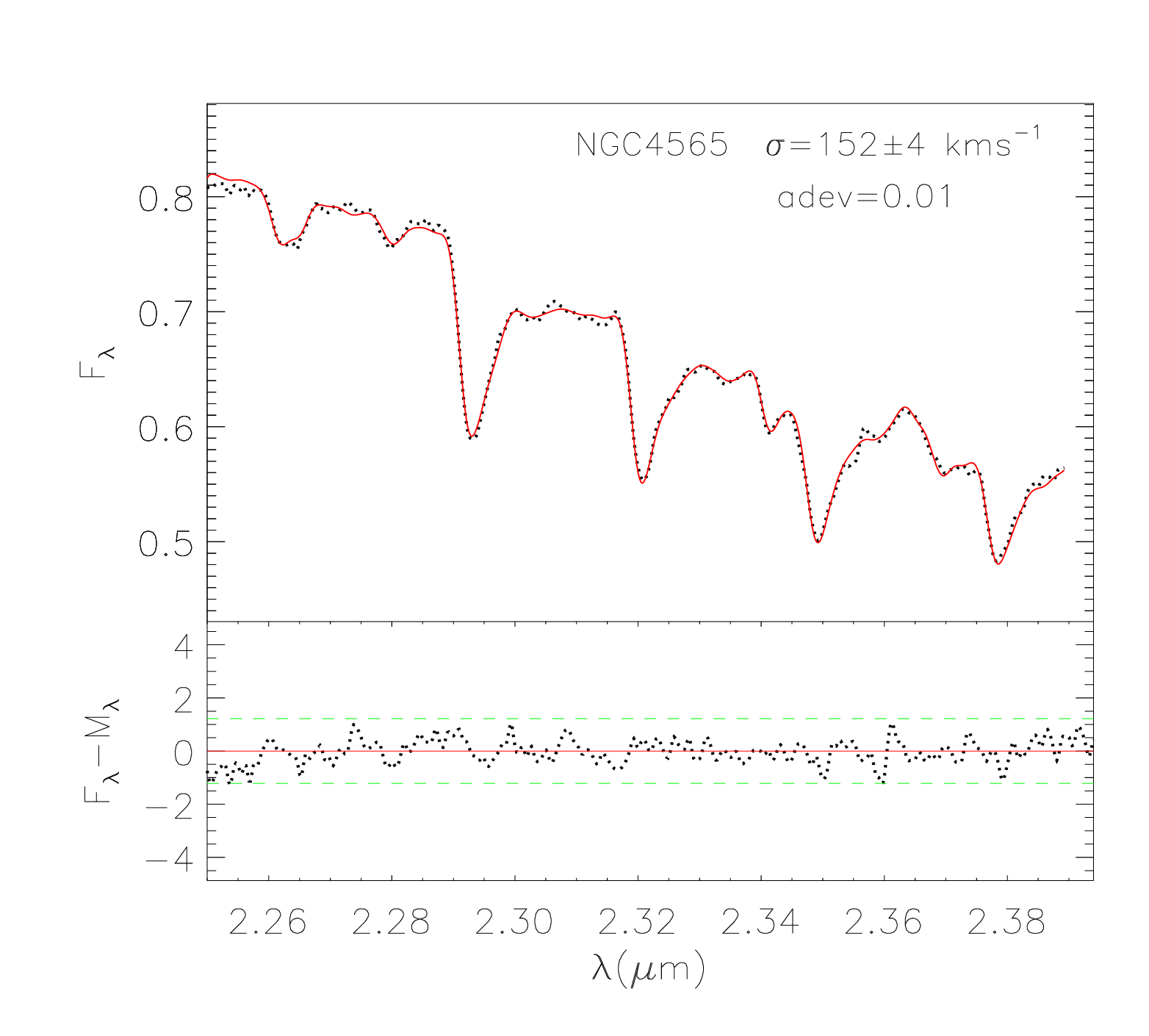} \\

    \includegraphics[scale=0.2]{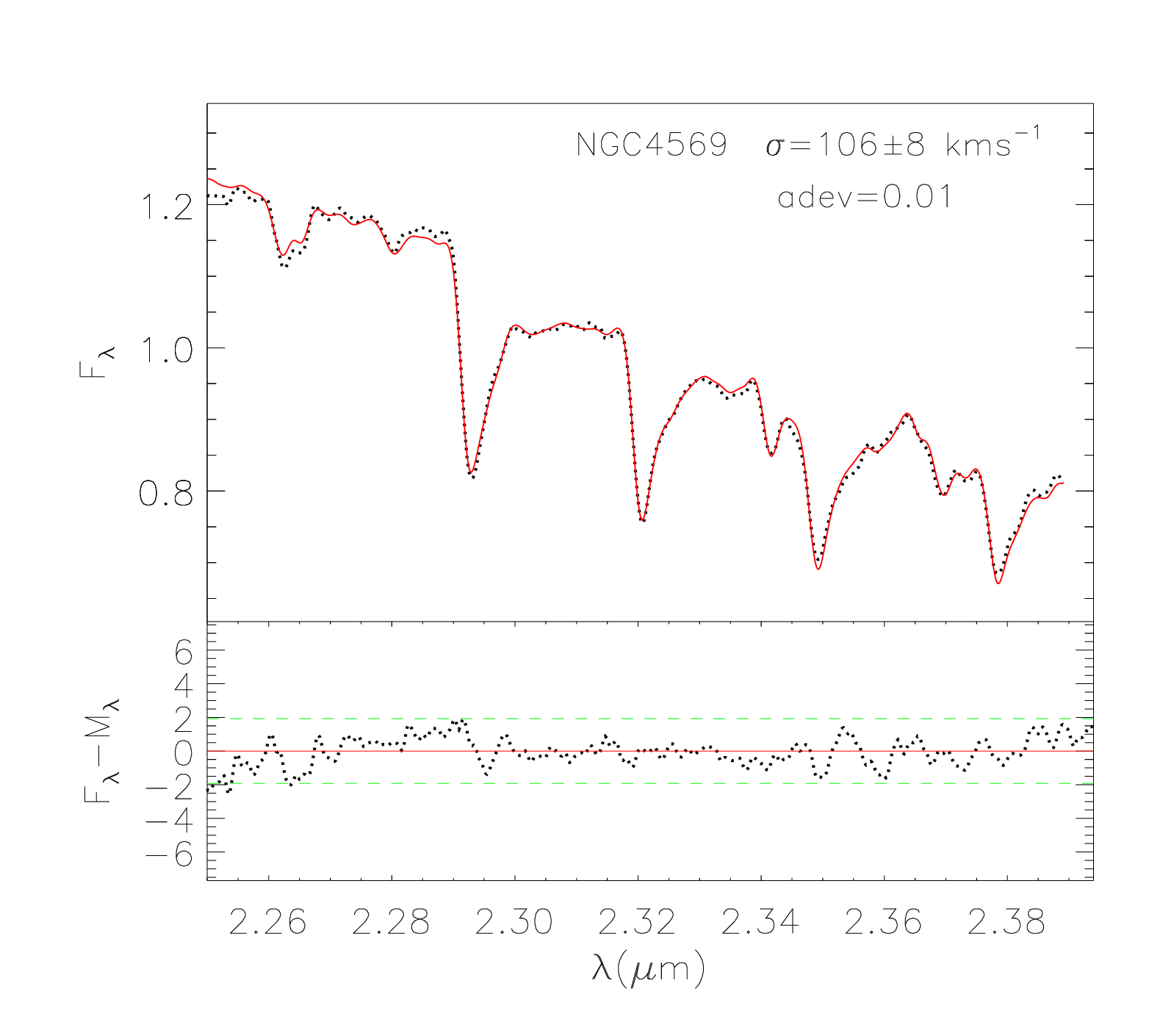}&
    \includegraphics[scale=0.2]{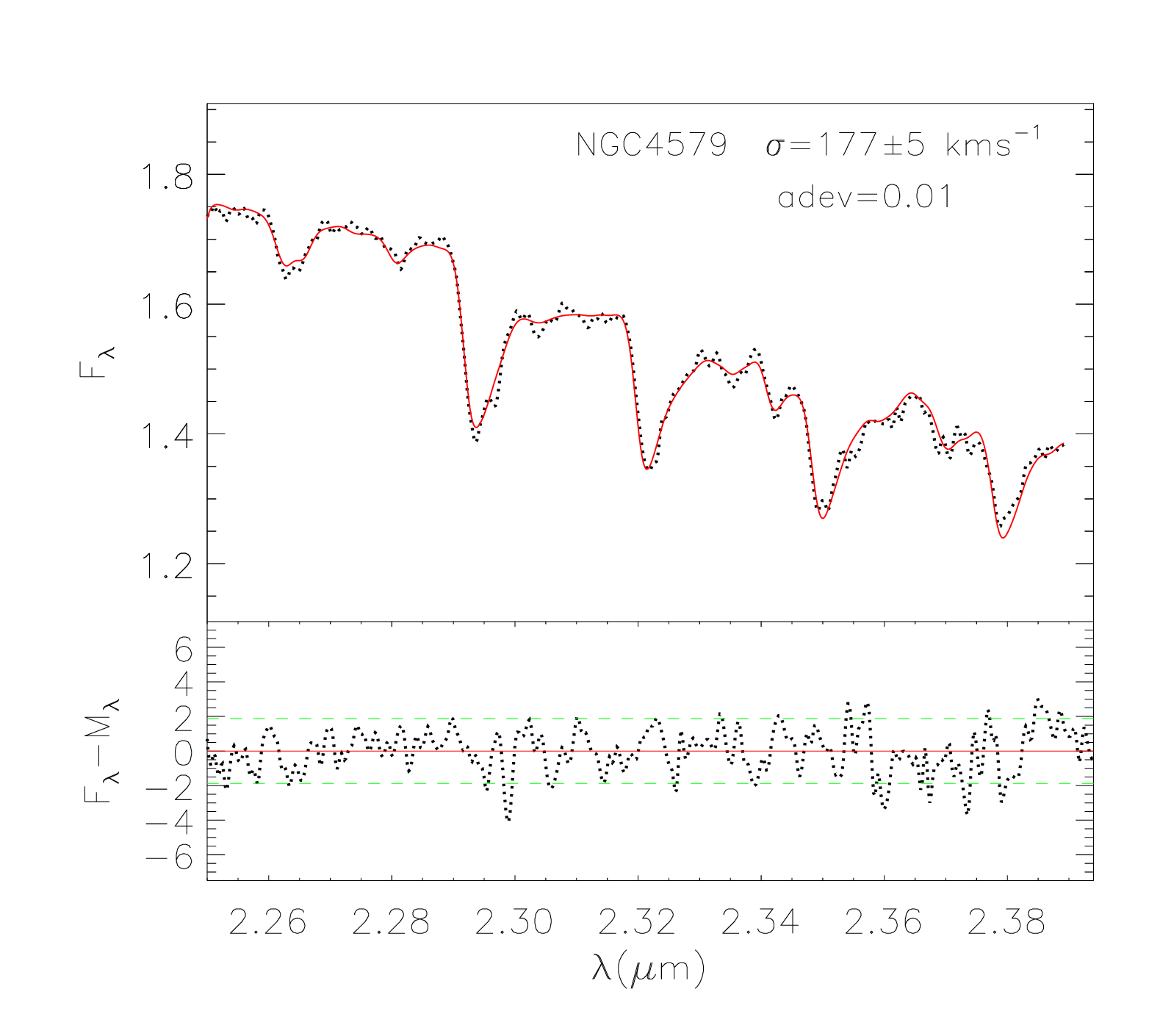}&
    \includegraphics[scale=0.2]{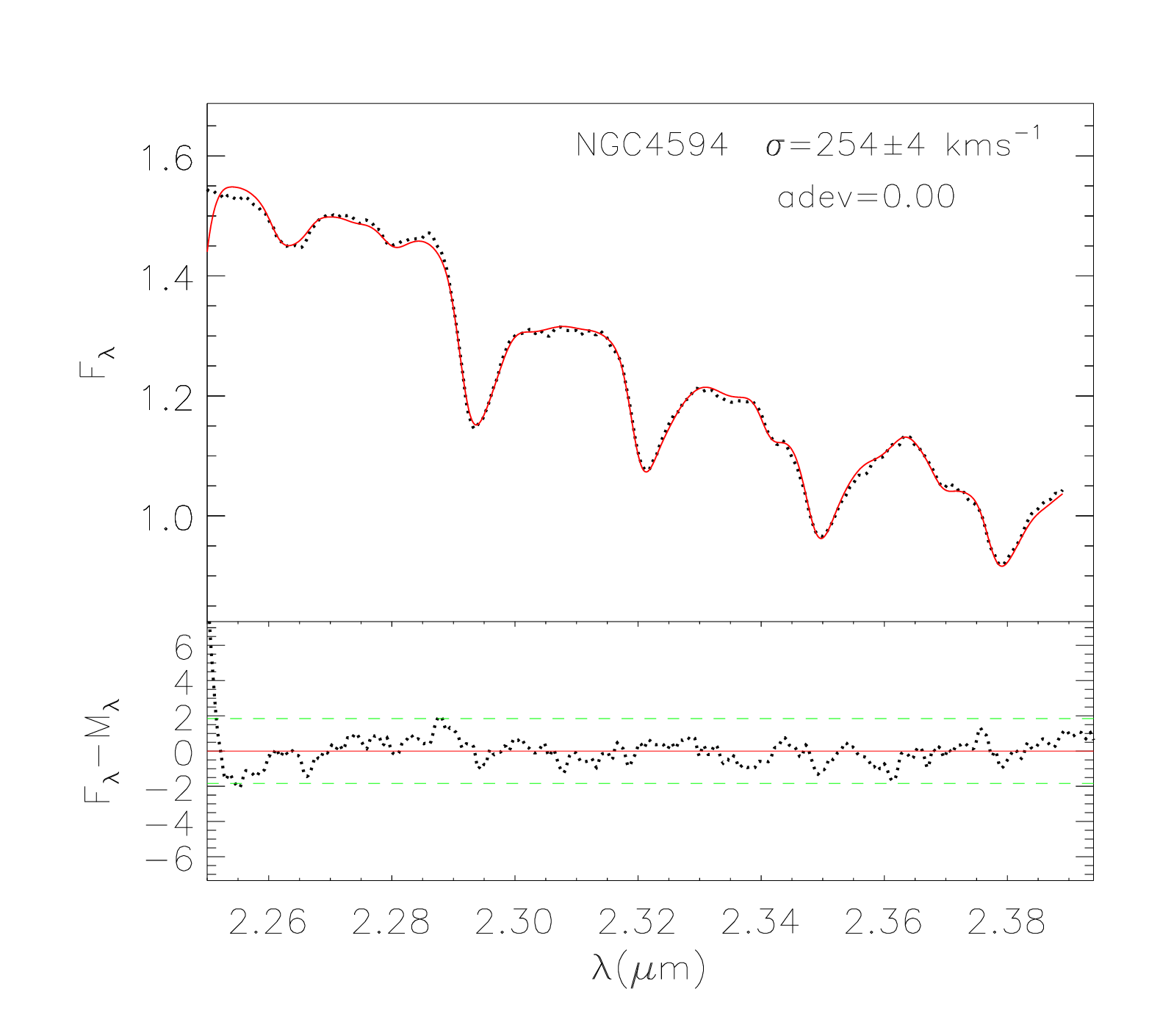}&    
    \includegraphics[scale=0.2]{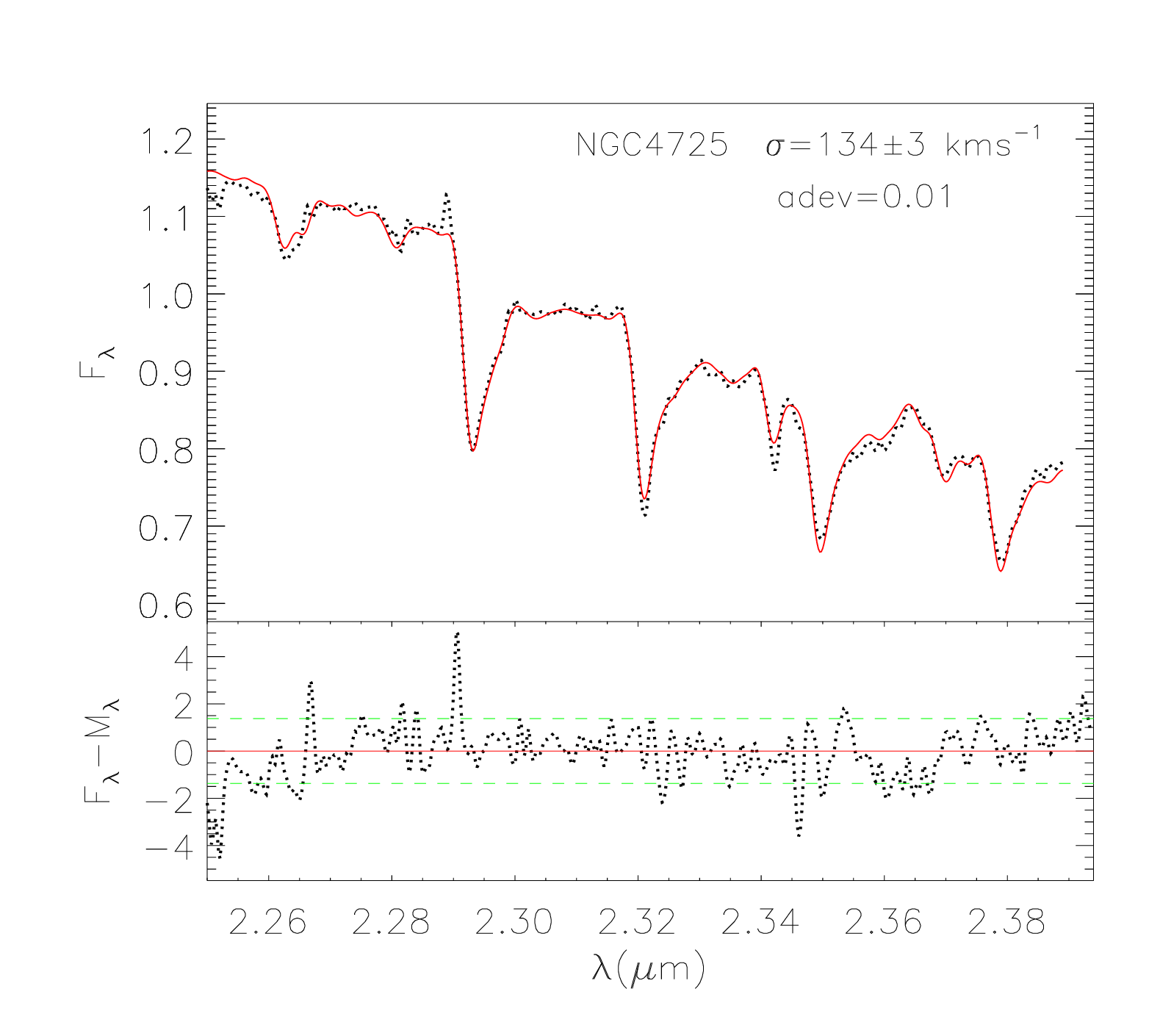} \\

    \includegraphics[scale=0.2]{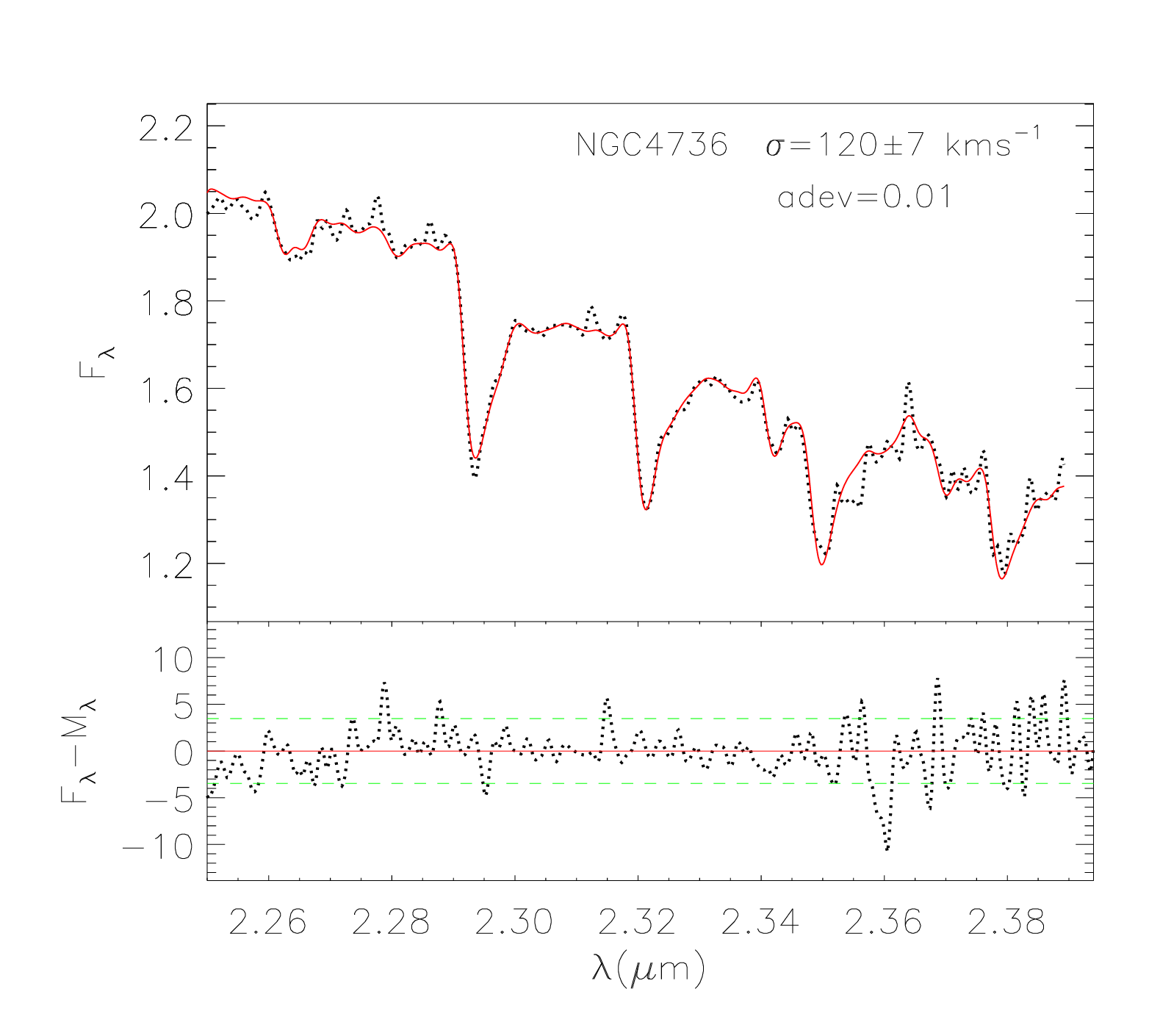}&
    \includegraphics[scale=0.2]{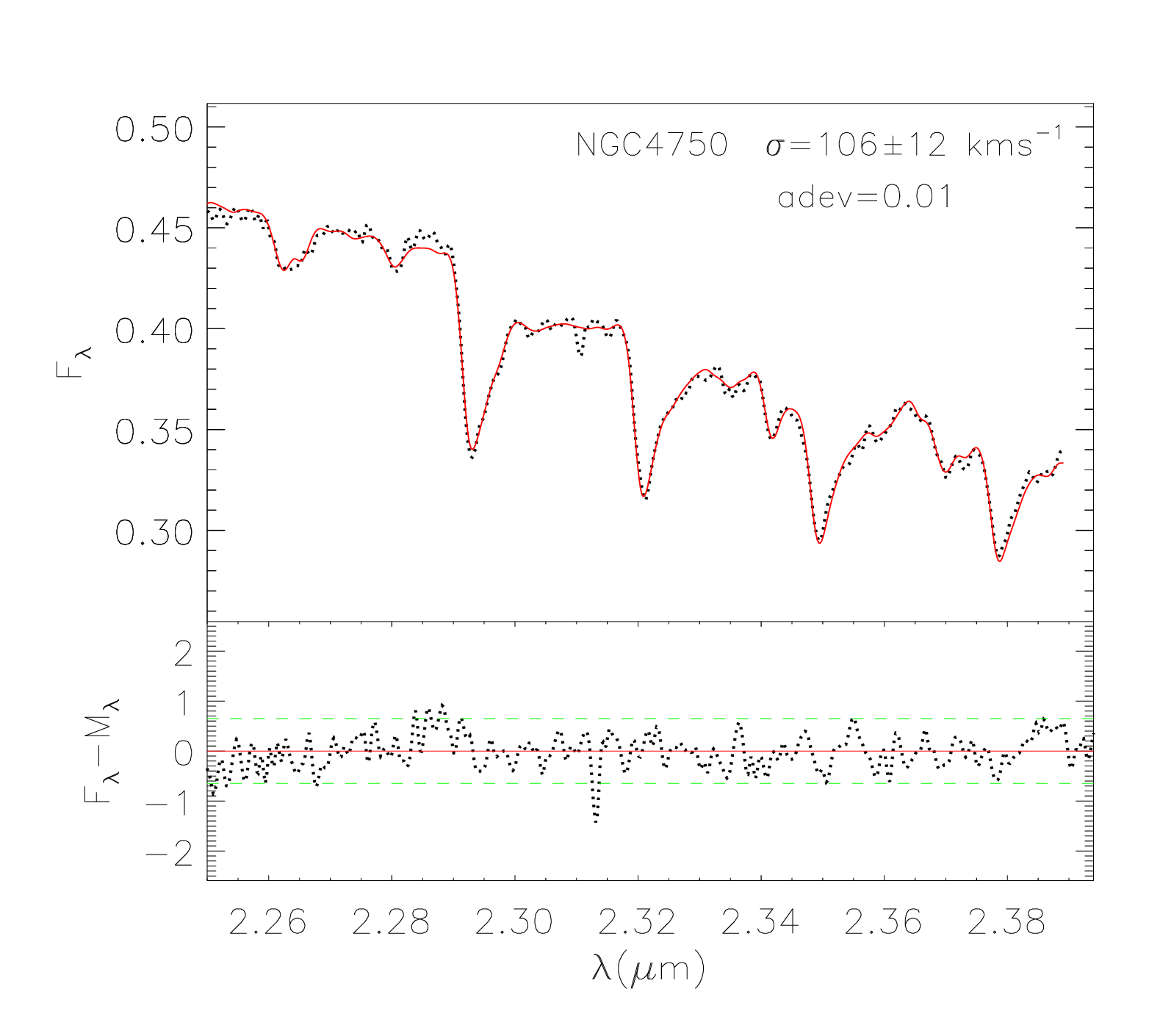}&
    \includegraphics[scale=0.2]{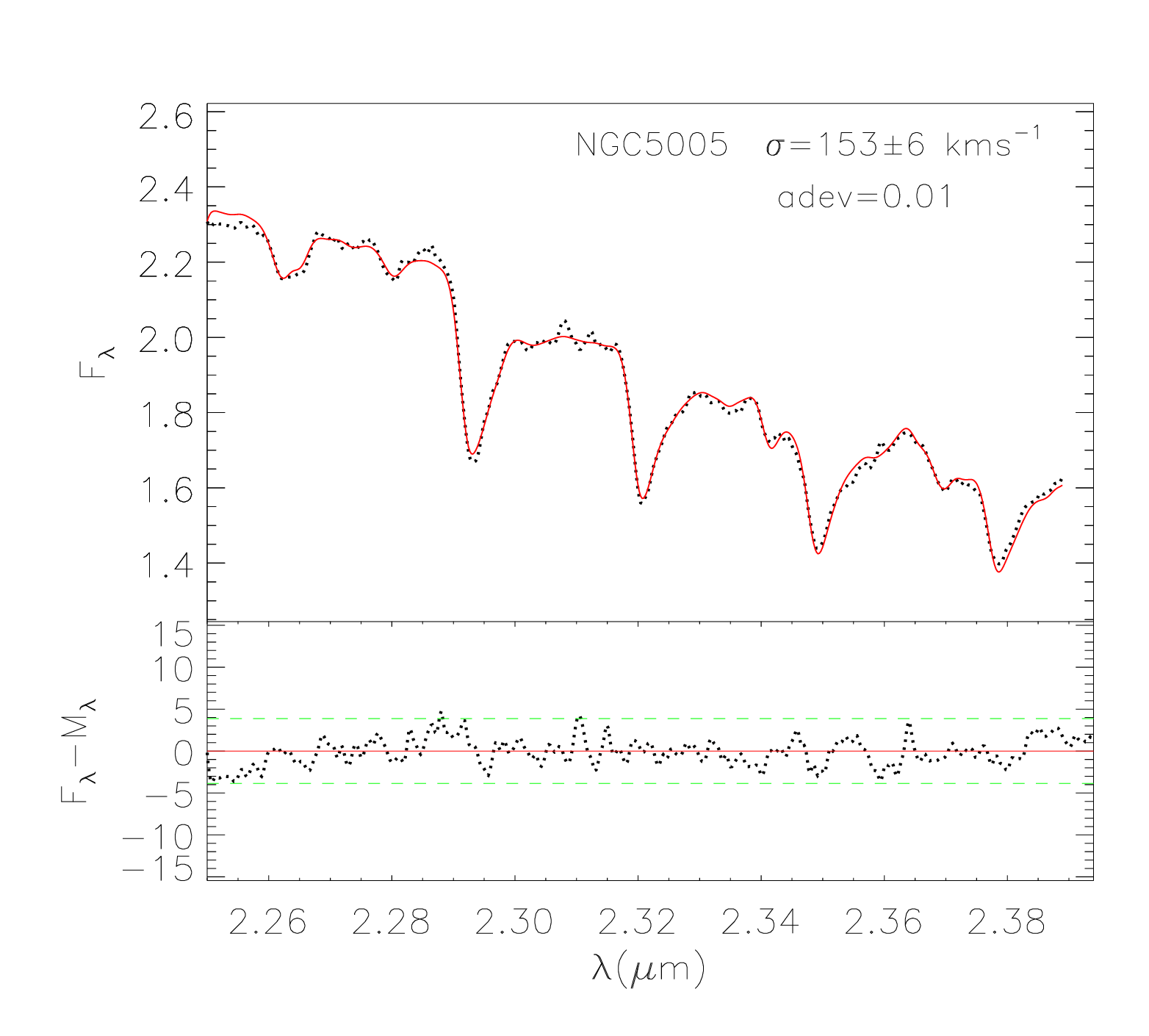}&    
    \includegraphics[scale=0.2]{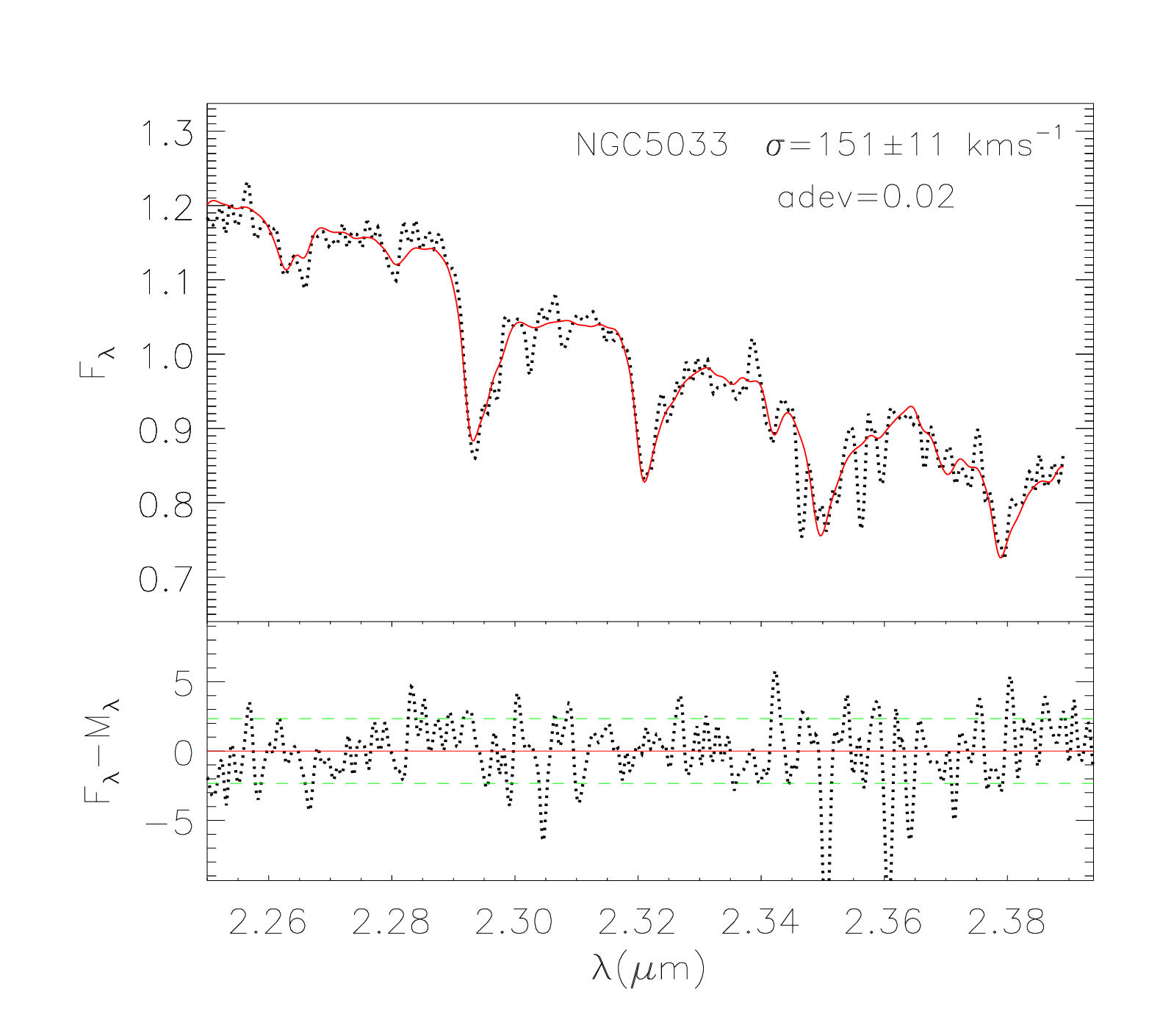} \\

    \includegraphics[scale=0.2]{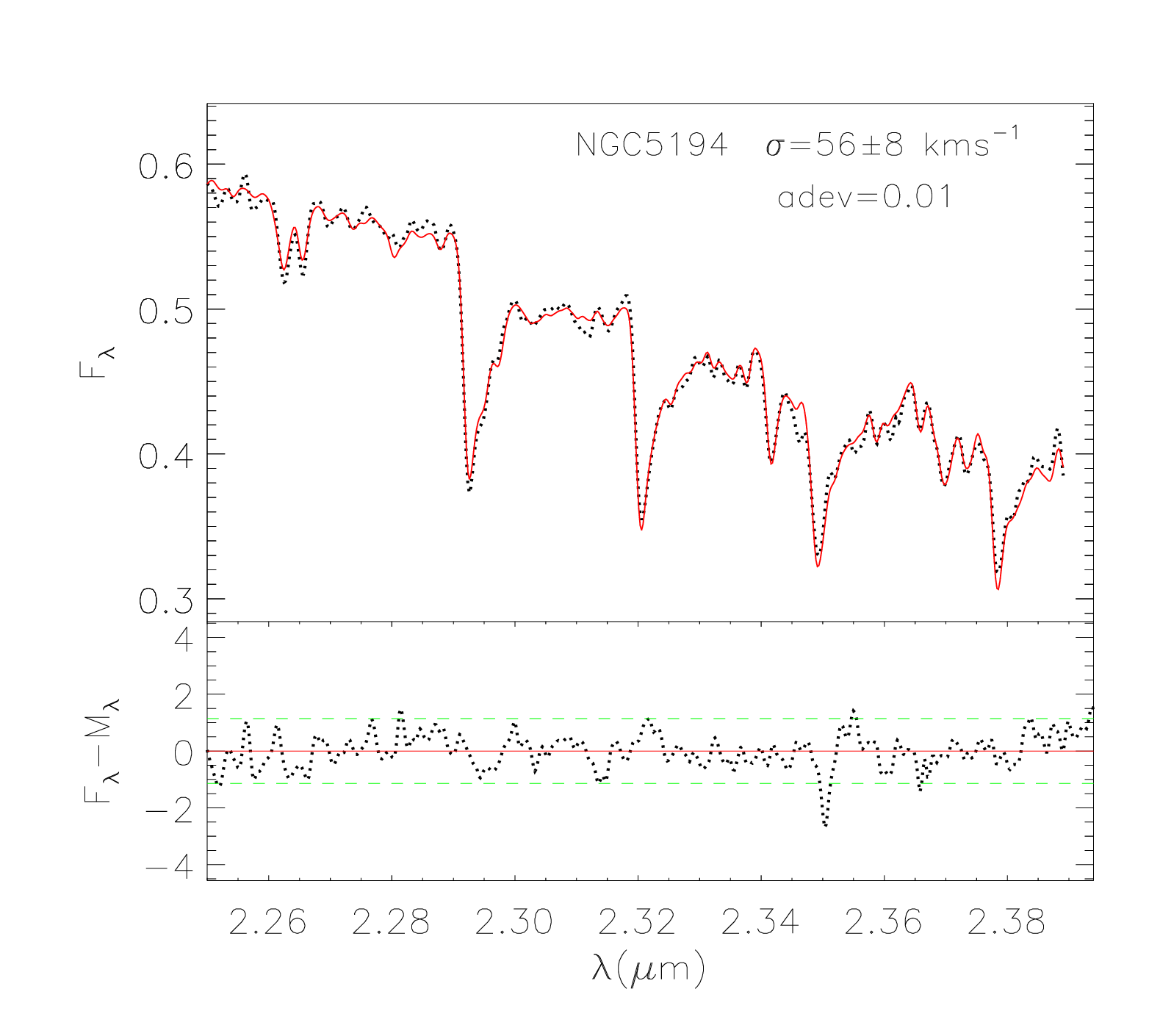}&
    \includegraphics[scale=0.2]{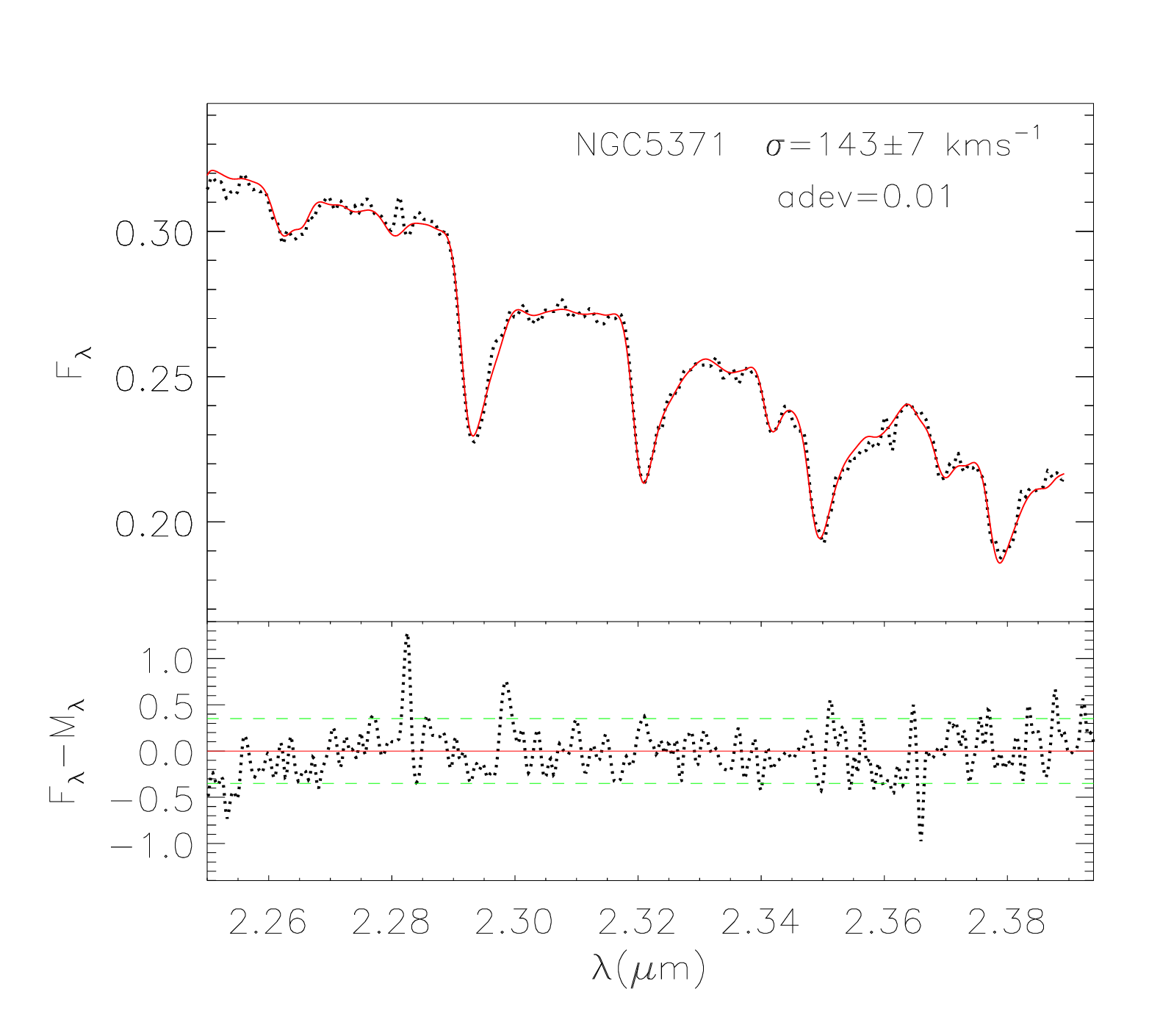}&
    \includegraphics[scale=0.2]{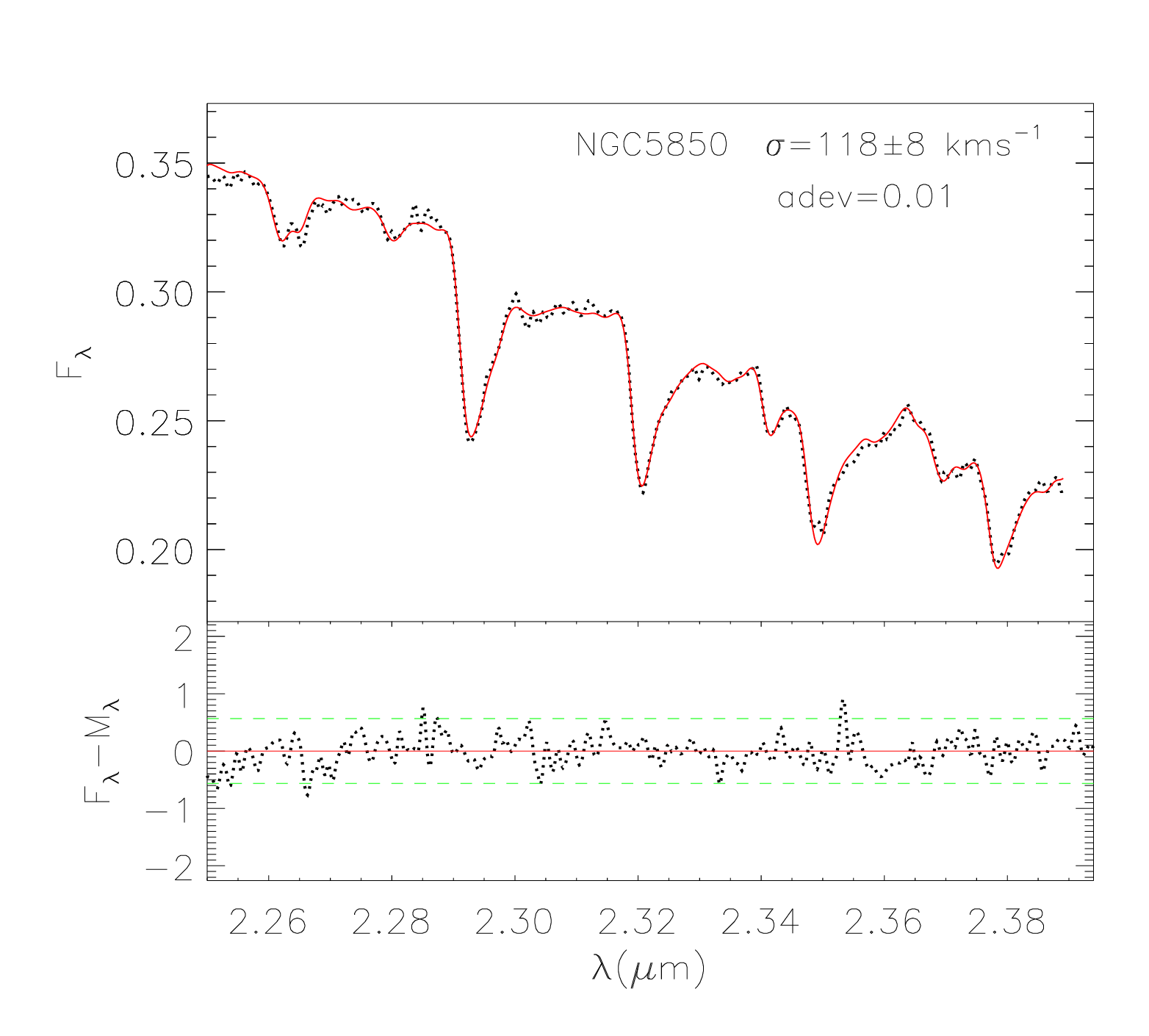}&    
    \includegraphics[scale=0.2]{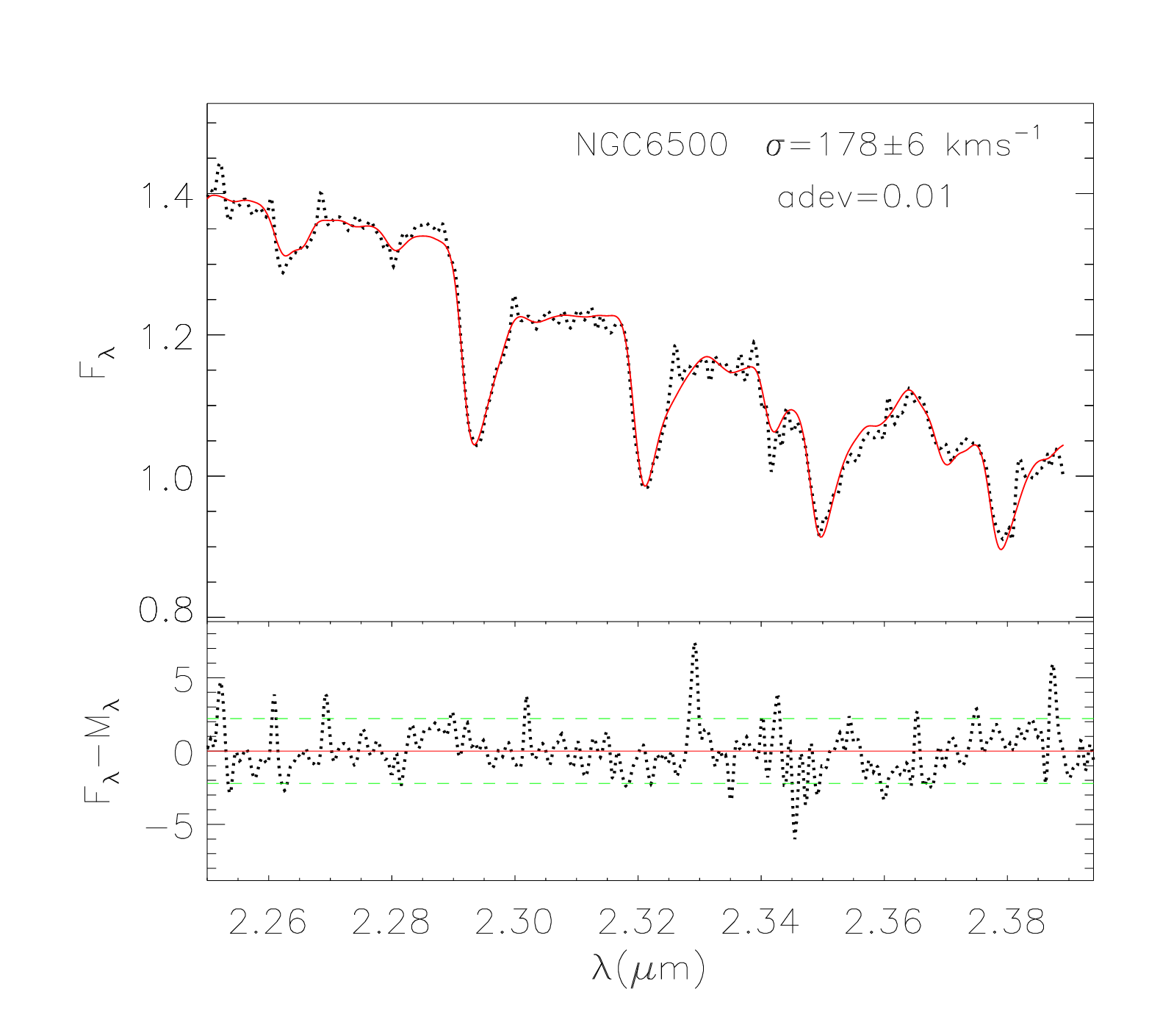} \\

    \includegraphics[scale=0.2]{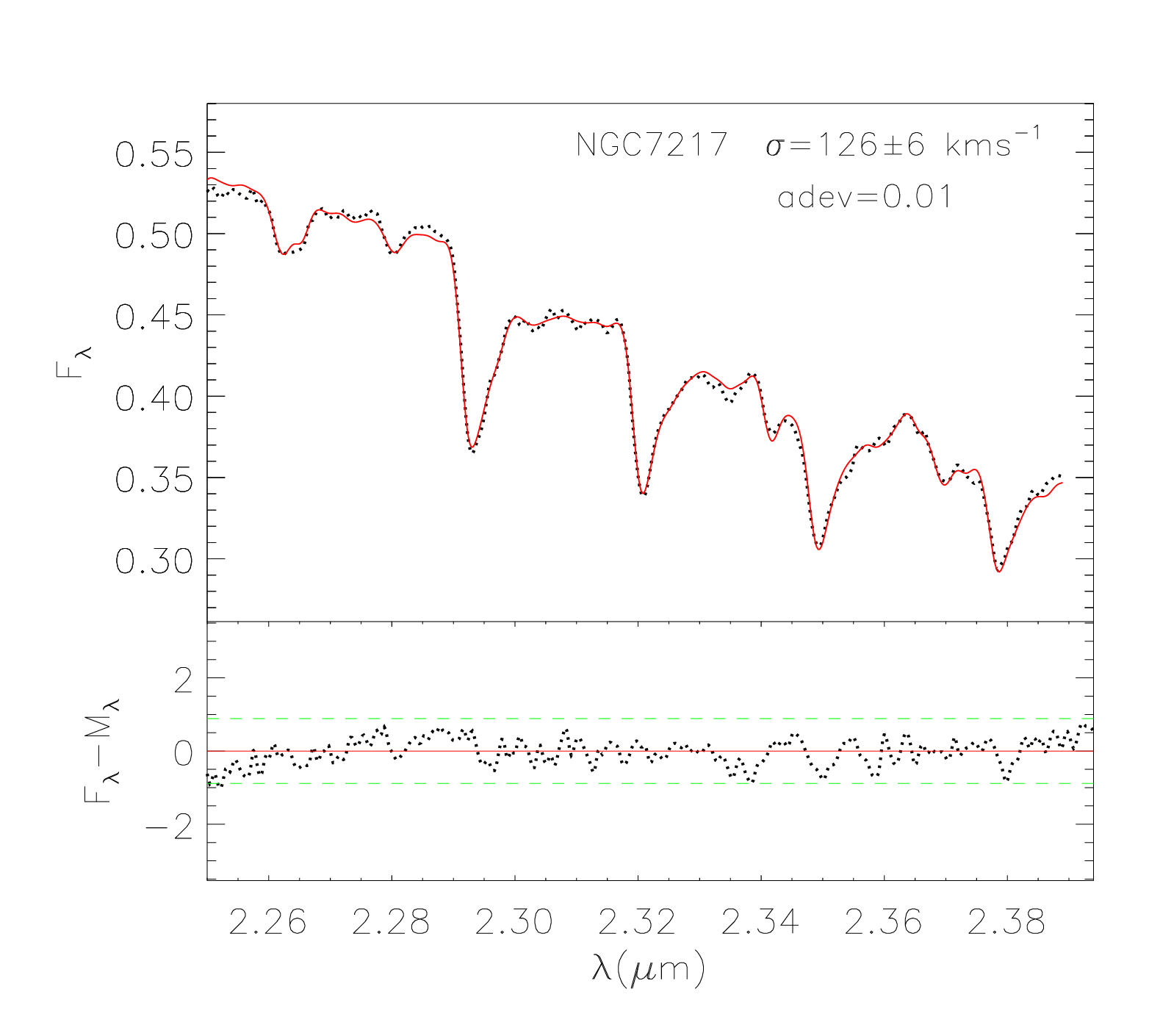}&
    \includegraphics[scale=0.2]{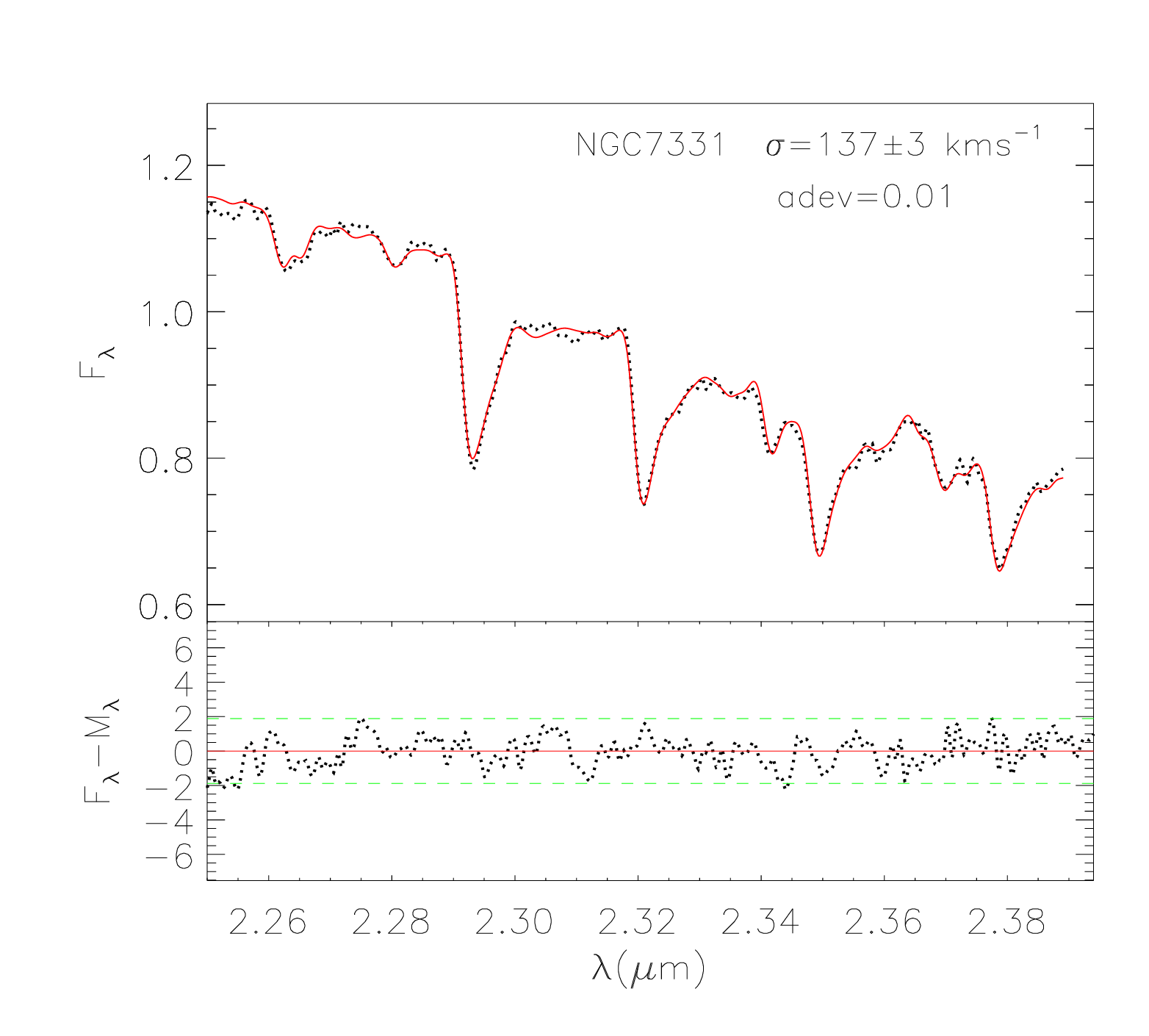}&
    \includegraphics[scale=0.2]{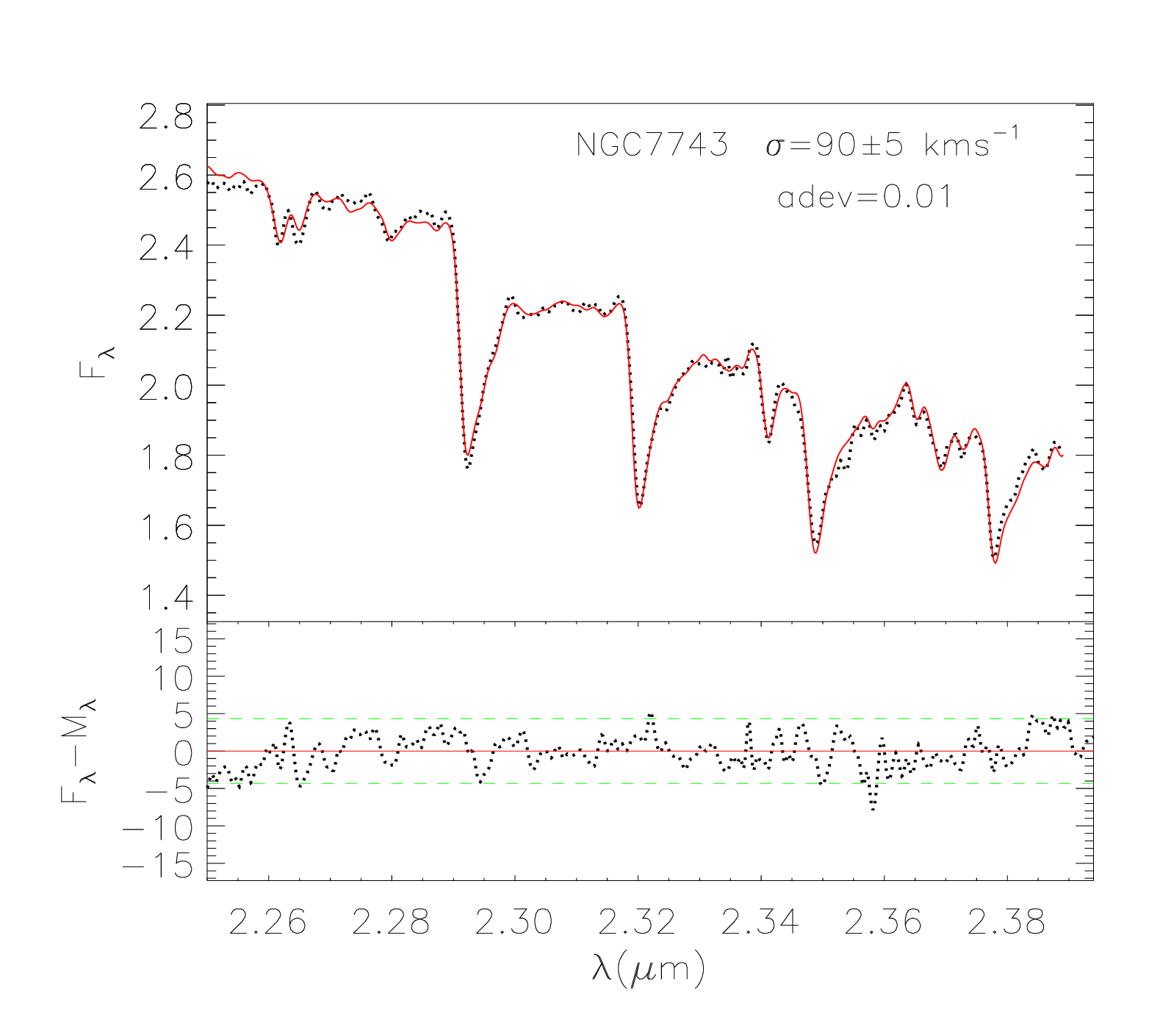}&
  \\

  \end{tabular}
  \caption{(continued)}
\end{figure*}

\begin{figure*}
\centering
  \begin{tabular}{cccc}
    \includegraphics[scale=0.2]{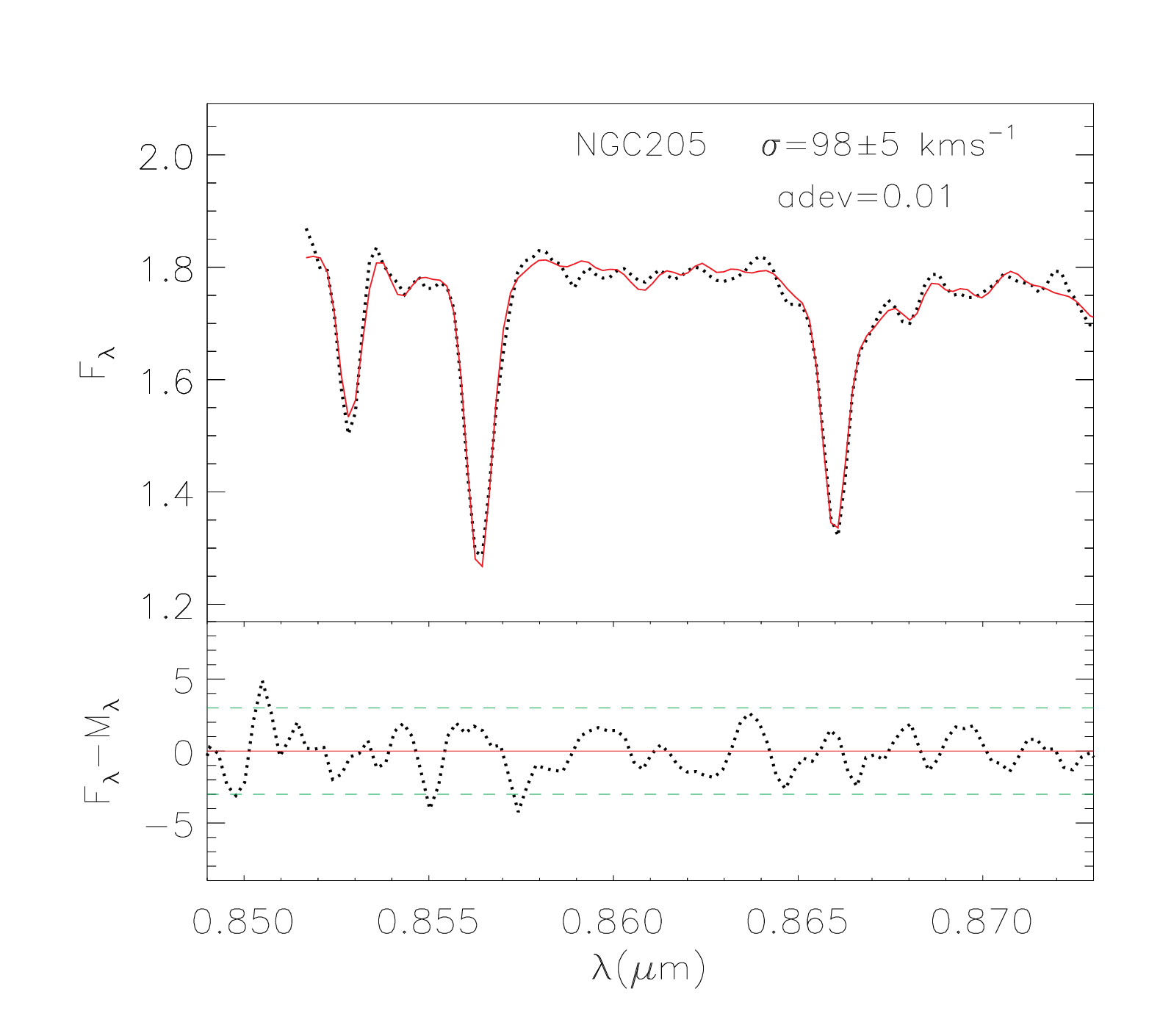}&
    \includegraphics[scale=0.2]{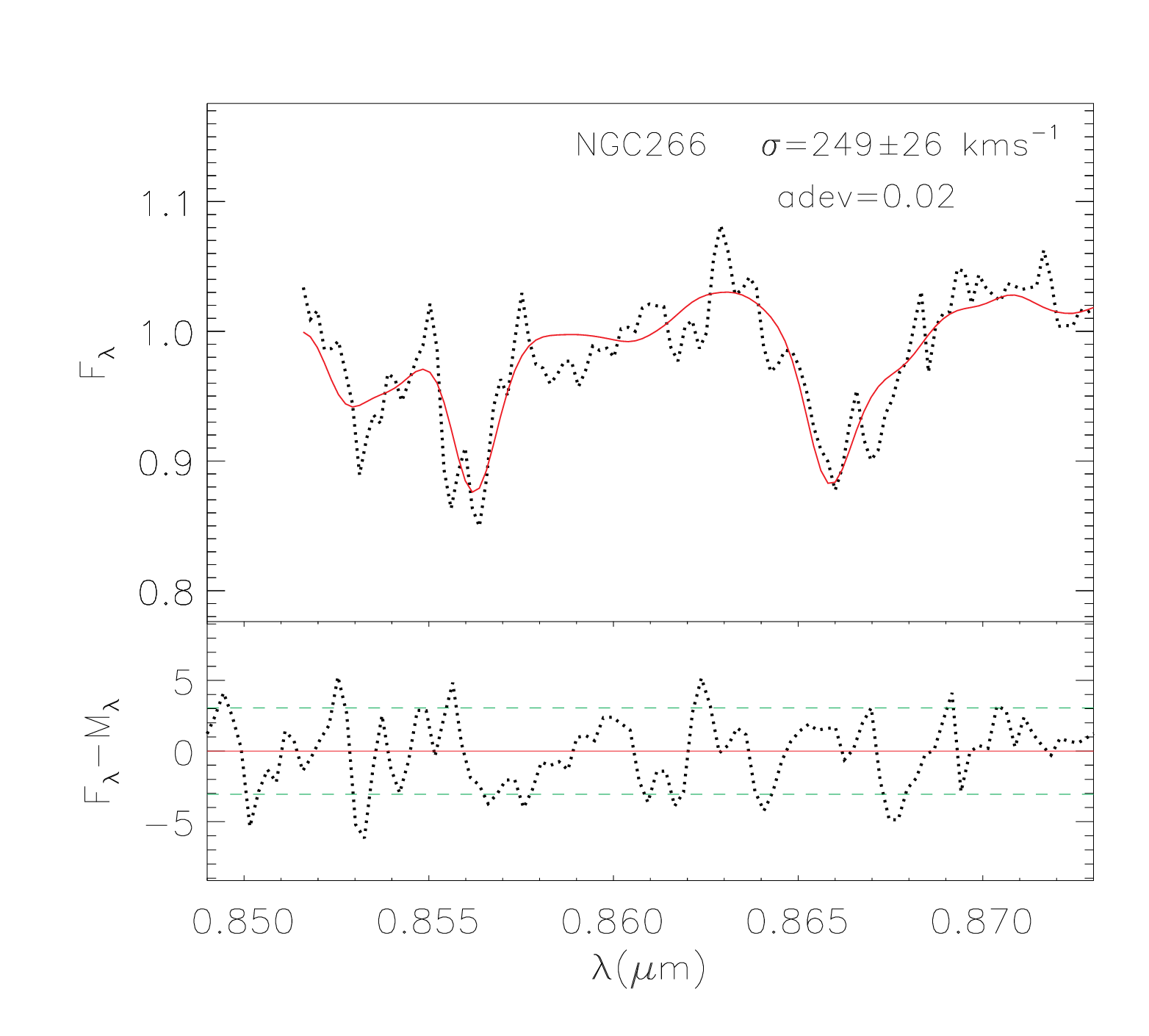}&
    \includegraphics[scale=0.2]{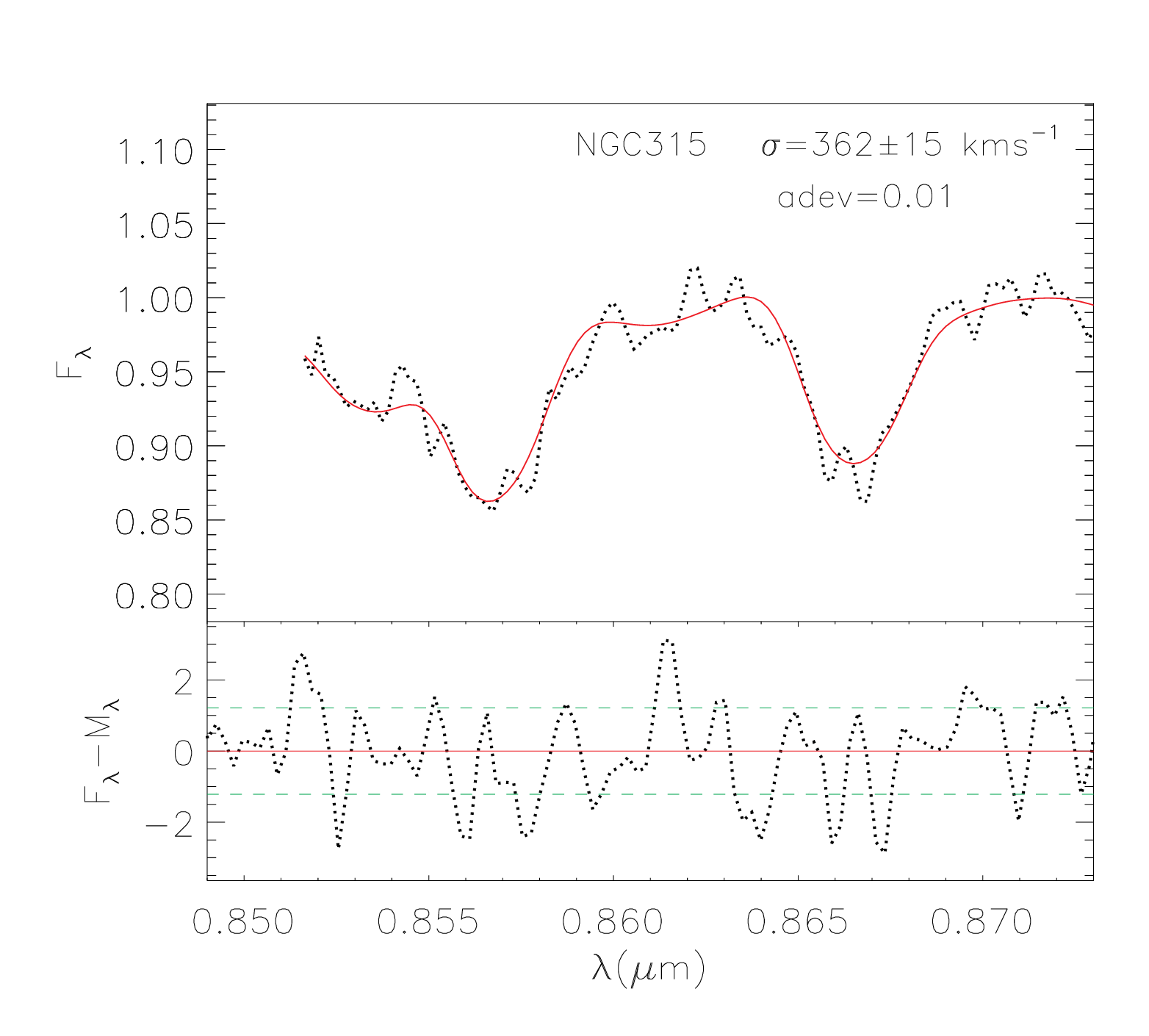}&    
    \includegraphics[scale=0.2]{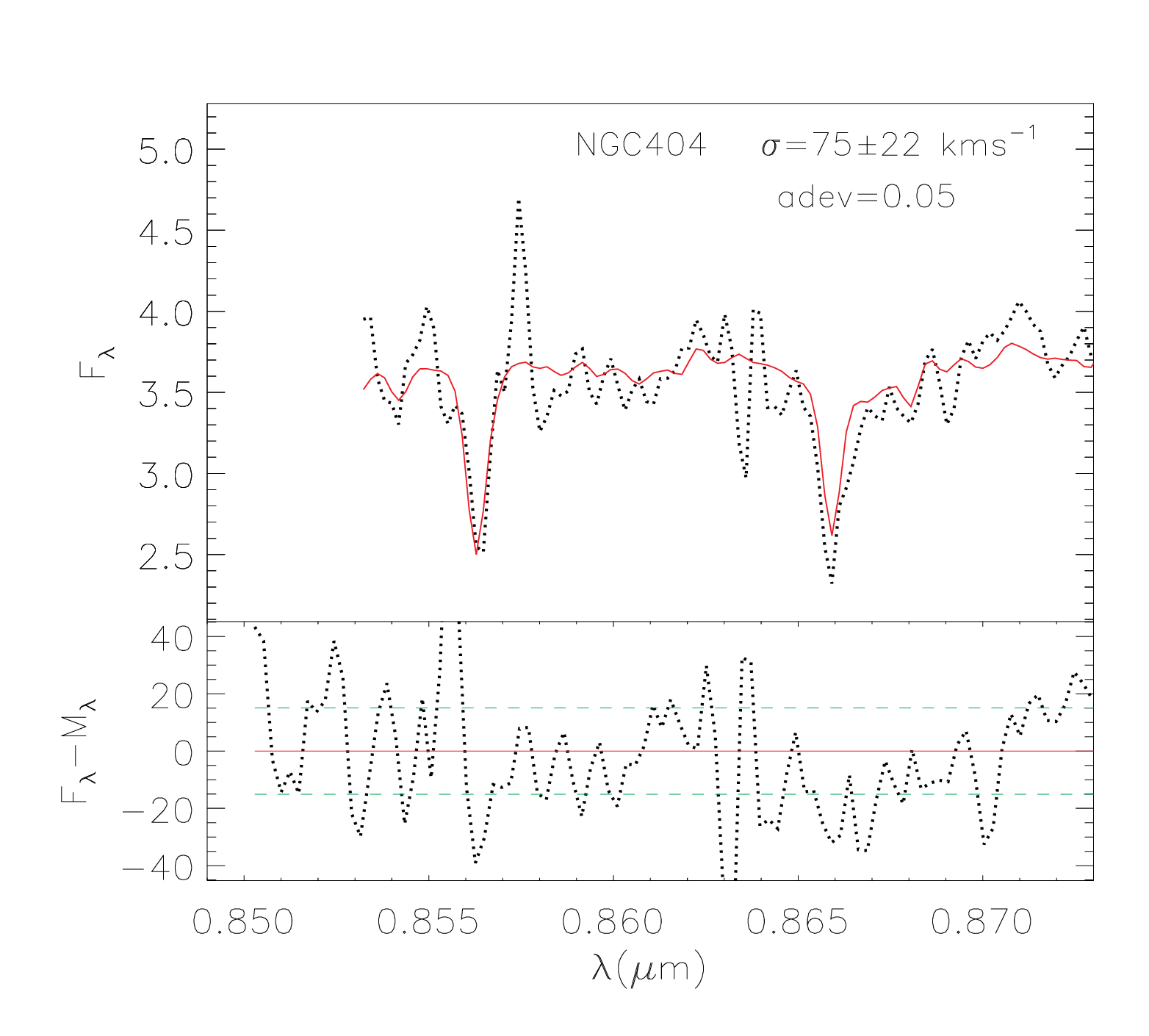} \\

    \includegraphics[scale=0.2]{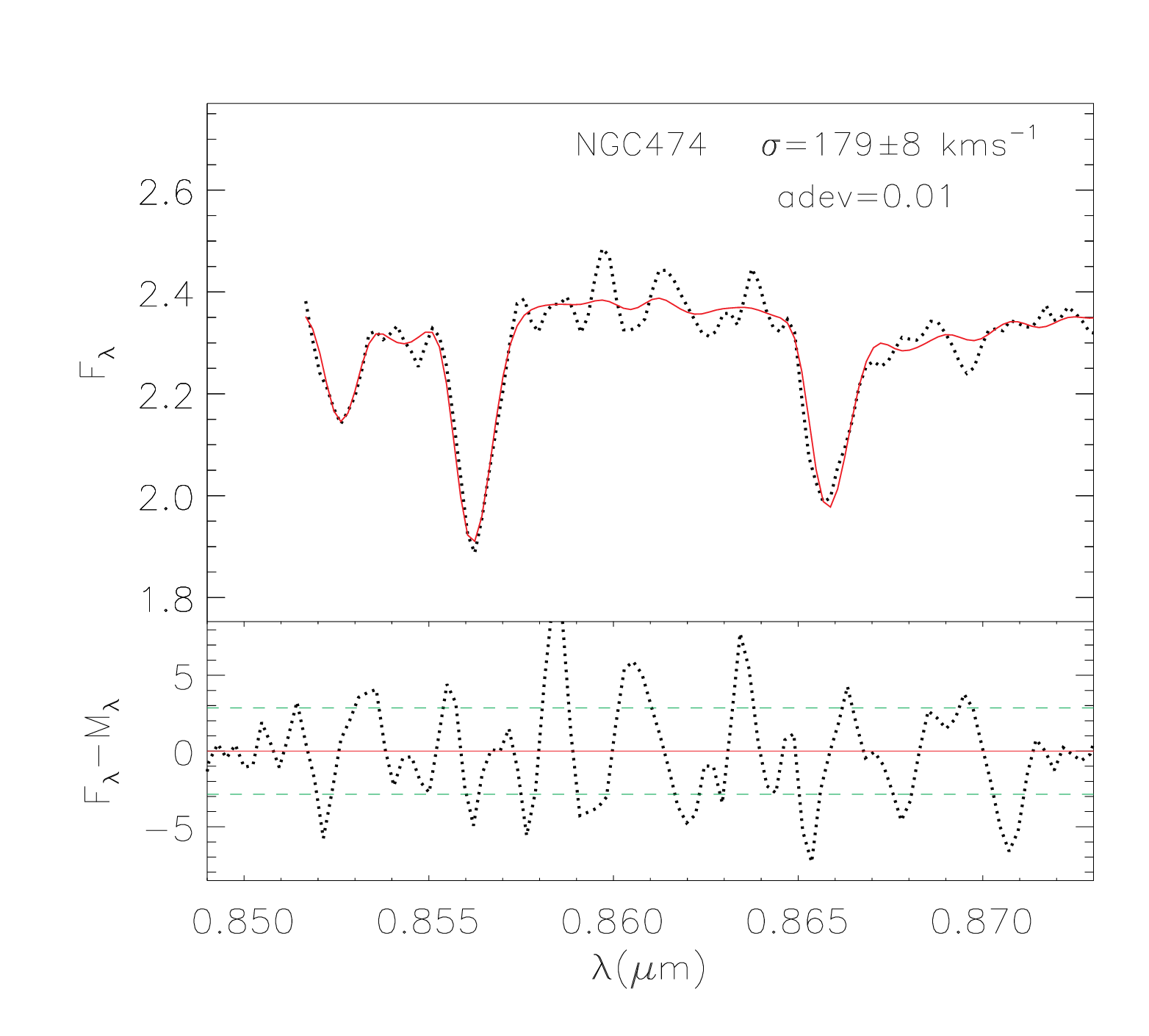}&
    \includegraphics[scale=0.2]{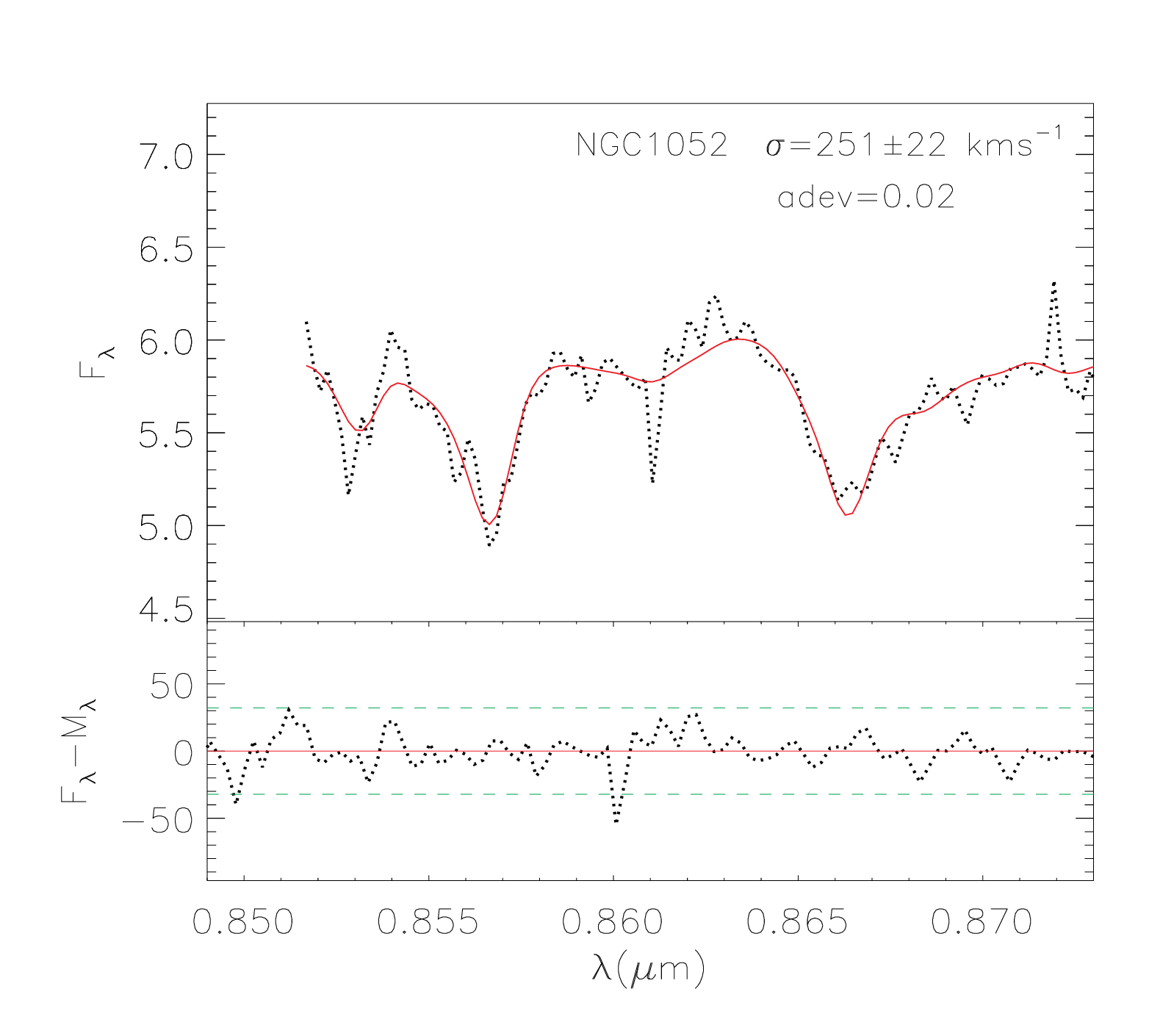}&
    \includegraphics[scale=0.2]{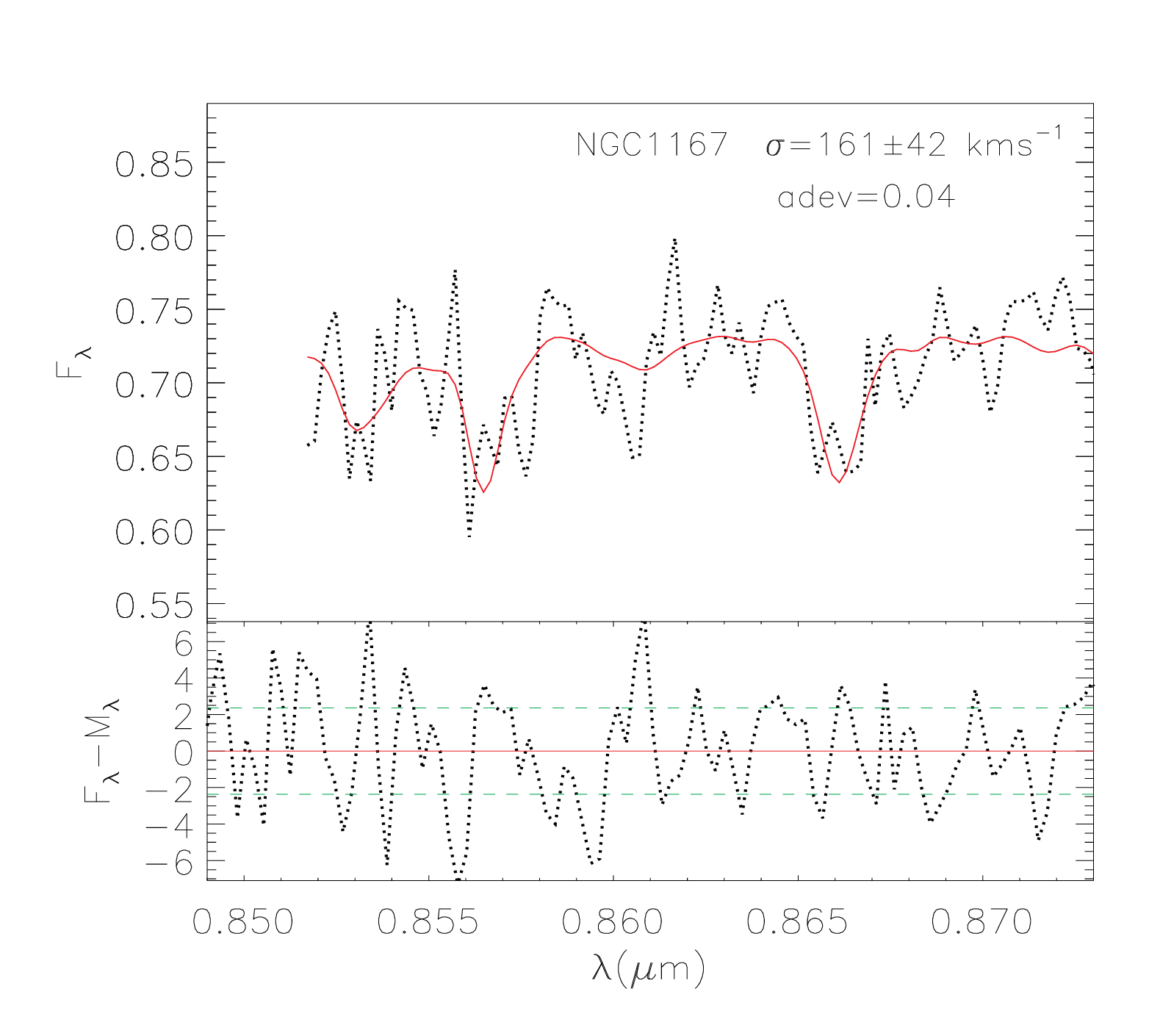}&    
    \includegraphics[scale=0.2]{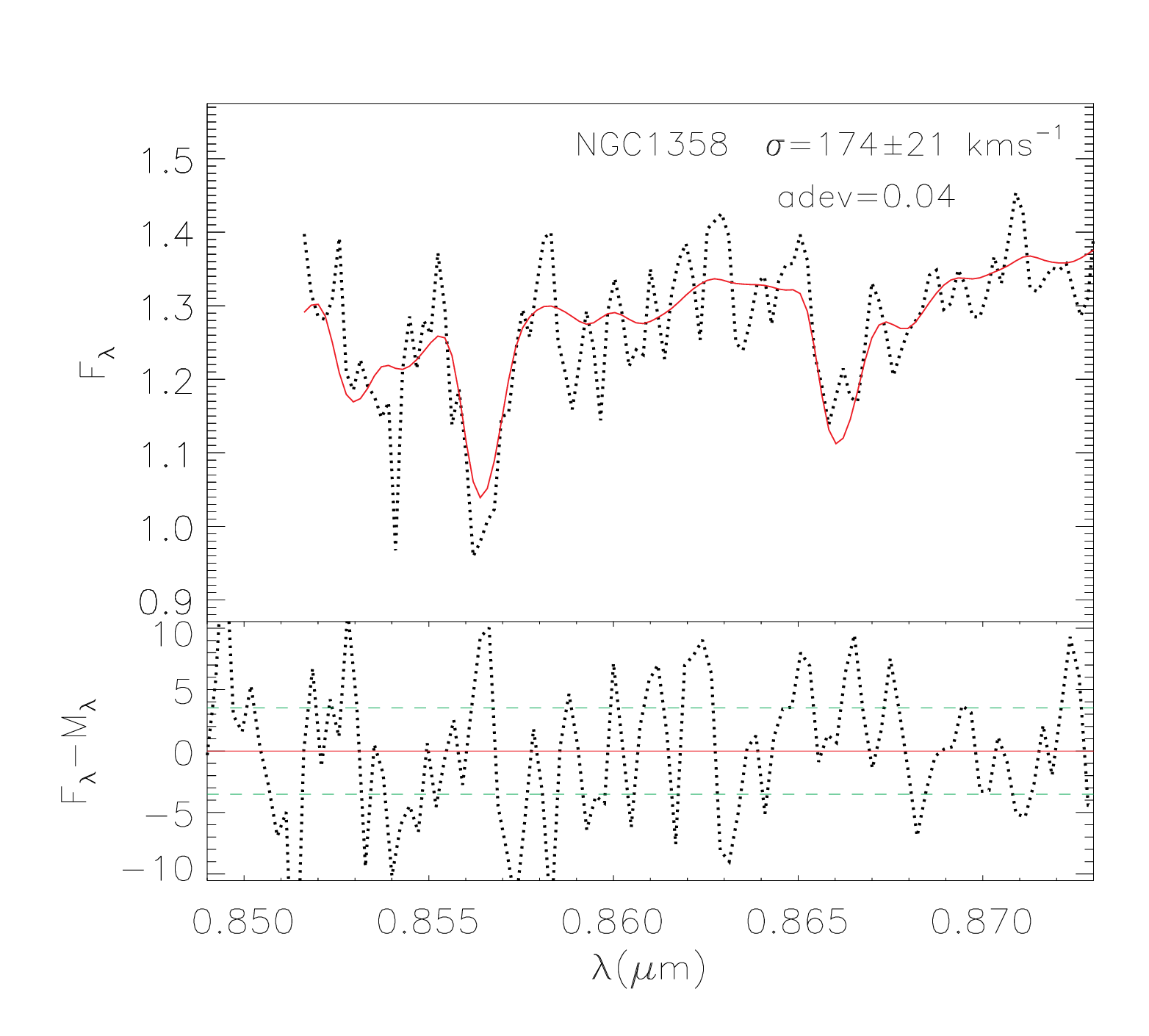} \\

    \includegraphics[scale=0.2]{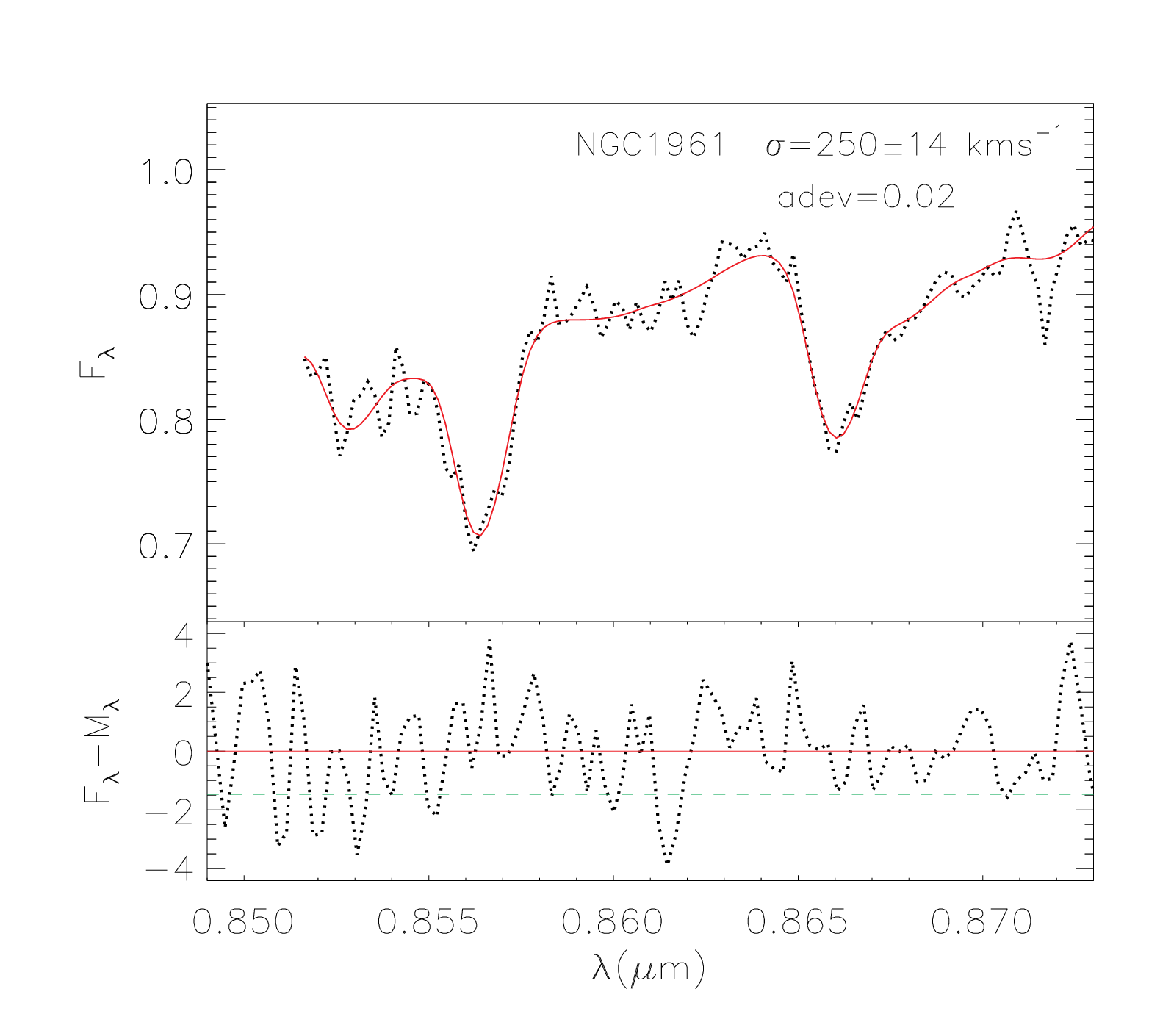}&
    \includegraphics[scale=0.2]{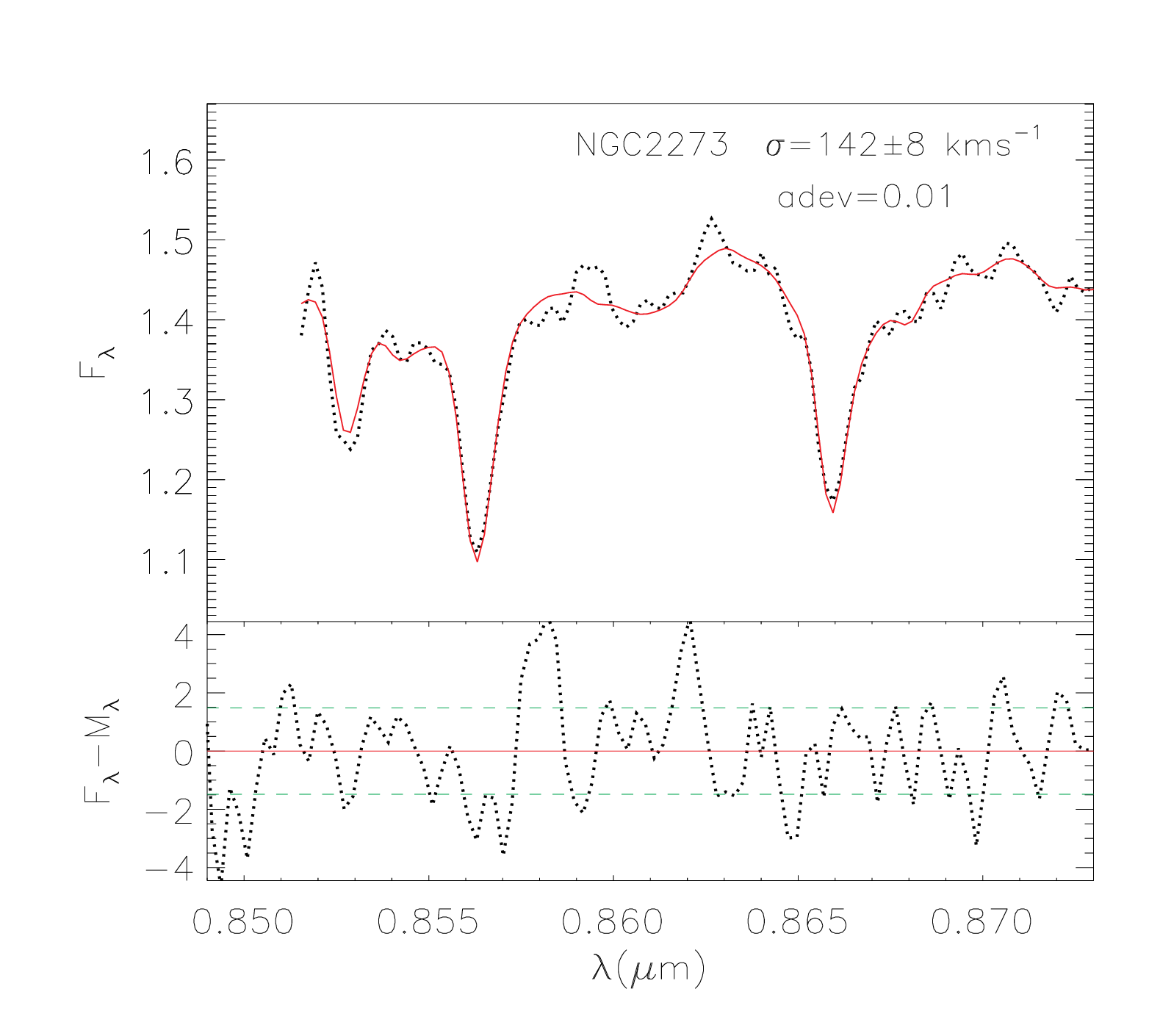}&
    \includegraphics[scale=0.2]{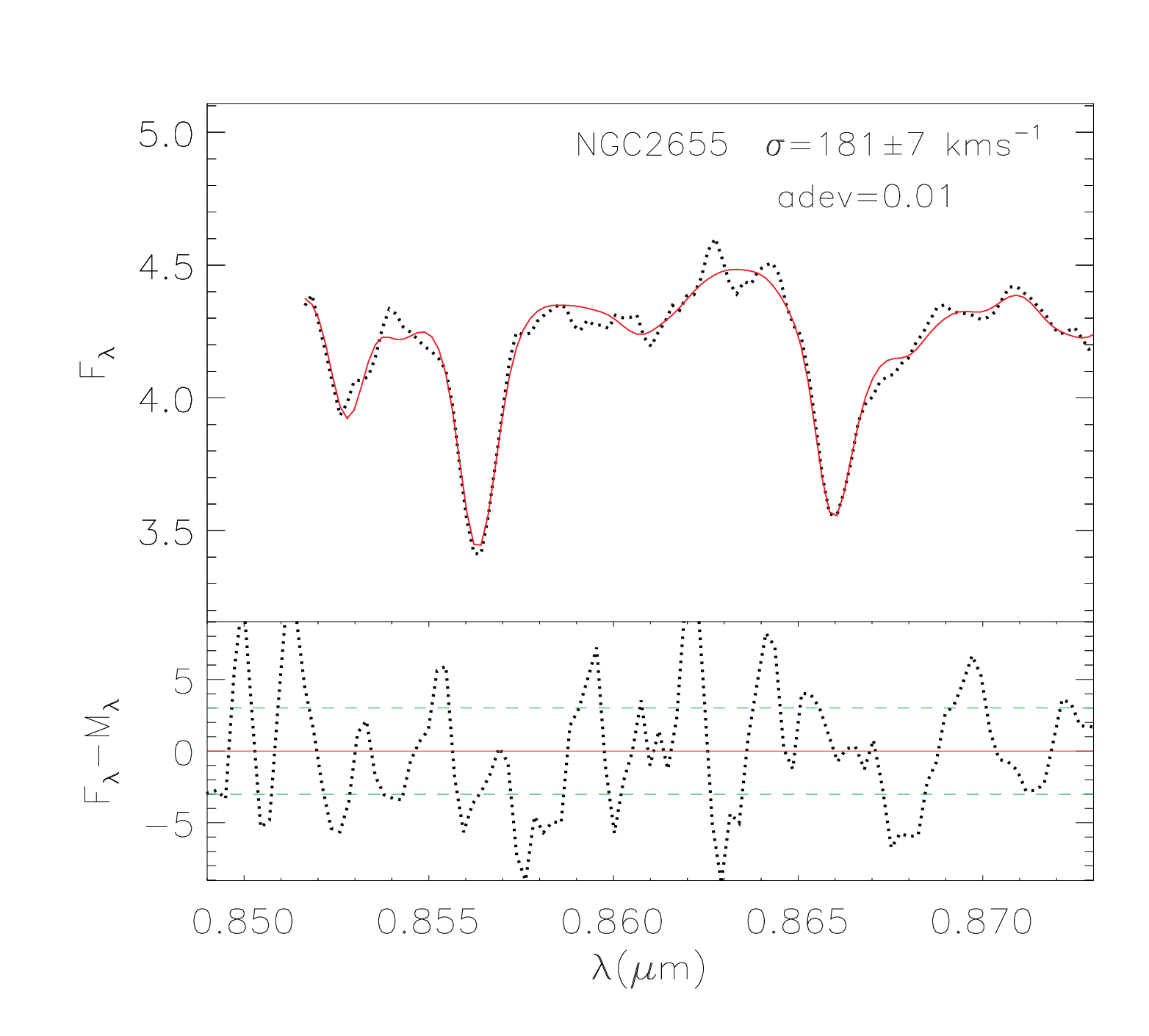}&    
    \includegraphics[scale=0.2]{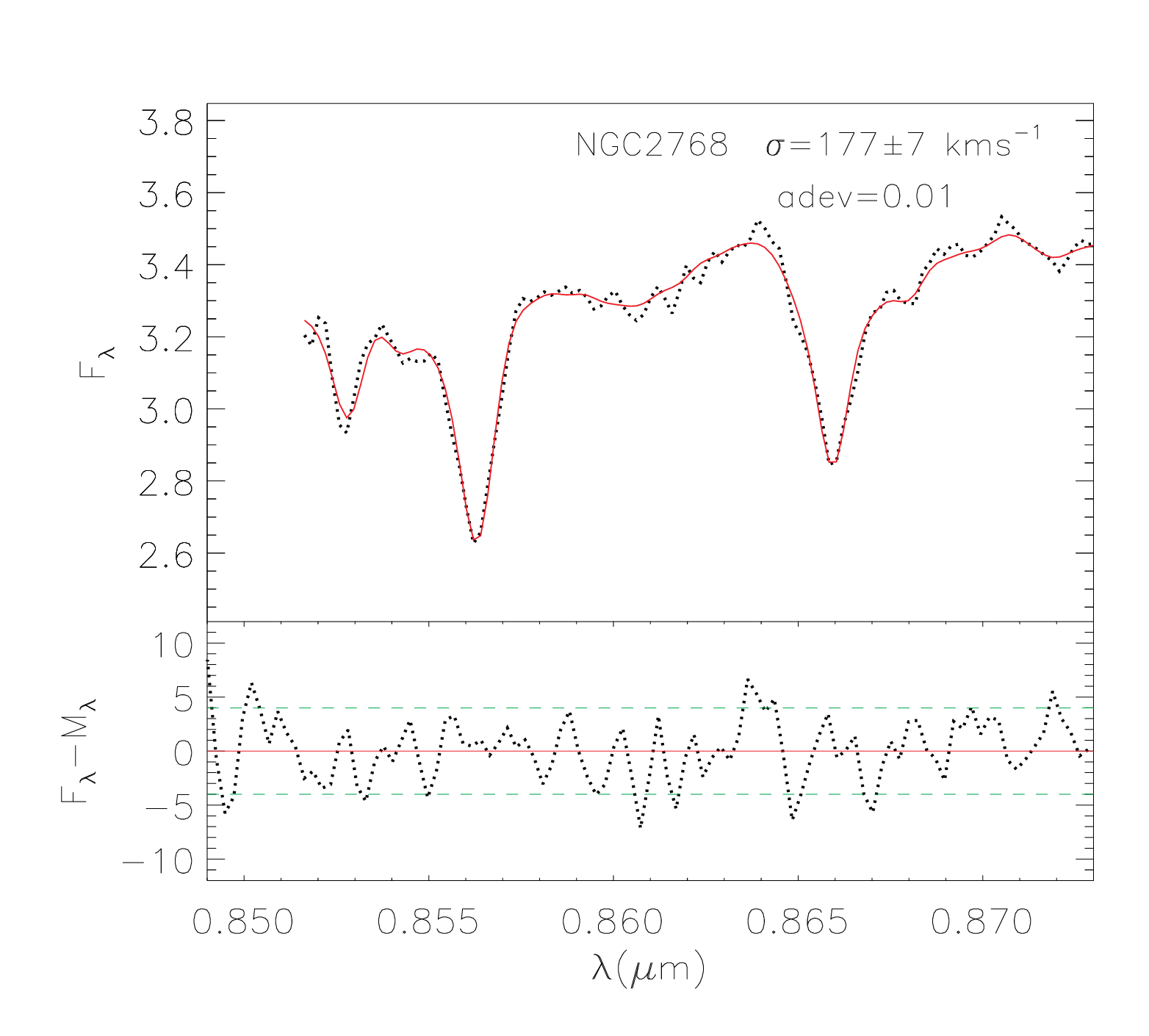} \\

    \includegraphics[scale=0.2]{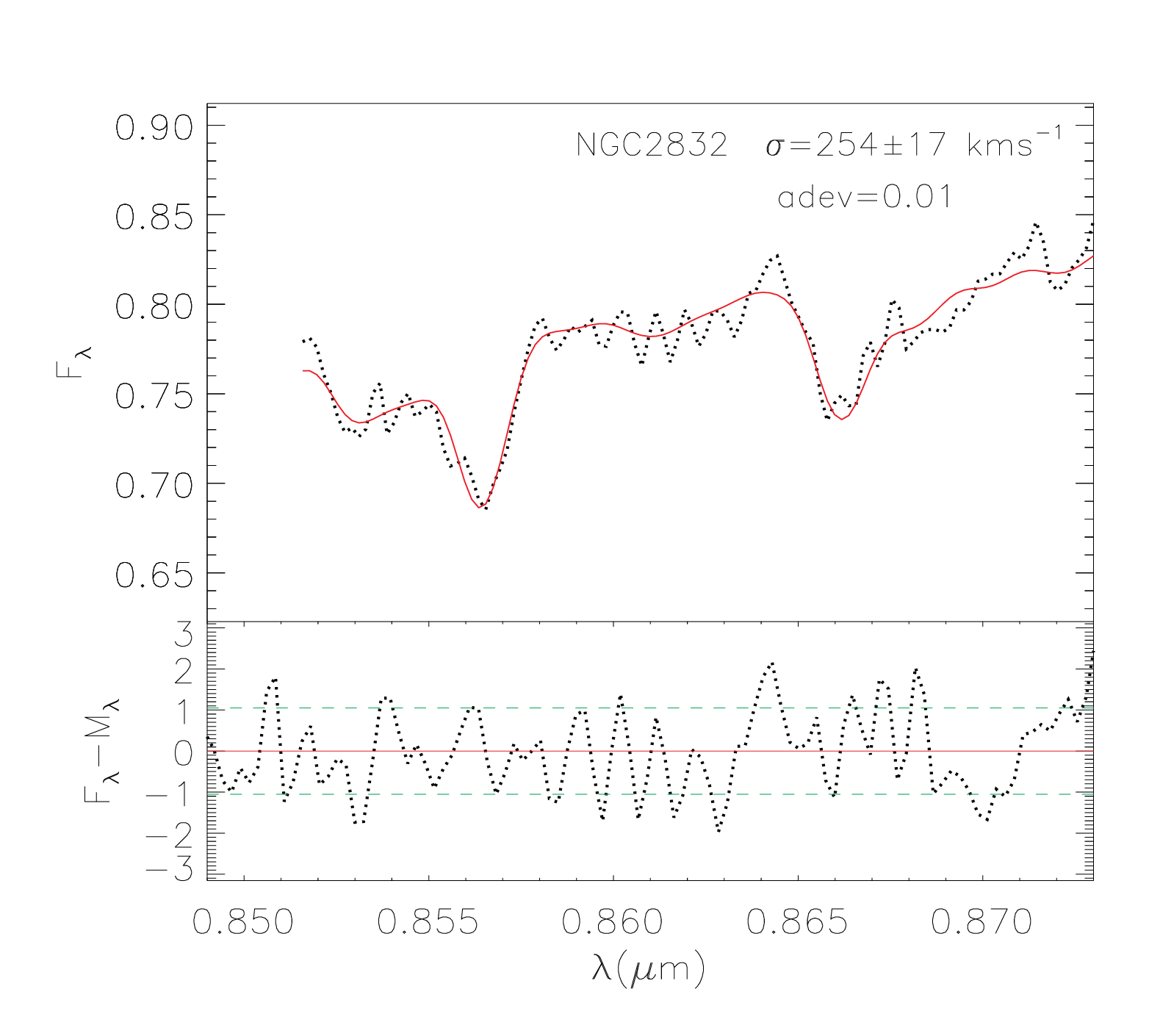} &
    \includegraphics[scale=0.2]{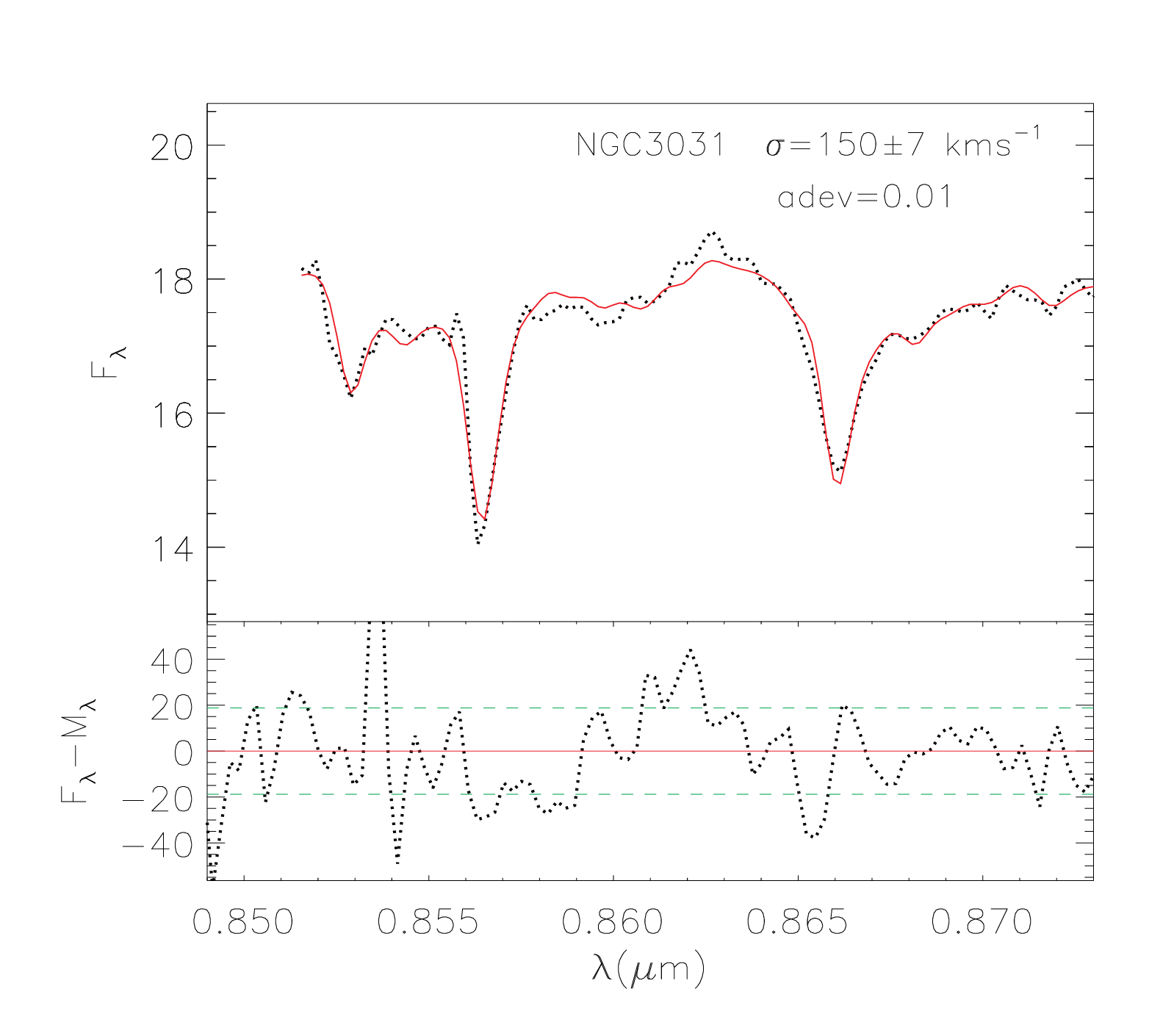}&
    \includegraphics[scale=0.2]{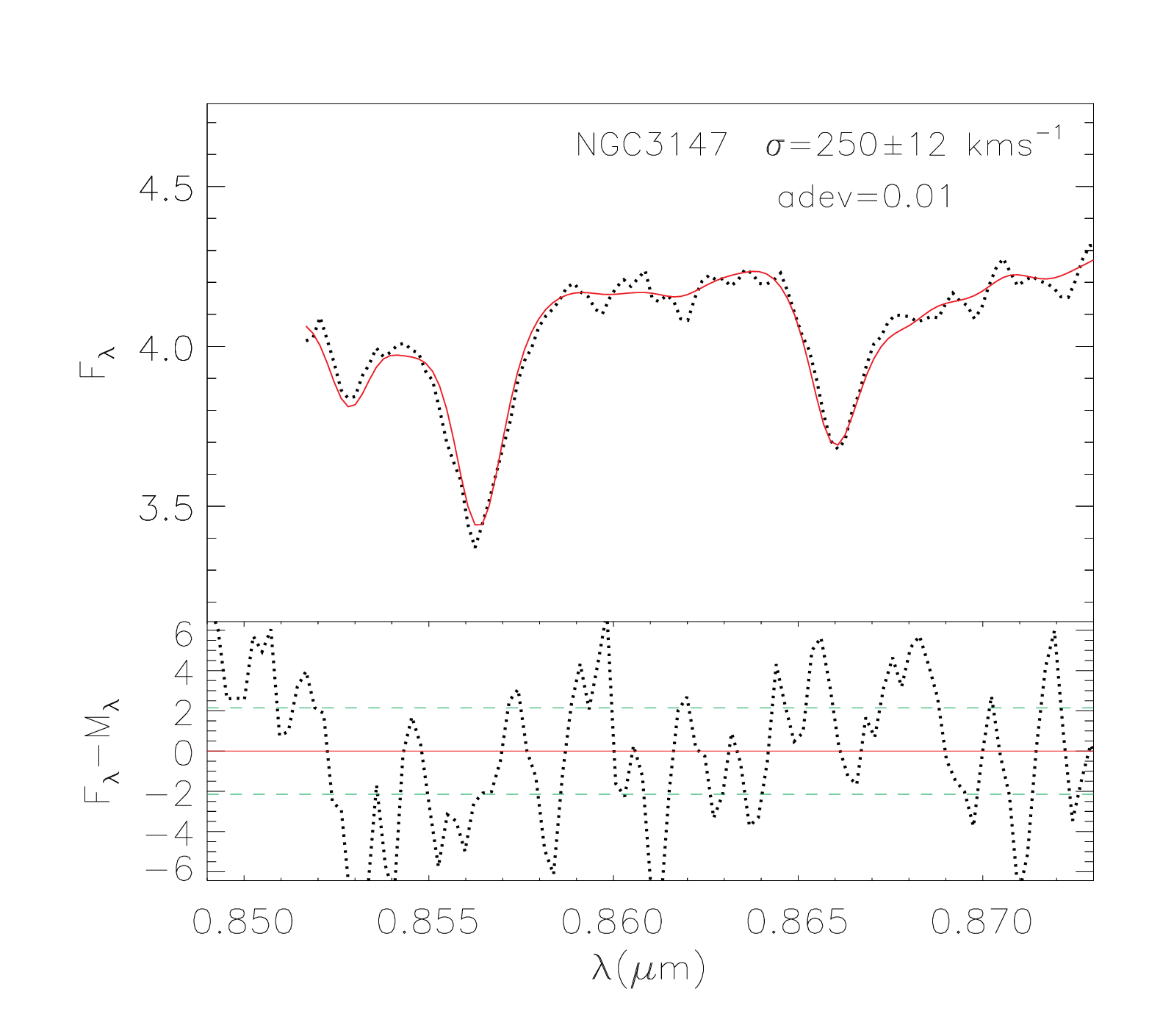} &
    \includegraphics[scale=0.2]{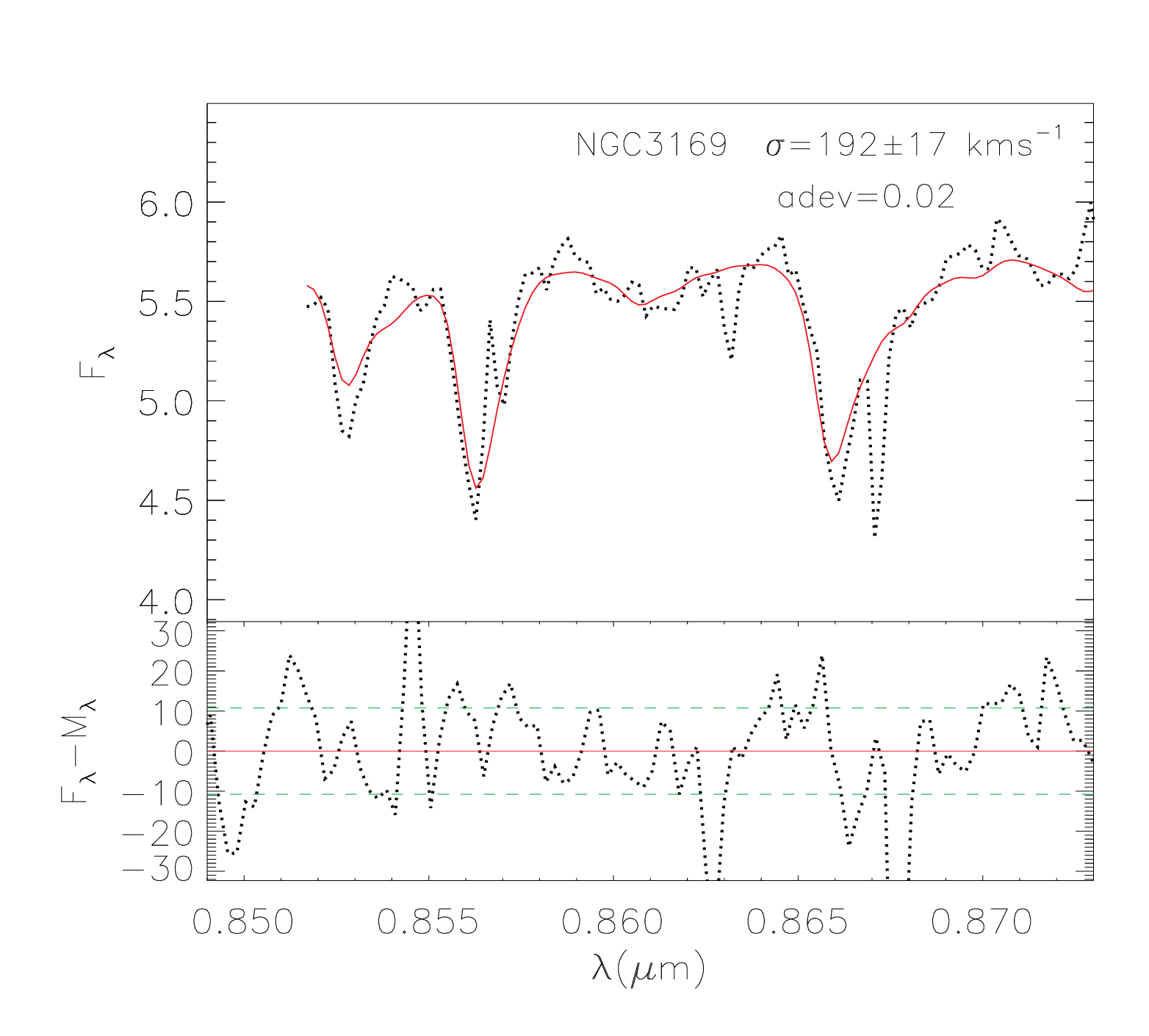} \\
    
    \includegraphics[scale=0.2]{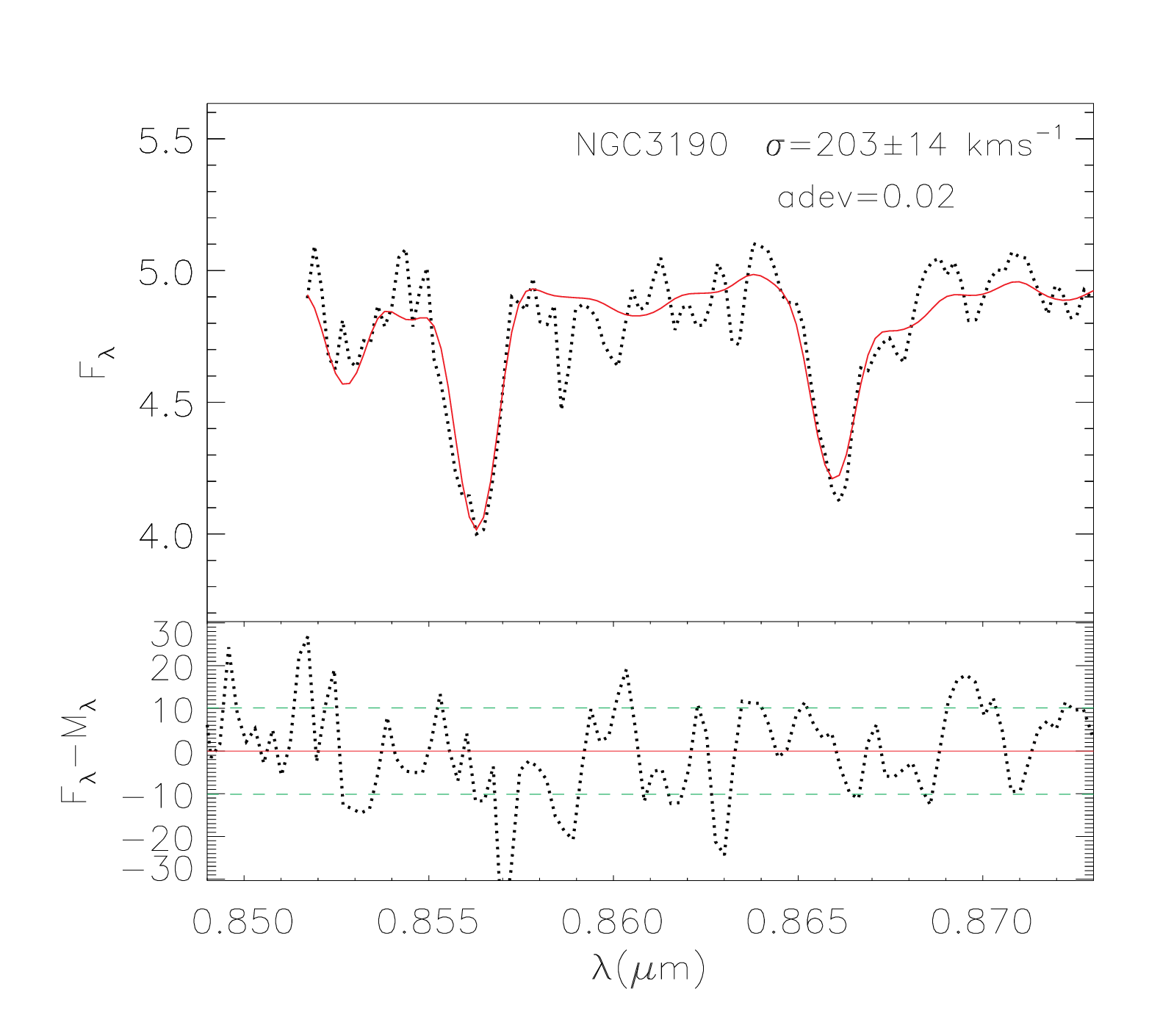}&
    \includegraphics[scale=0.2]{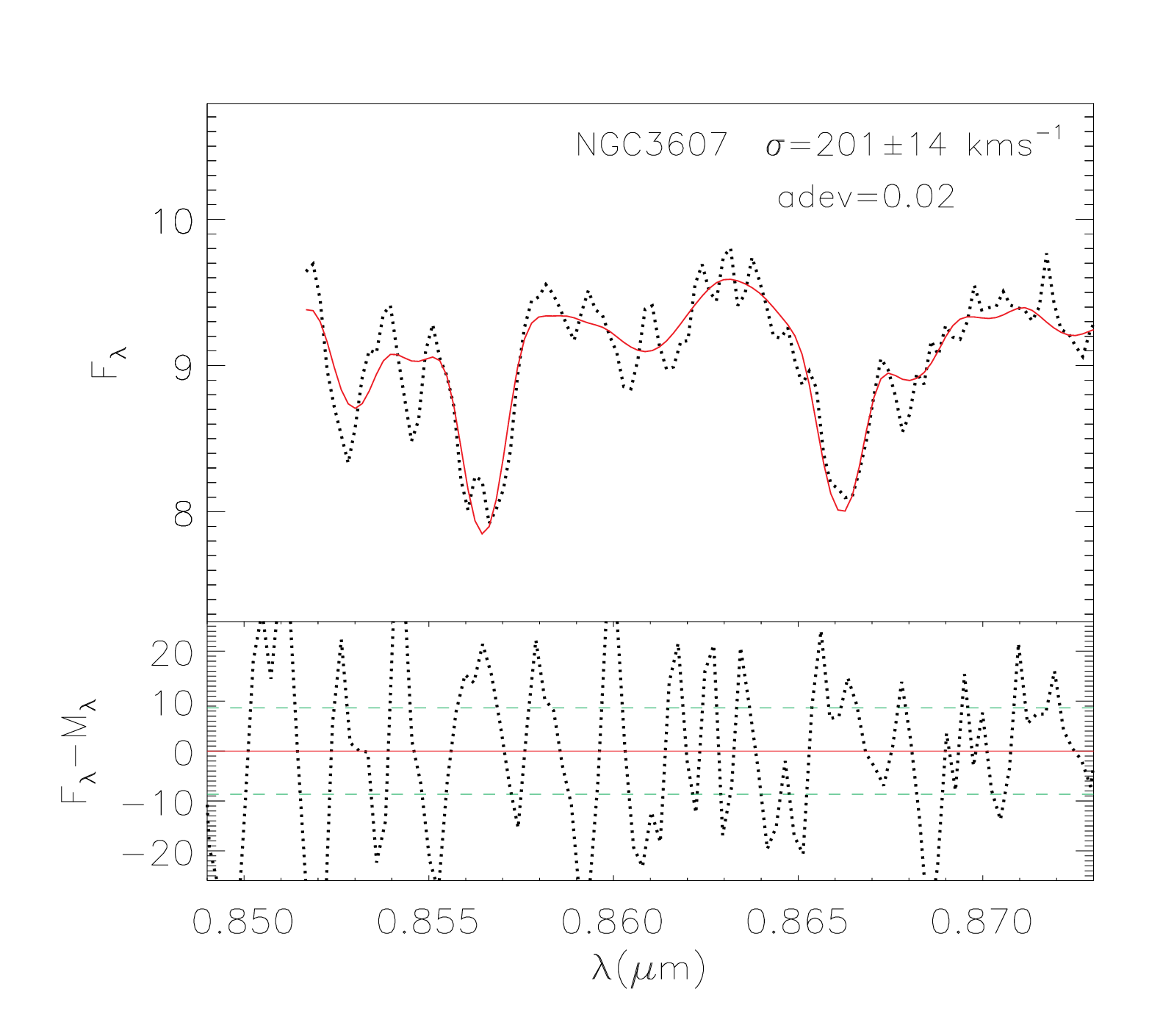}&
    \includegraphics[scale=0.2]{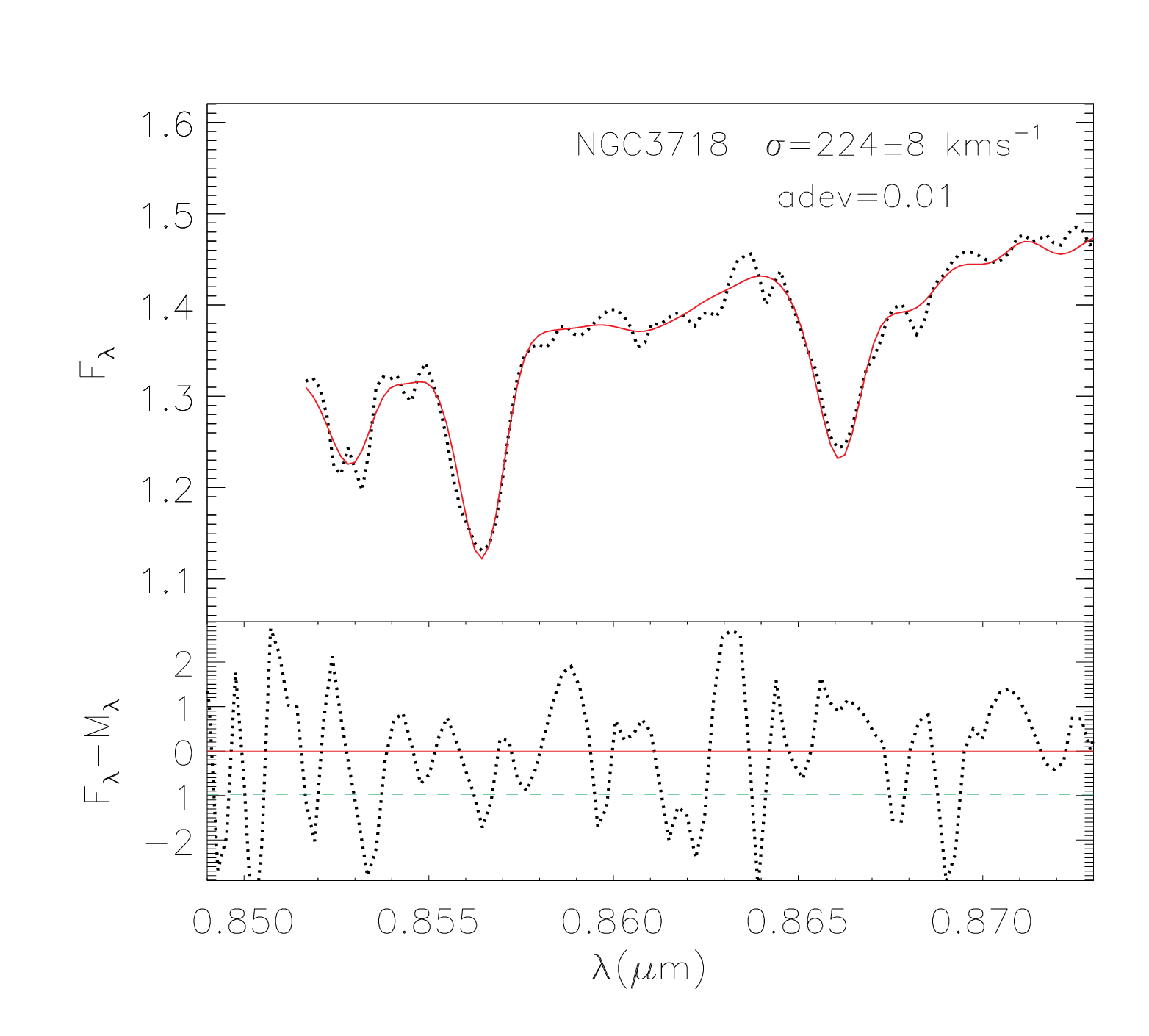}&
    \includegraphics[scale=0.2]{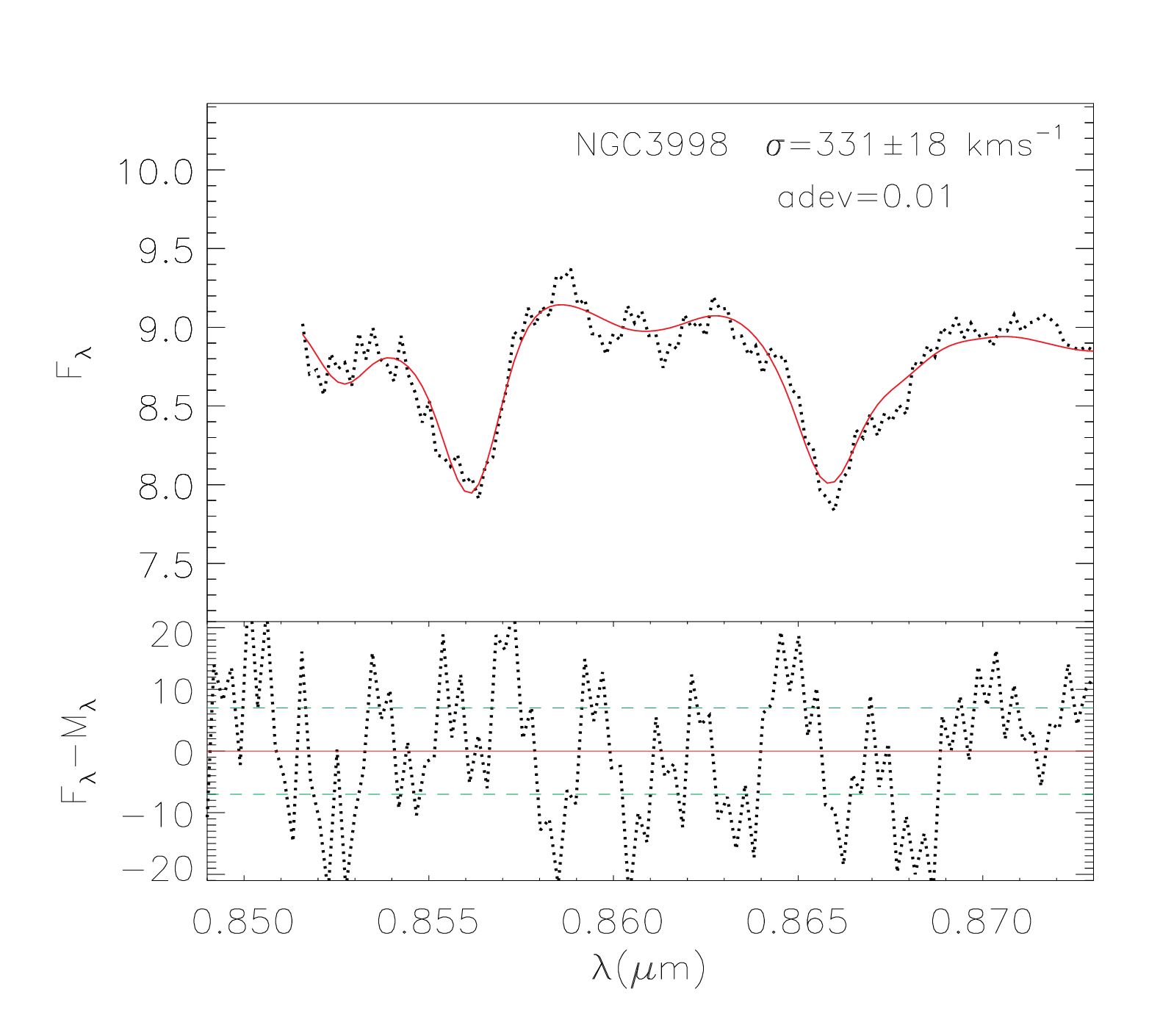} \\

    \includegraphics[scale=0.2]{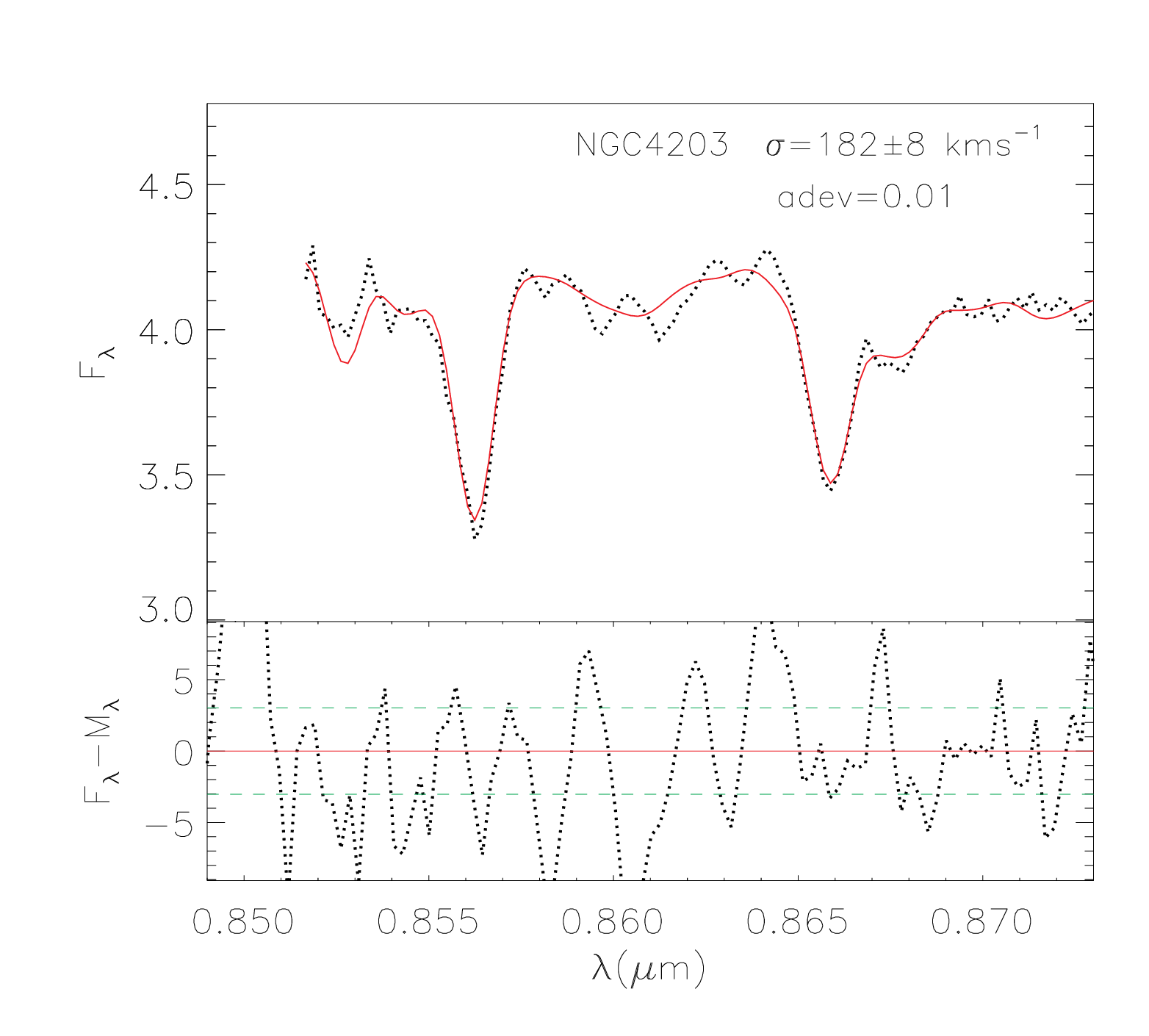}&
    \includegraphics[scale=0.2]{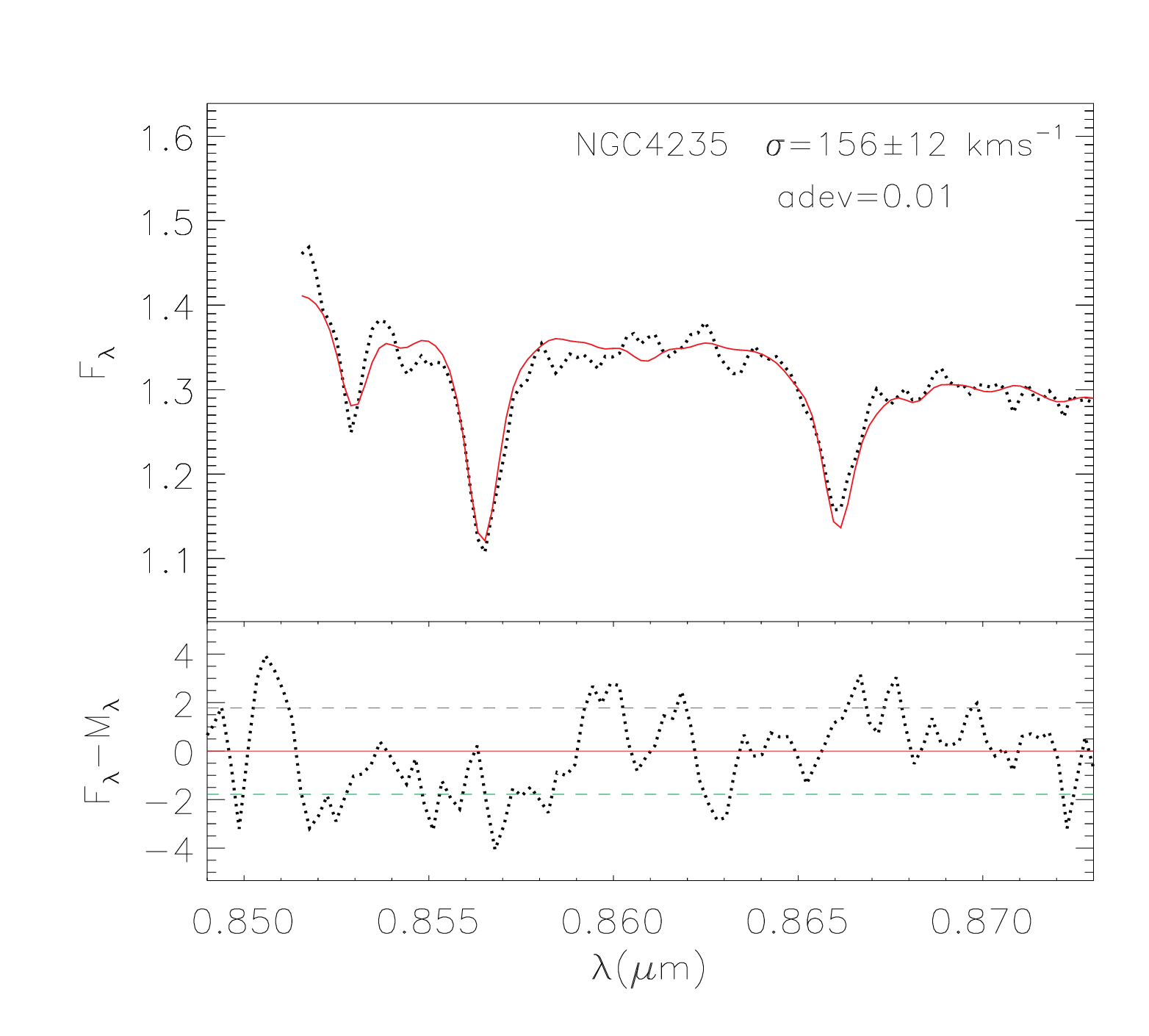}&
    \includegraphics[scale=0.2]{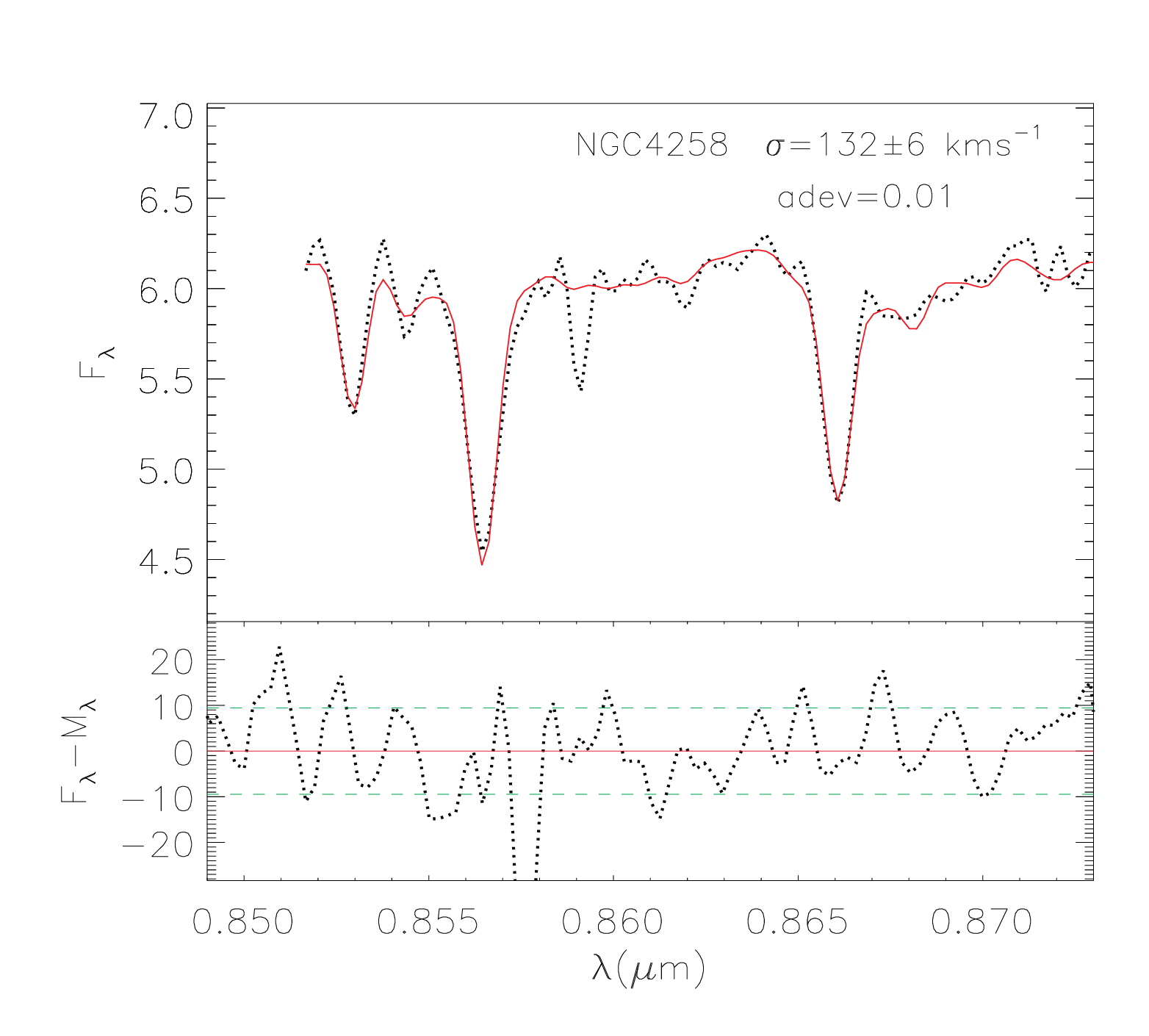} &
    \includegraphics[scale=0.2]{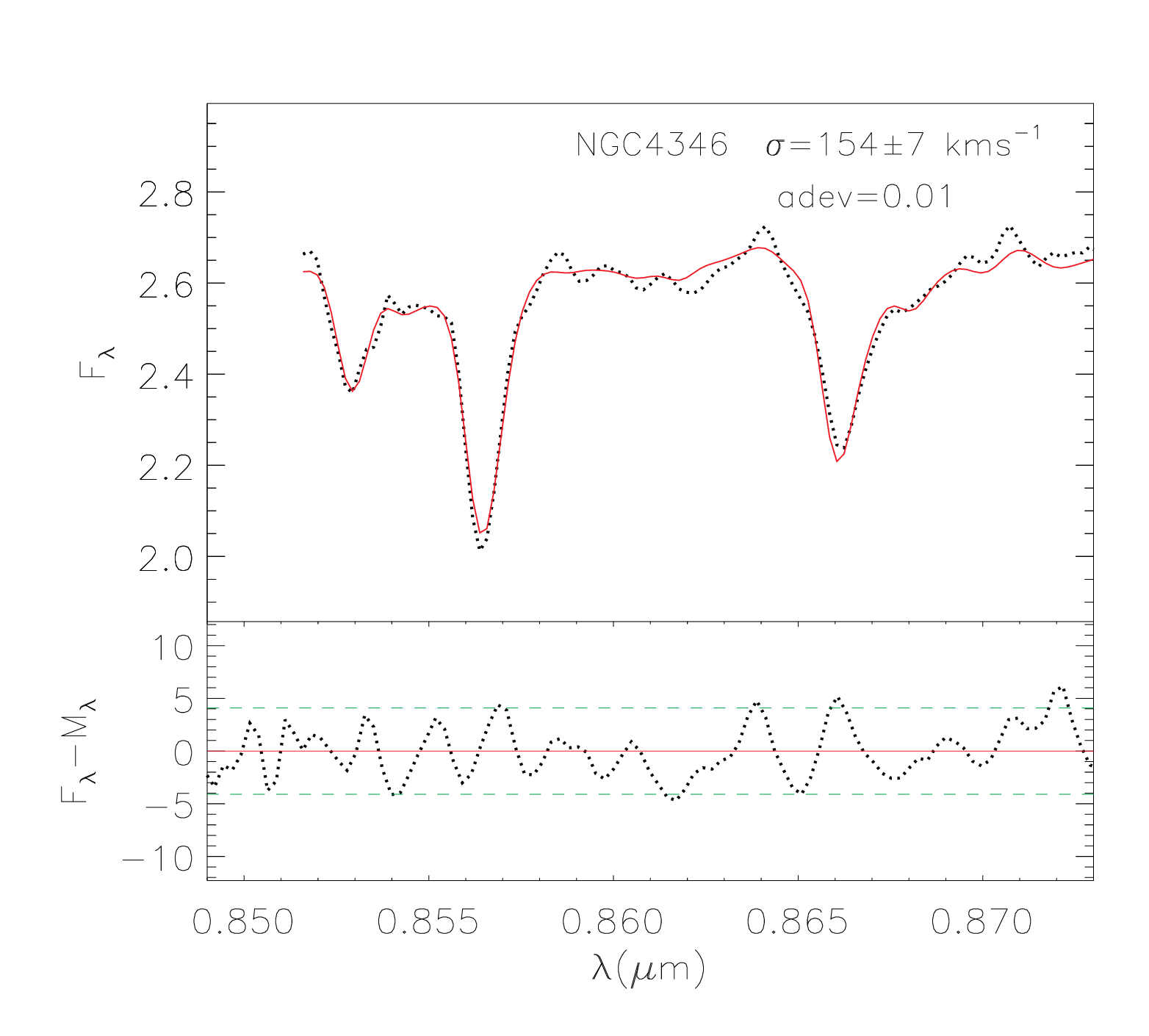} \\

  \end{tabular}
  \caption{Same as Fig.~\ref{fits-co} for the Ca triplet region.}
  \label{fits-ca}
\end{figure*}

\setcounter{figure}{1}
\begin{figure*}
\centering
\vspace{3 cm}
  \begin{tabular}{cccc}

    \includegraphics[scale=0.2]{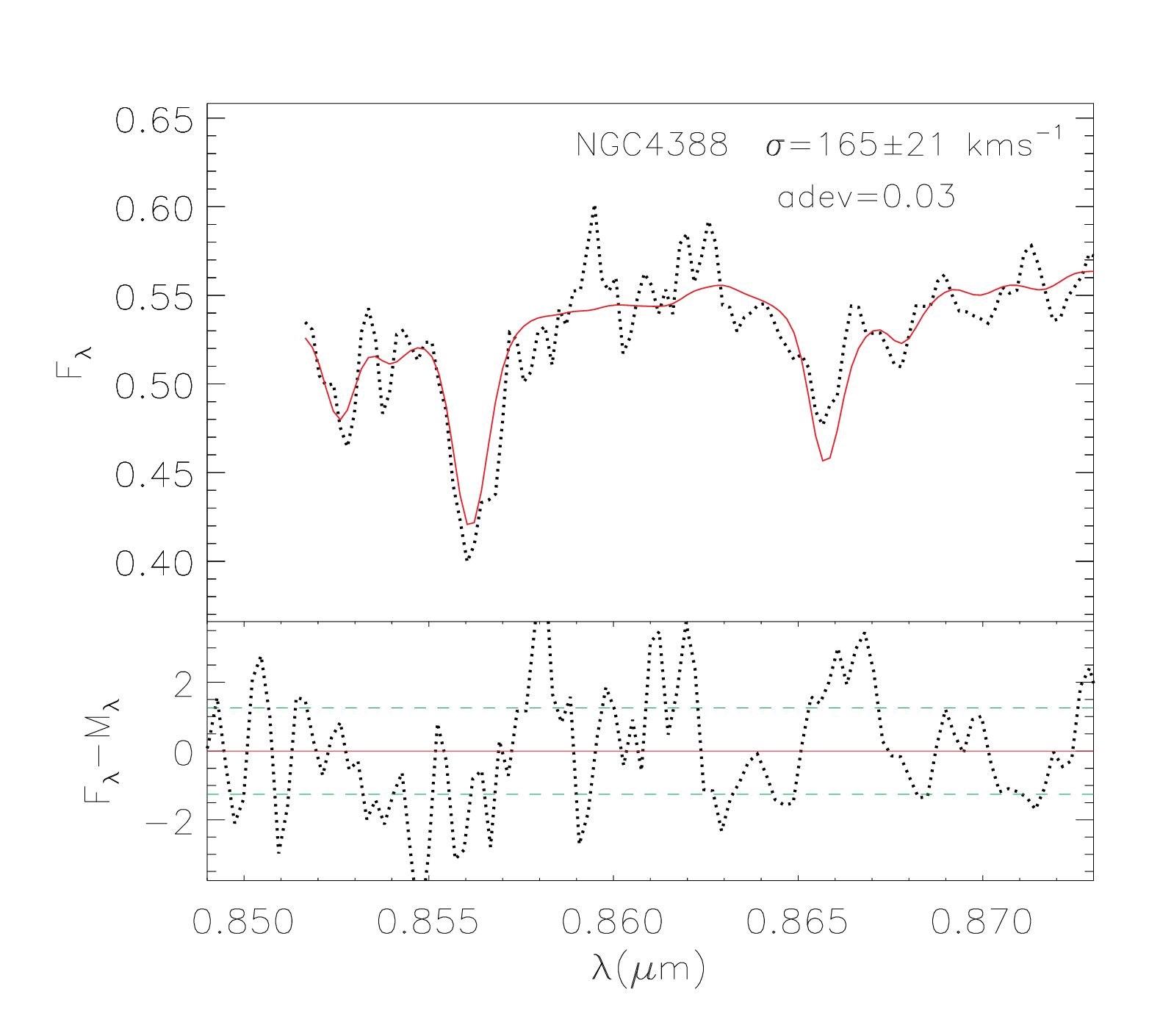}&
    \includegraphics[scale=0.2]{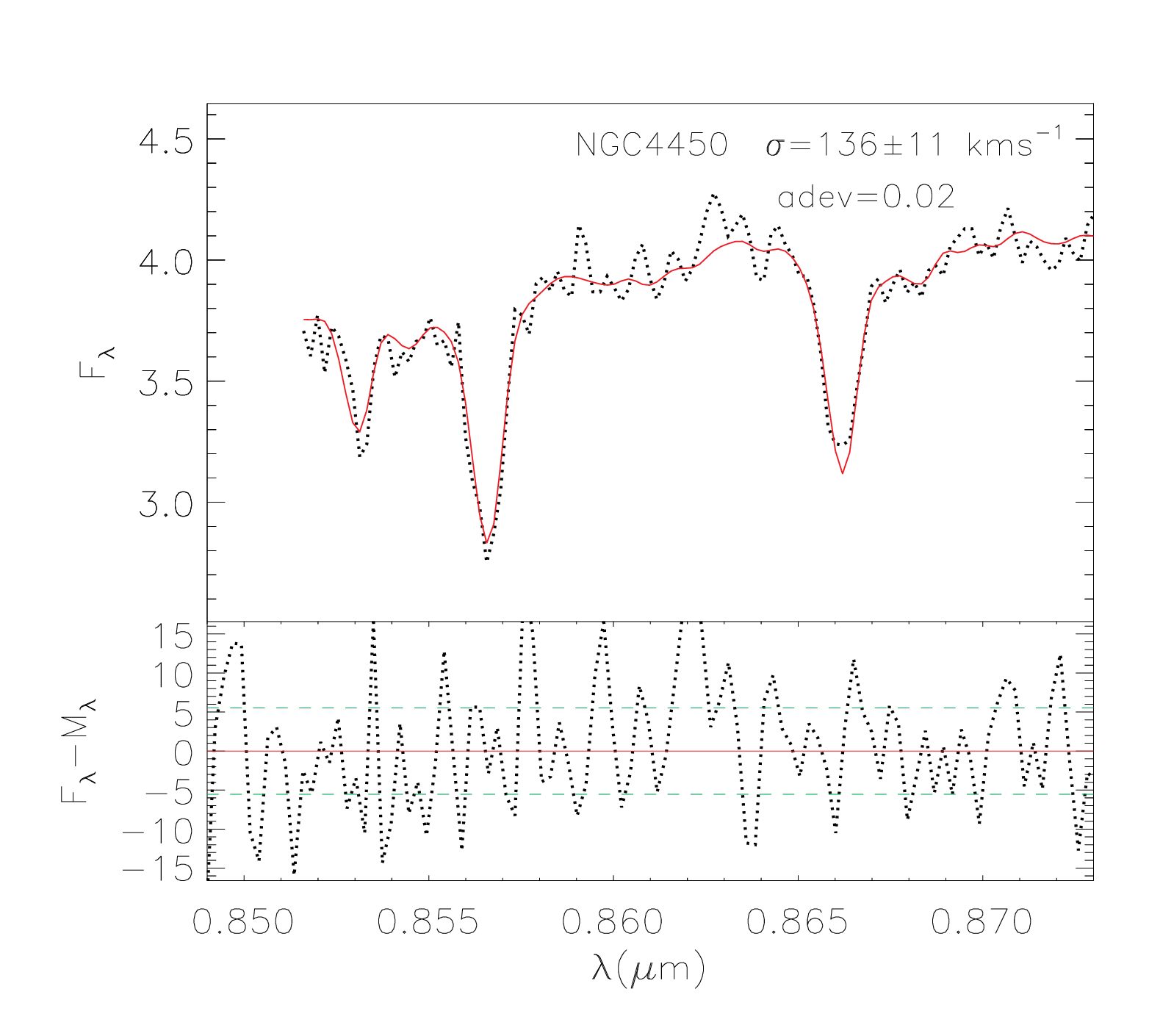}&
    \includegraphics[scale=0.2]{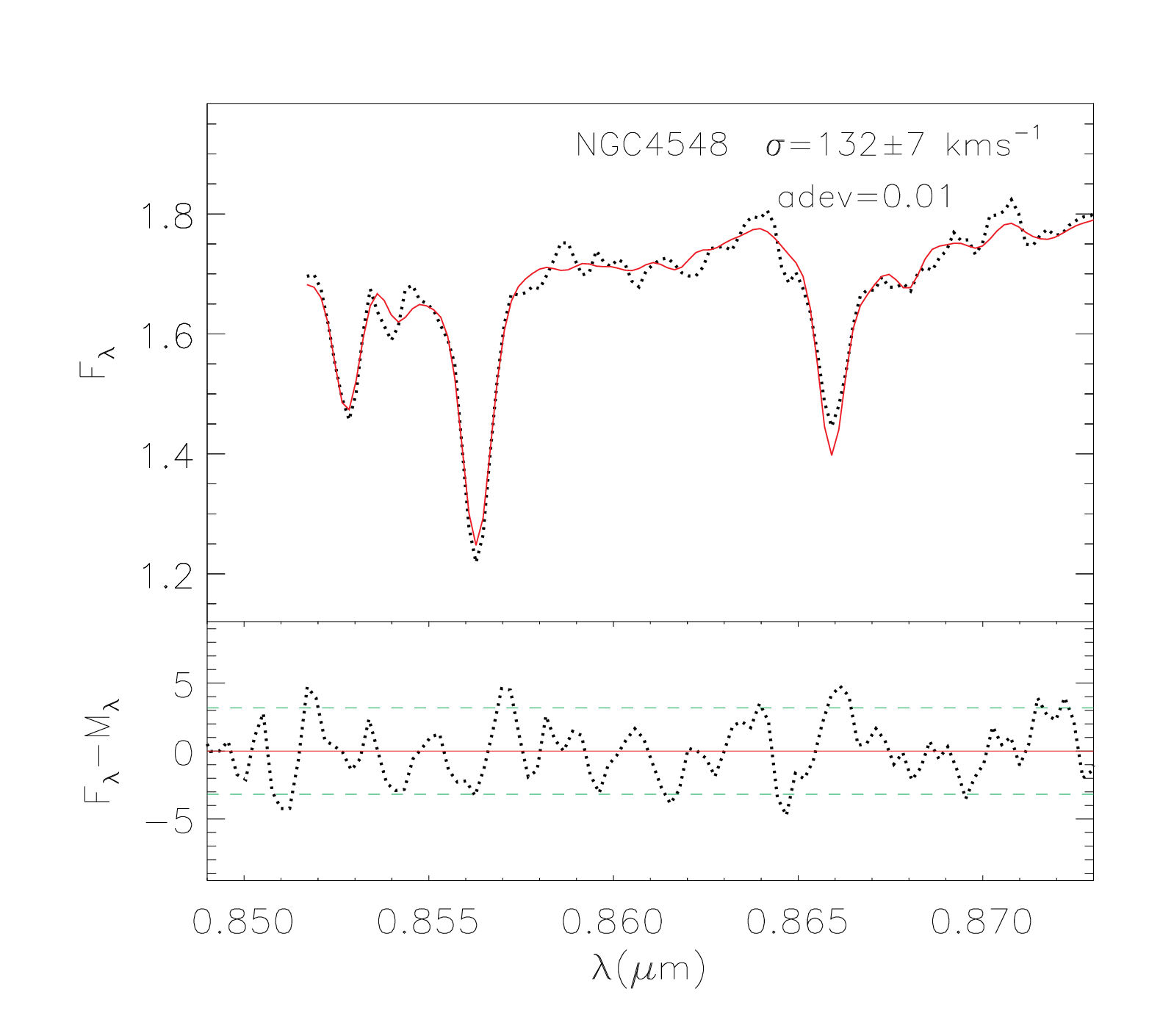}&    
    \includegraphics[scale=0.2]{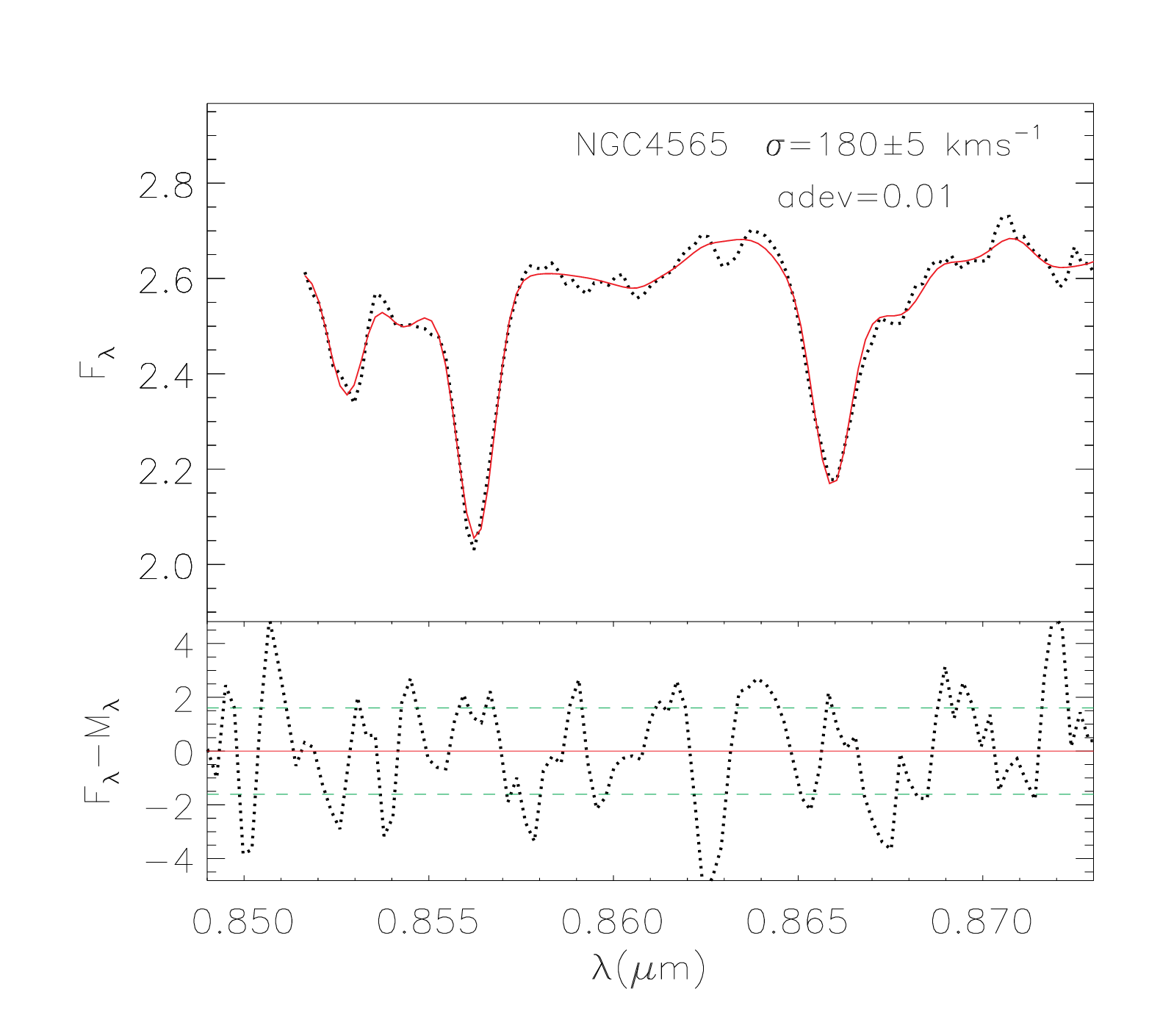} \\

    \includegraphics[scale=0.2]{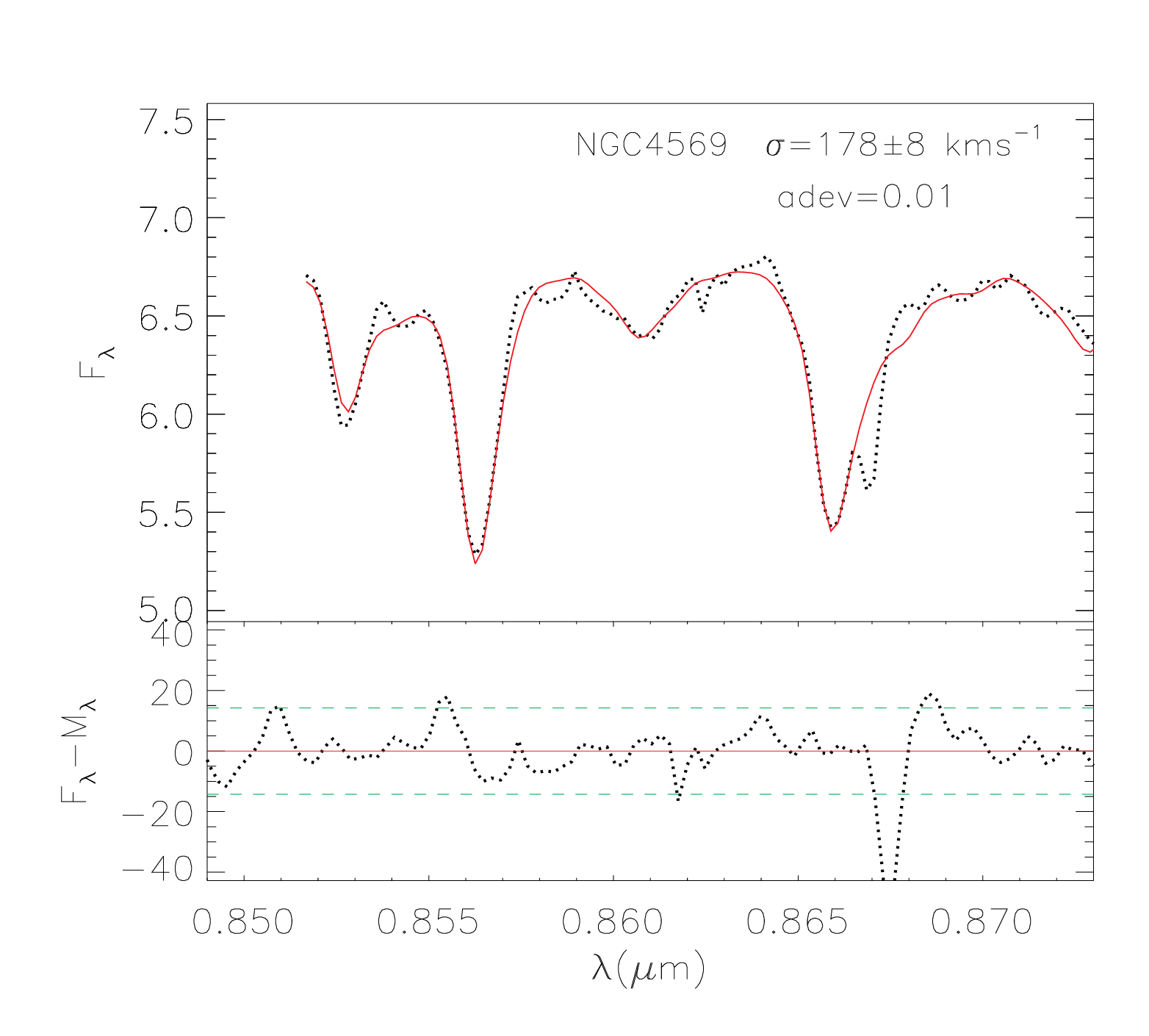}&
    \includegraphics[scale=0.2]{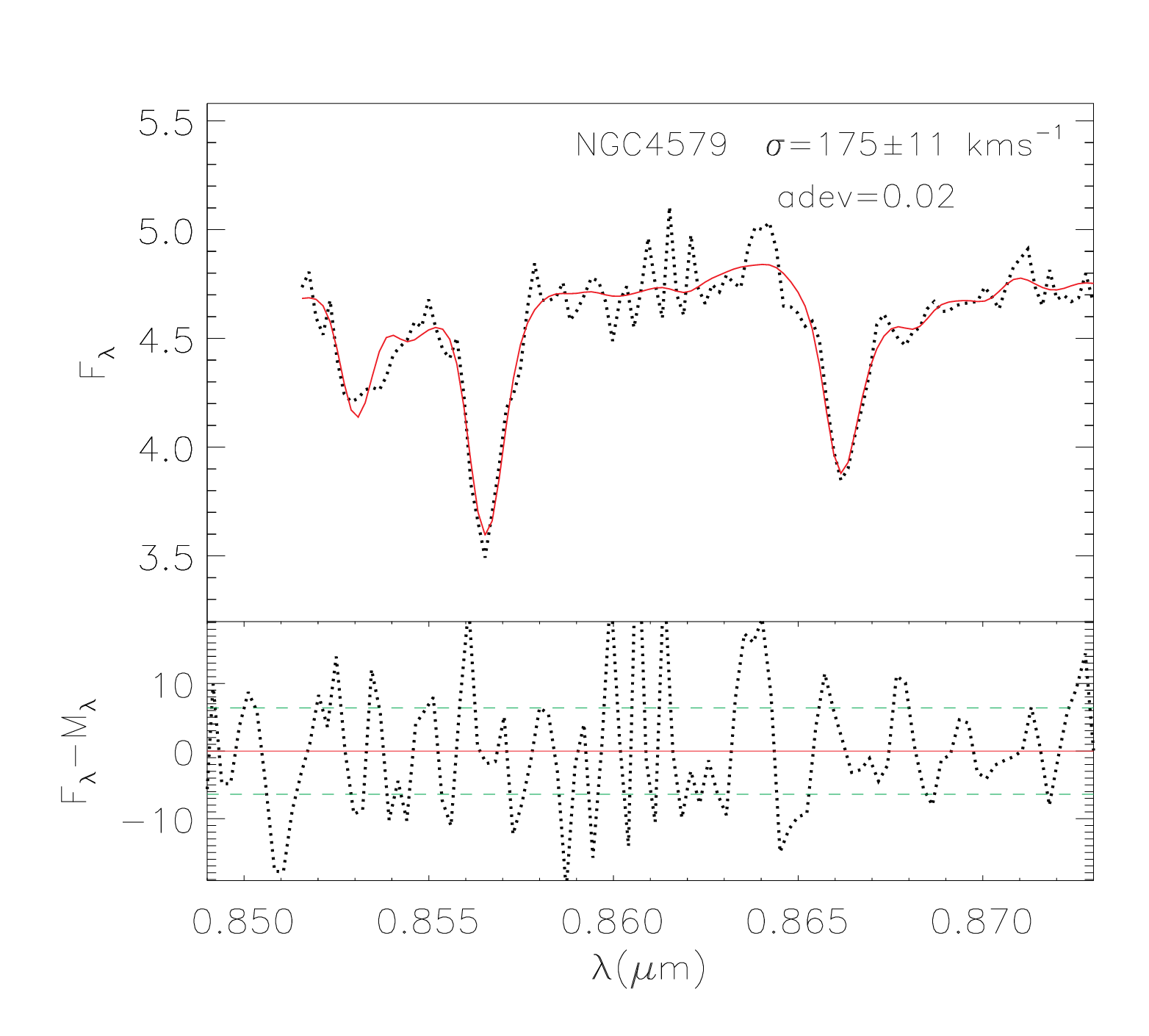}&
    \includegraphics[scale=0.2]{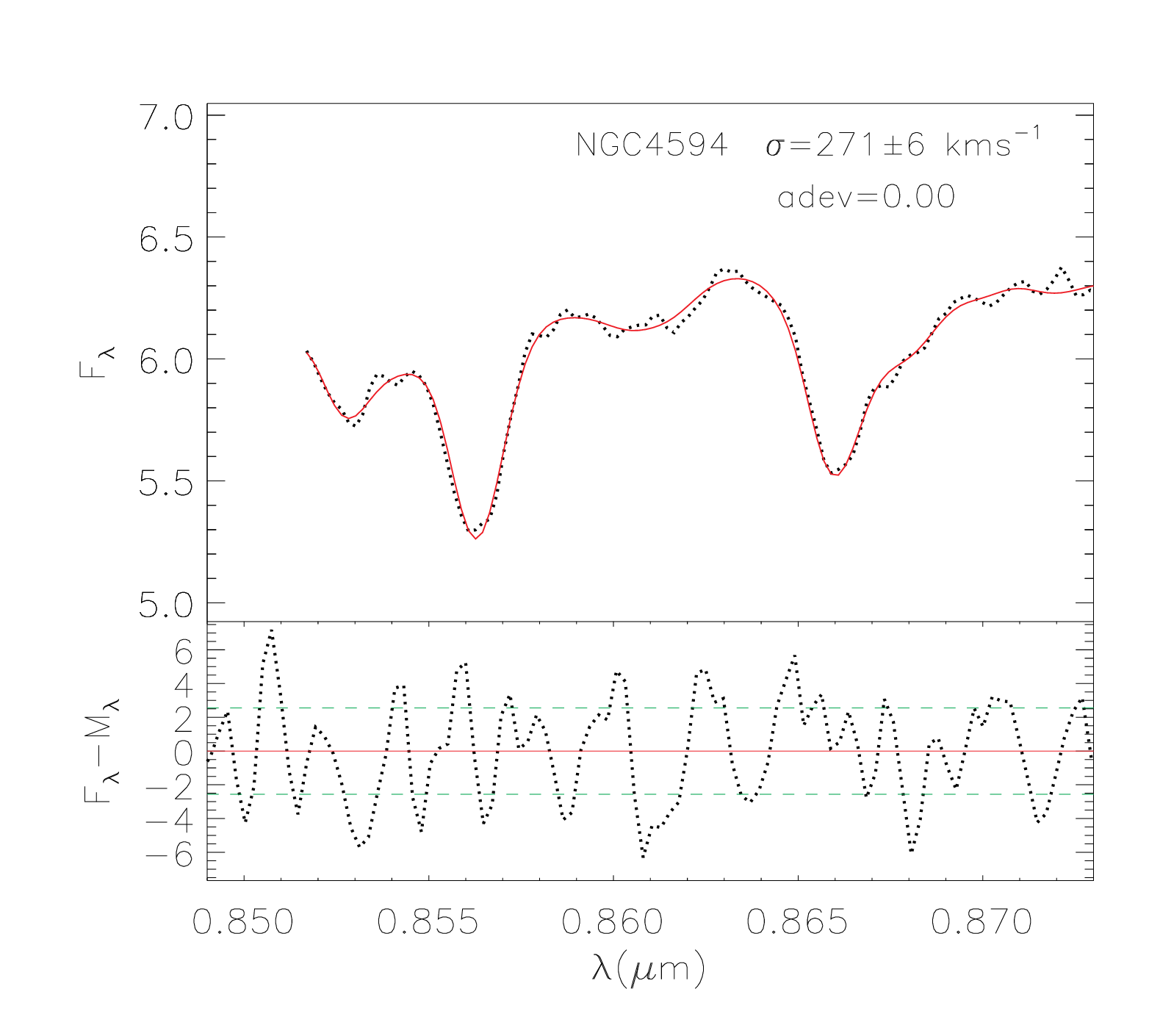}&    
    \includegraphics[scale=0.2]{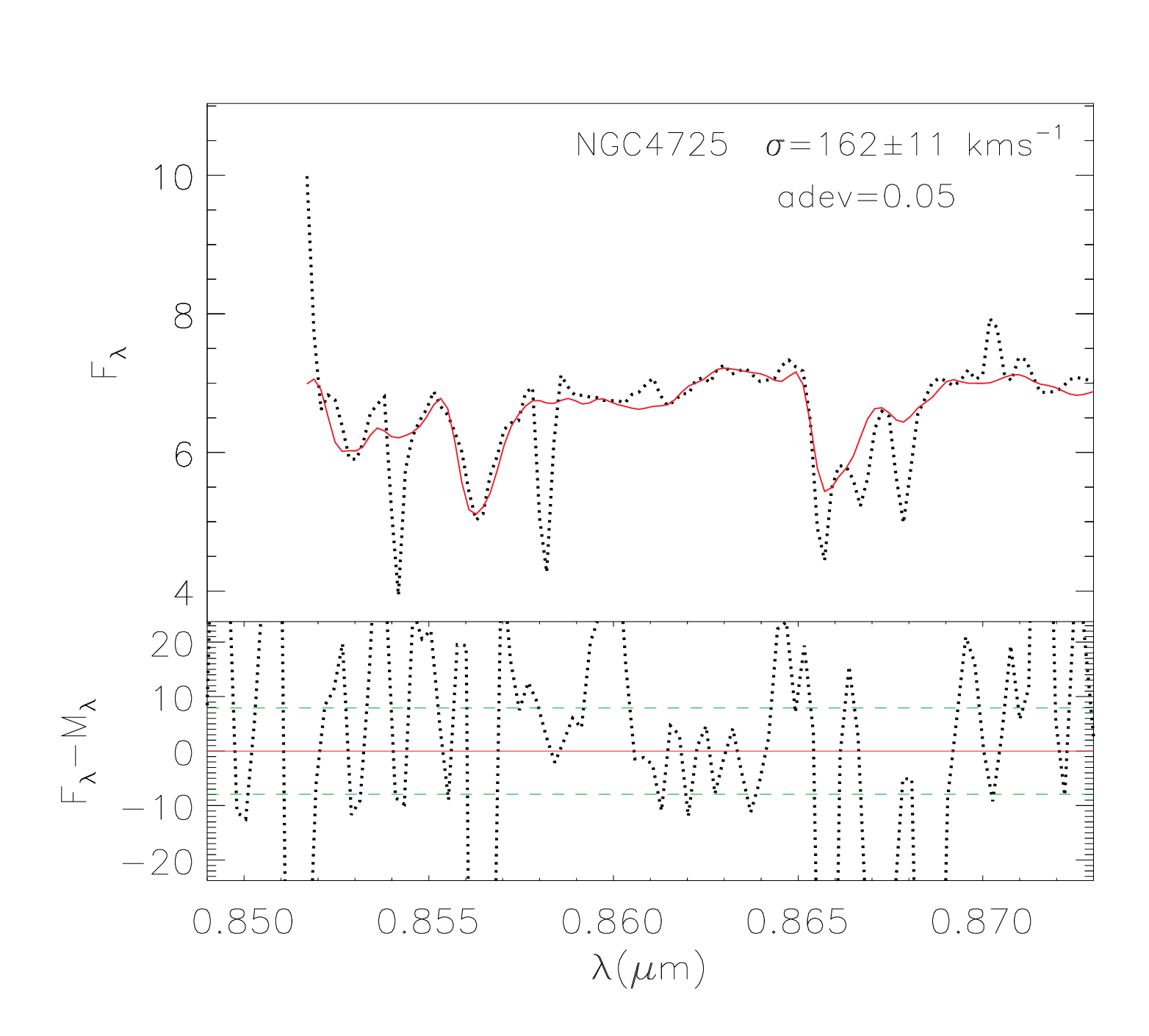} \\

    \includegraphics[scale=0.2]{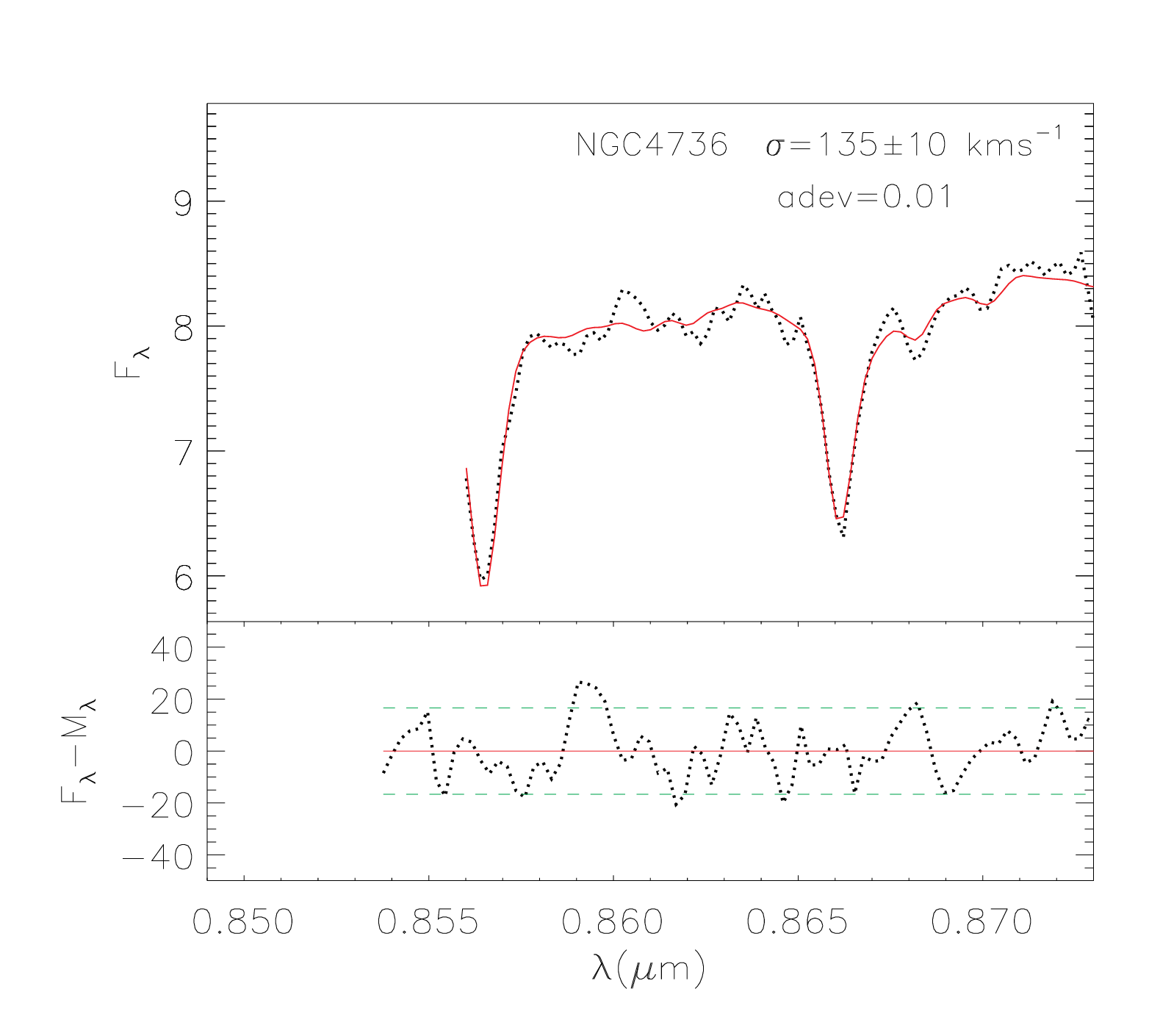}&
    \includegraphics[scale=0.2]{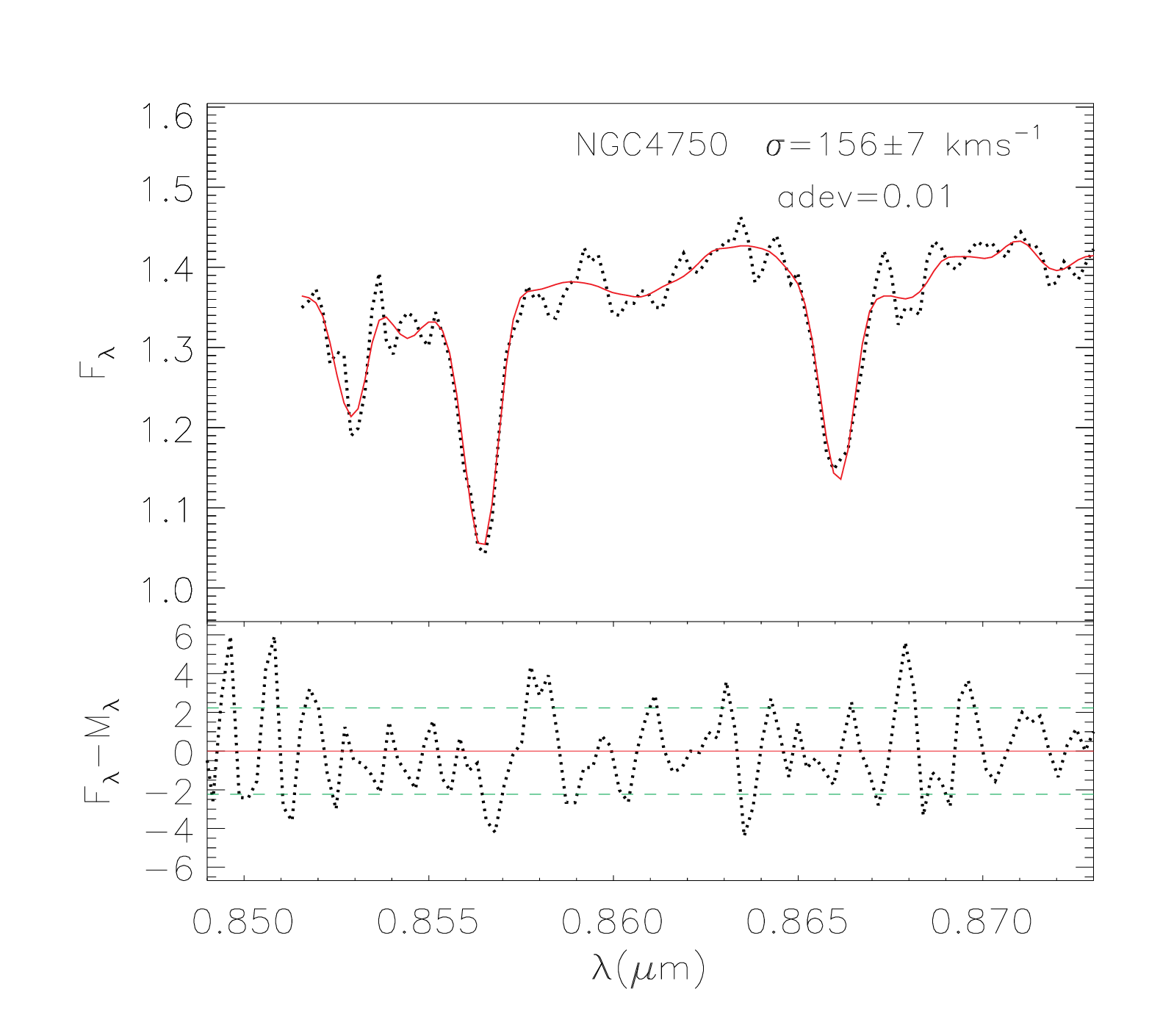}&
    \includegraphics[scale=0.2]{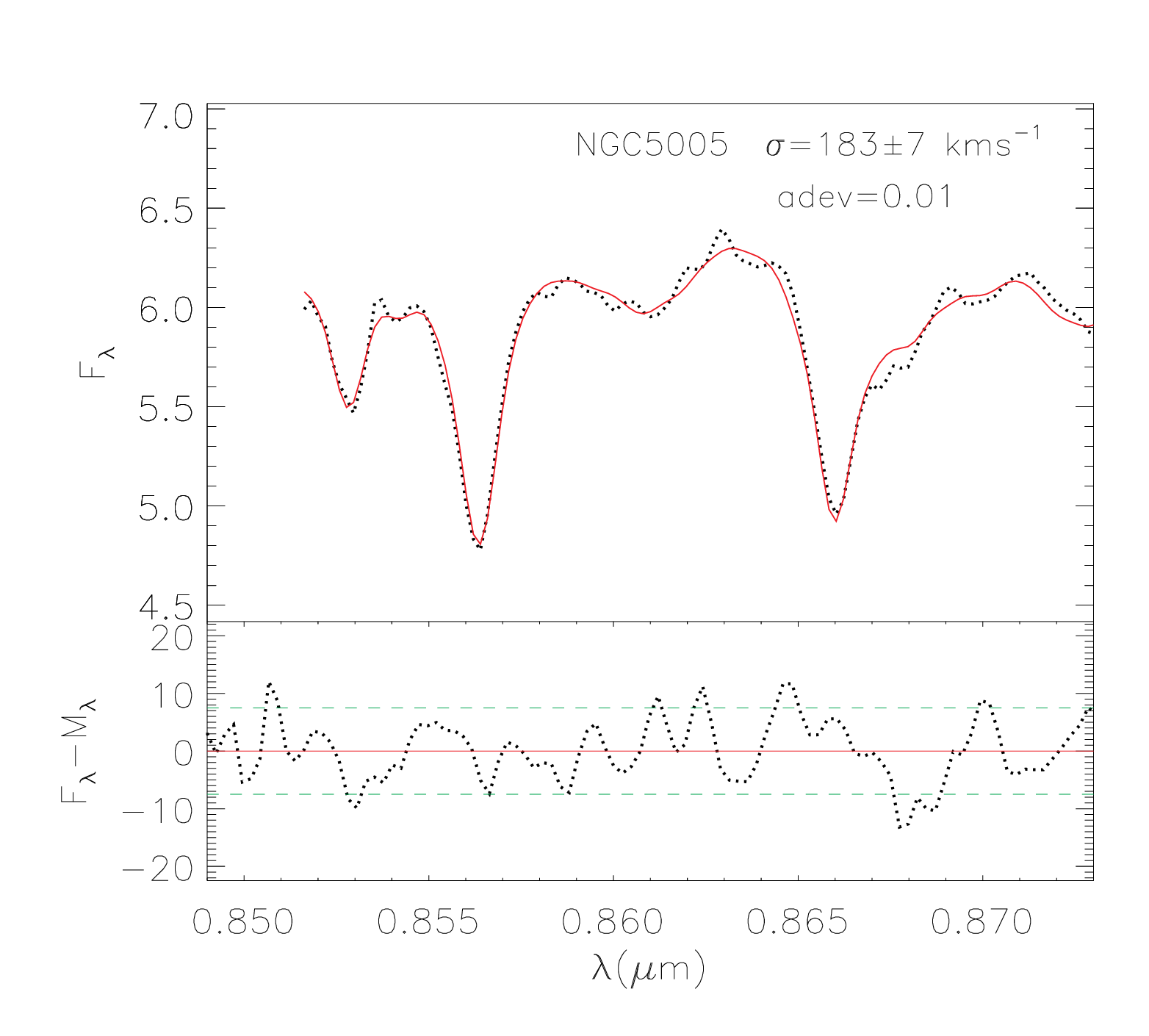}&    
    \includegraphics[scale=0.2]{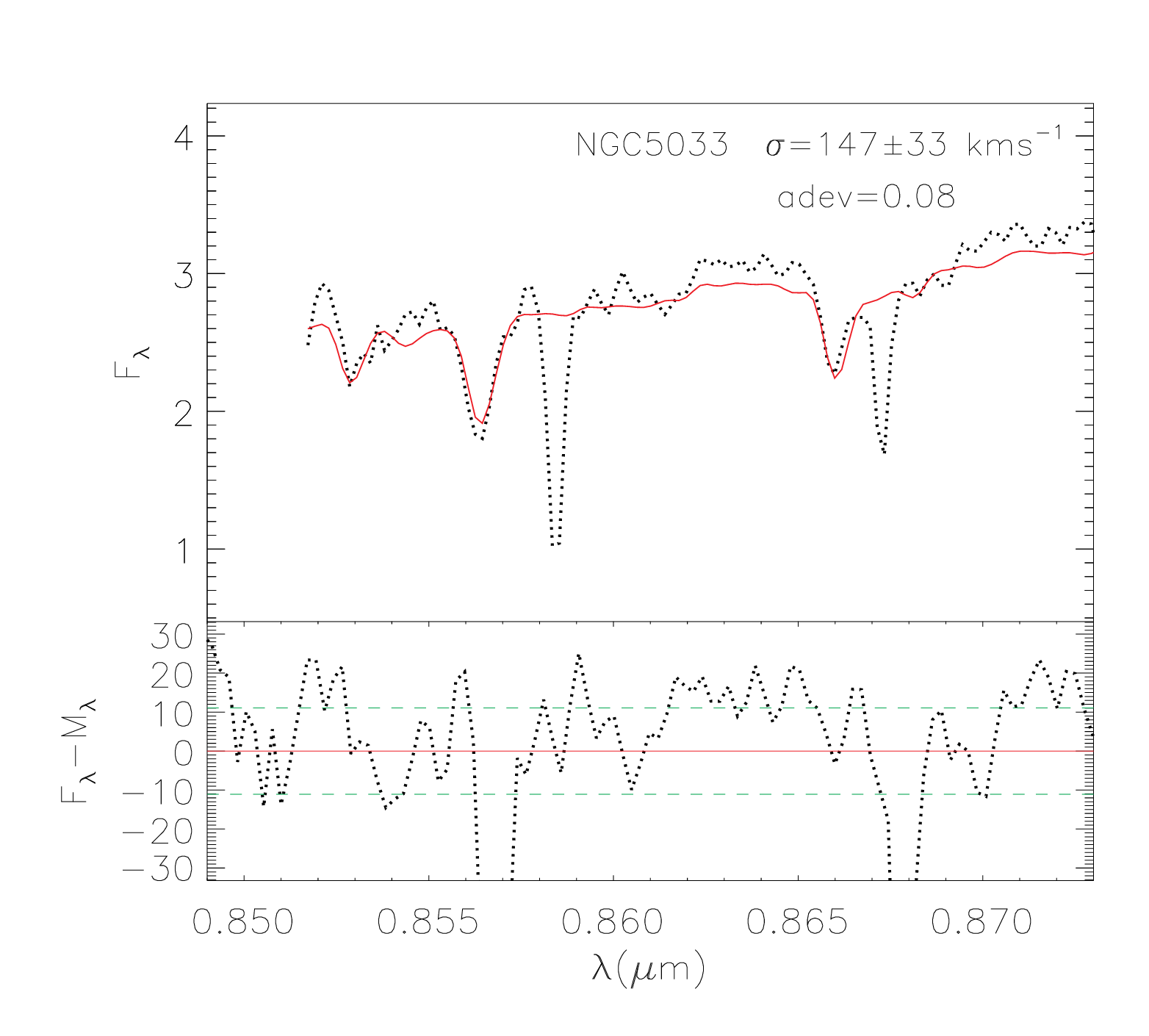} \\

    \includegraphics[scale=0.2]{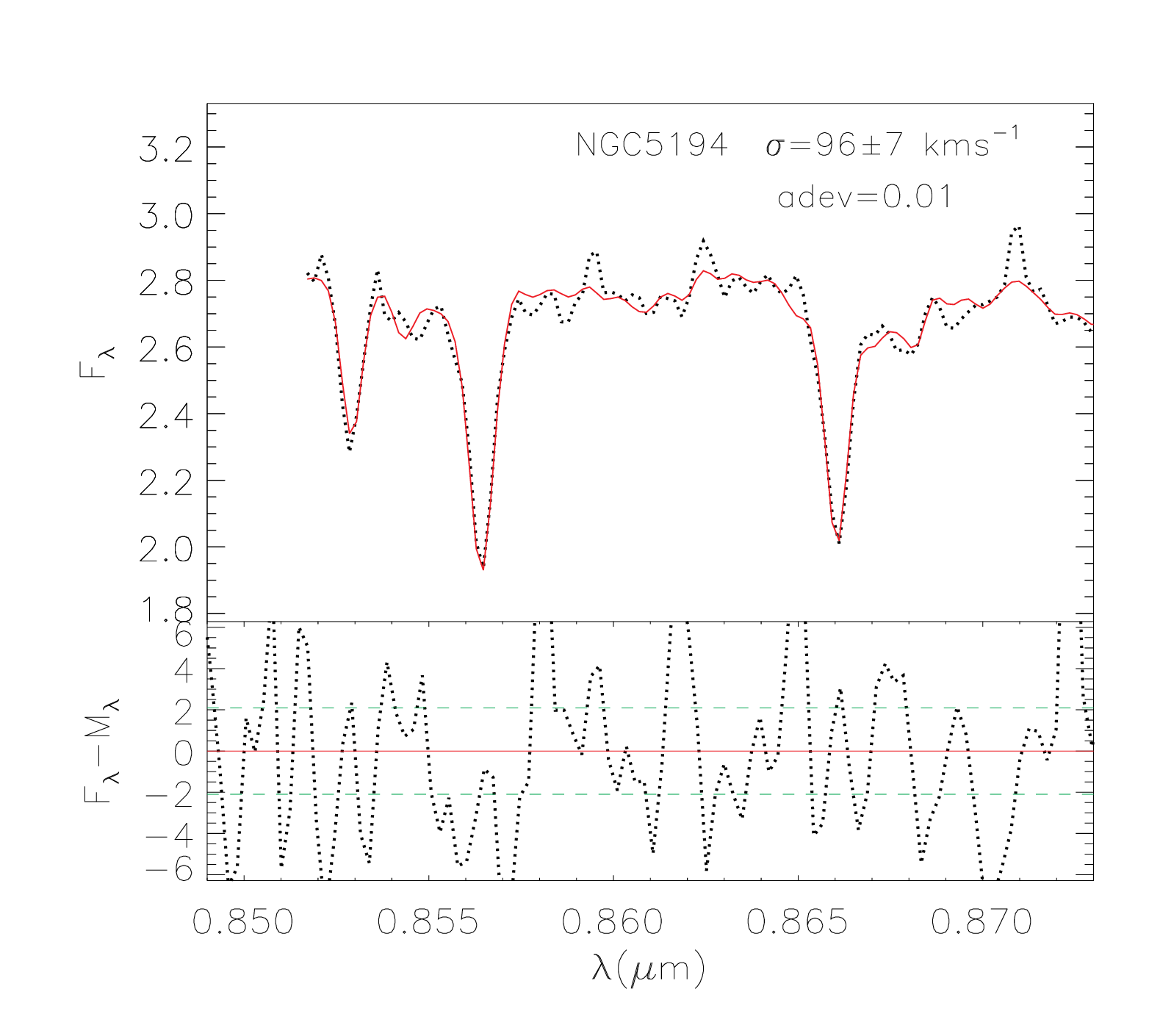}&
    \includegraphics[scale=0.2]{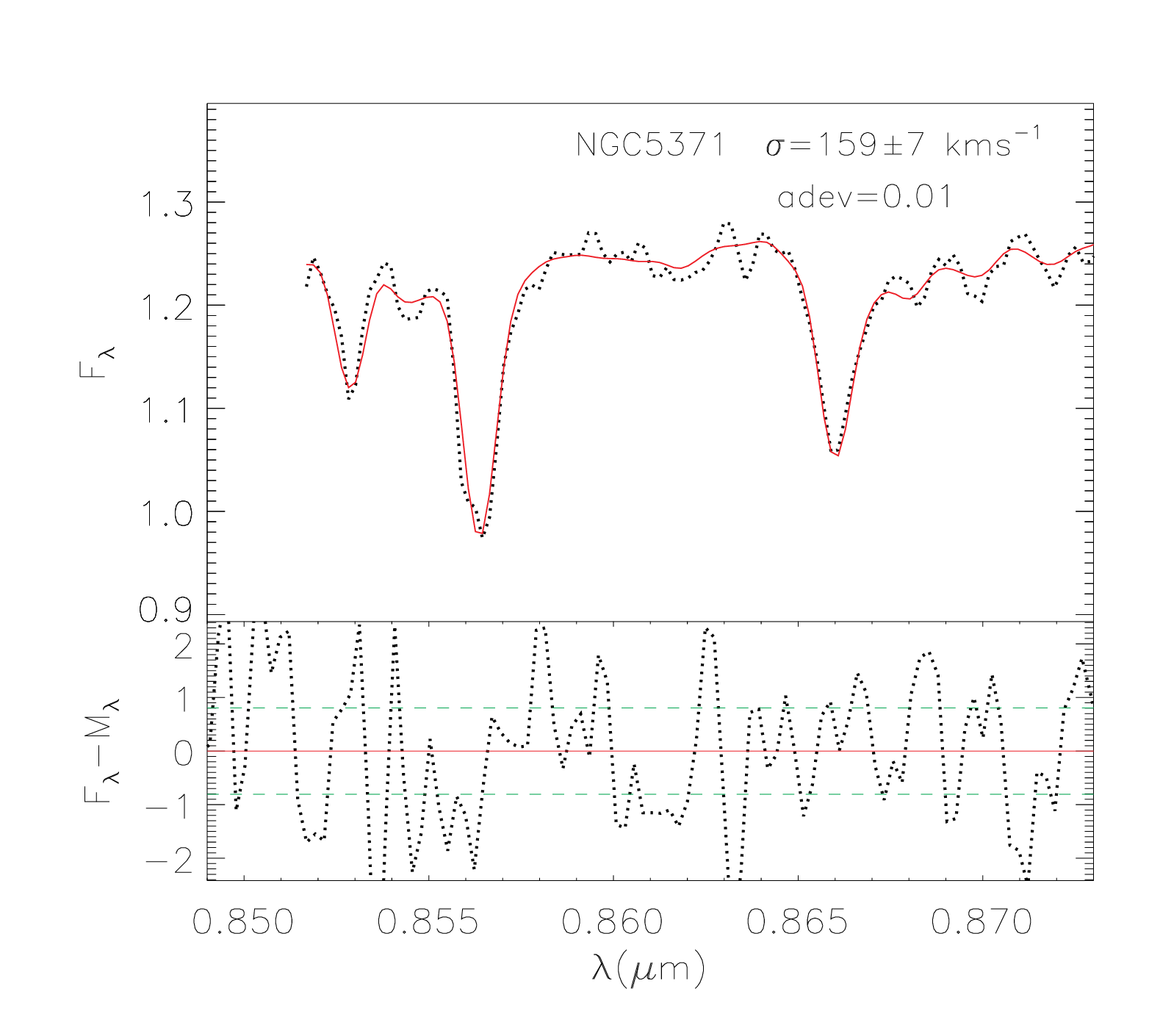}&
    \includegraphics[scale=0.2]{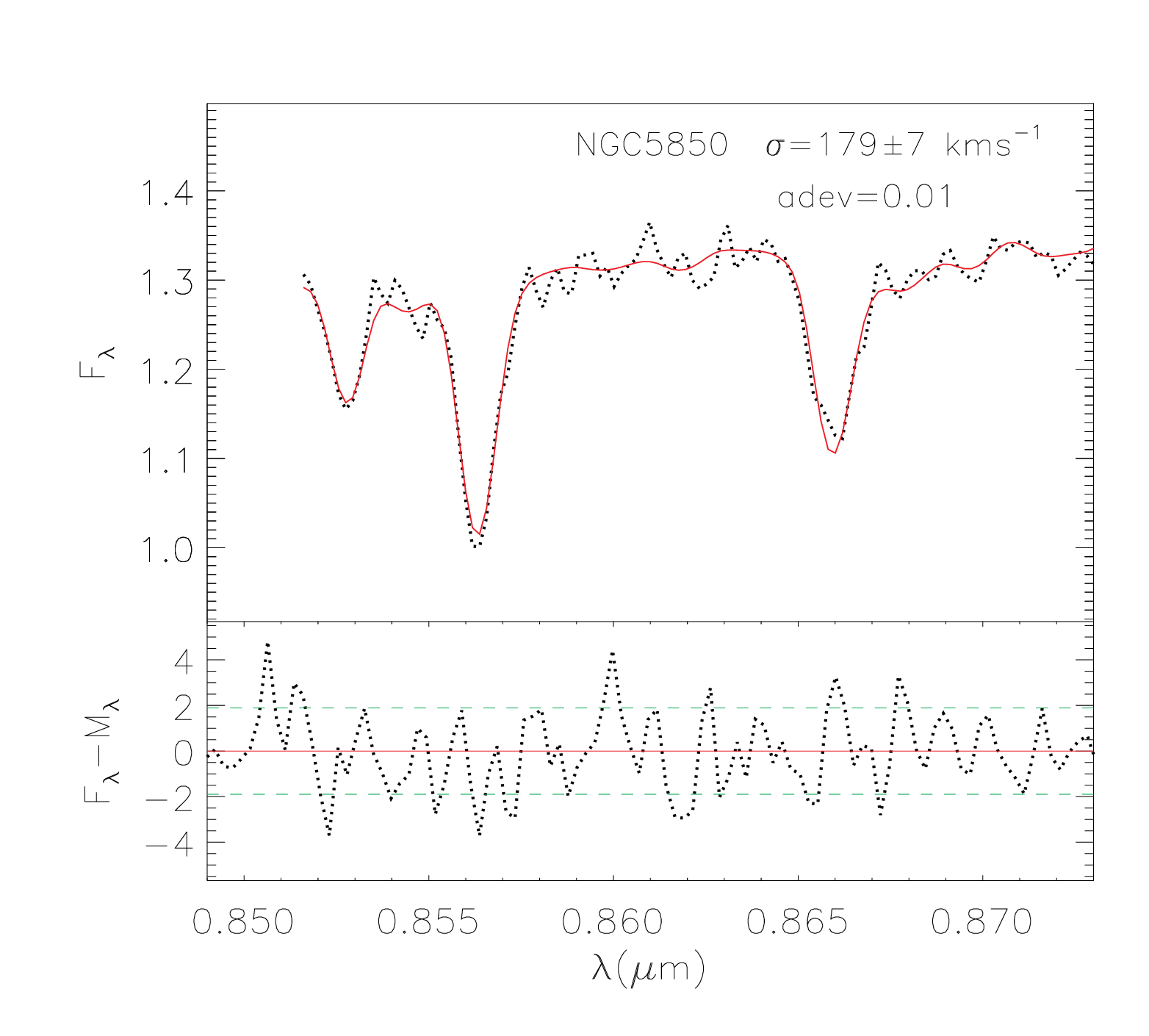}&    
    \includegraphics[scale=0.2]{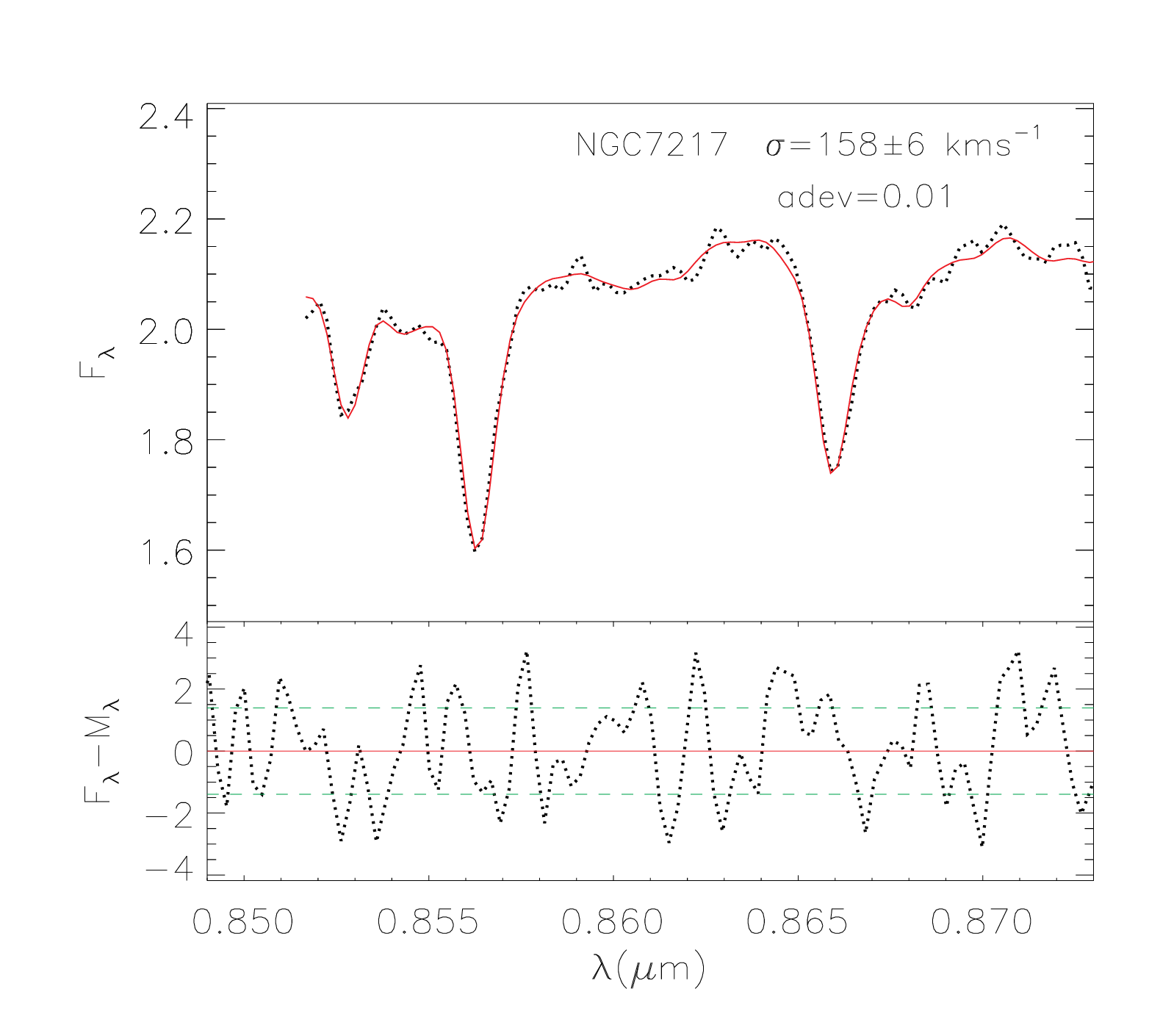} \\

    \includegraphics[scale=0.2]{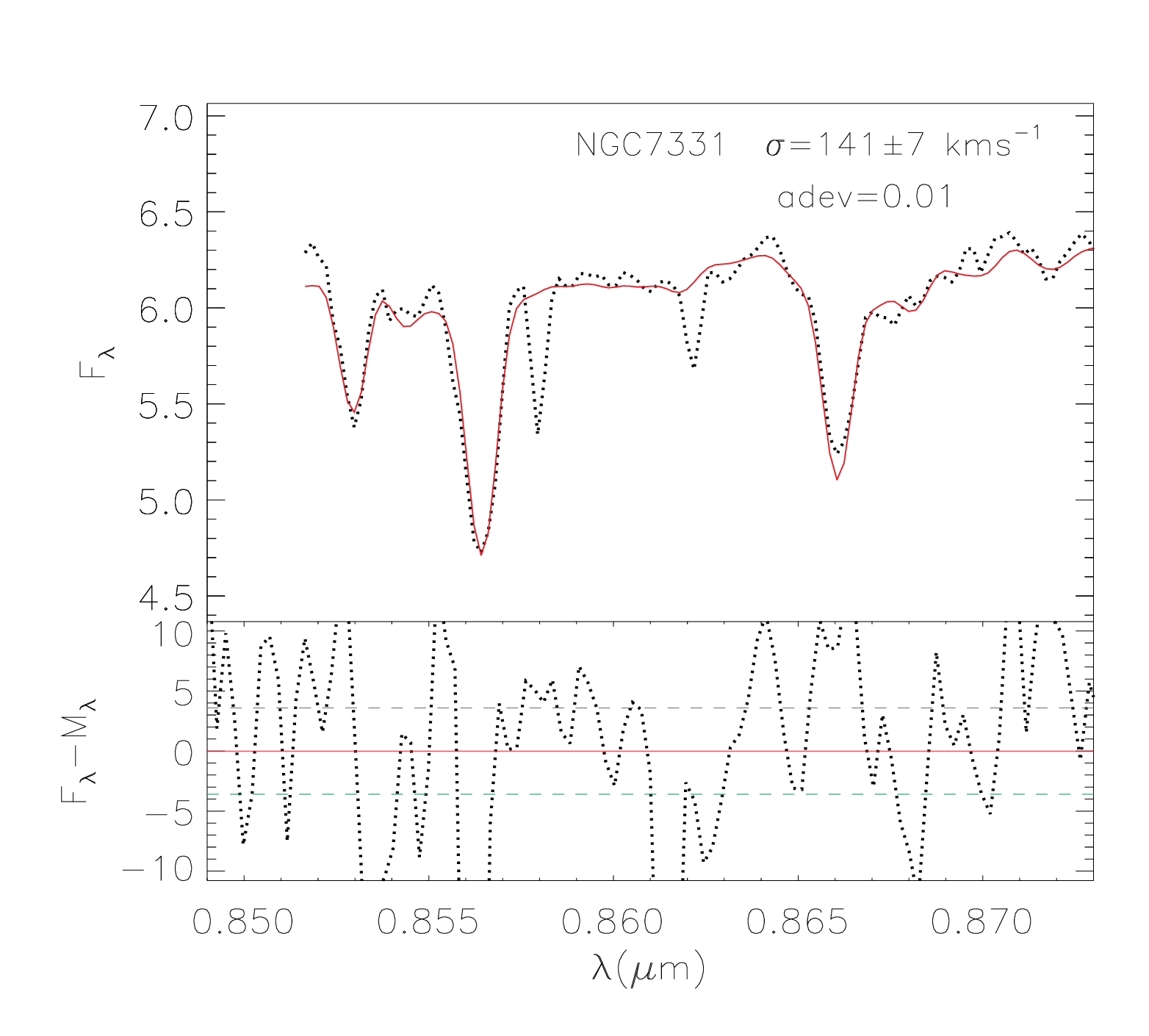}&
    \includegraphics[scale=0.2]{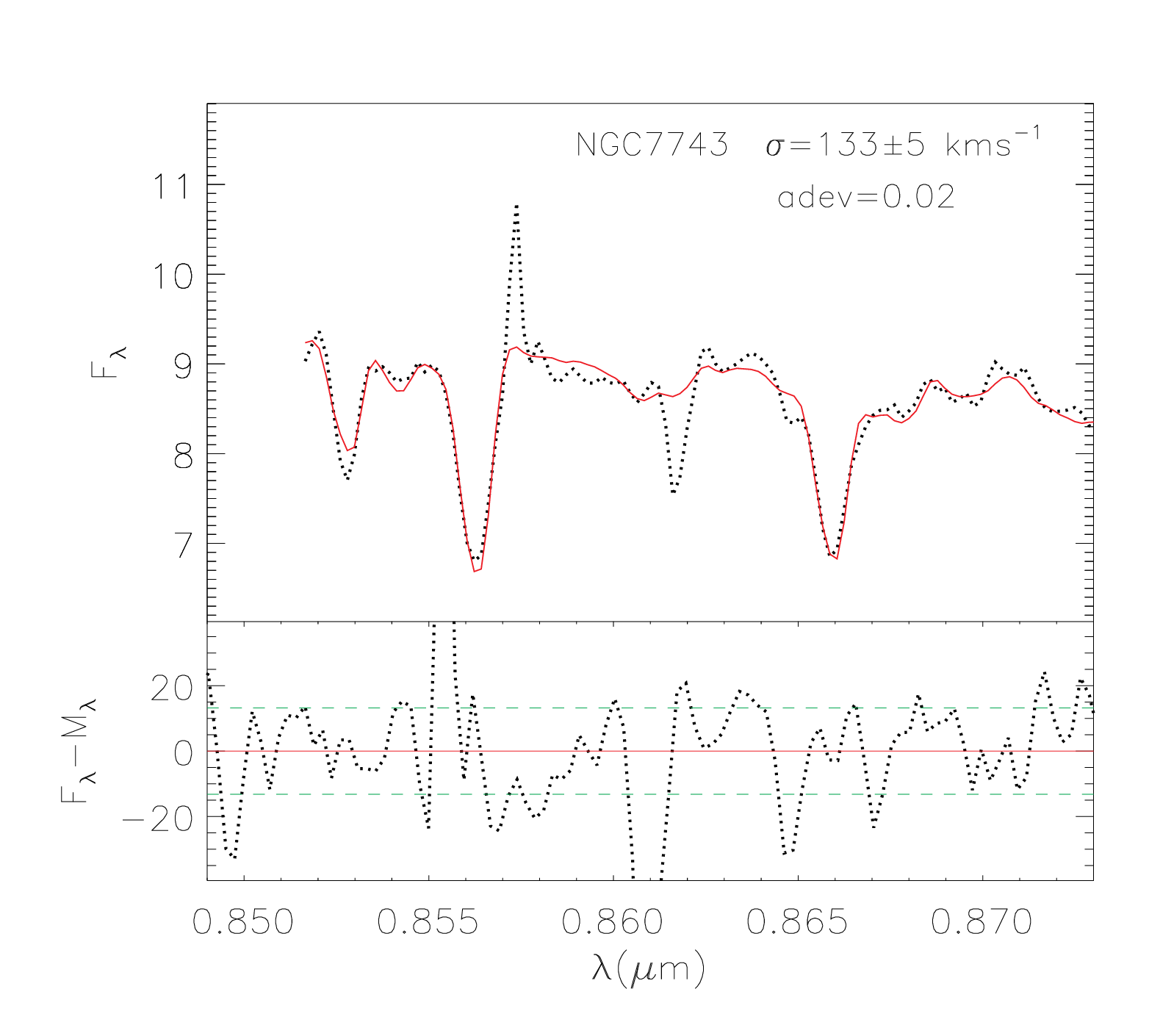}&
    &
  \\

  \end{tabular}
  \caption{(continued)}
\end{figure*}

\section{Stellar templates used to fit the stellar kinematics}\label{appendix}

Table \ref{table-apendix} shows the weights of each star (as well their spectral types) to the fit of the galaxy spectra for the CaT and CO spectral regions.

\setcounter{table}{0}
\begin{table*}
\caption{Stellar templates used for each galaxy to derive the stellar kinematics. Col 1: Galaxy name; Cols 2-4: Spectral and luminosity class, name and percentual contribution of the star to the fit of the galaxy spectrum for the Ca triplet region. Templates are from \citet{cenarro01} and for some cluster 
        stars, the authors list the positions of the stars in the HR diagram (SGB: subgiant branch; 
        GB: giant branch; HB: horizontal branch). Cols 5-7: Same as Cols 2-4 for the CO spectral region. Templates are from \citet{winge09}. }
\label{table-apendix}
\begin{tabular}{@{}lrlcrlc}
\hline
                    &\multicolumn{3}{|c|}{CaT region}								 & \multicolumn{3}{|c|}{CO region} \\
\hline
   Galaxy           &       Spectral Type   &             Star                      &   Weight (\%) &        Spectral Type		    &		  Star  		    &	Weight (\%) \\      
\hline
   NGC205           &        F3III          &             HD115604                  &    15	  &	K0III		    &		 HD105028		    &	  7 \\
                    &           GB          &            M67-F-108                  &    21	  &	 K2III		    &		  HD10598		    &	  5\\
                    &           HB          &             M71-1-41                  &     8	  &	G8III		    &		  HD107467		    &	 16\\
                    &           HB          &             M92-I-13                  &     3	  &	 M3III		    &		  HD27796		    &	  7\\
                    &       K0III           &             HD54810                   &    37	  &    M3III		    &		  HD112300		    &	  9\\
                    &        K3III          &             HD58972                   &     1	  &	M0III		    &		  HD2490		    &	 12\\
                    &     K4II              &             HD130705                  &     3	  &	K3III		    &		  HD4730		    &	 27\\
                    &        K5III          &             HD139669                  &     6	  &    K7III		    &		  HD63425B		    &	  1\\
                    &        K5III          &             HD149009                  &     1	  &	G3V		    &		  HD6461		    &	 11\\
                    &      M7.5III          &             HD126327                  &     2       &                         &                                       &  \\
   \hline  
    NGC266          &           GB          &            M67-F-108                  &    13       &	  M3III		    &  	   HD27796		            &     6  \\
                    &           HB          &             M5-II-76                  &    37       &	  M2III		    &  	   HD30354		            &    38  \\
                    &           HB          &             M71-C                     &     2       &    K8V		    &  	   HD113538		            &    41  \\
                    &      M5III            &             HD172816                  &    45    	  &	 M0III		    &  	   HD2490		            &     8  \\  
                    &        M6III          &             HD148783                  &     1       &    K7III		    &  	   HD63425B		            &     4  \\
   \hline     											 	    
   NGC315           &           GB          &            M67-F-108                  &    33       &	  M2III		    &  	   HD30354		            &    44 \\
                    &           GB          &        NGC188-II-122                  &    29       &    G2Ib		    &  	   HD209750		            &     6  \\
                    &           HB          &             M5-II-53                  &     4       &	 M0III		    &  	   HD2490		            &    21  \\
                    &        K5III          &             HD149009                  &    26       &	K0IV		    &  	   HD34642		            &    13  \\
                    &        M4III          &             HD17491                   &     2       &	G5II		    &  	   HD36079		            &    13  \\
                    &        M7III          &             HD207076                  &     3	  &                         &                                       &         \\
   \hline	            									  
   NGC404           &           GB          &            M67-F-108                  &    26	  &	   K0III	    &	    HD105028		            &    42  \\
                    &           HB          &             M5-II-76                  &    34	  &	  K3Iab		    &	     HD339034		            &     4  \\
                    &           HB          &           M92-XII-24                  &     2	  &	   M0III	    &	     HD2490		            &    42  \\
                    &        K5III          &             HD149009                  &    21   	  &	  K0IV		    &	     HD34642		            &    10 \\
                    &       M7III           &             HD114961                  &    15	  &                         &                                       &          \\
   \hline	       												 	   
   NGC410           &                       &                                       &     	  &	 M2III		    &		  HD30354		    &	 68  \\
                    &                       &                                       &    	  &	  K5II		    &		  HD3989		    &	 12  \\
                    &                       &                                       &    	  &    K8V		    &		  HD113538		    &	 17  \\
           	    &                       &                                       &             &     K6III		    &		  HD32440		    &	  2  \\
   \hline
   NGC474           &        F3III          &             HD115604                  &     1       &   M3III		    &  	   HD236791		            &     1        \\
                    &           GB          &          NGC7789-501                  &    25       &    M2III		    &  	   HD30354		            &    33        \\
                    &           HB          &             M92-I-13                  &    19       & K8V		            &  	   HD113538		            &     9        \\
                    &       K0III           &             HD54810                   &    37       & G2Ib		    &  	   HD209750		            &     3        \\
                    &          M6V          &         BD$+$19-5116-B                &    13       &   M0III		    &  	   HD2490		            &    29        \\
                    &          SGB          &            M67-IV-68                  &     4       &    K6III		    &  	   HD32440		            &    17        \\
   		    &                       &                                       &             &   K0IV		    &  	   HD34642		            &     5         \\
 	            &			    &				            &             &     K3Iab		    &		 HD339034		    &	 3  \\
  \hline
  NGC660	    &	             	    &	                		    &             &    G2Ib	            &		 HD209750		    &	 2   \\	  
   	            &	          	    &	                	            &             &    K7III		    &		 HD63425B		    &	15  \\
  		    &                       &                                       &             &     G8V		    &		 HD64606		    &	80  \\
   \hline
   NGC1052          &           HB          &             M5-II-53                  &     5       &      M3III 		    &	       HD236791 		    &    21    \\
                    &           HB          &             M92-I-13                  &    19       &       M3III 	    &	       HD27796  		    &     2    \\
                    &        K5III          &             HD139669                  &    34       &       M2III 	    &	       HD30354  		    &     4    \\
                    &       M7III           &             HD114961                  &    18       &    K8V 		    &	       HD113538 		    &    32    \\
                    &          SGB          &            M67-F-125                  &    22       &    K7III 		    &	       HD63425B 		    &    36   \\
           	    &                       &      				    &	          &      G3V 		    &	       HD6461			    &     3   \\
   \hline 
   NGC1167          &         F5VI          &             HD108177                  &    47       &       M2III 	    &	       HD30354  		    &    23  \\
                    &      G5IIIwe          &             HD88609                   &     1       &   M2III 	            &	     BD-01 3097 		    &     1  \\
                    &           GB          &        NGC188-II-122                  &    35       &    K8V 		    &	       HD113538 		    &    17 \\
                    &           HB          &           M92-XII-24                  &     6       &    G2Ib 		    &	       HD209750 		    &     4 \\
                    &       M7III           &             HD114961                  &     8       &      M0III 		    &	       HD2490			    &     8 \\
                    &                       &                                       &             &    K7III 		    &	       HD63425B 		    &    43 \\
  \hline  												
 \end{tabular}
   \end{table*}

\setcounter{table}{0}
\begin{table*}
\caption{(continued)}
\begin{tabular}{@{}lrlcrlc}  
\hline
                    &\multicolumn{3}{|c|}{CaT region}								 & \multicolumn{3}{|c|}{CO region} \\
\hline
   Galaxy           &       Spectral Type   &             Star                      &   Weight (\%) &        Spectral Type		    &		  Star  		    &	Weight (\%) \\      
\hline
   NGC1358          &           HB          &             M71-1-34                  &    44       &       M2III		    &		HD30354 		    &    57  \\
                    &           HB          &             M71-C                     &     9       &     M2	            &	       BD$+$59 274		    &	7  \\
                    &       K4III           &             HD149161                  &     6       &    K8V		    &		HD113538		    &    33  \\
                    &      M5III            &             HD172816                  &    24       &      M0III		    &		HD2490  		    &	1  \\
                    &       M7III           &             HD114961                  &    15       &                         &                                       &        \\
 \hline				
   NGC1961          &           GB          &          NGC188-I-57                  &    27       &	 M2III		    &		  HD30354		    &	 36  \\   
                    &          K0V          &             HD149661                  &     9       &    K8V		    &		  HD113538		    &	 24 \\
                    &        K5III          &             HD139669                  &    39       &	M0III		    &		  HD2490		    &	 39 \\
                    &      M5.5III          &             HD94705                   &    19       &                         &                                       &      \\
                    &          M6V          &         BD$+$19-5116-B                &     2       &                         &                                       &       \\   
                    &       M7III           &             HD114961                  &     1       &                         &                                       &        \\
   \hline  											
   NGC2273          &      HB             &             M5-II-53                  &     8         &	    M2III	       &	     HD30354		      &    18 \\
                    &     K4II            &             HD130705                  &    39         &	  K3Iab		       &	     HD339034		      &     1 \\
                    &        K5III        &             HD139669                  &     1     	  &	 K8V		       &	     HD113538		      &     1 \\
                    &        M4III        &             HD17491                   &    20     	  &	   M0III	       &	     HD2490		      &    19 \\
                    &      M5III          &             HD172816                  &     8     	  &	  K0IV		       &	     HD34642		      &     4 \\
                    &      M5.5III        &             HD94705                   &     1     	  &	  G5II		       &	     HD36079		      &    33 \\
                    &        M5III        &             HD175865                  &     7     	  &	 K7III		       &	     HD63425B		      &     3 \\
                    &          SGB        &            M67-F-115                  &    10     	  &	  G8V		       &	     HD64606		      &    11 \\
                    &          SGB        &            M67-F-125                  &     5     	  &	   G3V		       &	     HD6461		      &     6 \\
   \hline
      NGC2639	   &	         	  &	                		  &       	  &	  M2III		       &  	   HD30354		      &    42  \\
      		   &			  &					  &		  &    M3Iab		       &  	    BD$+$39 4208  	      &    11  \\
      		   &			  &					  &		  &    G2Ib		       &  	   HD209750		      &    29  \\
		   &		  	  &					  &		  &	 M0III		       &  	   HD2490		      &    16  \\
\hline												       												          
   NGC2655          &        F3III        &             HD115604                  &     6	  &	  M2III		       &  	   HD30354		      &     7  \\
                    &           GB        &            M67-F-108                  &     7	  &	K3Iab		       &  	   HD339034		      &     1  \\
                    &           HB        &             M5-II-53                  &     8	  &    K8V		       &  	   HD113538		      &    25  \\
                    &        K5III        &             HD139669                  &    31	  &	 M0III		       &  	   HD2490		      &    40  \\
                    &        M6III        &             HD18191                   &    22	  &	K0IV		       &  	   HD34642		      &     1  \\
                    &          SGB        &            M67-F-125                  &    24	  &    K7III		       &  	   HD63425B		      &    23  \\
  \hline	  
   NGC2768          &           GB        &          NGC188-I-85                  &     8	  &	 M2III		       &		  HD30354	      &	 23  \\
                    &           HB        &             M71-1-41                  &    16	  &     K3Iab		       &		  HD339034	      &	  1  \\
                    &           HB        &           M92-XII-24                  &     4	  &    K8V		       &		  HD113538	      &	 19  \\
                    &        K3III        &             HD102328                  &    37	  &    K1II		       &		  HD198700	      &	  3  \\
                    &        M6III        &             HD18191                   &    24	  &	M0III		       &		  HD2490	      &	 49  \\
                    &          SGB        &            M67-F-125                  &     7	  &     K0IV		       &		  HD34642	      &	  2  \\
  \hline         									
   NGC2832          &           F0        &           BD-01-2582                  &     1	  &                            &                                      &    1   \\
                    &           HB        &             M71-C                     &    41	  &	M2III		       &		 HD30354	      &	30  \\
                    &           HB        &           M92-XII-24                  &     2	  &    K8V		       &		 HD113538	      &	17  \\  
                    &      M5.5III        &             HD94705                   &    20	  &      M0III		       &		 HD2490 	      &	51  \\
                    &          M6V        &         BD$+$19-5116-B                &    16	  &                            &                                      &        \\
                    &        M7III        &             HD207076                  &     6         &                            &                                      &        \\
                    &          SGB        &            M67-F-125                  &    10         &                            &                                      &        \\
   \hline
    NGC3031	   &	       GB  	  &	     M67-F-231  		  &     1         &	    M3III		&	     HD27796		      &     5      \\
         	   &	       GB  	  &	     M92-XII-8  		  &     2         &	    M2III		&	     HD30354		      &    14      \\
         	   &	       GB  	  &	 NGC188-II-122  		  &     6         &	 K8V		        &      HD113538		              &    21     \\
         	   &	       HB  	  &	      M5-II-76  		  &     1         &	   M0III		&	     HD2490		      &    10     \\
         	   &	       HB  	  &	      M92-I-13  		  &     3         &	  K0IV		        &	     HD34642		      &     2     \\
         	   &	    K5III  	  &	      HD139669  		  &     5         &	  G5II		        &	     HD36079		      &     8     \\
         	   &	  M5III  	  &	      HD172816  		  &    25         &	 K7III		        &	     HD63425B		      &    25     \\
         	   &	   M7III 	  &	      HD114961  		  &    18         &	   G3V		        &	     HD6461		      &    10    \\
         	   &	      SGB  	  &	     M67-F-115  		  &     7         &	   K4III	        &	     HD9138		      &     1     \\
         	   &	      SGB  	  &	     M67-IV-68  		  &    29         &                             &                                     &          \\
  \hline
  NGC\,3079        &                      &                                       &               &	  M2III		        &  	   HD30354		      &    12    \\
 	           &			  &				          &	          &     K3Iab		        &		 HD339034	      &	 2   \\
 		    &			  &					  &               &    K8V		        &		HD113538	      &    18   \\
 		    &			  &					  &               &      M0III		        &		HD2490  	      &    39   \\
 		    &			  &					  &	          &      K7III		        &  	   HD63425B		      &       \\
\hline  												
\end{tabular}
\end{table*}

\setcounter{table}{0}
\begin{table*}
\caption{(continued)}
\begin{tabular}{@{}lrlcrlc}
\hline
                    &\multicolumn{3}{|c|}{CaT region}								 & \multicolumn{3}{|c|}{CO region} \\
\hline
   Galaxy           &       Spectral Type   &             Star                      &   Weight (\%) &        Spectral Type		    &		  Star  		    &	Weight (\%) \\      
\hline
   NGC3147          &          G2V        &            Hya-vB-64                  &     2         &	    M2III	        &	     HD30354		      &    26  \\
                    &           GB        &           NGC188-I-61                 &     8         &	 K8V		        &	     HD113538		      &    21  \\
                    &           HB        &             M71-C                     &     1         &	 G2Ib		        &	     HD209750		      &     6  \\
                    &        K0III        &             HD88284                   &    22         &	   M0III	        &	     HD2490		      &     5 \\\
                    &     K4II            &             HD130705                  &    46         &	  K0IV		        &	     HD34642		      &     1  \\
                    &          M6V        &         BD$+$19-5116-B                &    10         &	  G5II		        &	     HD36079		      &     7 \\
                    &      M7.5III        &             HD126327                  &     4         &	 K7III		        &	     HD63425B		      &    31 \\
                    &        M7III        &             HD207076                  &     3         &                             &                                     &         \\
   \hline												 	   
   NGC3169          &         F2II        &             HD164136                  &    15	  &	   M3III		& 	    HD27796		      &     6 \\
                    &           GB        &            M67-F-108                  &    19	  &	   M2III		& 	    HD30354		      &    33  \\
                    &           GB        &          NGC7789-501                  &    25	  &	 K3Iab		        &     HD339034		              &     4  \\
                    &           HB        &             M5-II-76                  &    14	  &	K8V		        & 	    HD113538		      &     6 \\
                    &           HB        &           M92-XII-24                  &     7	  &	G2Ib		        & 	    HD209750		      &     5  \\
                    &        M8III        &             HD113285                  &     4	  &	  M0III		        & 	    HD2490		      &    28  \\
                    &          SGB        &          NGC188-I-88                  &    12	  &	 K0IV		        & 	    HD34642		      &     4  \\
           	    &			  &					  &		  &	  G3V		        & 	    HD6461		      &    9 \\ 		
\hline
   NGC3190          &         F0V         &           Hya-vB-103                  &     3         &       M2III		  &		HD30354 		  &    14  \\
                    &         F6V         &             HD30652                   &     1         &     K3Iab		  &	        HD339034		  &	5  \\
                    &           GB        &            M67-F-108                  &     8         &    K8V		  &		HD113538		  &    23  \\
                    &           HB        &             M71-1-41                  &    32         &      M0III		  &		HD2490  		  &    24  \\
                    &        K3III        &             HD169191                  &    20         &    K7III		  &		HD63425B		  &    31 \\
                    &        M7III        &             HD207076                  &    16         &                       &                                       &       \\
                    &          SGB        &            M67-F-125                  &    17         &                       &                                       &       \\
   \hline	 	    
   NGC3607	    &		HB	  &	       M5-II-53 		  &     7         &       M2III		  &		HD30354 		  &    36  \\
          	    &		HB 	  &	       M92-I-13 		  &     6         &    K8V		  &		HD113538		  &    20  \\
          	    &	      K0V 	  &	      Coma-A-13 		  &    21         &      M0III		  &		HD2490  		  &    42  \\
          	   &	    K5III 	  &	      HD136028  		  &     4         &                       &                                       &        \\
         	   &	     K7V  	  &	      HD157881  		  &    10         &                       &                                       &        \\
         	   &	    M4III 	  &	      HD17491			  &    27         &                       &                                       &        \\
         	   &	   M7III 	  &	      HD114961  		  &    10         &                       &                                       &        \\
         	   &	      SGB 	  &	     M67-F-125  		  &    12         &                       &                                       &        \\
  \hline
 NGC3718	  &	      GB          &	    M67-F-108		          &     2         &                       &                                       &    1    \\
        	  &	      GB	  &	     M71-1-71		          &     1         &	M2III		  &		 HD30354		  &	25  \\
        	  &	      HB	  &	     M71-C		          &    14         &     K3Iab		  &		 HD339034		  &	 1  \\
        	  &	   K0III	  &	     HD63352		          &     6         &    K8V		  &		 HD113538		  &	22  \\
        	  &	   K5III	  &	     HD136028		          &    13         &      M0III		  &		 HD2490 		  &	33  \\
        	  &	   M5III	  &	     HD175865		          &    13         &    K7III		  &		 HD63425B		  &	14 \\
        	  &	     SGB	  &	    M67-F-125		          &    37         &      G3V		  &		 HD6461 		  &	 2  \\
        	  &	     SGB	  &	  NGC188-I-97		          &    11         &                       &                                       &        \\
\hline																				   
   NGC3998	    &		HB	  &	       M71-1-41 		  &     3         &       M2III		  &		HD30354 		  &    20   \\
          	    &		HB 	  &	       M71-C			  &    18         &    K8V		  &		HD113538		  &    10  \\
          	    &		HB 	  &	     M92-XII-24 		  &     9         &     K0IV		  &		HD34642 		  &    29  \\
          	    &	      K0V 	  &	      Coma-A-13 		  &    37         &    K7III		  &		HD63425B		  &    31 \\
          	    &	     K5III	  &	       HD136028 		  &    31         &      G3V		  &		HD6461  		  &	7  \\
   \hline	
 NGC4203	  &	      GB	  &	  NGC188-I-61		          &    30	  &       M2III		  & 	    HD30354		          &    17   \\
        	  &	      HB	  &	     M5-II-53		          &     4	  &    M3Iab		  & 	  BD+39 4208		          &     2   \\
        	  &	      HB	  &	     M71-1-41		          &    27	  &    K8V		  & 	    HD113538		          &    24   \\
        	  &	    K0V		  &	    Coma-A-13		          &     4	  &    G2Ib		  & 	    HD209750		          &     5   \\
        	  &	   K5III	  &	     HD139669		          &     4	  &     G5II		  & 	    HD36079		          &    14   \\
        	  &	   K5III	  &	     HD149009		          &     8	  &    K7III		  & 	    HD63425B		          &    33   \\
        	  &	    K7V		  &	     HD157881		          &     1	  &     G8V		  & 	    HD64606		          &     2    \\
        	  &	   M7III	  &	     HD114961		          &    19         &                       &                                       &          \\
\hline  														
 NGC4235	  &	  M1.5Vb	  &	     HD72905		          &    55         &     G5II		  &	 HD36079       			  &  77   \\
        	  &	   G5III	  &	     HD134063		          &    14         &     K7III		  &	 HD63425B               	  &  14     \\
        	  &	  K4III		  &	     HD131918		          &     2         &     G3V 		  &	 HD6461       			  &  7      \\
        	  &	   M7III	  &	     HD207076		          &    15         &			  &					  &	  \\
          	 &	   SGB		  &  	  M67-IV-68		          &    11         &			  &					  &	  \\
  \hline
   \end{tabular}
   \end{table*}

\setcounter{table}{0}
\begin{table*}
\caption{(continued)}
\begin{tabular}{@{}lrlcrlc}
\hline
                    &\multicolumn{3}{|c|}{CaT region}								 & \multicolumn{3}{|c|}{CO region} \\
\hline
   Galaxy           &       Spectral Type   &             Star                      &   Weight (\%) &        Spectral Type		    &		  Star  		    &	Weight (\%) \\      
\hline 				    
NGC4258 	 &	G5IIIwe		  & 	    HD2665		          &     3         &      M3III 		  &		 HD236791		  &	 5  \\
        	 &	     GB		  &        NGC188-II-187		  &     1 	  &       M3III  	  &		HD27796 		  &	4   \\
        	 &	 K1III		  & 	    HD185644		          &    62 	  &       M2III  	  &		HD30354 		  &    10   \\
        	 &	  K3III		  & 	    HD169191		          &    10 	  &     K3Iab		  &	        HD339034		  &	2  \\
        	 &	  M7III		  & 	    HD207076		          &    18 	  &    K8V		  &		HD113538		  &    10   \\
        	 &	    SGB		  & 	   M67-F-125		          &     3 	  &      M0III		  &		HD2490  		  &    10   \\
		 &			  & 				          &	          &     K0IV		  &		HD34642 		  &	7    \\
		 &			  & 				          &		  &    K7III		  &		HD63425B		  &    25    \\
		 &			  & 				          &	          &     G8V		  &		HD64606 		  &	1   \\
		 &			  & 				          &		  &      G3V		  &		HD6461  		  &    21   \\
\hline					 
 NGC4346	  &	  G9III		  &	     HD112989		          &     9         &       M2III  	  &		HD30354 		  &    32  \\
        	  &	      GB	  &	NGC188-II-122		          &     1         &    K8V		  &		HD113538		  &    19   \\
        	  &	      HB	  &	     M71-C		          &    21         &      M0III		  &		HD2490  		  &    25   \\
        	  &	   K3III	  &	     HD102328		          &    37         &    K7III		  &		HD63425B		  &    22    \\
        	  &	  M4III		  &	     HD17491		          &    13         &			  &  				          &  	     \\    
        	  &	     SGB	  &	    M67-F-125		          &     8         &			  &  				          &  	     \\   
        	  &	     SGB	  &	  NGC188-I-97		          &     8         &			  &  				          &   	     \\  
\hline
 NGC4388	  &	      GB	  &	    M67-F-108		          &     4         &       M2III  	  &		HD30354 		  &    23  \\
        	  &	    K7V		  &	     HD157881		          &    46         &    K8V		  &		HD113538		  &	4  \\
        	  &	   M1III	  &	     HD168720		          &    19         &     G5II		  &		HD36079 		  &    13  \\
        	  &	     M6V          &	 BD$+$19-5116-B 	          &     1         &     G8V		  &		HD64606 		  &    15  \\
        	  &	     SGB          &	    M67-F-115		          &    27         &      G3V		  &		HD6461  		  &    29   \\
        	  &	     SGB	  &	    M67-IV-68		          &     1         &      K4III		  &		HD9138  		  &    12   \\
\hline				
 NGC4450	   &	       GB  	 &	 NGC188-II-122  		&    13        &       M2III  		 &		HD30354 		  &    29   \\
        	   &	       HB  	 &	    M92-XII-24  		&     7        &    K8V			 &		HD113538		  &    19   \\
        	   &	    K5III  	 &	      HD120933  		&     1        &    K1II		 &		HD198700		  &    14   \\
        	   &	    K5III  	 &	      HD139669  		&    42        &      M0III		 &		HD2490  		  &	4   \\
        	   &	      M5V  	 &	      Gl-699			&    16        &    K7III		 &		HD63425B		  &    22   \\
        	   &	      SGB  	 &	     M67-F-115  		&    15        &      G3V		 &		HD6461  		  &	9   \\
                   &	      SGB  	 &	  NGC188-II-93  		&     2        &                         &                                        &         \\
\hline
 NGC4548	  &	      HB         &	     M71-C		        &    16        &	 M2III		 &		  HD30354		  &	 21  \\
        	  &	      HB	 &	 NGC188-I-105		        &     4        &    K8V  		 &		  HD113538		  &	 15  \\
        	  &	   K3III         &	     HD102328		        &    46        &	M0III  		 &		  HD2490		  &	 28  \\
        	  &	 M5.5III	 &	     HD94705		        &    10        &    K7III  		 &		  HD63425B		  &	 12  \\
        	  &	     SGB	 &	    M67-F-125		        &     4        &     G8V  		 &		  HD64606		  &	 21  \\
        	  &	     SGB	 &	  NGC188-I-97		        &    16        &			 &  				          &         \\
\hline
 NGC4565	   &	       GB 	 &	     M67-F-108  		&    16        &       M2III		 &	       HD30354  		  &    29  \\
        	   &	       GB  	 &	   NGC188-I-57  		&     3        &    K8V			 &	       HD113538 		  &    17  \\
        	   &	       HB 	 &	      M71-1-41  		&     7        &    G2Ib		 &	       HD209750 		  &     4  \\
        	   &	       HB  	 &	    M92-XII-24  		&     6        &      M0III		 &	       HD2490			  &    39  \\
        	   &	    K0III  	 &	      HD85503			&    11        &    K7III		 &	       HD63425B 		  &     4  \\
        	   &	    K5III  	 &	      HD139669  		&    17        &     G8V		 &	       HD64606  		  &     4  \\
        	   &	    M4III 	 &	      HD17491			&    22        &			 & 				          & 	\\
        	   &	   M7III 	 &	      HD114961  		&     4        &			 & 				          & 	\\
        	   &	      SGB 	 &	     M67-F-125  		&     8        &			 & 				          & 	\\
 \hline
 NGC4569	  &	   F3III         &	     HD115604		        &    26        &       M2III		 &	       HD30354  		  &    21 \\
        	  &	      GB	 &	    M67-F-108		        &    29        &    M3Iab		 &	     BD$+$39 4208		  &     1  \\
        	  &	      GB	 &	  NGC188-I-75		        &     1        &     K3Iab 		 &	       HD339034 		  &     4  \\
        	  &	      HB	 &	     M71-1-41		        &    18        &    K8V			 &	       HD113538 		  &    11  \\
        	  &	      HB	 &	   M92-XII-24		        &     7        &      M0III		 &	       HD2490			  &    59 \\
        	  &	   K5III	 &	     HD139669		        &     9        &			 &					  &	   \\
        	  &	   M7III	 &	     HD114961		        &     9        &			 &					  &	   \\
\hline													
 NGC4579	   &	    F3III  	 &	      HD115604  		&     2        &       M2III  		 &		HD30354 		  &    16  \\
        	   &	       GB  	 &	 NGC188-II-122  		&     1        &    G2Ib		 &	        HD209750		  &	1  \\
        	   &	       GB  	 &	   NGC7789-971  		&     8        &      M0III		 &		HD2490  		  &    35   \\
        	   &	       HB  	 &	    M92-XII-24  		&     4        &    K7III		 &		HD63425B		  &	4   \\
        	   &	K0III  		 &	      HD142091  		&    65        &     G8V		 &		HD64606 		  &    42  \\
        	   &	    K5III 	 &	      HD136028  		&     6        &			 &  					  &	     \\
        	   &	      SGB 	 &	     M67-F-125  		&    11        &			 &  					  &	     \\
\hline
  \end{tabular}
  \end{table*}

\setcounter{table}{0}
\begin{table*}
\caption{(continued)}
\begin{tabular}{@{}lrlcrlc}
\hline
                    &\multicolumn{3}{|c|}{CaT region}								 & \multicolumn{3}{|c|}{CO region} \\
\hline
   Galaxy           &       Spectral Type   &             Star                      &   Weight (\%) &        Spectral Type		    &		  Star  		    &	Weight (\%) \\      
\hline
  NGC4594	   &	       HB  	 &	      M71-C			&     7        &       M2III  		 &		HD30354 		  &    27  \\
         	   &	       HB  	 &	    M92-XII-24  		&     2        &    K8V			 &		HD113538		  &    19  \\
         	   &	     K0V  	 &	     Coma-A-13  		&    36        &      M0III		 &		HD2490  		  &    43  \\
         	   &	    K2III  	 &	      HD54719			&     8        &    K7III		 &		HD63425B		  &	9  \\
         	   &	    K3III  	 &	      HD102328  		&     2        &			 & 					  &	      \\
         	   &	    M4III 	 &	      HD17491			&     3        &			 & 					  &	      \\
         	   &	  M5III  	 &	      HD172816  		&    25        &			 &					  &	    \\
         	   &	      SGB 	 &	     M67-F-125  		&    14        &			 &					  &	    \\
  \hline
   NGC4725	   &	       GB 	 &	     M67-F-108  		&    24        &       M2III  		 &		HD30354 		  &    19 \\
         	   &	       GB  	 &	   NGC7789-971  		&     2        &    K8V			 &		HD113538		  &    19  \\
         	   &	       HB 	 &	      M5-II-53  		&    11        &    G2Ib		 &	        HD209750		  &	3  \\
         	   &	  M5III  	 &	      HD172816  		&    59        &      M0III		 &		HD2490  		  &    34  \\
         	   &	   M7III 	 &	      HD114961  		&     1        &    K7III		 &		HD63425B		  &    21  \\
		   &			 &					&	       &      G3V		 &		HD6461  		  &	2  \\
  \hline
  NGC4736	  &	      F5         &	     HD14938		        &     2        &       M2III  		 &		HD30354 		  &    28   \\
        	  &	     G2V	 &	     HD76932		        &    12        &    G2Ib		 &	        HD209750		  &    17    \\
        	  &	      GB	 &	    M67-F-108		        &    38        &      M0III		 &		HD2490  		  &    53   \\
        	  &	      GB	 &	  NGC7789-971		        &     8        &			 & 					  &	      \\
        	  &	      HB	 &	     M5-II-53		        &     3        &			 & 					  &	      \\
        	  &	      HB	 &	     M92-I-13		        &    10        &			 & 					  &	      \\
        	  &	   K0III	 &	     HD63352		        &    24        &			 & 					  &	      \\
\hline		       
   NGC4750	   &	       HB  	&	    M92-XII-24  		&     3      &       M2III			 &	       HD30354  		 &    22   \\
         	   &	       HB  	&	   NGC7789-676  		&     4      &     K3Iab 		 &	       HD339034 		 &     2   \\
         	   &	    K3III  	&	      HD102328  		&     6      &    K8V			 &	       HD113538 		 &    15    \\
         	   &	 K4II  		&	      HD130705  		&    14      &    G2Ib			 &	       HD209750 		 &    21   \\
         	   &	    K5III  	&	      HD149009  		&     9      &      M0III			 &	       HD2490			 &     8   \\
         	   &	    M4III 	&	      HD17491			&    16      &    K7III			 &	       HD63425B 		 &    28   \\
         	   &	  M5.5III  	&	      HD94705			&     7      &                                  &                                       &          \\
         	   &	      SGB  	&	     M67-F-115  		&     6      &				         &					  &         \\
         	   &	      SGB 	&	     M67-F-125  		&    13      &				         &					  &         \\
         	   &	      SGB.  	&	   NGC188-I-55  		&    16      &				         &					  &         \\
 \hline   
  NGC5005	   &	      HB	&	     M5-II-53		        &    10      &       M2III  		  &		HD30354 		  &    20  \\
        	  &	      HB	&	   M92-XII-24		        &     4      &    K8V			  &		HD113538		  &    24   \\
        	  &	    K0V		&	    Coma-A-13		        &     9      &      M0III			  &		HD2490  		  &    55   \\
        	  &	   K5III	&	     HD139669		        &    46      &				  &					   &	     \\
        	  &	 M5III		&	     HD172816		        &     3      &				  &					   &	     \\
        	  &	     SGB	&	    M67-F-125		        &    24      &				  &					   &	     \\
 \hline 												       
 NGC5033	  &	      GB	&	    M67-F-108		        &    64      &	M2III 		   &		 HD30354		   &	39  \\
        	  &	      HB	&	     M5-II-53		        &     7      &    K8V			   &		 HD113538		   &	14  \\
        	  &	 M5.5III	&	     HD94705		        &    18      &      M0III			   &		 HD2490 		   &	29  \\
        	  &	   M7III	&	     HD114961		        &     9      &     G8V			   &		 HD64606		   &	16  \\
 \hline
    NGC5194	   &	    F3III  	&	      HD115604  		&     2      &	 M2III		    &		  HD30354		    &	 27 \\
         	   &	       GB 	&	     M67-F-108  		&    33      &   M2III		    &		BD-01 3097		    &	 13  \\
         	   &	       HB  	&	    M92-XII-24  		&     7      &    K8V  		    &		  HD113538		    &	  8  \\
         	   &	    K3III  	&	      HD102328  		&     2      &    G2Ib  		    &		  HD209750		    &	 15  \\
         	   &	 K4II  		&	      HD130705  		&     5      &	M0III  		    &		  HD2490		    &	 34  \\
         	   &	    K5III  	&	      HD139669  		&    10      &                                  &                                       &         \\
         	   &	    M4III 	&	      HD17491			&    25      &				    &					     &         \\
         	   &	      SGB 	&	     M67-F-125  		&     8      &				    &					     &         \\
         	   &	      SGB  	&	     M67-IV-68  		&     2      &				    &					     &         \\
 \hline
 NGC5371 	 &	    F8V		& 	   Hya-vB-19		        &     8      &	 M2III		    &		  HD30354		    &	 38  \\
        	 &	    G2V		& 	   Hya-vB-64		        &    18      &     K3Iab		    &		  HD339034		    &	  4  \\
        	 &	     HB		& 	    M71-C		        &     4      &    K8V  		    &		  HD113538		    &	  4  \\
        	 &	  K0III		& 	    HD88284		        &    30      &    K1II  		    &		  HD198700		    &	  1   \\
        	 &     K4II		& 	    HD130705		        &    19      &    G2Ib  		    &		  HD209750		    &	 14  \\
        	 &	    M6V		& 	BD$+$19-5116-B  	        &     9      &    hd218594  		    &		  HD218594		    &	  2  \\
        	 &	M7.5III		& 	    HD126327		        &     9      &	M0III  		    &		  HD2490		    &	 11  \\
 		 &			& 				        &    	     &	K0IV 		    &  	    HD34642		           &     2  \\
 		 &			& 				        &    	     &   K7III			    &		 HD63425B		   &	16  \\
 		 &			& 				        &    	     &     G3V			   &		 HD6461 		   &	 2  \\
\hline 
  \end{tabular}
 \end{table*}

\setcounter{table}{0}
\begin{table*}
\caption{(continued)}
\begin{tabular}{@{}lrlcrlc}
\hline
                    &\multicolumn{3}{|c|}{CaT region}								 & \multicolumn{3}{|c|}{CO region} \\
\hline
   Galaxy           &       Spectral Type   &             Star                      &   Weight (\%) &        Spectral Type		    &		  Star  		    &	Weight (\%) \\      
\hline
NGC5850 	 &	    F0V		& 	    HD112412		        &     1      &       M2III  		  &		HD30354 		  &    30  \\
        	 &     G2V		& 	    HD76932		        &    10	     &     K3Iab		  &	        HD339034		  &	1  \\
        	 &	   F7V		& 	    HD102634		        &     2	     &    K8V			  &		HD113538		  &    19  \\
        	 &	    F8V		& 	    HD187691		        &     1	     &    G2Ib			  &	        HD209750		  &	2  \\
        	 &	     GB		& 	   M92-XII-8		        &    10	     &      M0III			  &		HD2490  		  &    33  \\
        	 &	  K0III		& 	    HD88284		        &    36	     &     K0IV			  &		HD34642 		  &	1  \\
        	 &	  K3III		& 	    HD169191		        &     1	     &    K7III			  &		HD63425B		  &    10  \\
        	 &     K4II		& 	    HD130705		        &    22      &                                  &                                       &         \\
        	 &	  M1III		& 	    HD168720		        &    12      &                                  &                                       &         \\
        	 &	M7.5III		& 	    HD126327		        &     1      &                                  &                                       &         \\
\hline
 NGC6500	  &	        	&	    	 		        &     	     &       M2III  		  &		HD30354 		  &    23  \\
        	  &	        	&	    	 		        &    	     &    K8V			  &		HD113538		  &    11   \\
        	  &	        	&	    	 		        &    	     &    G2Ib			  &	        HD209750		  &    18   \\
        	  &	        	&	    			        &    	     &    K7III			  &		HD63425B		  &    48    \\
\hline
NGC7217 	 &	  F3III		& 	    HD115604		        &     8	     &	M2III 		   &		 HD30354		   &	34  \\
        	 &	     GB		&        NGC188-II-122		        &     1	     &     K3Iab		   &		 HD339034		   &	 1  \\
        	 &	     HB		& 	  M92-XII-24		        &     2	     &    K8V			   &		 HD113538		   &	21  \\
        	 &	   K0V		& 	   Coma-A-13		        &    11	     &      M0III			   &		 HD2490 		   &	42  \\
        	 &	  K3III		& 	    HD102328		        &    52      &                                  &                                       &        \\
        	 &	 M4III		& 	    HD17491		        &    10	     &                                  &                                       &        \\
        	 &	  M5III		& 	    HD175865		        &     6	     &                                  &                                       &        \\
        	 &	    M6V		& 	BD$+$19-5116-B  	        &     6      &                                  &                                       &        \\
\hline
 NGC7331	  &	    F6V		&	     HD30652		        &     1	     &       M2III  		  &		HD30354 		  &    16  \\
        	  &	     G2V	&	    Hya-vB-64		        &    10	     &    K8V			  &		HD113538		  &    19  \\
        	  &	 G5IIIwe	&	     HD88609		        &    12	     &      M0III			  &		HD2490  		  &    37  \\
        	  &	   G8III	&	     HD38751		        &    36	     &    K7III			  &		HD63425B		  &    26  \\
        	  &	      HB	&	     M71-C		        &     9	     &				  &					  &	   \\
        	  &	   K3III	&	     HD102328		        &    11	     &				  &					  &	   \\
        	  &	   M7III	&	     HD207076		        &    16	     &				  &					  &	   \\
% \hline
%  NGC7469	  &	      GB	&	NGC188-II-122		        &    63	     &     G5II	      & 	   HD36079		     &    41   \\
%        	  &	   M7III	&	     HD207076		        &    36	     &    K7III	      & 	    HD63425B		      &     9	\\
%		  &                     &				        &	     &     G8V	      & 	    HD64606		      &    42	\\
%		  &                     &				        &	     &      G3V  		&	      HD6461			&     8  \\
\hline	 
 NGC7743	  &	      HB	&	     M71-C		        &    34	     &       M2III		     &  	   HD30354		     &     6  \\
        	  &	      HB	&	   M92-XII-24		        &    16	     &     K3Iab		     &  	   HD339034		     &     4  \\
        	  &	   K0III	&	     HD85503		        &    41	     &    K8V		     &  	   HD113538		     &     3  \\
        	  &	   M7III	&	     HD114961		        &     4	     &    G2Ib		     &  	   HD209750		     &     5  \\
        	  &	     SGB	&	    M67-F-125		        &     2	     &      M0III		     &  	   HD2490		     &    65  \\
		  &			&				        &	     &     K0IV		     &  	   HD34642		     &    12  \\
		  &			&				        &	     &      G3V		     &  	   HD6461		     &     2  \\
\hline
  \end{tabular} 
 \end{table*}

\label{lastpage}


\begin{thebibliography}{99}

\bibitem[\protect\citeauthoryear{Auld et al.}{}]{auld13} Auld, R. et al., 2013, 428, 1880.

\bibitem[\protect\citeauthoryear{Barth, Ho \& Sargent}{2002a}]{barth02a} Barth, A. J., Ho, L. C., Sargent, W. L. W., 2002, ApJ, 124, 2607.

\bibitem[\protect\citeauthoryear{Barth, Ho \& Sargent}{2002b}]{barth02} Barth, A. J., Ho, L. C., Sargent, W. L. W., 2002, ApJ, 566, L13.

\bibitem[\protect\citeauthoryear{Beers, Flynn \& Gebhardt}{1990}]{bootstrap} Beers, T. C., Flynn, K., Gebhardt, K., 1990, AJ, 100, 32.

\bibitem[\protect\citeauthoryear{Bellovary et al.}{2014}]{bellovary14} Bellovary, J., Holley-Bockelmann, K., G\"ultekin, K., Christensen, C., Governato, F., Brooks, A., Loebman, S., Munshi, F., 2014, {\it http://arxiv.org/abs/1405.0286}

\bibitem[\protect\citeauthoryear{Bender, Burstein \& Faber}{1993}]{bender93} Bender, R., Burstein, D., \& Faber, S. M. 1993, ApJ, 411, 153

\bibitem[\protect\citeauthoryear{Bernardi et al.}{2003}]{bernardi03} Bernardi, M. et al. 2003, AJ,125, 1866. 
 

\bibitem[\protect\citeauthoryear{Bourne et al.}{2013}]{bourne13} Bourne, N.et al. 2013, MNRAS, 436, 479.


\bibitem[\protect\citeauthoryear{Bower et al.}{2006}]{bower06} Bower, R. G. et al. 2006, MNRAS, 370, 645.
	
\bibitem[\protect\citeauthoryear{Bruzual \& Charlot}{2003}]{bruzual03} Bruzual, G., Charlot, S., 2003, MNRAS, 344, 1000.

\bibitem[\protect\citeauthoryear{Cappellari \& Emsellem}{2004}]{ppxf} Cappellari, M. \& Emsellem, E., 2004, PASP, 116, 138.

\bibitem[\protect\citeauthoryear{Cappellari et al.}{2007}]{cappellari07} Cappellari, M. et al., 2007, MNRAS, 379, 418.

\bibitem[\protect\citeauthoryear{Cenarro et. al.}{2001}]{cenarro01} Cenarro, A. J., Cardiel, N., Gorgas, J., Peletier, R. F., Vazdekis, A., Prada, F., 2001, MNRAS, 326, 959.

\bibitem[\protect\citeauthoryear{Cid Fernandes et al.}{2004}]{cid04}  Cid Fernandes, R., Gu, Q. Melnick, J., Terlevich, E., Terlevich, R., 
Kunth, D., Rodrigues Lacerda, R., Joguet, B., 2004, MNRAS, 355, 273.

 \bibitem[\protect\citeauthoryear{Ciesla et al.}{2012}]{ciesla12} Ciesla, L. et al., 2012, A\&A, 543, 161.

\bibitem[\protect\citeauthoryear{Cox et al.}{2006}]{cox06}  Cox, T. J., Dutta, S. N., Di Matteo, T., Hernquist, L., Hopkins, P. F., Robertson, B., Springel, V., 2006, ApJ, 650, 791.


\bibitem[\protect\citeauthoryear{Cushing, Rayner, \& Vacca}{2005}]{irtf05} Cushing, M.C., Rayner, J.T., \& Vacca, W.D., 2005, ApJ, 623, 1115.

\bibitem[\protect\citeauthoryear{Davies et al.}{2012}]{davies12} Davies J. I. et al., 2012,  MNRAS, 419, 3505. 

\bibitem[\protect\citeauthoryear{De Looze et al.}{2012}]{looze12} De Looze, I. et al., 2012, MNRAS, 423, 2359.

\bibitem[\protect\citeauthoryear{Di Matteo, Springel \& Hernquist}{2005}]{dimateo05} Di Matteo, T., Springel, V. \& Hernquist, L. 2005, Nature, 433, 604.

\bibitem[\protect\citeauthoryear{Djorgovski \& Davis}{1987}]{djorgovski87} Djorgovski, S., Davis, M., 1987, ApJ, 313, 59.

\bibitem[\protect\citeauthoryear{Draine}{2003}]{draine03}  Draine, B. T., 2003, ARAA, 41, 241.

\bibitem[\protect\citeauthoryear{Dressler}{1984a}]{dressler84a}  Dressler, A. 1984, ApJ, 281, 512

\bibitem[\protect\citeauthoryear{Dressler}{1984b}]{dressler84b}  Dressler, A. 1984, ApJ, 286, 97


\bibitem[\protect\citeauthoryear{Dressler et al.}{1987}]{dressler87} Dressler, A., Lynden-Bell, D., Burstein, D., Davies, R. L., Faber, S. M., Terlevich, R., Wegner, G., 1987, ApJ, 313, 42.
	
\bibitem[\protect\citeauthoryear{Emsellem et al.}{2001}]{emsellem01}	Emsellem, E., Greusard, D., Combes, F., Friedli, D., Leon, S., P\'econtal, E., Wozniak, H., 2001, A\&A, 368, 52.


\bibitem[\protect\citeauthoryear{Emsellem et al.}{2004}]{emsellem04} Emsellem, E., Cappellari, M.,
  Peletier, R. F., McDermid, R. M., Bacon, R., Bureau, M., Copin, Y., Davies,
  R. L., Krajnovi\'c, D., Kuntschner, H., Miller, B. W., \& de Zeeuw, P. T. 2004, MNRAS, 352, 721.

\bibitem[\protect\citeauthoryear{Falc\'on-Barroso, Peletier \& Balcells}{2002}]{falcon-barroso02} Falc\'on-Barroso, J. Peletier, R. F., Balcells, M., 2002, MNRAS, 335, 741.

\bibitem[\protect\citeauthoryear{Ferrarese \& Ford}{2005}]{ferrarese05} Ferrarese, L. \& Ford, H. C., 2005, Space Science Reviews, 116, 523.

\bibitem[\protect\citeauthoryear{Ferrarese \& Merrit}{2000}]{ferrarese00} Ferrarese, L. \& Merrit, D., 2000, ApJ, 547, 140.

\bibitem[\protect\citeauthoryear{Gebhardt et al.}{2003}]{gebhardt03}  Gebhardt, K., Richstone, D., Tremaine, S., Lauer, T. R., Bender, R., Bower,
G., Dressler, A., Faber, S. M., Filippenko, A. V., Green, R., Grillmair, C., Ho, L. C., Kormendy, J., Magorrian, J., \& Pinkney, J. 2003, ApJ, 583, 92

\bibitem[\protect\citeauthoryear{Gebhardt et al.}{2000}]{gebhardt00} Gebhardt, K. et al. 2000, ApJ, 539, 13.

\bibitem[\protect\citeauthoryear{Graham et al.}{2011}]{graham11} Graham, A. W., Onken, C. A., Athanassoula, E., Combes, F., 2011, MNRAS, 412, 2211.

\bibitem[\protect\citeauthoryear{G\"ultekin et al.}{2009}]{gultekin09} G\"ultekin, K., Cackett, E. M., Miller, J. M., Di Matteo, T., Markoff, S., Richstone, D. O., 2008, ApJ, 706, 404.

\bibitem[\protect\citeauthoryear{Heisler \& De Robertis}{1999}]{heisler99} Heisler C. A., De Robertis M. M., 1999, AJ, 118, 2038.

\bibitem[\protect\citeauthoryear{Ho, Filippenko \& Sargent}{1995}]{ho95} Ho, L. C., Filippenko, A. V., Sargent, W. L., 1995, ApJS, 98, 477.

\bibitem[\protect\citeauthoryear{Ho, Filippenko \& Sargent}{1997}]{ho97} Ho, L. C., Filippenko, A. V., Sargent, W. L., 1997, ApJS, 112, 31.

\bibitem[\protect\citeauthoryear{Ho et al.}{2009}]{ho09} Ho, L. C., Greene, J. E., Filippenko, A. V., Sargent, W. L., 2009, ApJS, 183, 1.

\bibitem[\protect\citeauthoryear{Hildebrand}{1983}]{hildebrand83} Hildebrand, R. H. 1983, Q. J. R. Astron. Soc., 24, 267

\bibitem[\protect\citeauthoryear{Ivanov et al.}{2000}]{ivanov00}  Ivanov, V. D., Rieke, G. H., Groppi, C. E., Alonso-Herrero, A., Rieke, M. J., Engelbracht, C. W., 2000, ApJ, 545, 190.

\bibitem[\protect\citeauthoryear{Kang et al.}{2013}]{kang13} Kang, W-R., Woo, J-H., Schulze, J., Riechers, D. A., Kim, S. C., Park, D., \& Smolcic, V., 2013, ApJ, 767, 26.



\bibitem[\protect\citeauthoryear{Kennicutt}{1998}]{kennicutt98} Kennicutt R. C., 1998, ARA\&A, 36, 189

\bibitem[\protect\citeauthoryear{Kennicutt \& Evans}{2012}]{kennicutt12} Kennicutt R. C., Evans N. J., 2012, ARA\&A, 50, 531.

\bibitem[\protect\citeauthoryear{Kormendy \& Kennicutt}{2004}]{kormendy04}  Kormendy, J., \& Kennicutt, R., ARA\&A, 2004, 42, 603.

\bibitem[\protect\citeauthoryear{Kotilainen et al.}{2012}]{kotilainen12} Kotilainen, J. K., Hyv\"onen, T., Reunanen, J., Ivanov, V. D., 2012, MNRAS, 425, 1057.

\bibitem[\protect\citeauthoryear{Kormendy, Bender \& Cornell}{2011}]{kormendy11}  Kormendy, J. Bender, R., \& Cornell, M. E., Nature, 2011, 469, 374.

\bibitem[\protect\citeauthoryear{Kormendy \& Ho}{2013}]{msigma} Kormendy, J., \& Ho, L. C. 2013, ARA\&A, 51, 511.

\bibitem[\protect\citeauthoryear{Kuntschner}{2000}]{kuntschner00} Kuntschner, H., MNRAS, 2000, 315, 184.

\bibitem[\protect\citeauthoryear{Maraston}{2005}]{maraston05} Maraston, C., 2005, MNRAS, 362, 799.

\bibitem[\protect\citeauthoryear{Magorrian et al.}{1998}]{magorrian98} Magorrian, J. et al. 1998, AJ, 115, 2285.

\bibitem[\protect\citeauthoryear{Nemmen et al.}{2007}]{nemmen07} Nemmen R., Bower, R., Babul, A. \& Storchi-Bergmann, T. 2007, MNRAS, 377, 1652

\bibitem[\protect\citeauthoryear{Marleau et al.}{2006}]{marleau06} Marleau, F. R. et al., 2006, ApJ, 646, 929.

\bibitem[\protect\citeauthoryear{Martins et al.}{2013a}]{martins13b} Martins, L. P., Rodr\'iguez-Ardila, A., Diniz, S., Gruenwald, R., de Souza, R., 2013a, MNRAS, 431, 1823.

\bibitem[\protect\citeauthoryear{Martins et al.}{2013b}]{martins13} Martins, L. P., Rodr\'iguez-Ardila, A., Diniz, S., Riffel, R., de Souza, R., 2013b, MNRAS, 435, 2861.


\bibitem[\protect\citeauthoryear{M\'arquez et al.}{2003}]{marquez03} M\'arquez, I. Masegosa, J., Durret, F., Gonz\'alez Delgado, R. M., Moles, M., Maza, J., P\'erez, E., Roth, M., 2003, A\&A, 409, 459.

\bibitem[\protect\citeauthoryear{Moorwood \&  Oliva}{1988}]{morwood88} Moorwood A. F. M., Oliva E., 1988, A\&A, 203, 278

\bibitem[\protect\citeauthoryear{Naab et al.}{2013}]{naab13} Naab, T. et al., 2013, arXiv:1311.0284

\bibitem[\protect\citeauthoryear{Origlia, Moorwood \& Oliva}{1993}]{origlia93} Origlia, L., Moorwood, A. F. M., Oliva, E., 1993, A\&A, 280, 5360.


%\bibitem[\protect\citeauthoryear{Ramos Almeida, P\'erez Garc\'ia \& Acosta-Pulido}{2009}]{ramos-almeida09} Ramos Almeida, C., P\'erez Garc\'ia, A. M., Acosta-Pulido, J. A., 2009, ApJ, 694, 1379.

\bibitem[\protect\citeauthoryear{Rayner, Cushing, \& Vacca}{2009}]{irtf09} Rayner, J.T., Cushing, M.C., \& Vacca, W.D., 2009, ApJS, 185, 289.

\bibitem[\protect\citeauthoryear{Richstone et al.}{1998}]{richstone98} Richstone, D. et al., 1998, Nature, 395, A14.


\bibitem[\protect\citeauthoryear{Rothberg \& Fischer}{2010}]{rothberg10} Rothberg, B., \& Fischer, J., 2010, ApJ, 712, 318.

\bibitem[\protect\citeauthoryear{Rothberg et al.}{2013}]{rothberg13} Rothberg, B., Fischer, J., Rodrigues, M., \& Sanders, D. B., 2013, ApJ, 767, 72.

\bibitem[\protect\citeauthoryear{Riffel, Rodr\'iguez-Ardila \& Pastoriza}{2006}]{rogerio06} Riffel, R., Rodr\'iguez-Ardila, A., Pastoriza, M. G., 2006, A\&A, 457, 61.

\bibitem[\protect\citeauthoryear{Riffel et al.}{2007}]{rogerio07}	Riffel, R., Pastoriza, M. G., Rodr\'iguez-Ardila, A., Maraston, C., 2007, ApJ, 659L, 103.

\bibitem[\protect\citeauthoryear{Riffel et al.}{2008}]{n4051} Riffel, Rogemar A., Storchi-Bergmann, T., Winge, C., McGregor, P. J., Beck, T., Schmitt, H., 2008, MNRAS, 385, 1129.

\bibitem[\protect\citeauthoryear{Riffel et al.}{2009}]{riffel09} Riffel, R., Pastoriza, M. G., Rodr\'iguez-Ardila, A., Bonatto, C., 2009, MNRAS, 400, 273.

\bibitem[\protect\citeauthoryear{Riffel}{2010}]{profit} Riffel, Rogemar A., 2010, Ap\&SS, 327, 239.

\bibitem[\protect\citeauthoryear{Riffel et al.}{2010}]{mrk1066-pop} Riffel, Rogemar A. \& Storchi-Bergmann, T., Riffel, R., \& Pastoriza, M. G., 2010, ApJ, 713, 469.

\bibitem[\protect\citeauthoryear{Riffel et al.}{2011}]{mrk1157-pop} Riffel, R., Riffel, Rogemar A., Ferrari, F., Storchi-Bergmann, T., 2011. MNRAS, 416, 493.

\bibitem[\protect\citeauthoryear{S\'anchez-Bl\'azquez et al.}{2006}]{miles} S\'anchez-Bl\'azquez, P., Peletier, R. F., Jim\'enez-Vicente, J., Cardiel, N., Cenarro, A. J., Falc\'on-Barroso, J., Gorgas, J., Selam, S., Vazdekis, A., 2006, MNRAS, 371, 703.

	
\bibitem[\protect\citeauthoryear{Sauvage et al.}{2013}]{sauvage10} Sauvage, M., et al. 2010, A\&A, 518, 64.

\bibitem[\protect\citeauthoryear{Silge \& Gebhardt}{2003}]{silge03} Silge, J. D., \& Gebhardt, K., 2003, ApJ, 125, 2809.

\bibitem[\protect\citeauthoryear{Silva et al.}{2008}]{silva08} Silva, D. R., Kuntschner, H., Lyubenova, M., 2008, ApJ, 674, 194.

\bibitem[\protect\citeauthoryear{Springel, Di Matteo \& Hernquist}{2005}]{springel05} Springel, V., Di Matteo, T. \& Hernquist, L., 2005, ApJ, 620, 79.

\bibitem[\protect\citeauthoryear{Storchi-Bergmann et al.}{2012}]{sb12} Storchi-Bergmann, T., Riffel, Rogemar A., Riffel, R., Diniz, M. R., Borges Vale, T., McGregor, P. J., 2012, ApJ, 755, 87.

\bibitem[\protect\citeauthoryear{Terlevich et al.}{1981}]{terlevich81}  Terlevich, R., Davies, R. L., Faber, S. M., Burstein, D. 1981, MNRAS, 196, 381

\bibitem[\protect\citeauthoryear{Tremaine et al.}{2002}]{tremaine02} Tremaine, S. et al. 2002, ApJ, 574, 740.

\bibitem[\protect\citeauthoryear{Thuan \& Sauvage}{1992}]{thuan92} Thuan, T. X., \& Sauvage, M. 1992, A\&AS, 92, 749


\bibitem[\protect\citeauthoryear{Valluri, Merrit \& Emsellem}{2004}]{valluri04} Valluri, M., Merritt, D., \& Emsellem, E. 2004, ApJ, 602, 66.

\bibitem[\protect\citeauthoryear{Vanderbeke et al.}{2011}]{vanderbeke11} Vanderbeke, J., Baes, M., Romanowsky, A. J., \& Schimidtobreick, L., 2011, MNRAS, 412, 2017.

\bibitem[\protect\citeauthoryear{van der Marel \& Franx}{1993}]{marel93} van der Marel, R. P., \& Franx, M. 1993, ApJ, 407, 525.

\bibitem[\protect\citeauthoryear{Weingartner \& Draine}{2001}]{weingartner01} Weingartner, J.C., \& Draine, B.T. 2001, ApJ, 548,296.

\bibitem[\protect\citeauthoryear{Winge, Riffel \& Storchi-Bergmann}{2009}]{winge09} Winge, C., Riffel, Rogemar A., \& Storchi-Bergmann, T., 2009, ApJS, 185, 186.

\bibitem[\protect\citeauthoryear{Woo et al.}{2013}]{woo13} Woo, J.-H., Schulze, A., Park, D., Kang, W.-R., Kim, S. C., \& Riechers, D. A. 2013, ApJ, 772, 49



\bibitem[\protect\citeauthoryear{Xiao et al.}{2011}]{xiao11} Xiao, T., Barth, A. J., Greene, J. E., Ho, L. C., Bentz, M., C., Ludwig, R. R., \& Jiang, Y., 2011, ApJ, 739, 28.

\end{thebibliography}
\end{document}